%% file: MSG_BS.tex
\documentclass[prx,reprint,twocolumn,showpacs,superscriptaddress]{revtex4-1}
\usepackage{amsmath,amssymb,bm,mathrsfs,graphicx, braket, times,amsthm,enumerate, scrextend}
\usepackage[colorlinks=true,citecolor=blue,linkcolor=blue]{hyperref}
\usepackage{longtable}
\usepackage{multirow} 
\usepackage[all,cmtip]{xy}
\usepackage[normalem]{ulem}
\usepackage[usenames,dvipsnames]{color}

\renewcommand{\vec}[1]{\bm{#1}}

\begin{document}
\title{Structure and Topology of Band Structures in the 1651 Magnetic Space Groups}

\begin{abstract}
The properties of electrons in magnetically ordered crystals are of interest both from the viewpoint of realizing novel topological phases, such as magnetic Weyl semimetals, and from the applications perspective of creating energy-efficient memories. A systematic study of symmetry and topology in magnetic materials has been challenging given that there are 1651 magnetic space groups (MSGs). Here, by using an efficient representation of allowed band structures, we obtain a systematic description of several basic properties of free electrons in all MSGs in three dimensions as well as in the 528 magnetic layer groups relevant to two dimensional magnetic materials. We compute constraints on electron fillings and band connectivity compatible with insulating behavior. Also, by contrasting with atomic insulators, we identify band topology entailed by the symmetry transformation of bands, as determined by the MSG alone. We give an application of our results to identifying topological semimetals arising in periodic arrangements of hedgehog-like magnetic textures.
\end{abstract}

\author{Haruki Watanabe}
\thanks{These two authors contributed equally to this work.} 
\affiliation{Department of Applied Physics, University of Tokyo, Tokyo 113-8656, Japan}

\author{Hoi Chun Po}
\thanks{These two authors contributed equally to this work.}
\author{Ashvin Vishwanath}
\affiliation{Department of Physics, Harvard University, Cambridge MA 02138}

\maketitle

\section{Introduction}
The recent discovery of topological insulators and other topological phases~\cite{Molenkamp} has revitalized the venerable subject of band theory. In addition to the explosion of understanding of  different  forms of band topology and how symmetries protect or prevent them, rapid progress is being made on several other fronts. For example, fundamental questions such as the connection between electron count and insulating behavior~\cite{Zak1999,Zak2000,Zak2001,Parameswaran, Roy,PNAS, SciAdv,PRL}, as well as the constraints imposed by crystal symmetries on the connectivity of bands \cite{Wigner,Herring,Zak1999,Zak2000,Zak2001,PRL,KaneDirac,Combinatorics, NC, Bernevig1, Graph}, have been resolved. New information has also been gleaned by contrasting  real-space and momentum-space descriptions in all 230 crystal space groups (SGs) of nonmagnetic materials~\cite{Zak1999,Zak2000,Zak2001,SciAdv,PNAS,PRL,NC,Bernevig1,Bernevig2}.  

Increasingly, attention is turning to electronic systems that combine magnetism with band topology. Here, a further panoply of novel topological phenomena is anticipated. Examples that have already been realized, to name just a few, include the quantized anomalous Hall effect in magnetic topological insulators~\cite{SCZhang,QiKunXue} and the topological Hall effect in skyrmion crystals~\cite{KimMillis,Tokura, Ong, BinzVishwanath,Pfliederer}. Promising magnetic Weyl semimetals candidates are also now being intensively studied~\cite{Nakatsuji,Felser,Yan}.  In addition to the fundamental  physics interest of these novel phases,  current-driven magnetic textures such as domain walls or skyrmions could be harnessed in technological applications, say in the development of energy-efficient memory devices~\cite{Fert}.

Despite these strong motivations, the pace of discovery of magnetic topological materials has been relatively slow compared to their nonmagnetic counterparts. There are at least two reasons for this: First, magnetic materials are necessarily correlated, making the  prediction of the magnetic structure and properties  more challenging. Consequently the set of well characterized magnetic materials is relatively small. Second, the sheer complexity of combining magnetic structures with SGs makes an exhaustive study of the theoretical possibilities daunting. Indeed, there are a total of 1651 different magnetic space groups (MSGs), which were only tabulated in the 1950s. Also, unlike for the 230 nonmagnetic SGs ~\cite{Bradley,Bilbao},  relevant  group-representation information is not always readily available.

Here, we will tackle the second problem by providing a systematic understanding of electronic band structures in the 1651 MSGs.  Inspired by the recent synthesis of atomically thin magnets ~\cite{Gong,Huang}, we also discuss the 528 magnetic layer groups (MLGs), relevant to two-dimensional magnetic materials.
We report  results for three key properties, tabulated in Appendix \ref{app:tables} and described below. 

First, we determine electron fillings that can be compatible with insulating behavior in all MSGs.  The electron count is a fundamental characteristic of an electronic crystal. The presence of nonsymmorphic symmetries, such as glides and screws, enforces connectivity of bands that raises the required fillings for realizing band insulators~\cite{Herring,Zak1999,Zak2000,Zak2001}. 
These conditions may be useful in the search for magnetic Weyl semimetals, as one can target fillings that, while forbidding band insulators, are nonetheless consistent with nodal-point Fermi surfaces~\cite{KaneDirac,RuChen,YoungWieder}.  

Next, it has long been known that representations of energy levels at high-symmetry points must connect in specific ways in obtaining a set of isolated bands~\cite{Wigner}. This can be viewed as a refinement of the filling condition, which  imposes  additional constraints on band structures (BSs)~\cite{Combinatorics,NC,Bernevig1,Graph}. The solutions to these constraints are most conveniently represented as a vector space (but with integer coefficients, so more accurately a `lattice' in mathematical terminology)~\cite{Ari,Combinatorics,NC}, and are described using only a handful of basis vectors whose precise number $d_{\rm BS}$ depends on the MSG.

Lastly, we contrast the general BSs defined in the momentum space with the subset of those obtained from `atomic insulators' (AI), in which electrons are tightly bound to sites in the lattice furnishing orbitals with different symmetries. Significantly, we find that this produces a vector space of the {\em same} dimension, which in mathematical terms is summarized by the equality $d_{\rm AI}=d_{\rm BS}$. However, the bases for BSs and AIs do not generally coincide, and we determine the classes of band structures that cannot be reduced to any atomic limit. 
This leads to an obstruction to finding symmetric localized Wannier states ~\cite{Thouless, Vanderbilt}, and corresponds to band topology that can be diagnosed without detailed information about the electronic wave-functions. We give an example of how this can be used to diagnose topological semimetals, and also an example of a nodal-line semimetal diagnosed through its filling. The latter consists of hedgehog-antihedgehog magnetic order with a lone electron at the core of these defects, leading to gapless behavior.  

We note that much progress has recently been made on related problems for nonmagnetic SGs~\cite{SciAdv,PRL,NC,Bernevig1,Bernevig2}. However, in the case of MSGs these problems have only been attacked in certain restricted settings, for example in generalizing the parity criterion~\cite{FuKane} to identify Weyl semimetals, Chern insulators, and axion insulators~\cite{Ari10, Ari, PhysRevB.83.205101, Bernevig, Chern_Rotation}.  
Here, we furnish a systematic study of all MSGs.
In particular, it is important to emphasize that, although our approach has several features in common with K-theory-based classifications~\cite{MooreFreed, Ktheory, Combinatorics, Ken2017}, we do not seek to fully classify crystalline electronic phases here. As discussed in a recent insightful work~\cite{Combinatorics} for two dimensional wallpaper groups, part of such a general classification includes band-structure connectivities, a problem that has been posed since the early days of band theory~\cite{Wigner}. Utilizing a result we prove below, we construct a framework that greatly simplifies such computations, which enables an extension of previous results \cite{SciAdv,PRL,NC} to the physically important case of MSGs.

In addition, topological distinctions revealed using our symmetry-based indicators fit naturally into well-established frameworks~\cite{MooreFreed,Combinatorics,Ken2017,NC}, and remain stable upon addition of trivial degrees of freedom. 
Yet, our approach is inherently symmetry-based, and so for crystals with low symmetry, say with only lattice translations, topological insulators cannot be diagnosed without further knowledge of the wavefunctions.
A similar caveat pertains to the complementary quantum chemistry approach of Ref.\ \onlinecite{Bernevig1}, which elaborates on the theory developed in Ref.\ \onlinecite{Zak1999}, and takes as input a specific set of orbitals at fixed locations in the crystal. While convenient for representing quantum chemistry information, it is less suited to capturing stable topological distinctions that survive the inclusion of additional trivial bands. 
 Also, one forgoes the simplification arising from the vector space like representation of energy bands that allowed us to generate results for all MSGs, while Ref. \onlinecite{Bernevig1} is currently confined to just the 230 SGs.

\section{Magnetic Band structures}
Let us begin by reviewing some background materials concerning electronic band structures arising from a magnetic material. There are in total 1651 MSGs and 528 MLGs~\cite{Bradley, Bilbao, Litvin, IUC, BNSdatabase}.
Among the 1651 MSGs, 230 of them are identical to SGs in which only unitary spatial symmetries are considered (type I MSGs).
All other MSGs have an equal number of unitary and anti-unitary elements: $\mathcal{M}=\mathcal{G}+\mathcal{A}$. The unitary part $\mathcal{G}$ is identical to one of the 230 SGs and the anti-unitary part $\mathcal{A}$ can be generally written in the form $\tilde{\mathcal{T}}\mathcal{G}$, where $\tilde{\mathcal{T}}\equiv\mathcal{T}g_0$ is the product of a spatial operation $g_0$ and time-reversal (TR) $\mathcal{T}$.  When $g_0$ belongs to $\mathcal{G}$, the MSG is simply the direct product of a SG $\mathcal{G}$ and $\mathbb Z_2^{\mathcal{T}}$. 
This leads to another 230 MSGs, one for each SG, that are called type II.
When $g_0$ is not an element of $\mathcal{G}$, there are two types further differentiated by whether $g_0$ is a pure translation (type IV) or not (type III).
For type II-IV MSGs, the little group of $\vec{k}$ and the site-symmetry group of $\vec{x}$ may also have an anti-unitary part, in addition to the usual unitary part.   Note that, in the literature (e.g.\ \cite{PennState}), our definition of type III and IV MSGs are sometimes called type IIIa and IIIb respectively.
In addition, there are two common labeling schemes for MSGs: Opechowski-Guccione (OG) and Belov-Neronova-Smirnova (BNS). In this work we follow the BNS notation, where an MSG is labeled by a pair of integers written as $S.L$, with $S$ corresponding to one of the 230 SG and $L$ an extra label to differentiate between different magnetic descendants. (Please refer to, e.g., Ref.\ \onlinecite{BilbaoMagnData}, for the precise meaning of these numbers.)

Next, we introduce a general formalism for the efficient analysis of band structure properties based on representations of the little group~\cite{Ari,Combinatorics,NC}.
 Our main focus here is to address how the formalism developed for SGs can be readily applied to MSGs.
We define a band structure (BS) as a set of bands isolated from others by band gaps above and below at all high-symmetry momenta, where by a high-symmetry momentum we refer to a $\vec{k}$ in the Brillouin zone at which the \emph{unitary} part of the little group, $\mathcal{G}_{\vec{k}}$, is strictly larger than the translation subgroup $T$. In 3D, they can be high-symmetry points, lines or planes.
A BS can be characterized by the set of nonnegative integers $\vec{n}=\{n_{\vec{k}}^\alpha\}$ that count the number of times an irreducible representation (irrep) $u_{\vec{k}}^\alpha$ of $\mathcal{G}_{\vec{k}}$ appears.  As we are interested in systems of spinful electrons, the irreps $u_{\vec{k}}^\alpha$ here are generally projective due to the spin-1/2 nature of electrons (also called `double-valued'). Since $\{n_{\vec{k}}^\alpha\}$ cannot be changed smoothly without gap closing or symmetry breaking, $\vec{n}=\{n_{\vec{k}}^\alpha\}$ serves as `topological invariants' defined for each BS.

The integers $\vec{n}=\{n_{\vec{k}}^\alpha\}$ cannot be chosen freely, since symmetries demand that they satisfy a collection of compatibility relations~\cite{Wigner, Bradley}.  Since $\mathcal{G}$ is a subgroup of $\mathcal{M}=\mathcal{G}+\mathcal{A}$, the full list of compatibility relation $\mathcal{C}$ imposed on an $\mathcal{M}$-symmetric BS can be split into two sets, $\mathcal{C}_{\mathcal{G}}$ arising from $\mathcal G$ and $\tilde{\mathcal{C}}_{\mathcal{A}}$ from $\mathcal A$. 
Let us denote by $\{{\rm BS}\}_{\rm phys}^{\mathcal{G}}$ and $\{{\rm BS}\}_{\rm phys}$ the set of all $\vec{n}$'s satisfying $\mathcal{C}_{\mathcal{G}}$ and $\mathcal{C}$, respectively. 
Here, the subscript `phys' indicates that all $n_{\vec{k}}^\alpha$'s in $\vec{n}=\{n_{\vec{k}}^\alpha\}$ are nonnegative, which is required for interpreting them as the multiplicities of irreps in a physical band structure.
We will introduce another set $\{{\rm BS}\}^{(\mathcal{G})}$ that relaxes this nonnegative condition later.
Note that $\{{\rm BS}\}_{\rm phys}^{\mathcal{G}}$ and $\{{\rm BS}\}_{\rm phys}$ differ only in the imposition of $\tilde{\mathcal{C}}_{\mathcal{A}}$.
In general, the anti-unitary part $\mathcal{A}$ requires a pairing of $\vec{b}\in\{{\rm BS}\}_{\rm phys}^{\mathcal{G}}$ with another $\vec{b}'\in\{{\rm BS}\}_{\rm phys}^{\mathcal{G}}$, unless $\vec{b}$ itself is already symmetric under $\tilde{\mathcal{T}}$.  The pairing type can be easily determined using the Herring rule~\cite{Bradley}. 
(In Appendix~\ref{app:compatibility}, we provide a more elaborated review on the compatibility relations and the Herring rule.)

\section{Band Topology}
Having described some generalities about BSs, we now review how knowledge about the real space can inform band topology \cite{SciAdv,NC}.
We define the trivial class of BSs by the AIs, which are band insulators that are smoothly connected to a limit of vanishing hopping, and hence are deformable to product states in real space. Equivalently, an AI admits symmetric, exponentially localized Wannier functions.

To specify an AI, one should choose a position $\vec{x}$ in real space at which electrons are localized, and the type of the orbital put on that site.  
All inequivalent choices of the position $\vec{x}$ are classified by Wyckoff positions~\cite{ITC}. The orbital can be chosen from the (co-)irreps of the site-symmetry group of $\vec{x}$ (Appendix~\ref{app:AI}). Given these choices, an $\mathcal{M}$-invariant AI can be constructed by placing a symmetry-related orbital on each site of the $\mathcal{M}$-symmetric lattice and filling them by electrons.  The AI has a specific combination of irreps in the momentum space, which automatically satisfies $\mathcal{C}=\mathcal{C}_{\mathcal{G}}+\tilde{\mathcal{C}}_{\mathcal{A}}$.  We list up all distinct $\vec{n}$'s corresponding to an AI by varying $\vec{x}$ and the orbital type.  We list up all distinct $\vec{n}$'s corresponding to an AI by varying the position $\vec{x}$ and the orbital type and we obtain $\{{\rm AI}\}_{\rm phys}$, a subset of $\{{\rm BS}\}_{\rm phys}$.   If one replaces $\mathcal{M}$ above with $\mathcal{G}$, one gets the set of $\mathcal{G}$-symmetric AIs, $\{{\rm AI}\}_{\rm phys}^{\mathcal{G}}$.

Now we are ready to tell which elements of $\{{\rm BS}\}_{\rm phys}$ must be topologically nontrivial and which elements can be trivial.  This can be judged by contrasting the elements of $\{{\rm BS}\}_{\rm phys}$ with those in $\{{\rm AI}\}_{\rm phys}$. Namely, any $\vec{b}\in\{{\rm BS}\}_{\rm phys}$ not belonging to $\{{\rm AI}\}_{\rm phys}$ necessarily features nontrivial band topology, because, by definition, there does not exist any atomic limit of the BS with the same combinations of irreps.  This is a sufficient (but not necessary) condition to be topologically nontrivial --- here we exclusively focus on the band topology that can be diagnosed by the set of irreps at high-symmetry momenta.  

The simplest way in exploring the nontrivial elements of $\{{\rm BS}\}_{\rm phys.}$ is thus to consider the complement of $\{{\rm AI}\}_{\rm phys.}$ in $\{{\rm BS}\}_{\rm phys.}$, as in Refs.~\onlinecite{NC, fragile}.
However, this set has a complicated mathematical structure. To simplify the analysis, we allow for the formal \emph{subtraction} of bands, and extend the values of $n_{\vec{k}}^\alpha$ to any integer, including the negative ones, {\it {\` a} la} a K-theory analysis.  $\{{\rm BS}\}_{\rm phys}$ then becomes an abelian group $\{{\rm BS}\}=\mathbb{Z}^{d_{{\rm BS}}}$ (known as a `lattice' in the mathematical nomenclature)~\cite{Ari,Combinatorics,NC}.   In other words, there are $d_{{\rm BS}}$ basis `vectors' $\{\vec{b}_i\}_{i=1}^{d_{{\rm BS}}}$, and $\{{\rm BS}\}$ can be expressed as $ \left \{\sum_{i=1}^{d_{{\rm BS}}}m_i\vec{b}_i\,|\,m_i\in\mathbb{Z} \right\}$.  Similarly, by allowing negative integers when taking superposition of AIs, we get another abelian group $\{{\rm AI}\}=\mathbb{Z}^{d_{{\rm AI}}}$, which is a subgroup of $\{{\rm BS}\}$. The band topology we are interested in is now encoded in the quotient group 
\begin{equation}
X_{\rm BS} \equiv \{ {\rm BS}\}/\{{\rm AI} \},
\end{equation}
dubbed the symmetry-based indicator of band topology~\cite{NC}.  As we will see shortly, the quotient group is always a finite abelian group and hence must be a product of the form $\prod_i {\mathbb Z}_{n_i}$.

\section{Constructing band structures from atomic insulators}
To compute $X_{\rm BS}$, the natural first step is to identify $\tilde{\mathcal C}_{\mathcal A}$, the extra compatibility relations enforced by the anti-unitary symmetries. Contrary to this expectation, we now show that, based on our earlier results on SGs, one can directly compute $\{{\rm BS}\}$ and $X_{\rm BS}$ for any MSG $\mathcal M$ without deriving $\tilde {\mathcal C}_{\mathcal A}$. This serves to demonstrate the power of the present approach: symmetry content and connectivity of BSs can be readily extracted without the large overheads mandated by the conventional approach.

To this end, we first revisit the relevant aspects of the theory for an SG $\mathcal G$.
By definition, $\{{\rm AI}\}^{\mathcal{G}}$ is a subgroup of $\{{\rm BS}\}^{\mathcal{G}}$, and therefore, a priori, it could be the case that $d_{{\rm AI}}^{\mathcal{G}}<d_{{\rm BS}}^{\mathcal{G}}$ strictly.  However, by an explicit computation for all the 230 SGs, Ref.\ \onlinecite{NC} found that $d_{{\rm BS}}^{\mathcal{G}}=d_{{\rm AI}}^{\mathcal{G}}$ always holds.  This statement has an important implication --- every $\vec{b}\in\{ {\rm BS} \}^{\mathcal{G}}$ can, in fact, be expanded as $\vec{b} = \sum_iq_i \, \vec{a}_i$ with \emph{rational} coefficients $q_i\in\mathbb{Q}$, using a basis $\{\vec{a}_i\}_{i=1}^{d_{{\rm AI}}^{\mathcal{G}}}$ of $\{ {\rm AI} \}^{\mathcal{G}}$.
In other words, full knowledge on the group $\{ {\rm BS} \}^{\mathcal{G}}$ can be obtained from that on $\{ {\rm AI} \}^{\mathcal{G}}$.

Based on this result for SGs, we will now prove the same statement, namely $d_{{\rm BS}}=d_{{\rm AI}}$, for any MSG $\mathcal M$.
This result enables us to efficiently compute $\{{\rm BS}\}$ and $X_{\rm BS}$ for all MSGs and MLGs using only information contained in $\{{\rm AI}\}$, which can be readily extracted from the tabulated Wyckoff positions~\cite{Litvin, IUC, BNSdatabase}.
 
Our proof is centered on the following observation: Recall $\mathcal M = \mathcal G + \mathcal A$, where $\mathcal G$ is unitary and $\mathcal A=\tilde{\mathcal{T}}\mathcal G$ is the anti-unitary part generated by $\tilde{\mathcal{T}} = \mathcal T g_0$ for some spatial symmetry $g_0$. Note that for type II MSGs, $g_0 \in \mathcal G$ and can be chosen to be the identity, whereas for types III and IV, $g_0 \not \in \mathcal G$.
Now consider a $\mathcal G$-symmetric collection of fully-filled local orbitals in real space, which defines an AI $\vec{a} \in \{ {\rm AI}\}_{\text{phys}}^{\mathcal{G}}$. This AI is not generally symmetric under $\mathcal M$, i.e., it may not be invariant under the action of $\tilde{\mathcal{T}}$. However, if we stack it together with its $\tilde{\mathcal{T}}$-transformed copy, we will arrive at an $\mathcal M$-symmetric AI. Algebraically, this means $\vec a + \tilde{\mathcal{T}}\vec a\in \{ {\rm AI}\}_{\text{phys}}$. 

In the momentum space, a similar symmetrization procedure can be performed on the representation content.  
Suppose that $\{|\vec{k},i\rangle\}$ is a basis of an irrep $u_{\vec{k}}^\alpha$ of $\mathcal{G}_{\vec{k}}$. Then the $\tilde{\mathcal{T}}$-transformed copy, $\{\tilde{\mathcal{T}}|\vec{k},i\rangle\}$, forms a basis of an irrep $u_{\vec{k}'}^{\alpha'}$ of $\mathcal{G}_{\vec{k}'}$ (Appendix~\ref{app:compatibility}).  When $\vec{b}$ represents a BS that contains $u_{\vec{k}}^{\alpha}$ $n_{\vec{k}}^\alpha$-times, we denote by $\tilde{\mathcal{T}}\vec{b}$ a BS that contains $u_{\vec{k}'}^{\alpha'}$ the same number of times.   For any $\vec b \in \{ {\rm BS}\}^{\mathcal{G}}$, $\vec{b}+\tilde{\mathcal{T}}\vec{b}$ satisfies the compatibility conditions $\tilde{\mathcal{C}}_{\mathcal{A}}$, and hence belongs to $\{{\rm BS}\}$.   We present the explicit form of $\tilde{\mathcal{T}}\vec{b}$ in Appendix~\ref{app:themapf}.

\begin{table}
\begin{center}
\caption{\textbf{Characterization of band structures (BSs) in a magnetic space group (MSG); excerpt from Tables~\ref{tab:spinful_Tri}--\ref{tab:SpinfulCubic}.}
\label{tab:ex1}}
\begin{tabular}{lc|ccc} 
\hline \hline 
\multicolumn{2}{c|}{MSG\footnote{MSG number in the Belov-Neronova-Smirnova notation, followed by a Roman numeral ${\rm I}, \ldots, {\rm IV}$ indicating its type.}} &
$d$\footnote{Number of linearly independent BSs.}
&
$X_{\rm BS}$\footnote{Symmetry-based indicator of band topology, which takes the form $\prod_i {\mathbb Z}_{n_i}$; denoted by the collection of positive integers $(n_1,n_2, \cdots)$.}
&
$\nu_{\rm BS}$\footnote{For most of the MSGs, the set of physical BS fillings $\{\nu\}_{\rm BS}$ and the set of AI fillings $\{\nu\}_{\rm AI}$ agree with each other, and they take the form $\{\nu\}_{\rm BS}=\{\nu\}_{\rm AI}=\nu_{\rm BS} \,\mathbb{N}$.  The asterisk indicates violation to this rule, detailed in Table~\ref{tab:spinfulnuEx}.}
\\ \hline
2.4 & I & $9$ & $(2, 2, 2, 4)$ & $1$\\
2.7 & IV & $5$ & $(2)$ & $2$\\
3.4 & IV & $3$ & $(2)$ & $2$ \\
209.51 & IV & $3$ & $(1)$ & $2^*$ \\
\hline \hline 
\end{tabular}
\end{center}
\end{table}

We are now ready to prove the statement. Observe any $\vec{B}\in\{{\rm BS}\}$ also belongs to $\{{\rm BS}\}^{\mathcal{G}}$, and as $d_{{\rm BS}}^{\mathcal{G}}=d_{{\rm AI}}^{\mathcal{G}}$, it can be expanded in the basis of $\{{\rm AI}\}^{\mathcal{G}}$:
\begin{equation}
\vec{B}=\textstyle\sum_{i}q_i\vec{a}_i,\quad q_i\in\mathbb{Q}.\label{eq:expand}
\end{equation}
Now we symmetrize both sides of Eq.~\eqref{eq:expand} by $\tilde{\mathcal{T}}$.  Since $\vec{B}$ is $\mathcal{M}$-invariant, $\vec{B}+\tilde{\mathcal{T}}\vec{B}=2\vec{B}$, so
\begin{equation}
\vec{B}=\textstyle\frac{1}{2}(\vec{B}+\tilde{\mathcal{T}}\vec{B})=\sum_{i}\frac{1}{2}q_i\big(\vec{a}_i+\tilde{\mathcal{T}}\vec{a}_i\big).
\end{equation}
As argued, $\vec{a}_i+\tilde{\mathcal{T}}\vec{a}_i \in \{ {\rm AI}\}$.  (Note that linearity was invoked when we extend the argument form $\{ {\rm AI}\}_{\text{phys}}^{\mathcal{G}}$ to a general element of $\{ {\rm AI}\}^{\mathcal{G}}$.) This proves that any $\vec{B}\in\{{\rm BS}\}$ can be expanded in terms of $\{{\rm AI}\}$ (using rational coefficients), an equivalent statement of $d_{{\rm AI}}=d_{{\rm BS}}$.  Furthermore, it implies that the quotient group $X_{\rm BS}=\{{\rm BS}\}/\{{\rm AI}\}$ does not contain any $\mathbb{Z}$-factor, and hence is a finite abelian group of the form $\prod_i {\mathbb Z}_{n_i}$.

To summarize, we have shown that the set of AIs and BSs are identical as far as their dimensionality goes.  This is a powerful statement, since it means we could simply focus on AIs, study their symmetry representations in momentum space, and then take rational combinations to generate all BSs. We caution, however, that one has to properly rescale the entries of $\vec n$ when an irrep is paired with another copy of itself according to the Herring rule. For full details of the treatment, we refer the interested readers to Ref.\ \onlinecite{NC}.

Following this strategy, we perform the first calculation of  $d_{\rm BS}$, $X_{\rm BS}$, and $\nu_{\rm BS}$ for all of the 1651 MSGs and 528 MLGs. The full list of the computation results are tabulated in Tables~\ref{tab:spinful_Tri}--\ref{tab:spinful_MLG} in Appendix~\ref{app:tables}. 
For readers' convenience, we reproduce a few examples from these tables in Tables~\ref{tab:ex1} and \ref{tab:ex2}.

\begin{table}
\begin{center}
\caption{\textbf{Characterization of magnetic layer groups through the corresponding magnetic space group (MSG); excerpt from Table~\ref{tab:spinful_MLG}.}
\label{tab:ex2}
}
\begin{tabular}{cc|ccc} 
\hline \hline 
\multicolumn{2}{c|}{MSG\footnote{The numbers in parentheses label the different ways to project the MSG down to 2D planes (Appendix~\ref{app:MLG}).}} & $d$\footnote{\label{FootTableI} Defined as in Table \ref{tab:ex1}.
}
& $X_{\rm BS}$\footref{FootTableI} & $\nu_{\rm BS}$\footref{FootTableI}
\\\hline
2.5 (1) & II & $5$ & (2) & $2$ \\
2.5 (2) & II & $5$ & (2) & $2$ \\
2.5 (3) & II & $5$ & (2) & $2$ \\
3.4 (1) & IV & $3$ & $(2)$ & $2$ \\
3.4 (2) & IV & $2$ & $(1)$ & $2$ \\
\hline\hline 
\end{tabular}
\end{center}
\end{table}

\section{Representation-enforced semimetals}
As an application of our results, here we introduce new classes of representation-enforced semimetals (reSM).  A BS is said to be a reSM when (i) there are gap closings at some generic (i.e., not high-symmetry) momenta in the Brillouin zone, and (ii) the gaplessness is mandated by the combination of irreps at high-symmetry momenta. 
Note that only unitary symmetries are incorporated in defining a high-symmetry momentum, and the notion of a BS only requires a continuous gap at all high-symmetry points, lines and planes. ReSMs arise when the topological band-gap closings are buried in the low-symmetry regions of the Brillouin zone, rendering them hard to diagnose in conventional electronic structure calculations, but are nonetheless readily detectable and are robust against numerical uncertainty due to their representation-enforced nature.
All previously known examples of reSMs concern TR-broken systems in 3D with either the inversion~\cite{Ari, Bernevig} or $\bar{4}$ rotation symmetry~\cite{Chern_Rotation}.  The new classes we propose here are realized in spin-orbit coupled systems in 2D or 3D, and are associated to type III or IV MSGs.

Our example is the type IV MSG 3.4 ($P_a112$) is generated by $C_{2z}$ (the $\pi$ rotation about $z$-axis) and $\tilde{\mathcal{T}}\equiv\mathcal{T}T_{\frac{1}{2}\vec{a}_1}$ (half translation by $\frac{1}{2}\vec{a}_1$ followed by TR) apart from lattice translations.  (The type III MSG 13.69 can also host a Dirac semimetal through a nearly identical mechanism. See Appendix \ref{app:tbmodel}.) According to Table~\ref{tab:ex1}, this MSG has two classes of combinations of $C_{2z}$ eigenvalues, as indicated by $X_{\rm BS}=\mathbb{Z}_2$.  As we shall see now, the 3D band structure belonging to this nontrivial class hosts at least four Weyl points, where the Weyl points with opposite charities are maximally separated in the $k_3$ direction.

To understand this band topology step by step, let us first remove the lattice translation in $z$ and take a single layer parallel to the $xy$ plane.
The MSG then reduces to the MLG $p_a112$, which corresponds to the entry 3.4 (1) in Table~\ref{tab:ex2} with $X_{\rm BS}=\mathbb{Z}_2$.  The rotation $C_{2z}$ is a symmetry at the four TR-invariant momenta of the 2D BZ, whose eigenvalues are either $\pm i$ for spinful electrons. When $k_1=0$, the anti-unitary symmetry $\tilde{\mathcal{T}}$ with $\tilde{\mathcal{T}}^2=(-1)^{\hat{F}}\hat{T}_{\vec{a}_1}=-1$ demands the pairing of the two eigenvalues of $C_{2z}$.  In contrast, when $k_1=\pi$, $\tilde{\mathcal{T}}^2=+1$ and no pairing is required.  For this MLG, $X_{\rm BS}=\mathbb{Z}_2$ and the topological index distinguishing the two class is given by
\begin{equation}
\label{eq:topoinv}
\eta=\prod_{n: \text{occupied}}\prod_{\vec{k}=(\pi,0),(\pi,\pi)}\,\eta_{n, \vec{k}},
\end{equation}
where $\eta_{n,\vec{k}}$ is the $C_{2z}$-eigenvalue of the $n$-th band.  
$\eta=+1$ for any AI, while $\eta=-1$ for nontrivial BSs.
Note that this invariant is similar to those identified in some earlier works \cite{Ari, Bernevig, Chern_Rotation, FuKane,Rappe}.

\begin{figure*}[tbh!!]
\begin{center}
{\includegraphics[width=0.8 \textwidth]{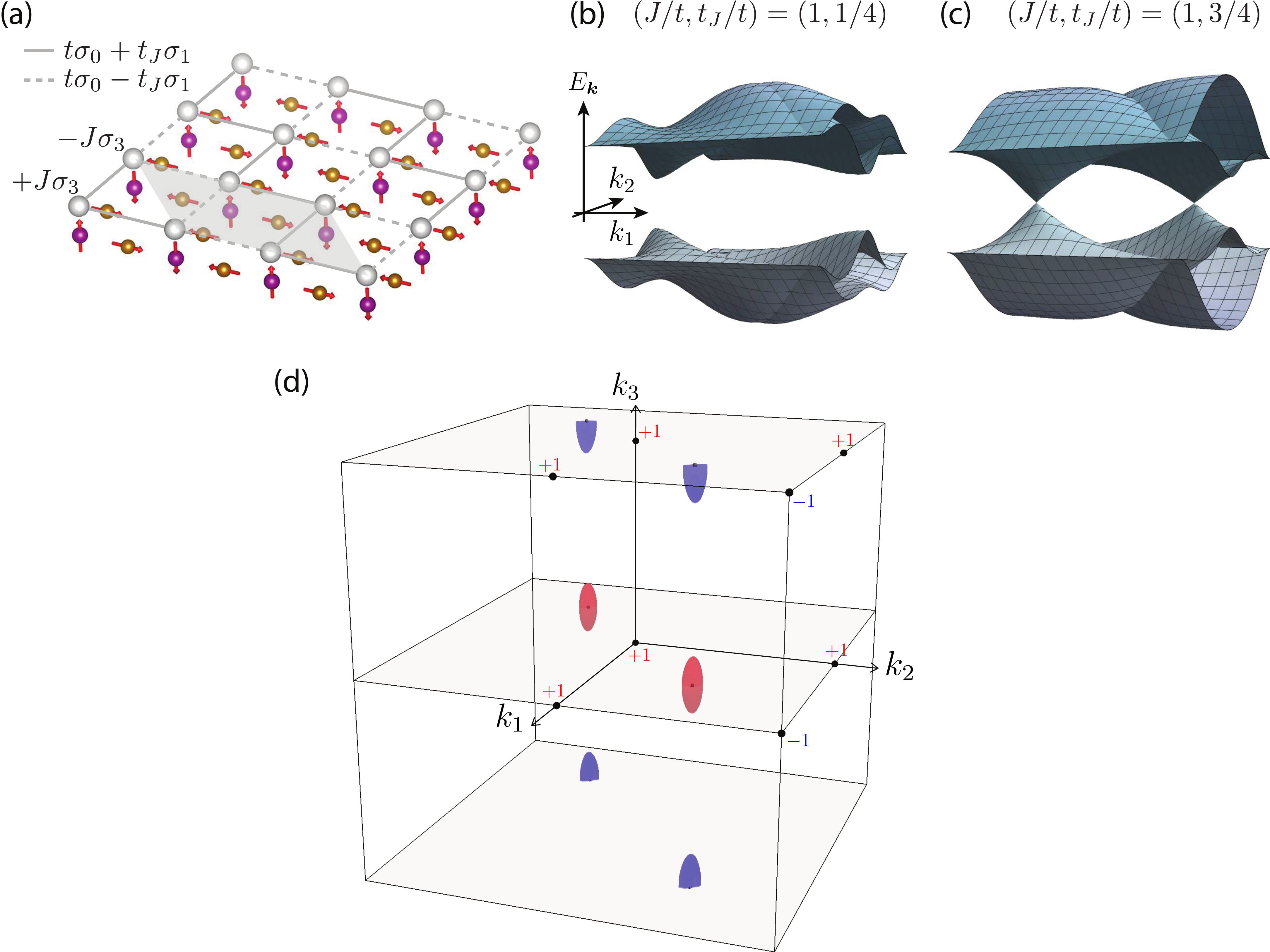}}
\caption{
{\bf Magnetic representation-enforced semimetal.} 
(a) An example tight-binding model for representation-enforced semimetal.  There are two sub-lattices per unit cell (shaded) due to an anti-ferromagnetic order $\vec{m}_{\vec{x}}$, producing in total four bands.  On-site potential $J$ stands for the exchange coupling $-J\vec{m}_{\vec{x}}\cdot c_{\vec{x}}^\dagger\vec{\sigma}c_{\vec{x}}$.  In addition to the standard nearest-neighbor (NN) hopping $t$, a spin-dependent hopping $\pm t_J\sigma_x$ is included, which can be viewed as originating from an exchange coupling to a magnetic moment in the middle of the NN bonds (Appendix~\ref{app:tbmodel}). (b) The dispersion relation at $J/t=1$ and $t_J/t=1/4$. In this case, $\eta$ in Eq.~\eqref{eq:topoinv} is $+1$ and the dispersion is gapped between the second and the third band. (c) The dispersion relation at $J/t=1$ and $t_J/t=3/4$. Now $\eta=-1$ and a pair of Dirac nodes exists as predicted. (d) The Fermi surface of the 3D version of the representation-enforced semimetal. The two Weyl points on the $k_3=0$ ($k_3=\pi$) plane have the chirality $+1$ ($-1$), indicating a huge Fermi arc on some 2D surfaces.  The signs on the TRIMs indicate the product of the $C_{2z}$ rotation eigenvalues of occupied bands in this model.}
\label{fig:reSM}
\end{center}
\end{figure*}

Next, we show that the nontrivial BS with $\eta=-1$ corresponds to 2D magnetic Dirac reSMs (two-fold degeneracy at the gapless point) enabled by a strong spin-rotation symmetry breaking, as illustrated in Fig.\ \ref{fig:reSM}. 
Denoting the standard Pauli matrices by $\sigma_{0,\dots,3}$, the combined symmetry $\tilde{\mathcal{T}}C_{2z}$ dictates that the local $2$ by $2$ Hamiltonian $h(\vec{k})$ near a  gapless point takes the form
\begin{equation}
h(\vec{k}) = \sum_{i=0}^{2} g_i(\vec{k})\sigma_i,
\end{equation}
i.e., the $\sigma_3$ term is missing.
Since there are the same number of the tunable parameters $k_1$ and $k_2$ as the number of the relevant coefficients ($g_1(\vec{k})$ and $g_2(\vec{k})$), the gaplessness is stable against perturbations respecting the symmetries~\cite{BJY}.  
To see this more explicitly, we note that the symmetry $\tilde{\mathcal{T}}C_{2z}$ quantizes the Berry phase $B(k_1)=\int_{-\pi}^{\pi} dk_2 A_2(k_1)$ to be $0$ or $\pi$ at any $k_1$.  The topological index $\eta=-1$ dictates that $B(0)\neq B(\pi)$, implying the presence of a gap closing somewhere in $0<k_1<\pi$.
Using a $\vec k \cdot \vec p$ analysis, one can confirm that the bands disperse linearly, leading to a Dirac node.
We note that, compared to the protection of Dirac node by an eigenvalue exchange of nonsymmorphic symmetries~\cite{YoungKane,YoungWieder,JingWang,Vanderbilt2}, the present Dirac node has a more topological character.

\begin{figure*}[tbh!!]
\begin{center}
{\includegraphics[width=0.98 \textwidth]{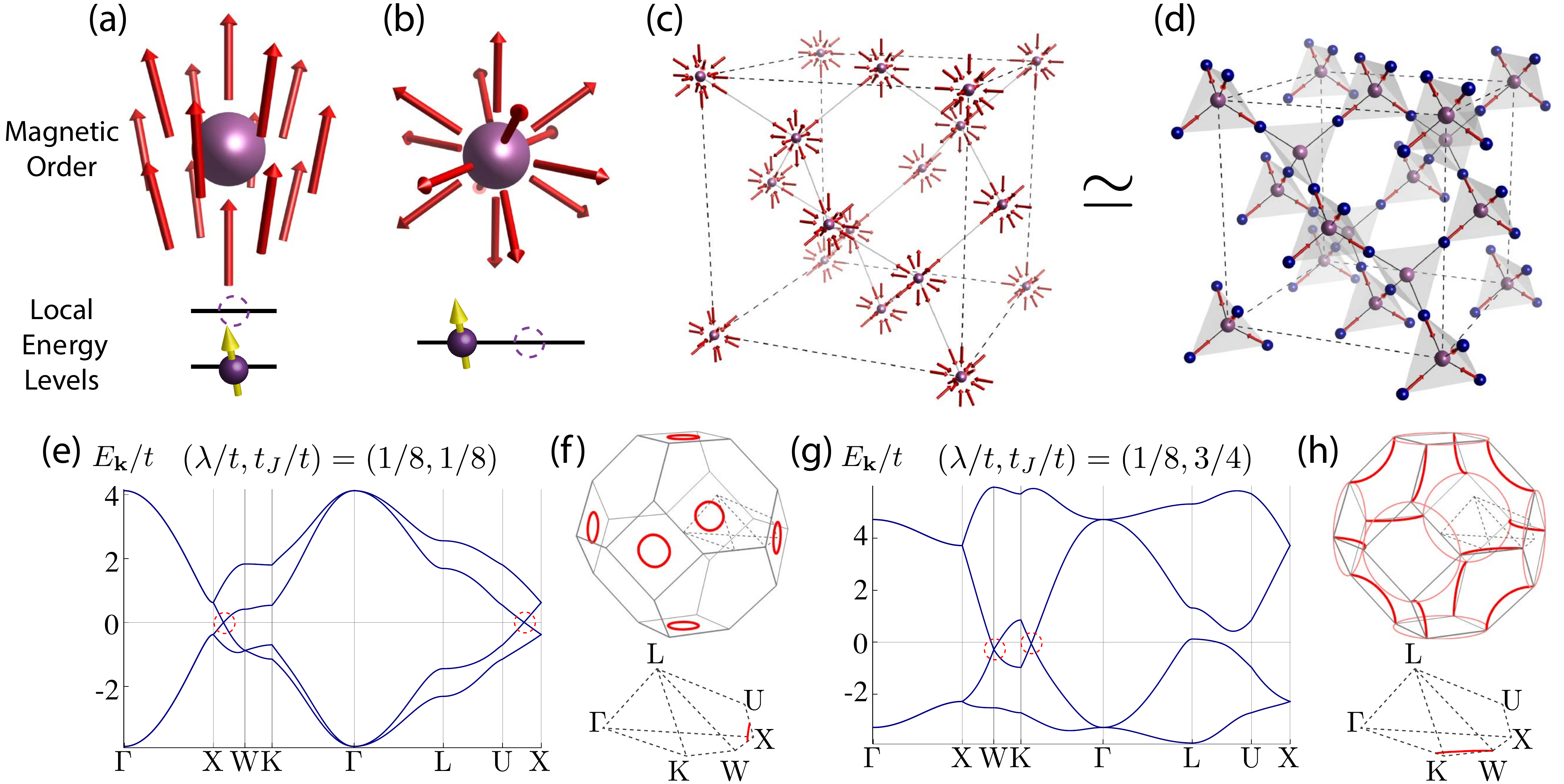}}
\caption{
{\bf Magnetic filling-enforced semimetals.} 
(a,b) Symmetries of a magnetic order can prohibit atomic insulators (AIs) at odd site fillings. 
(a) The magnetic-point-group symmetries of a ferromagnetic arrangement is compatible with nondegenerate local energy levels. (b) Those of the depicted hedgehog defect, however, force all the energy levels to exhibit even degeneracies, which forbids AIs when a lone electron is localized to the purple site.
(c) When hedgehog and anti-hedgehog defects are arranged into a diamond lattice, the previous argument suggests that no AI is allowed whenever the site fillings at the defect cores are odd.
(d) The hypothetical magnetic structure in (c) could be realizable in spinel structures if the diamond sites are occupied by atoms with odd atomic numbers (purple), and the magnetic atoms (blue) at the pyrochlore sites exhibit  an all-in-all-out magnetic order.
(e--h) Filling-enforced semimetals arising form the magnetically-ordered Fu-Kane-Mele model, Eq.\ \eqref{eq:JFKM}.
(e) When the magnetically modulated hopping $t_J$ is weak compared to the spin-orbit coupling $\lambda$, the fermiology is governed by rings of gap-closing (circled in red), growing out from the original Dirac points at $X$ when $t_J=0$. (f) The positions of the rings in the Brillouin zone are shown in red.
(g,h) For $t_J \geq 2\sqrt{2} \lambda$, the nodal rings become connected at the momentum W. (Thin lines indicate copies of the gapless momenta in the repeated zone scheme, included to illustrate the connectivity of the rings.)
}
\label{fig:feSM}
\end{center}
\end{figure*}

Given this understanding of the 2D layer with MLG 3.4, let us now recover the third dimension by stacking the copies of 2D layers with generic inter-layer couplings.   The resulting band structure, fully symmetric under MSG 3.4, host in total four Weyl points; two on the $k_3=0$ plane and the other two on the $k_3=\pi$ plane, as depicted in Fig.~\ref{fig:reSM}d.  Note that the two Weyl points on the $k_3=0$ plane are related by $C_{2z}$ rotation (or, equivalently in this setup, $\tilde{\mathcal T}$) symmetry, and so must have the same chirality.  Consequently, the two Weyl points on the $k_3=\pi$ plane have the opposite chirality. This large separation of the Weyl points with opposite chiralities dictates the presence of long Fermi arcs on the surface.  Let us stress again that this nontrivial band structure is indicated by the rotation eigenvalues at TRIMs, via the formula in Eq.~\eqref{eq:topoinv} applied to the $k_3=0$ plane of the 3D band structure.This illustrates how the ideas  described in the present work can guide the search for ideal magnetic Weyl semimetals.
We leave a full analysis of magnetic reSMs to future works.

\section{Filling-enforced semimetals}
As another application of our theory, we attack the following problem: Given an MSG $\mathcal M$, what are the electron fillings $\nu$ at which band insulators are allowed? One of the primary interests in solving this problem lies in the search of topological semimetals --- if a material has an electron filling incompatible with any band insulator, it must possess symmetry-protected gaplessness near the Fermi energy, provided a band-theory description is applicable. We refer to such systems as filling-enforced (semi-)metals. 
Once the MSG describing the magnetic order is identified, such systems can be readily predicted using our results on $\nu_{\rm BS}$. For stoichiometric compounds with perfect crystalline order and commensurate magnetism, one can further show that spatial symmetries quantize the physically allowed fillings, which in turns allows one to further diagnose the fermiology of the filling-enforced (semi-)metals. We relegate a detailed discussion to Appendix \ref{app:feSM}, and summarize the results on the predicted fermiology in Tables \ref{tab:feSM-p}--\ref{tab:feSMSOC}.

While a systematic analysis of filling-enforced semimetals (feSMs) can be performed using the tables mentioned above, it is instructive to discuss a physical picture linking a real-space description of the magnetic ordering to the momentum-space obstruction to realizing any band insulator. To this end, we first contrast the effect of topologically trivial and nontrivial magnetic orders on local electronic energy levels. 
When the surrounding magnetic moments order ferromagnetically, as in Fig.\ \ref{fig:feSM}a, the electronic levels are Zeeman-split and nondegenerate. In contrast, when the moments are arranged into the hedgehog defect shown in Fig.\ \ref{fig:feSM}b, the magnetic-point-group symmetries at the defect core demand that all single-particle local energy levels exhibit even degeneracies, i.e., such a magnetic order forbids a gapped state when a lone electron is localized to the defect core.

To be compatible with lattice translations, however, there must be a balance between defects and anti-defects in each unit cell. An interesting situation arises when the positions of the defects and anti-defects are further related by symmetries prohibiting them from annihilation. 
As an example, consider the diamond lattice with one sublattice occupied by hedgehog defects, and the other sublattice, symmetry-related to the first by a glide mirror, occupied by anti-hedgehogs (Fig.\ \ref{fig:feSM}c). The symmetries of this system are described by MSG 227.131 (type III). 
Suppose the defect core (i.e., the diamond sites) corresponds to an atom with an odd atomic number, which leads to an electron filling of  $\nu = 2 \mod 4$ per primitive unit cell. By the previous argument, it should be impossible to obtain an AI by localizing all the electrons. This suggests the electron filling might lead to an obstruction to forming a band insulator and hence enforcing a (semi-)metallic behavior.
Indeed, our result of $\nu_{\rm BS} = 4$ for this MSG (Table \ref{tab:SpinfulCubic}) implies all band insulators are ruled out at the specified filling of $\nu = 2 \mod 4$. From Table \ref{tab:feSM-m},  we further see that this corresponds to a feSM with movable nodal features.

Curiously, the hypothetical structure depicted in Fig.\ \ref{fig:feSM}c could be relevant for materials with the chemical formula $A B_2 O_4$ taking the spinel structure, where the $A$ atoms, occupying the diamond sites, are surrounded by the $B$ atoms sitting at the pyrochlore positions. If the $B$ atoms are magnetically ordered into the all-in-all-out configuration (Fig.\ \ref{fig:feSM}d), it can be viewed as a lattice realization of the described hedgehog-anti-hedgehog lattice in MSG 227.131.
This suggests the following tight-binding model as an example of the feSM described:
\begin{equation}\begin{split}\label{eq:JFKM}
\hat H = & \sum_{\langle ij\rangle} \hat c_{i}^\dagger \left( t \, \sigma_0 + 4\, t_J\, \vec d_{ij} \cdot \vec \sigma \right) \hat c_j\\
&~~~~~~~~~~~~ + i\, 8 \lambda \sum_{\langle \langle ij \rangle \rangle} \hat c_{i}^\dagger \left(\vec d_{li}  \times \vec d_{lj}\right)\cdot \vec \sigma\,\hat c_j,
\end{split}\end{equation}
where $\hat c_i$ and $\hat c_i^\dagger$ represent the fermion operators corresponding to an $s$-orbital localized to the diamond site $i$ (spin indices are suppressed), $\langle i j \rangle$ and $\langle \langle ij \rangle \rangle$ respectively denote first- and second-nearest-neighbor bonds, and $\vec d_{ij}$ is the vector connecting the site $j$ to $i$ (the lattice constant of the conventional cell is set to $1$).
In the term $\propto \lambda$, we let $l$ denote the common nearest neighbor of $i$ and $j$.
Note that $t_J$ can be physically interpreted as a modulation of the nearest-neighbor hopping by exchange coupling with the magnetic moment at pyrochlore sites, and $\lambda$ parameterizes the symmetry-allowed spin-orbit coupling between the second-nearest-neighbor sites.

When $t_J=0$, Eq.~\eqref{eq:JFKM} is TR-symmetric (corresponding to the type II MSG 227.129), and reduces to the undistorted Fu-Kane-Mele model defined on the diamond lattice hosting three symmetry-related Dirac points (Table \ref{tab:feSM-p}) \cite{Z2_3D}. Unlike lattice distortions, which lead to topological insulators~\cite{Z2_3D}, TR-symmetry breaking described by $t_J\neq 0$ leads to a feSM.
Interestingly, the Dirac points in the original Fu-Kane-Mele model are unstable towards such perturbation, and our results indicate that the semimetallic behavior is enforced by a more delicate band connectivity (Table \ref{tab:feSM-m}).
This is explicitly verified in the example band structures plotted in Figs.\ \ref{fig:feSM}e-h, showing how the Dirac points split into (doubly degenerate) nodal rings as $t_J$ increases from $0$.

This example serves only as a particular instance of the 421 MSGs we identified to be compatible with feSMs (Appendix \ref{app:feSM}). A full list of such MSGs is presented in Tables \ref{tab:feSM-p} and \ref{tab:feSM-m}. 
Armed with this list, we studied the magnetic structures listed on the Bilbao Crystallographic Server \cite{BilbaoMagnData}, and found that the experimentally-characterized magnetic materials YFe$_4$Ge$_2$ and  LuFe$_4$Ge$_2$ (and related compounds) in the type III MSG 58.399 are realistic magnetic feSM candidates \cite{YFG, LFG}. 
For these compounds, a nonmagnetic atom with an odd atomic number (Lu or Y) sits at a maximal-symmetry site of multiplicity 2, whereas their  surrounding magnetic moments (Fe) ordered into a pattern symmetric under the combined symmetry of spatial inversion and TR.
Such a magnetic ordering falls into our broad description of nontrivial defect lattices, and these compounds are expected to feature Dirac points, pinned to high-symmetry momenta, near the Fermi energy (Table \ref{tab:feSM-p}).

We remark that the method described here can be readily applied to any other commensurate magnetic crystals where both the chemical formula and the MSG describing the magnetic order have been identified. In particular, a local-moment description is inessential --- the filling criterion established based on our computation of $\nu_{\rm BS}$ applies equally well to systems exhibiting itinerant magnetism, as long as a band-theory description remains applicable.

\begin{figure*}[tbh!!]
\begin{center}
{\includegraphics[width=0.60 \textwidth]{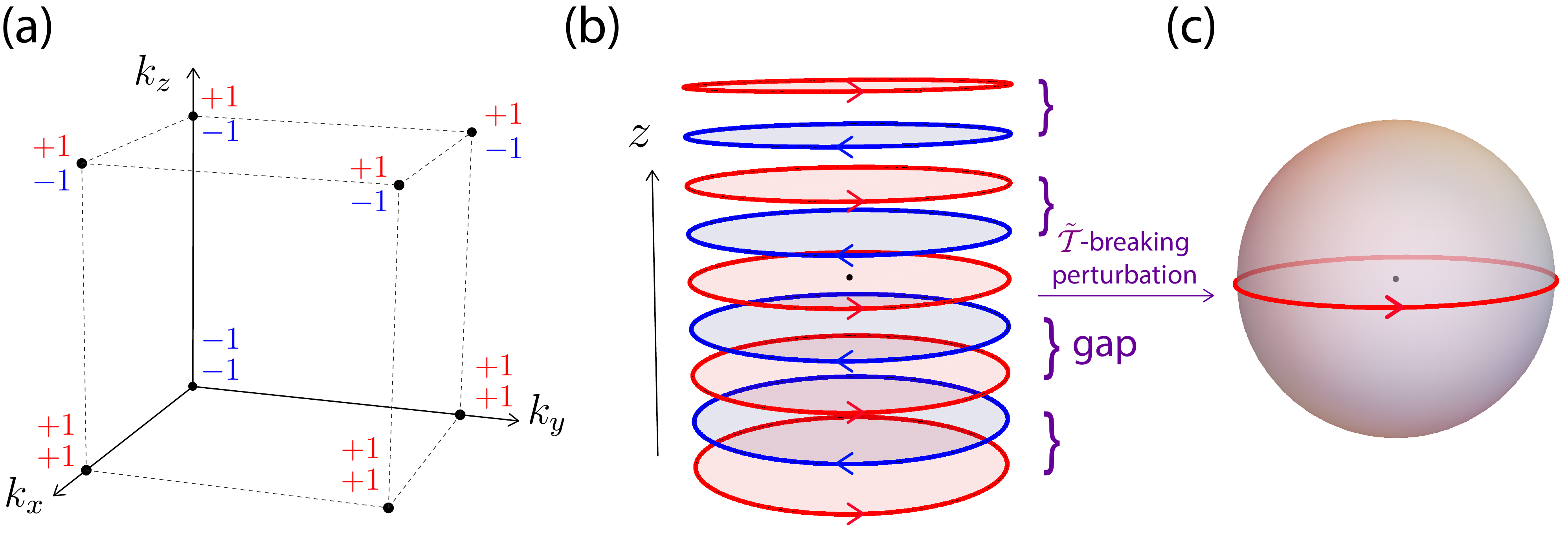}}
\caption{
{\bf Symmetry indicators for anti-ferromagnetic topological insulators.}
(a) An example of the inversion parity combination of valence bands in the $\mathbb{Z}_2$ nontrivial phase of MSG 2.7. (b) The realization of the $\mathbb{Z}_2$ nontrivial phase by staggered stacking of Chern insulators. The red (blue) disks represent a Chern insulator with $C=+1$ $(C=-1)$. (c) Breaking the $\tilde{T}$-symmetry (the half translation in $z$ followed by the time-reversal) leads to a `higher-order' state with a 1D equatorial chiral mode on the surface \cite{SitteRoschAltmanFritz, ZhangKaneMele,Teo,  Benalcazar17,  Hashimoto2017, Song17, Schindler17,  Benalcazar17Aug, Fang17, Langbehn17}.
}
\label{fig:AFTI}
\end{center}
\end{figure*}

\section{Indicator for Antiferromagnetic Topological Insulator}
As the last application of our results, let us make a connection of symmetry-based indicators of MSGs, fully computed in this work, to some previously studied topological insulators in TR broken settings.

As a canonical example, we discuss the type IV MSG 2.7 generated by the inversion symmetry $I$ and the half translation followed by TR, $\tilde{\mathcal{T}}\equiv\mathcal{T}T_{\frac{1}{2}\vec{a}_z}$.  This MSG has two classes of the combination of inversion parities as indicated by $X_{\rm BS}=\mathbb{Z}_2$ in Table~\ref{tab:ex1}.  The topological index distinguishing the two classes is given by
\begin{equation}
\label{eq:topoinv2}
\xi=(-1)^{\frac{1}{4}\sum_{n: \text{occupied}}\sum_{\text{TRIMs}}\,\xi_{n, \vec{k}}}=\pm1
\end{equation}
Here, $\xi_{n, \vec{k}}$ is the parity of the $n$-th occupied band.  To understand this formula, note first that even- and odd-parity bands form a Kramers pair at the TRIMs with $k_z=\pi$. This is because $I$ and $\tilde{\mathcal{T}}$ anti-commute when $k_z=\pi$. Thus, the four TRIMs with $k_z=\pi$ never contribute to the index. On the other hand, when $k_z=0$, $I$ and $\tilde{\mathcal{T}}$ commute, and $\tilde{\mathcal{T}}^2=-1$. This means that the same parity eigenvalue appears twice at TRIMs with $k_z=0$. In Fig.~\ref{fig:AFTI}a, we illustrated an example of the parity combination with $\xi=-1$.  

As demonstrated in Ref.~\cite{MongEssinMoore}, the index $\xi=-1$ implies a nontrivial band topology, the so-called `antiferromagnetic topological insulator' that supports gapless surface Dirac modes. The anti-ferromagnetic topological insulator can be understood as a staggered stacking of Chern insulators, in which a Chern insulator with $C=+1$ (described by red disks in Fig.~\ref{fig:AFTI}b) is on every $z=n$ ($n\in\mathbb{Z}$) plane and the one with $C=-1$ (blue disks) is on $z=n+\frac{1}{2}$ ($n\in\mathbb{Z}$) plane. All of these Chern insulators are related by the $\tilde{ \mathcal T}$ symmetry.

If we introduce a perturbation that breaks the $\tilde{\mathcal T}$ symmetry, the $C=+1$ Chern insulator at $z=n$ ($n\neq0$) and the $C=-1$  Chern insulator at $z=n\mp\text{sign}(n)\frac{1}{2}$ can pair annihilate with each other in an inversion symmetric manner, leaving only the $C=+1$ Chern insulator on the $z=0$ plane as shown in Fig.~\ref{fig:AFTI}c. Such one-dimensional chiral edge state on the surface of 3D topological insulators has been studied in Refs.\ \onlinecite{ZhangKaneMele,SitteRoschAltmanFritz}, and their counterparts in many other symmetry settings have been discovered recently \cite{Teo,  Benalcazar17,  Hashimoto2017, Song17, Schindler17,  Benalcazar17Aug, Fang17, Langbehn17}.
Assuming that the inversion symmetry is unbroken, we see that this phase can be diagnosed by another symmetry-based indicator for the type-I MSG 2.4, which is simply the SG with only the inversion $I$ apart from the lattice translation. According to the Table~\ref{tab:ex1}, this MSG has $\mathbb{Z}_2\times\mathbb{Z}_2\times\mathbb{Z}_2\times\mathbb{Z}_4$ classification, and the insulator hosting the 1D chiral edge mode constructed here belongs to the class $(0,0,0,2)$ in this classification.

\section{Conclusion}
In this work, we revisit the old problem of assigning a global irrep label to each connected branch of a band structure~\cite{Wigner}, generalizing our recent results on time-reversal symmetric systems \cite{NC} to all 1651 magnetic space groups. 
Our central results are the computation of three fundamental quantities associated with magnetic band structures, which are tabulated in Appendix \ref{app:tables}: (i) $d_{\rm BS}$, which characterizes the number of independent building blocks of energy bands; (ii) $X_{\rm BS}$, which, akin to the Fu-Kane parity criterion \cite{FuKane}, serves as a symmetry-based indicator of band topology; and (iii) $\nu_{\rm BS}$, which dictates the electron fillings at which band insulators are possible. We further demonstrate the utility of these results by applying them to the study of topological semimetals, focusing on cases where the absence of a band gap is either diagnosed through the symmetry representations at isolated high-symmetry points, or is mandated by the electron filling. In particular, we identify the exhaustive list of magnetic space groups capable of hosting filling-enforced semimetals (Appendix \ref{app:feSM}).

Although a full database on our computation of the basis `vectors' of the groups $\{{\rm BS}\}$ and $\{{\rm AI}\}$ is not included, we note that such a database can be readily generated \cite{NC}: Our proof of $d_{\rm BS} = d_{\rm AI}$ for all magnetic space groups implies full knowledge on $\{{\rm BS}\}$ is encoded in that of $\{{\rm AI} \}$, and the latter can be computed by analyzing, for instance, hypothetical structures constructed by combining irreps of the site-symmetry groups to each of the Wyckoff positions. As long as all high-symmetry momenta are included in the analysis and the doubling of any irrep (according to the Herring rule) is incorporated, $\{{\rm BS}\}$ can be recovered by forming linear superpositions of the entries in $\{ {\rm AI} \}$ using rational coefficients, subjected only to the constraint that the resultant irrep multiplicities must all be integer-valued.
Such calculations can be performed without deriving the compatibility relations, and this circumvents the careful convention-fixing required in both tabulating and utilizing such results.  In view of this, we refrain from providing a database on them.

We close by highlighting several interesting open questions that we leave for future studies: (i) performing a comprehensive study on the identification and characterization of representation-enforced semimetals in all 2D and 3D magnetic space groups; (ii) developing a theory of quantized physical responses, or a proof of the absence thereof, for each nontrivial class in the exhaustive list of $X_{\rm BS}$ we computed; and (iii) discovering realistic topological materials based on systematic searches using the diagnostics our theory provides. Finally, we remark that some aspects of our filling criterion generalize to the interacting setting \cite{MSGInt}, similar to the corresponding relation between the results for time-reversal symmetric crystals \cite{PRL, PNAS}.

\begin{acknowledgments}
HW thank Shuichi Murakami and Chen Fang for useful discussion. AV and HCP were supported by NSF DMR-1411343. AV acknowledges support from a Simons Investigator Award. 
HW acknowledges support from JSPS KAKENHI Grant Number JP17K17678.
\end{acknowledgments}

\clearpage
\bibliography{references}

\clearpage
\appendix

\onecolumngrid

\section{Tables for spinful electrons
\label{app:tables}}
Here we include the following tables for spinful electrons: 
\begin{itemize}
\item Tables \ref{tab:spinful_Tri}--\ref{tab:SpinfulCubic}:  $d$, $X_{\rm BS}$, and $\nu_{\rm BS}$ for MSGs.
\item Table \ref{tab:spinful_MLG}:  $d$, $X_{\rm BS}$, and $\nu_{\rm BS}$ for MLGs.
\item Table \ref{tab:spinfulnuEx}: $\{\nu\}_{\rm BS}$ and $\{\nu\}_{\rm AI}$ for exceptional MSGs.
\end{itemize}
Instructions on how to read the tables can be found in Tables \ref{tab:ex1} and \ref{tab:ex2}.

We remark that the results on  $d$, $X_{\rm BS}$, and $\nu_{\rm BS}$ for types I and II MSGs, which account for 460 out of the 1651 entries, were already published in Ref.\ \onlinecite{NC}. For the readers' convenience, we reproduce them here alongside with the new results on the remaining 1191 type III and IV MSGs. Similarly, the data on exceptional filling patterns among the type II MSGs (i.e., for time-reversal symmetric systems) were reported in Refs.\ \onlinecite{SciAdv, PRL}, but are also reproduced below.

\begin{center}
\begin{table}[h]
\caption{Characterization of magnetic space groups (MSGs) in the triclinic family for spinful electrons.
\label{tab:spinful_Tri}}
\input{spinful_Tri}\\
\input{spinful_Tri_foot}
\end{table}
\end{center}

\begin{center}
\begin{table}
\caption{Characterization of magnetic space groups (MSGs) in the monoclinic family for spinful electrons.
\label{tab:}}
\input{spinful_Mono}\\
\input{spinful_Mono_foot}
\end{table}
\end{center}

\begin{center}
\begin{table}
\caption{Characterization of magnetic space groups (MSGs)  in the orthorhombic family for spinful electrons
\label{tab:}}
\input{spinful_Ortho}
\end{table}
\end{center}

\begin{center}
\begin{table*}
(Continued from the previous page)\\
\input{spinful_OrthoII}
\end{table*}
\end{center}

\begin{center}
\begin{table*}
(Continued from the previous page)\\
\input{spinful_OrthoIII}\\
\input{spinful_OrthoIII_foot}
\end{table*}
\end{center}

\begin{center}
\begin{table}
\caption{Characterization of magnetic space groups (MSGs)  in the tetragonal family for spinful electrons
\label{tab:}}
\input{spinful_Tetra}
\end{table}
\end{center}

\begin{center}
\begin{table*}
(Continued from the previous page)\\
\input{spinful_TetraII}
\end{table*}
\end{center}

\begin{center}
\begin{table*}
(Continued from the previous page)\\
\input{spinful_TetraIII}\\
\input{spinful_TetraIII_foot}
\end{table*}
\end{center}

\begin{center}
\begin{table}
\caption{Characterization of magnetic space groups (MSGs)  in the hexagonal family for spinful electrons
\label{tab:}}
\input{spinful_Hexa}
\end{table}
\end{center}

\begin{center}
\begin{table*}
(Continued from the previous page)\\
\input{spinful_HexaII}\\
\input{spinful_HexaII_foot}
\end{table*}
\end{center}

\begin{center}
\begin{table}
\caption{Characterization of magnetic space groups (MSGs)  in the cubic family for spinful electrons
\label{tab:SpinfulCubic}}
\input{spinful_Cubic}\\
\input{spinful_Cubic_foot}
\end{table}
\end{center}


\begin{center}
\begin{table}
\caption{Characterization of the projections of magnetic space groups (MSGs) which correspond to magnetic layer group, assuming spinful electrons.
\label{tab:spinful_MLG}}
\input{spinful_MLG}
\end{table}
\end{center}

\begin{center}
\begin{table*}
(Continued from the previous page)\\
\input{spinful_MLGII}
\end{table*}
\end{center}

\begin{center}
\begin{table*}
(Continued from the previous page)\\
\input{spinful_MLGIII}\\
\input{spinful_MLGIII_foot}
\end{table*}
\end{center}


\begin{center}
\begin{table}
\caption{
Magnetic space groups (MSGs) for which spinful electrons exhibit exceptional filling patterns
\label{tab:spinfulnuEx}}
\input{spinful_nuSkip}\\
\input{spinful_nuSkip_foot}
\end{table}
\end{center}

\clearpage

\twocolumngrid
\section{Filling-enforced semimetals in stoichiometric compounds
\label{app:feSM}}
While the electron filling $\nu$ of a physical system can, in principle, take any real value upon doping, in perfectly ordered stoichiometric crystals the realizable fillings are further constrained by crystalline symmetries. 
As the atoms in such a crystal form $\mathcal M$-symmetric lattices, the allowed fillings for an MSG $\mathcal M$ are determined by the multiplicities of the Wyckoff positions \cite{RuChen}. Mathematically, this is captured by demanding $\nu \in \nu_{\rm P} \,\mathbb N$, where $\nu_{\rm P}$ is the gcd of the multiplicities of the Wyckoff positions and $\mathbb N$ denotes the set of natural number.
Whenever $\nu_{\rm P} < \nu_{\rm BS}$, one can find systems with a filling $\nu$ that is incompatible with a band gap, and therefore $\mathcal M$ admits feSMs. We also remark that for the four  `Wyckoff-mismatched' SGs identified in Ref.\ \onlinecite{SciAdv}, as well as their magnetic descendants, the set of physically realizable filling does not take the simple form of $\nu_{\rm P} \mathbb N$. The exhaustive list of such Wyckoff-mismatched MSGs can be found in Table \ref{tab:spinlessnuEx}, since the set physically realizable fillings coincides with the $\{ \nu\}_{\rm AI}$ of \emph{spinless} fermions.

For TR symmetric systems described by the 230 type II MSGs, $\nu_{\rm P} < \nu_{\rm BS}$ holds generally, and the filling criterion for filling-enforced (semi-)metals simplifies into whether or not the filling is an odd multiple of $\nu_{\rm P}$. For a general MSG, however, $\nu_{\rm P} \leq \nu_{\rm BS}$, and to identify magnetic settings which can host filling-enforced (semi-)metals one must first isolate the list of MSGs where $\nu_{\rm P} \neq \nu_{\rm BS}$. This is easily achieved using our theory, and we found that 647 out of the 1651 MSGs can host filling-enforced (semi-)metals (this includes the 230 type II MSGs, i.e., there are are 417 intrinsically magnetic MSGs). For all these MSGs, $\nu_{\rm BS} = 2 \nu_{\rm P}$, and therefore gaplessness near the Fermi energy is enforced whenever the filling of the system is an odd multiple of $\nu_{\rm P}$.

Some of these MSGs, however, are still incompatible with feSMs due to `band-sticking.' For instance, whenever the MSG contains the combination of spatial inversion and TR as a symmetry, each band is doubly degenerate. For such MSGs, if $\nu_{\rm P}$ is also odd, the filling-enforced gaplessness always manifests in the form of a large Fermi-surface enclosing half of the Brillouin zone, resulting in a conventional metal. A similar situation arise for band-sticking along high-symmetry line. We refer to such systems as filling-enforced metals. We found that 226 of the identified MSGs fall into this category, and they are listed in Table \ref{tab:feM}.

This leaves behind 421 MSGs, 250 of which are intrinsically magnetic, which can potentially host feSMs. Among them, we further characterize their minimal fermiology into two types. If the irrep dimensions at any high-symmetry point is incompatible with a band gap at filling $\nu_{\rm P}$, the fermi surface of the feSM consists of pinned nodal points. These MSGs are tabulated in Table \ref{tab:feSM-p}. If all the high-symmetry momenta can be in principle gapped, then the semimetallic behavior must be enforced by a more intricate band connectivity, resulting in nodal features that are movable, but irremovable. (Such fermiology is exemplified by that of simple nonsymmorphic SGs, as discussed in, e.g., Ref.\ \onlinecite{Zak2001}.) These MSGs are tabulated in Table \ref{tab:feSM-m}. Note that these are not mutually exclusive characterization of feSMs --- certain MSGs in Table \ref{tab:feSM-p} can also display such movable nodal features.

Lastly, we also include results specific to TR-symmetric systems (type II MSGs) in Table \ref{tab:feSMSOC}, which describes the effect of spin-orbit coupling on the fermiology of the feSMs. 
\clearpage

\onecolumngrid
\begin{center}
\begin{table}[h]
\caption{Magnetic space groups (MSGs) which can host filling-enforced semimetals with nodal-point Fermi surfaces pinned at high-symmetry momenta.
\label{tab:feSM-p}}
\input{spinful_pfeSM}\\
\input{spinful_pfeSM_foot}
\end{table}
\end{center}

\begin{center}
\begin{table}
\caption{Magnetic space groups (MSGs) which can host filling-enforced semimetals with movable nodal Fermi surfaces.
\label{tab:feSM-m}}
\input{spinful_mfeSM}\\
\input{spinful_mfeSM_foot}
\end{table}
\end{center}

\begin{center}
\begin{table}
\caption{Magnetic space groups (MSGs) which can host filling-enforced metals.
\label{tab:feM}}
\input{spinful_feM}\\
\input{spinful_feM_foot}
\end{table}
\end{center}

\begin{center}
\begin{table}
\caption{Effect of turning on spin-orbit coupling on the fermiology of time-reversal symmetric filling-enforced (semi-)metals
\label{tab:feSMSOC}}
\input{SOC_feSM}\\
\input{SOC_feSM_foot}
\end{table}
\end{center}


\clearpage

\twocolumngrid

\section{Notations}
For a MSG $\mathcal{M}=\mathcal{G}+\mathcal{A}$, $g\in\mathcal{G}$ maps the position $\vec{x}$ and the momentum $\vec{k}$ to
\begin{eqnarray}
g(\vec{x})\equiv p_g\vec{x}+\vec{t}_g,\,\,\, g(\vec{k})\equiv p_g(\vec{k}),\,\,\, p_g\in O(3),
\end{eqnarray}
while $a\in\mathcal{A}$ acts on them as
\begin{eqnarray}
a(\vec{x})=p_{a}\vec{x}+\vec{t}_{a},\quad a(\vec{k})=-p_{a}\vec{k},\,\,\, p_a\in O(3).
\end{eqnarray}

The little group of $\vec{k}$ is defined by
\begin{eqnarray}
\mathcal{M}_{\vec{k}}=\{m\in\mathcal{M}\,|\,m(\vec{k})=\vec{k}\}=\mathcal{G}_{\vec{k}}+\mathcal{A}_{\vec{k}},
\end{eqnarray}
where $\mathcal{G}_{\vec{k}}$ is the unitary part and $\mathcal{A}_{\vec{k}}$ is the anti-unitary part:
\begin{eqnarray}
\mathcal{G}_{\vec{k}}&=&\{g\in\mathcal{G}\,|\,g(\vec{k})=\vec{k}+\vec{G}\},\\
\mathcal{A}_{\vec{k}}&=&\{a\in\mathcal{A}\,|\,a(\vec{k})=\vec{k}+\vec{G}\}
\end{eqnarray}
and $\vec{G}$ is a reciprocal lattice vector.  $\mathcal{G}_{\vec{k}}$ is a subgroup of $\mathcal{M}_{\vec{k}}$ but $\mathcal{A}_{\vec{k}}$ is not even a group. 

Similarly, the site symmetry group of $\vec{x}$ is defined by
\begin{eqnarray}
\mathcal{M}_{\vec{x}}=\{m\in\mathcal{G}\,|\,m(\vec{x})=\vec{x}\}=\mathcal{G}_{\vec{x}}+\mathcal{A}_{\vec{x}}. 
\label{eq:sitesym}
\end{eqnarray}
where $\mathcal{G}_{\vec{x}}$ is the unitary part and $\mathcal{A}_{\vec{x}}$ is the anti-unitary part:
\begin{eqnarray}
\mathcal{G}_{\vec{x}}&=&\{g\in\mathcal{G}\,|\,g(\vec{x})=\vec{x}\},\\
\mathcal{A}_{\vec{x}}&=&\{a\in\mathcal{A}\,|\,a(\vec{x})=\vec{x}\}.
\end{eqnarray}

Since we are interested in systems of spinful electrons with spin-orbit coupling, we assume a projective representation, i.e.,
\begin{eqnarray}
u(g)u(g')&=&z_{g,g'}u(gg'),\\
u(g)u(a')&=&z_{g,a'}u(ga'),\\
u(a)u(g')^{*}&=&z_{a,g'}u(ag'),\\
u(a)u(a')^{*}&=&z_{a,a'}u(aa').  
\end{eqnarray}
For example, $z_{\mathcal{T},\mathcal{T}}=-1$ for the time-reversal symmetry $\mathcal{T}$.  For spinless electrons, all $z$ should be set to be unity in the following discussions.

\section{Compatibility relations}
\label{app:compatibility}
Here we review the compatibility relations among the integers $\vec{n}=\{n_{\vec{k}}^\alpha\}$ to be consistent with $\mathcal{M}=\mathcal{G}+\mathcal{A}$.  

\subsection{Basics}
Suppose that $\{|\vec{k},i\rangle\}$ is a basis of an irrep $u_{\vec{k}}^\alpha$ of $\mathcal{G}_{\vec{k}}$, i.e.,
\begin{eqnarray}
\hat{h}|\vec{k},i\rangle=|\vec{k},j\rangle[u_{\vec{k}}^\alpha(h)]_{ji},\quad h\in\mathcal{G}_{\vec{k}}.
\label{eq:gki}
\end{eqnarray}
Here we show that $\{\hat{g}|\vec{k},i\rangle\}$ ($g\in\mathcal{G}$) forms a basis of the irrep 
\begin{eqnarray}
u_{g(\vec{k})}^{\alpha'}(h')\equiv \rho_{h',g} u_{\vec{k}}^\alpha(g^{-1}h'g),\quad h'\in \mathcal{G}_{g(\vec{k})}.
\end{eqnarray}
Here, $\rho_{h,g}$ is related to the projective factor $z_{g,g'}=\pm1$ as 
\begin{eqnarray}
\rho_{h,g}=\frac{z_{h,g}}{z_{g,g^{-1}hg}}=\pm1.
\end{eqnarray}
To see this, note that $h'\in \mathcal{G}_{g(\vec{k})}$ can be written as $h'=ghg^{-1}$ using $h\in\mathcal{G}_{\vec{k}}$. Then,
\begin{eqnarray}
&&\hat{h}'(\hat{g}|\vec{k},i\rangle)=z_{h',g}(\hat{h'g})|\vec{k},i\rangle\notag\\
&&=z_{h',g}(\hat{gh})|\vec{k},i\rangle=\frac{z_{h',g}}{z_{g,h}}\hat{g}(\hat{h}|\vec{k},i\rangle)\notag\\
&&=\rho_{h',g}\hat{g}(|\vec{k},j\rangle [u_{\vec{k}}^\alpha(h)]_{ji})\notag\\
&&=(\hat{g}|\vec{k},j\rangle) \rho_{h',g}[u_{\vec{k}}^\alpha(g^{-1}h'g)]_{ji}.
\end{eqnarray}

Similarly, $\{\hat{a}|\vec{k},i\rangle\}$ ($a\in\mathcal{A}$) forms a basis of the irrep 
\begin{eqnarray}
u_{a(\vec{k})}^{\alpha'}(h')\equiv \rho_{h',a}[u_{\vec{k}}^\alpha(a^{-1}h'a)]_{ji}^*,\quad h'\in \mathcal{G}_{g(\vec{k})}.
\label{eq:aki}
\end{eqnarray}
where
\begin{eqnarray}
\rho_{h,a}=\frac{z_{h,a}}{z_{a,a^{-1}ha}}=\pm1.
\end{eqnarray}
Again writing $h'\in \mathcal{G}_{a(\vec{k})}$ as $h'=aha^{-1}$ using $h\in\mathcal{G}_{\vec{k}}$,
\begin{eqnarray}
&&\hat{h}'(\hat{a}|\vec{k},i\rangle)=z_{h',a}(\hat{h'a})|\vec{k},i\rangle\notag\\
&&=z_{h',a}(\hat{ah})|\vec{k},i\rangle=\frac{z_{h',a}}{z_{a,h}}\hat{a}(\hat{h}|\vec{k},i\rangle)\notag\\
&&=\rho_{h',a}\hat{a}(|\vec{k},j\rangle [u_{\vec{k}}^\alpha(h)]_{ji})\notag\\
&&=(\hat{a}|\vec{k},j\rangle) \rho_{h',a}[u_{\vec{k}}^\alpha(a^{-1}h'a)]_{ji}^*.
\end{eqnarray}

\subsection{Compatibility relations from the unitary part $\mathcal{C}_{\mathcal{G}}$}
Here we summarize the compatibility relations arising from the unitary part $\mathcal{G}$ of $\mathcal{M}$.  There are three major categories: 

(i) The representations used at symmetry-related momenta are related to each other.  Suppose that an irrep $u_{\vec{k}}^\alpha$ of $\mathcal{G}_{\vec{k}}$ is used $n_{\vec{k}}^\alpha$ times.  When $g\notin\mathcal{G}_{\vec{k}}$, $\vec{k}$ and $g(\vec{k})$ are symmetry-related but distinct momenta.  Let $\{|\vec{k},i\rangle\}$ be the basis of the representation.  As explained above, the $g$-transformed copy, $\{\hat{g}|\vec{k},i\rangle\}$ forms a basis of the irrep $u_{g(\vec{k})}^{\alpha'}$ in Eq.~\eqref{eq:gki}. It implies
\begin{equation}
n_{g(\vec{k})}^{\alpha'}=n_{\vec{k}}^\alpha.
\end{equation}

(ii) At a high-symmetry momenta slightly off from a higher-symmetry point $\vec{k}$, the continuity of the band structure requires that 
\begin{equation}
n_{\vec{k}+\delta(\vec{k})}^\beta=\sum_\alpha n_{\vec{k}}^\alpha m^{\alpha\beta},
\end{equation}
where nonnegative integers $m^{\alpha\beta}$ are the coefficients appearing in the decomposition $u_{\vec{k}}^\alpha=\sum_{\beta}c^{\alpha\beta}u_{\vec{k}+\delta(\vec{k})}^{\beta}$ regarding $u_{\vec{k}}^\alpha$ as a representation of $\mathcal{G}_{\vec{k}+\delta(\vec{k})}<\mathcal{G}_{\vec{k}}$. 

(iii) For nonsymmorphic operations $g$ with a `fractional translation', such as glide reflections or screw rotations, 
\begin{equation}
\sum_{\alpha}n_{\vec{k}}^{\alpha}\,\text{tr}\, u_{\vec{k}}^\alpha(g)=0,
\end{equation}
again due to the continuity of the band structure.

\subsection{Compatibility relations from the anti-unitary part  $\tilde{\mathcal{C}}_{\mathcal{A}}$}
The anti-unitary part $\mathcal{A}$ adds two more constraints. 

(iv) $\mathcal{A}$ may relate two momenta that were not related by $\mathcal{G}$.  This occurs when the anti-unitary part $\mathcal{A}_{\vec{k}}$ is empty.
Suppose that an irrep $u_{\vec{k}}^\alpha$ of $\mathcal{G}_{\vec{k}}$ is used $n_{\vec{k}}^\alpha$ times.  Let $\{|\vec{k},i\rangle\}$ be the basis of the representation.  The $a$-transformed copy, $\{\hat{a}|\vec{k},i\rangle\}$ forms a basis of the irrep $u_{a(\vec{k})}^{\alpha'}$ in Eq.~\eqref{eq:aki}. It requires a constraint
\begin{equation}
n_{a(\vec{k})}^{\alpha'}=n_{\vec{k}}^\alpha.
\end{equation}

(v) When the anti-unitary part $\mathcal{A}_{\vec{k}}$ is not empty, $u_{a(\vec{k})}^{\alpha'}$ in Eq.~\eqref{eq:aki} is also an irrep at $\vec{k}$ if $a\in\mathcal{A}_{\vec{k}}$.  The nature of paring due to $\mathcal{A}_{\vec{k}}$ can be judged by the Herring rule~\cite{Bradley}:  
\begin{eqnarray}
\xi_{\vec{k}}^\alpha=\frac{1}{|\mathcal{A}_{\vec{k}}/T|}\sum_{a\in\mathcal{A}_{\vec{k}}/T}\delta_{a(\vec{k}),\vec{k}}\,z_{a,a}\,\chi^\alpha_{\vec{k}}(a^2),
\end{eqnarray}
where $\chi_{\vec{k}}^{\alpha}(h)\equiv\text{tr}\,u_{\vec{k}}^{\alpha}(h)$ is the character of the irrep.  When $\xi_{\vec{k}}^\alpha=0$, $u_{\vec{k}}^{\alpha'}$ and $u_{\vec{k}}^{\alpha}$ are inequivalent and $n_{\vec{k}}^{\alpha'}=n_{\vec{k}}^{\alpha}$ ($\alpha'\neq\alpha$) is required.  When $\xi_{\vec{k}}^\alpha=-1$ or $+1$, $u_{\vec{k}}^{\alpha'}$ and $u_{\vec{k}}^{\alpha}$ are equivalent ($\alpha'=\alpha$).  When $\xi_{\vec{k}}^\alpha=-1$, $\{\hat{a}|\vec{k},i\rangle\}$ and $\{|\vec{k},i\rangle\}$ are orthogonal to each other and $n_{\vec{k}}^\alpha$ must be an even integer.  When $\xi_{\vec{k}}^\alpha=+1$, $\{\hat{a}|\vec{k},i\rangle\}$ and $\{|\vec{k},i\rangle\}$ are linearly dependent and there is no extra constraint on $n_{\vec{k}}^{\alpha}$.  

\section{Details of Atomic insulators}
\label{app:AI}
As explained in the main text, an AI can be specified by choosing the position $\vec{x}$ in real space at which electrons are localized and the type of orbital on the site. The type of orbital is a (co-)representation of the site symmetry group $\mathcal{M}_{\vec{x}}$ in Eq.~\eqref{eq:sitesym}.

Let $u_{\vec{x}}^r$ ($r=1,2,\ldots$) be an irrep of $\mathcal{G}_{\vec{x}}$.  When $\mathcal{A}_{\vec{x}}$ is empty, $u_{\vec{x}}^r$ is already an irrep of $\mathcal{M}_{\vec{x}}$.  We put an orbital $u_{\vec{x}}^r$ on the site $\vec{x}$.  One can construct an AI by placing a symmetry-related copy on each site of an $\mathcal{M}$-symmetric lattice $\mathcal{M}\vec{x}\equiv\{m\vec{x}\,|\,m\in\mathcal{M}\}$.  In this case the lattice $\mathcal{M}\vec{x}$ contains twice as many number of site as $\mathcal{G}\vec{x}$.

On the other hand, when $\mathcal{A}_{\vec{x}}$ is not empty, $\mathcal{A}_{\vec{x}}$ dictates a certain pairing among $u_{\vec{x}}^r$'s, leading to a co-irrep of $\mathcal{M}_{\vec{x}}$. The details of paring can be again judged by the Herring rule~\cite{Bradley}:
\begin{eqnarray}
\xi_{\vec{x}}^r=\frac{1}{|\mathcal{A}_{\vec{x}}|}\sum_{a\in\mathcal{A}_{\vec{x}}}\,z_{a,a}\,\chi^r_{\vec{x}}(a^2).
\end{eqnarray}
When $\xi_{\vec{x}}^r=0$, $u_{\vec{x}}^{r'}$ and $u_{\vec{k}}^{r}$ are inequivalent ($r'\neq r$) and must be paired to form a co-irrep of $\mathcal{M}_{\vec{x}}$.
When $\xi_{\vec{x}}^r=-1$ or $+1$, $u_{\vec{x}}^{r'}$ and $u_{\vec{x}}^{r}$ are equivalent ($r'= r$). When $\xi_{\vec{x}}^r=-1$, $\{\hat{a}|\vec{x},i\rangle\}$ and $\{|\vec{x},i\rangle\}$ are orthogonal to each other, and $u_{\vec{x}}^{r}$ is `paired with itself'.  Finally, when $\xi_{\vec{x}}^r=+1$, $\{\hat{a}|\vec{x},i\rangle\}$ and $\{|\vec{x},i\rangle\}$ are linearly dependent and $u_{\vec{x}}^{r}$ alone stands a co-irrep of $\mathcal{M}_{\vec{x}}$.  We put a co-irrep on the site $\vec{x}$ and generate AI by placing symmetry-related copy of the orbital on each site of the lattice $\mathcal{M}\vec{x}=\mathcal{G}\vec{x}$.

\section{The details of $\tilde{\mathcal{T}}b$}
\label{app:themapf}

Let us write $\vec{b}=\{n_{\vec{k}}^\alpha\}$ and $\tilde{\mathcal{T}}\vec{b}=\{\tilde{n}_{\vec{k}}^\alpha\}$.  By definition, $\vec{b}$ has the representation $\oplus_{\alpha}n_{\vec{k}}^\alpha u_{\vec{k}}^{\alpha}$ of $\mathcal{G}_{\vec{k}}$ at $\vec{k}$. Then $a=\tilde{\mathcal{T}}$ maps it to the representation $\oplus_{\alpha}n_{\vec{k}}^\alpha u_{a(\vec{k})}^{\alpha'}$ of $\mathcal{G}_{a(\vec{k})}$ at $a(\vec{k})$ [see Eq.~\eqref{eq:aki}].  By definition of $\tilde{\mathcal{T}}\vec{b}$, this coincides with $\oplus_{\alpha}\tilde{n}_{a(\vec{k})}^\alpha u_{a(\vec{k})}^{\alpha}$, i.e.,
\begin{eqnarray}
\oplus_{\alpha}\tilde{n}_{a(\vec{k})}^\alpha u_{a(\vec{k})}^{\alpha}(h)=\oplus_{\alpha}n_{\vec{k}}^\alpha u_{a(\vec{k})}^{\alpha'}(h).
\end{eqnarray}
Solving for $\tilde{n}_{\vec{k}}^\alpha$, we get
\begin{eqnarray}
\tilde{n}_{\vec{k}}^\beta = \frac{1}{|\mathcal{G}_{\vec{k}}/T|}\sum_{h\in\mathcal{G}_{\vec{k}}/T}\chi_{\vec{k}}^\beta(h)^*\sum_{\alpha} n_{a^{-1}(\vec{k})}^{\alpha}\,\chi_{\vec{k}}^{\alpha'}(h).
\end{eqnarray}

\section{Magnetic layer groups}
\label{app:MLG}
For each MLG, there is a corresponding MSG that can be achieved by stacking of 2D layers. Conversely, if an MSG $\mathcal{M}$ is such that every element of $m\in\mathcal{M}$ maps, say, $z$ to either $n\pm z$ ($n\in\mathbb{Z}$) (i.e., no mixing of $z$ with $x$ or $y$ or no fractional translation along $z$), the MSG can be reduced to an MLG by projecting down to $xy$ plane.  (In the actual calculation, it suffices to set $k_z=0$).  The projection down to $yz$ and $zx$ planes, if allowed, may also lead to an MSG.  These three possible ways of projections lead to sometimes distinct and sometimes the identical MLGs.  We present the computation results in in Appendix~\ref{app:tables} in a redundant way --- for each `parent' MSG, we list the computed numbers for all possible directions of projections among the three, regardless of their dependence/independence as MLGs.  For example, for MSG 3.4, two projections down to $xy$ and $xz$ planes are possible and correspondingly we have two rows 3.4 (1) and 3.4 (2).  For this MSG, the two projections correspond to two different MLGs.  On the other hand, for the type II MSG 2.5, the three ways of projection lead to an identical MLG.  Even in this case the redundant presentation has an advantage: e.g. it tells us that the three $\mathbb{Z}_2$ factor of $X_{\rm BS}=(\mathbb{Z}_2)^3\times\mathbb{Z}_4$ for MSG 2.5 can actually be attributed to the thee orthogonal 2D planes, implying the presence of `weak' phases.

\section{The tight-binding model for representation-enforced semimetal}
\label{app:tbmodel}
Here we present a tight-binding model that realizes the 2D reSM described by the main text.
We assume the MLG $p_a112$.  There are two sites in a unit cell. The A-sublattice is at $\vec{x}=n_1\vec{a}_1+n_2\vec{a}_2$ ($n_1,n_2\in\mathbb{Z}$) and the B-sublattice is at $\vec{x}=(n_1+\frac{1}{2})\vec{a}_1+n_2\vec{a}_2$ ($n,m\in\mathbb{Z}$).  For simplicity let us set $\vec{a}_1=(2,0)$, $\vec{a}_2=(-1,1)$.  In the following, four by four matrices are in the basis $(|A,\uparrow\rangle,|A,\downarrow\rangle,|B,\uparrow\rangle,|B,\downarrow\rangle)$.  We assume the standard symmetry representation:
\begin{eqnarray}
U_{\vec{k}}(C_{2z})&=&i
\begin{pmatrix}
1&0&0&0\\
0&-1&0&0\\
0&0&e^{-i k_1}&0\\
0&0&0&-e^{-i k_1}
\end{pmatrix},\\
U_{\vec{k}}(\mathcal{T}T_{\frac{1}{2}\vec{a}_1})&=&
\begin{pmatrix}
0&0&0&-e^{i k_1}\\
0&0&e^{i k_1}&0\\
0&-1&0&0\\
1&0&0&0
\end{pmatrix}
\end{eqnarray}
satisfying
\begin{eqnarray}
&&U_{-\vec{k}}(C_{2z})U_{\vec{k}}(C_{2z})=-1,\\
&&U_{-\vec{k}}(\mathcal{T}T_{\frac{1}{2}\vec{a}_1})U_{\vec{k}}(\mathcal{T}T_{\frac{1}{2}\vec{a}_1})^{*}=-e^{-ik_1}.
\end{eqnarray}

An example of the tight-binding Hamiltonian in the momentum space reads
\begin{eqnarray}
h(\vec{k})&=&\sum_{\mu,\nu=0,4}g_{\mu,\nu}(\vec{k})\,\sigma_\mu\otimes\sigma_\nu,\\
g_{1,0}(\vec{k})&=&t[1+\cos k_1+\cos k_2+\cos (k_1+k_2)],\\
g_{2,0}(\vec{k})&=&t[\sin k_1-\sin k_2+\sin (k_1+k_2)]\\
g_{1,1}(\vec{k})&=&t_J[1-\cos k_1+\cos k_2-\cos (k_1+k_2)],\\
g_{2,1}(\vec{k})&=&-t_J[\sin k_1+\sin k_2+\sin (k_1+k_2)],\\
g_{3,3}(\vec{k})&=&J.
\end{eqnarray}
It is easy to check the symmetry of the Hamiltonian:
\begin{eqnarray}
&&U_{\vec{k}}(C_{2z})h(\vec{k})=h(-\vec{k})U_{\vec{k}}(C_{2z}),\\
&&U_{\vec{k}}(\mathcal{T}T_{\frac{1}{2}\vec{a}_1})h(\vec{k})^*=h(-\vec{k})U_{\vec{k}}(\mathcal{T}T_{\frac{1}{2}\vec{a}_1}).
\end{eqnarray}

When $|t_J|<\frac{1}{2}J$, $\eta=+1$ and the system is a band insulator.  When $|t_J|>\frac{1}{2}J$, $\eta=-1$ and the system realizes a Dirac semimetal.

The MLG corresponding to the type III MSG 13.69 ($P112'/a'$) can also host a Dirac semimetal through an almost identical mechanism.  The MLG is generated by $T_{\vec{a}_2}$, $I$ (the 3D inversion), and $\mathcal{T}G$ ($G\equiv T_{\frac{1}{2}\vec{a}_1}IC_{2z}$ is the glide symmetry). The corresponding topological invariant $\eta$ is defined by 
\begin{equation}
\eta=\prod_{n: \text{occupied}}\prod_{\vec{k}=(0,0),(0,\pi)}\,\eta_{n, \vec{k}},
\end{equation}
where $\eta_{n, \vec{k}}$ represents the inversion eigenvalue $\pm1$.  

\clearpage
\onecolumngrid

\section{Tables for spinless electrons}
Here we include the following tables for spinless electrons:
\begin{itemize}
\item Tables \ref{tab:spinless_Tri}--\ref{tab:spinless_Cubic}:  $d$, $X_{\rm BS}$, and $\nu_{\rm BS}$ for MSGs.
\item Table \ref{tab:spinlessnuEx}: $\{\nu\}_{\rm BS}$ and $\{\nu\}_{\rm AI}$ for exceptional MSGs.
\end{itemize}

Similar to the remarks in Appendix \ref{app:tables}, some previously published results are reproduced below for completeness.

\begin{center}
\begin{table}[h]
\caption{Characterization of magnetic space groups (MSGs) in the triclinic family for spinless electrons.
\label{tab:spinless_Tri}}
\input{spinless_Tri}\\
\input{spinless_Tri_foot}
\end{table}
\end{center}

\begin{center}
\begin{table}[h]
\caption{Characterization of magnetic space groups (MSGs) in the monoclinic family for spinless electrons.
\label{tab:}}
\input{spinless_Mono}\\
\input{spinless_Mono_foot}
\end{table}
\end{center}

\begin{center}
\begin{table}
\caption{Characterization of magnetic space groups (MSGs)  in the orthorhombic family for spinless electrons
\label{tab:}}
\input{spinless_Ortho}
\end{table}
\end{center}

\begin{center}
\begin{table*}
(Continued from the previous page)\\
\input{spinless_OrthoII}
\end{table*}
\end{center}

\begin{center}
\begin{table*}
(Continued from the previous page)\\
\input{spinless_OrthoIII}\\
\input{spinless_OrthoIII_foot}
\end{table*}
\end{center}

\begin{center}
\begin{table}
\caption{Characterization of magnetic space groups (MSGs)  in the tetragonal family for spinless electrons
\label{tab:}}
\input{spinless_Tetra}
\end{table}
\end{center}

\begin{center}
\begin{table*}
(Continued from the previous page)\\
\input{spinless_TetraII}
\end{table*}
\end{center}

\begin{center}
\begin{table*}
(Continued from the previous page)\\
\input{spinless_TetraIII}\\
\input{spinless_TetraIII_foot}
\end{table*}
\end{center}

\begin{center}
\begin{table}
\caption{Characterization of magnetic space groups (MSGs)  in the hexagonal family for spinless electrons
\label{tab:}}
\input{spinless_Hexa}
\end{table}
\end{center}

\begin{center}
\begin{table*}
(Continued from the previous page)\\
\input{spinless_HexaII}\\
\input{spinless_HexaII_foot}
\end{table*}
\end{center}

\begin{center}
\begin{table}
\caption{Characterization of magnetic space groups (MSGs)  in the cubic family for spinless electrons
\label{tab:spinless_Cubic}}
\input{spinless_Cubic}\\
\input{spinless_Cubic_foot}
\end{table}
\end{center}


\begin{center}
\begin{table}
\caption{
Magnetic space groups (MSGs) for which spinless electrons exhibit exceptional filling patterns
\label{tab:spinlessnuEx}}
\input{spinless_nuSkip}\\
\input{spinless_nuSkip_foot}
\end{table}
\end{center}

\end{document}

%% file: spinful_Tri.tex
\begin{tabular}{cc|ccc||cc|ccc||cc|ccc||cc|ccc} 
\hline \hline 
\multicolumn{2}{c|}{MSG} & $d$ & $X_{\rm BS}$ & $\nu_{\rm BS}$ & \multicolumn{2}{c|}{MSG} & $d$ & $X_{\rm BS}$ & $\nu_{\rm BS}$ & \multicolumn{2}{c|}{MSG} & $d$ & $X_{\rm BS}$ & $\nu_{\rm BS}$ & \multicolumn{2}{c|}{MSG} & $d$ & $X_{\rm BS}$ & $\nu_{\rm BS}$\\ 
\hline 
1.1 & I & $1$ & $(1)$ & $1$ & 1.3 & IV & $1$ & $(1)$ & $2$ & 2.5 & II & $9$ & $(2, 2, 2, 4)$ & $2$ & 2.7 & IV & $5$ & $(2)$ & $2$ \\ 
1.2 & II & $1$ & $(1)$ & $2$ & 2.4 & I & $9$ & $(2, 2, 2, 4)$ & $1$ & 2.6 & III & $1$ & $(1)$ & $2$ & ~ & ~ & ~ & ~ \\ 
\hline \hline 
\end{tabular}

%% file: spinful_Tri_foot.tex
\newlength{\tabLspinfulTri} 
\settowidth{\tabLspinfulTri}{\input{spinful_Tri}} 
\begin{minipage}{\tabLspinfulTri} 
\begin{flushleft} 
{\footnotesize $d$: Rank of the band structure group $\{{\rm BS}\}$\\ 
$X_{\rm BS}$: Symmetry-based indicators of band topology\\ 
$\nu_{\rm BS}$: Set of $\nu$ bands are symmetry-forbidden from being isolated by band gaps if $\nu \not \in \nu_{\rm BS}\, \mathbb Z$ }
\end{flushleft}\end{minipage}

%% file: spinful_Mono.tex
\begin{tabular}{cc|ccc||cc|ccc||cc|ccc||cc|ccc} 
\hline \hline 
\multicolumn{2}{c|}{MSG} & $d$ & $X_{\rm BS}$ & $\nu_{\rm BS}$ & \multicolumn{2}{c|}{MSG} & $d$ & $X_{\rm BS}$ & $\nu_{\rm BS}$ & \multicolumn{2}{c|}{MSG} & $d$ & $X_{\rm BS}$ & $\nu_{\rm BS}$ & \multicolumn{2}{c|}{MSG} & $d$ & $X_{\rm BS}$ & $\nu_{\rm BS}$\\ 
\hline 
3.1 & I & $5$ & $(2)$ & $1$ & 7.24 & I & $1$ & $(1)$ & $2$ & 10.47 & IV & $7$ & $(2, 2)$ & $2$ & 13.70 & IV & $4$ & $(2)$ & $4$ \\ 
3.2 & II & $1$ & $(1)$ & $2$ & 7.25 & II & $1$ & $(1)$ & $4$ & 10.48 & IV & $6$ & $(2)$ & $2$ & 13.71 & IV & $3$ & $(2)$ & $4$ \\ 
3.3 & III & $1$ & $(1)$ & $1$ & 7.26 & III & $1$ & $(1)$ & $2$ & 10.49 & IV & $8$ & $(2)$ & $2$ & 13.72 & IV & $5$ & $(2)$ & $2$ \\ 
3.4 & IV & $3$ & $(2)$ & $2$ & 7.27 & IV & $1$ & $(1)$ & $4$ & 11.50 & I & $6$ & $(2)$ & $2$ & 13.73 & IV & $3$ & $(2)$ & $4$ \\ 
3.5 & IV & $1$ & $(1)$ & $2$ & 7.28 & IV & $1$ & $(1)$ & $2$ & 11.51 & II & $5$ & $(2, 2, 4)$ & $4$ & 13.74 & IV & $4$ & $(2)$ & $4$ \\ 
3.6 & IV & $3$ & $(1)$ & $2$ & 7.29 & IV & $1$ & $(1)$ & $4$ & 11.52 & III & $2$ & $(1)$ & $2$ & 14.75 & I & $5$ & $(2)$ & $2$ \\ 
4.7 & I & $1$ & $(1)$ & $2$ & 7.30 & IV & $1$ & $(1)$ & $4$ & 11.53 & III & $1$ & $(1)$ & $2$ & 14.76 & II & $5$ & $(2, 4)$ & $4$ \\ 
4.8 & II & $1$ & $(1)$ & $4$ & 7.31 & IV & $1$ & $(1)$ & $2$ & 11.54 & III & $5$ & $(2, 2, 4)$ & $2$ & 14.77 & III & $1$ & $(1)$ & $4$ \\ 
4.9 & III & $1$ & $(1)$ & $2$ & 8.32 & I & $2$ & $(1)$ & $1$ & 11.55 & IV & $3$ & $(2)$ & $4$ & 14.78 & III & $1$ & $(1)$ & $4$ \\ 
4.10 & IV & $1$ & $(1)$ & $4$ & 8.33 & II & $1$ & $(1)$ & $2$ & 11.56 & IV & $5$ & $(2)$ & $2$ & 14.79 & III & $5$ & $(2, 4)$ & $2$ \\ 
4.11 & IV & $1$ & $(1)$ & $2$ & 8.34 & III & $1$ & $(1)$ & $1$ & 11.57 & IV & $3$ & $(2)$ & $4$ & 14.80 & IV & $3$ & $(2)$ & $4$ \\ 
4.12 & IV & $1$ & $(1)$ & $2$ & 8.35 & IV & $1$ & $(1)$ & $2$ & 12.58 & I & $10$ & $(2, 2)$ & $1$ & 14.81 & IV & $3$ & $(2)$ & $4$ \\ 
5.13 & I & $3$ & $(1)$ & $1$ & 8.36 & IV & $1$ & $(1)$ & $2$ & 12.59 & II & $7$ & $(2, 2, 4)$ & $2$ & 14.82 & IV & $3$ & $(2)$ & $4$ \\ 
5.14 & II & $1$ & $(1)$ & $2$ & 9.37 & I & $1$ & $(1)$ & $2$ & 12.60 & III & $1$ & $(1)$ & $2$ & 14.83 & IV & $5$ & $(2)$ & $2$ \\ 
5.15 & III & $1$ & $(1)$ & $1$ & 9.38 & II & $1$ & $(1)$ & $4$ & 12.61 & III & $1$ & $(1)$ & $2$ & 14.84 & IV & $3$ & $(2)$ & $4$ \\ 
5.16 & IV & $2$ & $(1)$ & $2$ & 9.39 & III & $1$ & $(1)$ & $2$ & 12.62 & III & $7$ & $(2, 2, 4)$ & $1$ & 15.85 & I & $6$ & $(2, 2)$ & $2$ \\ 
5.17 & IV & $1$ & $(1)$ & $2$ & 9.40 & IV & $1$ & $(1)$ & $2$ & 12.63 & IV & $5$ & $(2)$ & $2$ & 15.86 & II & $5$ & $(2, 4)$ & $4$ \\ 
6.18 & I & $3$ & $(1)$ & $1$ & 9.41 & IV & $1$ & $(1)$ & $4$ & 12.64 & IV & $5$ & $(2)$ & $2$ & 15.87 & III & $1$ & $(1)$ & $2$ \\ 
6.19 & II & $1$ & $(1)$ & $2$ & 10.42 & I & $15$ & $(2, 2, 2)$ & $1$ & 13.65 & I & $7$ & $(2, 2)$ & $2$ & 15.88 & III & $2$ & $(1)$ & $2$ \\ 
6.20 & III & $1$ & $(1)$ & $1$ & 10.43 & II & $9$ & $(2, 2, 2, 4)$ & $2$ & 13.66 & II & $5$ & $(2, 2, 4)$ & $4$ & 15.89 & III & $5$ & $(2, 4)$ & $2$ \\ 
6.21 & IV & $1$ & $(1)$ & $2$ & 10.44 & III & $1$ & $(1)$ & $2$ & 13.67 & III & $1$ & $(1)$ & $2$ & 15.90 & IV & $4$ & $(2)$ & $2$ \\ 
6.22 & IV & $2$ & $(1)$ & $2$ & 10.45 & III & $1$ & $(1)$ & $2$ & 13.68 & III & $3$ & $(1)$ & $2$ & 15.91 & IV & $3$ & $(2)$ & $4$ \\ 
6.23 & IV & $2$ & $(1)$ & $2$ & 10.46 & III & $9$ & $(2, 2, 2, 4)$ & $1$ & 13.69 & III & $5$ & $(2, 2, 4)$ & $2$ & ~ & ~ & ~ & ~ \\ 
\hline \hline 
\end{tabular}

%% file: spinful_Mono_foot.tex
\newlength{\tabLspinfulMono} 
\settowidth{\tabLspinfulMono}{\input{spinful_Mono}} 
\begin{minipage}{\tabLspinfulMono} 
\begin{flushleft} 
{\footnotesize $d$: Rank of the band structure group $\{{\rm BS}\}$\\ 
$X_{\rm BS}$: Symmetry-based indicators of band topology\\ 
$\nu_{\rm BS}$: Set of $\nu$ bands are symmetry-forbidden from being isolated by band gaps if $\nu \not \in \nu_{\rm BS}\, \mathbb Z$ }
\end{flushleft}\end{minipage}

%% file: spinful_Ortho.tex
\begin{tabular}{cc|ccc||cc|ccc||cc|ccc||cc|ccc} 
\hline \hline 
\multicolumn{2}{c|}{MSG} & $d$ & $X_{\rm BS}$ & $\nu_{\rm BS}$ & \multicolumn{2}{c|}{MSG} & $d$ & $X_{\rm BS}$ & $\nu_{\rm BS}$ & \multicolumn{2}{c|}{MSG} & $d$ & $X_{\rm BS}$ & $\nu_{\rm BS}$ & \multicolumn{2}{c|}{MSG} & $d$ & $X_{\rm BS}$ & $\nu_{\rm BS}$\\ 
\hline 
16.1 & I & $1$ & $(1)$ & $2$ & 23.51 & III & $3$ & $(1)$ & $1$ & 29.101 & III & $1$ & $(1)$ & $4$ & 33.151 & IV & $1$ & $(1)$ & $4$ \\ 
16.2 & II & $1$ & $(1)$ & $2$ & 23.52 & IV & $1$ & $(1)$ & $2$ & 29.102 & III & $1$ & $(1)$ & $4$ & 33.152 & IV & $1$ & $(1)$ & $4$ \\ 
16.3 & III & $5$ & $(2)$ & $1$ & 24.53 & I & $4$ & $(1)$ & $2$ & 29.103 & III & $1$ & $(1)$ & $4$ & 33.153 & IV & $1$ & $(1)$ & $4$ \\ 
16.4 & IV & $1$ & $(1)$ & $2$ & 24.54 & II & $1$ & $(1)$ & $4$ & 29.104 & IV & $1$ & $(1)$ & $4$ & 33.154 & IV & $1$ & $(1)$ & $4$ \\ 
16.5 & IV & $1$ & $(1)$ & $4$ & 24.55 & III & $2$ & $(1)$ & $2$ & 29.105 & IV & $1$ & $(1)$ & $8$ & 33.155 & IV & $1$ & $(1)$ & $4$ \\ 
16.6 & IV & $1$ & $(1)$ & $4$ & 24.56 & IV & $3$ & $(1)$ & $2$ & 29.106 & IV & $1$ & $(1)$ & $4$ & 34.156 & I & $3$ & $(1)$ & $2$ \\ 
17.7 & I & $5$ & $(1)$ & $2$ & 25.57 & I & $1$ & $(1)$ & $2$ & 29.107 & IV & $1$ & $(1)$ & $4$ & 34.157 & II & $1$ & $(1)$ & $4$ \\ 
17.8 & II & $1$ & $(1)$ & $4$ & 25.58 & II & $1$ & $(1)$ & $2$ & 29.108 & IV & $1$ & $(1)$ & $4$ & 34.158 & III & $1$ & $(1)$ & $2$ \\ 
17.9 & III & $1$ & $(1)$ & $2$ & 25.59 & III & $3$ & $(1)$ & $1$ & 29.109 & IV & $1$ & $(1)$ & $4$ & 34.159 & III & $3$ & $(1)$ & $2$ \\ 
17.10 & III & $3$ & $(2)$ & $2$ & 25.60 & III & $5$ & $(2)$ & $1$ & 29.110 & IV & $1$ & $(1)$ & $4$ & 34.160 & IV & $2$ & $(1)$ & $4$ \\ 
17.11 & IV & $2$ & $(1)$ & $4$ & 25.61 & IV & $1$ & $(1)$ & $4$ & 30.111 & I & $3$ & $(1)$ & $2$ & 34.161 & IV & $1$ & $(1)$ & $4$ \\ 
17.12 & IV & $5$ & $(1)$ & $2$ & 25.62 & IV & $1$ & $(1)$ & $2$ & 30.112 & II & $1$ & $(1)$ & $4$ & 34.162 & IV & $2$ & $(1)$ & $4$ \\ 
17.13 & IV & $4$ & $(1)$ & $2$ & 25.63 & IV & $1$ & $(1)$ & $4$ & 30.113 & III & $1$ & $(1)$ & $2$ & 34.163 & IV & $3$ & $(2)$ & $4$ \\ 
17.14 & IV & $3$ & $(1)$ & $4$ & 25.64 & IV & $1$ & $(1)$ & $4$ & 30.114 & III & $1$ & $(1)$ & $2$ & 34.164 & IV & $3$ & $(1)$ & $2$ \\ 
17.15 & IV & $3$ & $(1)$ & $4$ & 25.65 & IV & $1$ & $(1)$ & $4$ & 30.115 & III & $3$ & $(1)$ & $2$ & 35.165 & I & $2$ & $(1)$ & $2$ \\ 
18.16 & I & $3$ & $(1)$ & $2$ & 26.66 & I & $1$ & $(1)$ & $2$ & 30.116 & IV & $2$ & $(1)$ & $4$ & 35.166 & II & $1$ & $(1)$ & $2$ \\ 
18.17 & II & $1$ & $(1)$ & $4$ & 26.67 & II & $1$ & $(1)$ & $4$ & 30.117 & IV & $3$ & $(2)$ & $4$ & 35.167 & III & $2$ & $(1)$ & $1$ \\ 
18.18 & III & $3$ & $(2)$ & $2$ & 26.68 & III & $1$ & $(1)$ & $2$ & 30.118 & IV & $1$ & $(1)$ & $4$ & 35.168 & III & $4$ & $(2)$ & $1$ \\ 
18.19 & III & $1$ & $(1)$ & $2$ & 26.69 & III & $3$ & $(1)$ & $2$ & 30.119 & IV & $3$ & $(1)$ & $2$ & 35.169 & IV & $1$ & $(1)$ & $4$ \\ 
18.20 & IV & $2$ & $(1)$ & $4$ & 26.70 & III & $1$ & $(1)$ & $2$ & 30.120 & IV & $2$ & $(1)$ & $4$ & 35.170 & IV & $2$ & $(1)$ & $2$ \\ 
18.21 & IV & $1$ & $(1)$ & $4$ & 26.71 & IV & $1$ & $(1)$ & $4$ & 30.121 & IV & $2$ & $(1)$ & $4$ & 35.171 & IV & $2$ & $(1)$ & $4$ \\ 
18.22 & IV & $2$ & $(1)$ & $4$ & 26.72 & IV & $1$ & $(1)$ & $4$ & 30.122 & IV & $2$ & $(1)$ & $4$ & 36.172 & I & $1$ & $(1)$ & $2$ \\ 
18.23 & IV & $3$ & $(1)$ & $2$ & 26.73 & IV & $1$ & $(1)$ & $2$ & 31.123 & I & $2$ & $(1)$ & $2$ & 36.173 & II & $1$ & $(1)$ & $4$ \\ 
18.24 & IV & $3$ & $(1)$ & $2$ & 26.74 & IV & $1$ & $(1)$ & $2$ & 31.124 & II & $1$ & $(1)$ & $4$ & 36.174 & III & $1$ & $(1)$ & $2$ \\ 
19.25 & I & $1$ & $(1)$ & $4$ & 26.75 & IV & $1$ & $(1)$ & $4$ & 31.125 & III & $1$ & $(1)$ & $2$ & 36.175 & III & $2$ & $(1)$ & $2$ \\ 
19.26 & II & $1$ & $(1)$ & $8$ & 26.76 & IV & $1$ & $(1)$ & $4$ & 31.126 & III & $2$ & $(1)$ & $2$ & 36.176 & III & $1$ & $(1)$ & $2$ \\ 
19.27 & III & $1$ & $(1)$ & $4$ & 26.77 & IV & $1$ & $(1)$ & $4$ & 31.127 & III & $1$ & $(1)$ & $2$ & 36.177 & IV & $1$ & $(1)$ & $2$ \\ 
19.28 & IV & $1$ & $(1)$ & $4$ & 27.78 & I & $1$ & $(1)$ & $2$ & 31.128 & IV & $2$ & $(1)$ & $4$ & 36.178 & IV & $1$ & $(1)$ & $4$ \\ 
19.29 & IV & $1$ & $(1)$ & $4$ & 27.79 & II & $1$ & $(1)$ & $4$ & 31.129 & IV & $1$ & $(1)$ & $4$ & 36.179 & IV & $1$ & $(1)$ & $2$ \\ 
19.30 & IV & $1$ & $(1)$ & $4$ & 27.80 & III & $1$ & $(1)$ & $2$ & 31.130 & IV & $1$ & $(1)$ & $4$ & 37.180 & I & $2$ & $(1)$ & $2$ \\ 
20.31 & I & $3$ & $(1)$ & $2$ & 27.81 & III & $5$ & $(2)$ & $2$ & 31.131 & IV & $1$ & $(1)$ & $4$ & 37.181 & II & $1$ & $(1)$ & $4$ \\ 
20.32 & II & $1$ & $(1)$ & $4$ & 27.82 & IV & $1$ & $(1)$ & $2$ & 31.132 & IV & $2$ & $(1)$ & $2$ & 37.182 & III & $1$ & $(1)$ & $2$ \\ 
20.33 & III & $1$ & $(1)$ & $2$ & 27.83 & IV & $1$ & $(1)$ & $4$ & 31.133 & IV & $2$ & $(1)$ & $4$ & 37.183 & III & $4$ & $(2)$ & $2$ \\ 
20.34 & III & $2$ & $(1)$ & $2$ & 27.84 & IV & $1$ & $(1)$ & $4$ & 31.134 & IV & $2$ & $(1)$ & $2$ & 37.184 & IV & $1$ & $(1)$ & $2$ \\ 
20.35 & IV & $3$ & $(1)$ & $2$ & 27.85 & IV & $1$ & $(1)$ & $4$ & 32.135 & I & $3$ & $(1)$ & $2$ & 37.185 & IV & $2$ & $(2)$ & $4$ \\ 
20.36 & IV & $1$ & $(1)$ & $4$ & 27.86 & IV & $1$ & $(1)$ & $4$ & 32.136 & II & $1$ & $(1)$ & $4$ & 37.186 & IV & $2$ & $(1)$ & $2$ \\ 
20.37 & IV & $3$ & $(1)$ & $2$ & 28.87 & I & $4$ & $(1)$ & $2$ & 32.137 & III & $1$ & $(1)$ & $2$ & 38.187 & I & $1$ & $(1)$ & $2$ \\ 
21.38 & I & $2$ & $(1)$ & $2$ & 28.88 & II & $1$ & $(1)$ & $4$ & 32.138 & III & $3$ & $(2)$ & $2$ & 38.188 & II & $1$ & $(1)$ & $2$ \\ 
21.39 & II & $1$ & $(1)$ & $2$ & 28.89 & III & $1$ & $(1)$ & $2$ & 32.139 & IV & $1$ & $(1)$ & $4$ & 38.189 & III & $2$ & $(1)$ & $1$ \\ 
21.40 & III & $4$ & $(2)$ & $1$ & 28.90 & III & $2$ & $(1)$ & $2$ & 32.140 & IV & $2$ & $(1)$ & $4$ & 38.190 & III & $3$ & $(1)$ & $1$ \\ 
21.41 & III & $3$ & $(1)$ & $1$ & 28.91 & III & $3$ & $(2)$ & $2$ & 32.141 & IV & $3$ & $(1)$ & $2$ & 38.191 & III & $3$ & $(1)$ & $1$ \\ 
21.42 & IV & $1$ & $(1)$ & $2$ & 28.92 & IV & $4$ & $(1)$ & $2$ & 32.142 & IV & $2$ & $(1)$ & $4$ & 38.192 & IV & $1$ & $(1)$ & $2$ \\ 
21.43 & IV & $2$ & $(1)$ & $2$ & 28.93 & IV & $2$ & $(1)$ & $4$ & 32.143 & IV & $3$ & $(1)$ & $4$ & 38.193 & IV & $1$ & $(1)$ & $2$ \\ 
21.44 & IV & $2$ & $(1)$ & $2$ & 28.94 & IV & $1$ & $(1)$ & $4$ & 33.144 & I & $1$ & $(1)$ & $4$ & 38.194 & IV & $1$ & $(1)$ & $4$ \\ 
22.45 & I & $1$ & $(1)$ & $2$ & 28.95 & IV & $2$ & $(1)$ & $4$ & 33.145 & II & $1$ & $(1)$ & $8$ & 39.195 & I & $1$ & $(1)$ & $2$ \\ 
22.46 & II & $1$ & $(1)$ & $2$ & 28.96 & IV & $4$ & $(1)$ & $4$ & 33.146 & III & $1$ & $(1)$ & $4$ & 39.196 & II & $1$ & $(1)$ & $4$ \\ 
22.47 & III & $3$ & $(1)$ & $1$ & 28.97 & IV & $3$ & $(1)$ & $2$ & 33.147 & III & $1$ & $(1)$ & $4$ & 39.197 & III & $2$ & $(1)$ & $2$ \\ 
22.48 & IV & $1$ & $(1)$ & $2$ & 28.98 & IV & $3$ & $(1)$ & $4$ & 33.148 & III & $1$ & $(1)$ & $4$ & 39.198 & III & $1$ & $(1)$ & $2$ \\ 
23.49 & I & $1$ & $(1)$ & $2$ & 29.99 & I & $1$ & $(1)$ & $4$ & 33.149 & IV & $1$ & $(1)$ & $4$ & 39.199 & III & $3$ & $(2)$ & $2$ \\ 
23.50 & II & $1$ & $(1)$ & $2$ & 29.100 & II & $1$ & $(1)$ & $8$ & 33.150 & IV & $1$ & $(1)$ & $8$ & 39.200 & IV & $1$ & $(1)$ & $4$ \\ 
\hline \hline 
\end{tabular}

%% file: spinful_OrthoII.tex
\begin{tabular}{cc|ccc||cc|ccc||cc|ccc||cc|ccc} 
\hline \hline 
39.201 & IV & $1$ & $(1)$ & $2$ & 47.251 & III & $1$ & $(1)$ & $2$ & 51.301 & IV & $4$ & $(2)$ & $4$ & 54.351 & IV & $3$ & $(2)$ & $4$ \\ 
39.202 & IV & $1$ & $(1)$ & $2$ & 47.252 & III & $15$ & $(2, 2, 2)$ & $1$ & 51.302 & IV & $4$ & $(2, 2)$ & $4$ & 54.352 & IV & $3$ & $(2)$ & $8$ \\ 
40.203 & I & $3$ & $(1)$ & $2$ & 47.253 & III & $1$ & $(1)$ & $2$ & 51.303 & IV & $5$ & $(2)$ & $4$ & 55.353 & I & $7$ & $(2, 2)$ & $2$ \\ 
40.204 & II & $1$ & $(1)$ & $4$ & 47.254 & IV & $5$ & $(2, 2, 4)$ & $4$ & 51.304 & IV & $4$ & $(2)$ & $4$ & 55.354 & II & $5$ & $(2, 4)$ & $4$ \\ 
40.205 & III & $1$ & $(1)$ & $2$ & 47.255 & IV & $5$ & $(2, 4)$ & $4$ & 52.305 & I & $5$ & $(2)$ & $4$ & 55.355 & III & $1$ & $(1)$ & $4$ \\ 
40.206 & III & $2$ & $(1)$ & $2$ & 47.256 & IV & $5$ & $(4)$ & $4$ & 52.306 & II & $3$ & $(4)$ & $8$ & 55.356 & III & $1$ & $(1)$ & $4$ \\ 
40.207 & III & $2$ & $(1)$ & $2$ & 48.257 & I & $3$ & $(2)$ & $4$ & 52.307 & III & $2$ & $(1)$ & $4$ & 55.357 & III & $9$ & $(2, 2, 2)$ & $2$ \\ 
40.208 & IV & $3$ & $(1)$ & $2$ & 48.258 & II & $3$ & $(2, 4)$ & $4$ & 52.308 & III & $1$ & $(1)$ & $4$ & 55.358 & III & $5$ & $(2)$ & $2$ \\ 
40.209 & IV & $1$ & $(1)$ & $4$ & 48.259 & III & $3$ & $(1)$ & $2$ & 52.309 & III & $2$ & $(1)$ & $4$ & 55.359 & III & $1$ & $(1)$ & $4$ \\ 
40.210 & IV & $3$ & $(1)$ & $2$ & 48.260 & III & $5$ & $(2)$ & $2$ & 52.310 & III & $4$ & $(2)$ & $4$ & 55.360 & IV & $4$ & $(2, 2)$ & $4$ \\ 
41.211 & I & $2$ & $(1)$ & $2$ & 48.261 & III & $1$ & $(1)$ & $4$ & 52.311 & III & $4$ & $(2)$ & $4$ & 55.361 & IV & $3$ & $(2)$ & $4$ \\ 
41.212 & II & $1$ & $(1)$ & $4$ & 48.262 & IV & $2$ & $(2)$ & $4$ & 52.312 & III & $3$ & $(2)$ & $4$ & 55.362 & IV & $4$ & $(2)$ & $4$ \\ 
41.213 & III & $1$ & $(1)$ & $2$ & 48.263 & IV & $3$ & $(2)$ & $4$ & 52.313 & III & $3$ & $(1)$ & $4$ & 55.363 & IV & $7$ & $(2, 2)$ & $2$ \\ 
41.214 & III & $1$ & $(1)$ & $2$ & 48.264 & IV & $2$ & $(2)$ & $4$ & 52.314 & IV & $3$ & $(2)$ & $4$ & 55.364 & IV & $5$ & $(2)$ & $4$ \\ 
41.215 & III & $2$ & $(2)$ & $2$ & 49.265 & I & $6$ & $(2)$ & $2$ & 52.315 & IV & $4$ & $(2)$ & $4$ & 56.365 & I & $3$ & $(2)$ & $4$ \\ 
41.216 & IV & $2$ & $(1)$ & $4$ & 49.266 & II & $5$ & $(2, 2, 4)$ & $4$ & 52.316 & IV & $3$ & $(2)$ & $8$ & 56.366 & II & $3$ & $(4)$ & $8$ \\ 
41.217 & IV & $1$ & $(1)$ & $4$ & 49.267 & III & $3$ & $(1)$ & $2$ & 52.317 & IV & $4$ & $(2)$ & $4$ & 56.367 & III & $1$ & $(1)$ & $4$ \\ 
41.218 & IV & $2$ & $(1)$ & $2$ & 49.268 & III & $1$ & $(1)$ & $2$ & 52.318 & IV & $3$ & $(2)$ & $4$ & 56.368 & III & $1$ & $(1)$ & $4$ \\ 
42.219 & I & $1$ & $(1)$ & $2$ & 49.269 & III & $10$ & $(2, 2)$ & $2$ & 52.319 & IV & $4$ & $(2)$ & $4$ & 56.369 & III & $5$ & $(2, 2)$ & $4$ \\ 
42.220 & II & $1$ & $(1)$ & $2$ & 49.270 & III & $7$ & $(2, 2)$ & $2$ & 52.320 & IV & $4$ & $(2)$ & $4$ & 56.370 & III & $3$ & $(2)$ & $4$ \\ 
42.221 & III & $2$ & $(1)$ & $1$ & 49.271 & III & $1$ & $(1)$ & $4$ & 53.321 & I & $9$ & $(2)$ & $2$ & 56.371 & III & $3$ & $(1)$ & $4$ \\ 
42.222 & III & $3$ & $(2)$ & $1$ & 49.272 & IV & $3$ & $(2)$ & $4$ & 53.322 & II & $5$ & $(2, 4)$ & $4$ & 56.372 & IV & $2$ & $(2)$ & $8$ \\ 
42.223 & IV & $1$ & $(1)$ & $2$ & 49.273 & IV & $6$ & $(2)$ & $2$ & 53.323 & III & $1$ & $(1)$ & $4$ & 56.373 & IV & $2$ & $(2)$ & $4$ \\ 
43.224 & I & $2$ & $(1)$ & $2$ & 49.274 & IV & $4$ & $(2)$ & $4$ & 53.324 & III & $2$ & $(1)$ & $4$ & 56.374 & IV & $3$ & $(2)$ & $4$ \\ 
43.225 & II & $1$ & $(1)$ & $4$ & 49.275 & IV & $3$ & $(2)$ & $4$ & 53.325 & III & $1$ & $(1)$ & $4$ & 56.375 & IV & $3$ & $(2)$ & $4$ \\ 
43.226 & III & $1$ & $(1)$ & $2$ & 49.276 & IV & $4$ & $(2)$ & $4$ & 53.326 & III & $5$ & $(2)$ & $2$ & 56.376 & IV & $2$ & $(2)$ & $4$ \\ 
43.227 & III & $2$ & $(1)$ & $2$ & 50.277 & I & $3$ & $(2)$ & $4$ & 53.327 & III & $8$ & $(2, 2)$ & $2$ & 57.377 & I & $4$ & $(2)$ & $4$ \\ 
43.228 & IV & $1$ & $(1)$ & $4$ & 50.278 & II & $3$ & $(2, 4)$ & $4$ & 53.328 & III & $6$ & $(2, 2)$ & $2$ & 57.378 & II & $3$ & $(2, 4)$ & $8$ \\ 
44.229 & I & $1$ & $(1)$ & $2$ & 50.279 & III & $3$ & $(1)$ & $2$ & 53.329 & III & $2$ & $(1)$ & $4$ & 57.379 & III & $3$ & $(1)$ & $4$ \\ 
44.230 & II & $1$ & $(1)$ & $2$ & 50.280 & III & $3$ & $(1)$ & $2$ & 53.330 & IV & $5$ & $(2)$ & $4$ & 57.380 & III & $1$ & $(1)$ & $4$ \\ 
44.231 & III & $2$ & $(1)$ & $1$ & 50.281 & III & $5$ & $(2, 2)$ & $2$ & 53.331 & IV & $4$ & $(2)$ & $4$ & 57.381 & III & $1$ & $(1)$ & $4$ \\ 
44.232 & III & $3$ & $(1)$ & $1$ & 50.282 & III & $5$ & $(2)$ & $2$ & 53.332 & IV & $6$ & $(2)$ & $4$ & 57.382 & III & $4$ & $(2)$ & $4$ \\ 
44.233 & IV & $1$ & $(1)$ & $4$ & 50.283 & III & $1$ & $(1)$ & $4$ & 53.333 & IV & $5$ & $(2)$ & $4$ & 57.383 & III & $4$ & $(2, 2)$ & $4$ \\ 
44.234 & IV & $1$ & $(1)$ & $2$ & 50.284 & IV & $3$ & $(2)$ & $4$ & 53.334 & IV & $8$ & $(2)$ & $2$ & 57.384 & III & $3$ & $(2)$ & $4$ \\ 
45.235 & I & $1$ & $(1)$ & $2$ & 50.285 & IV & $2$ & $(2)$ & $4$ & 53.335 & IV & $6$ & $(2)$ & $4$ & 57.385 & III & $2$ & $(1)$ & $4$ \\ 
45.236 & II & $1$ & $(1)$ & $4$ & 50.286 & IV & $2$ & $(2)$ & $8$ & 53.336 & IV & $5$ & $(2)$ & $4$ & 57.386 & IV & $2$ & $(2)$ & $8$ \\ 
45.237 & III & $1$ & $(1)$ & $2$ & 50.287 & IV & $3$ & $(2)$ & $4$ & 54.337 & I & $4$ & $(2)$ & $4$ & 57.387 & IV & $3$ & $(2)$ & $4$ \\ 
45.238 & III & $3$ & $(2)$ & $2$ & 50.288 & IV & $2$ & $(2)$ & $8$ & 54.338 & II & $3$ & $(2, 4)$ & $8$ & 57.388 & IV & $4$ & $(2)$ & $4$ \\ 
45.239 & IV & $1$ & $(1)$ & $2$ & 51.289 & I & $7$ & $(2, 2)$ & $2$ & 54.339 & III & $1$ & $(1)$ & $4$ & 57.389 & IV & $4$ & $(2)$ & $4$ \\ 
45.240 & IV & $1$ & $(1)$ & $4$ & 51.290 & II & $5$ & $(2, 2, 4)$ & $4$ & 54.340 & III & $2$ & $(1)$ & $4$ & 57.390 & IV & $3$ & $(2)$ & $8$ \\ 
46.241 & I & $2$ & $(1)$ & $2$ & 51.291 & III & $1$ & $(1)$ & $2$ & 54.341 & III & $1$ & $(1)$ & $4$ & 57.391 & IV & $2$ & $(2)$ & $4$ \\ 
46.242 & II & $1$ & $(1)$ & $4$ & 51.292 & III & $2$ & $(1)$ & $2$ & 54.342 & III & $5$ & $(2, 2)$ & $4$ & 57.392 & IV & $3$ & $(2)$ & $4$ \\ 
46.243 & III & $1$ & $(1)$ & $2$ & 51.293 & III & $1$ & $(1)$ & $4$ & 54.343 & III & $3$ & $(2)$ & $4$ & 58.393 & I & $8$ & $(2)$ & $2$ \\ 
46.244 & III & $2$ & $(1)$ & $2$ & 51.294 & III & $7$ & $(2, 2)$ & $2$ & 54.344 & III & $4$ & $(2, 2)$ & $4$ & 58.394 & II & $5$ & $(4)$ & $4$ \\ 
46.245 & III & $2$ & $(1)$ & $2$ & 51.295 & III & $6$ & $(2)$ & $2$ & 54.345 & III & $4$ & $(1)$ & $4$ & 58.395 & III & $1$ & $(1)$ & $4$ \\ 
46.246 & IV & $1$ & $(1)$ & $2$ & 51.296 & III & $9$ & $(2, 2, 2)$ & $2$ & 54.346 & IV & $4$ & $(2)$ & $4$ & 58.396 & III & $1$ & $(1)$ & $4$ \\ 
46.247 & IV & $2$ & $(1)$ & $2$ & 51.297 & III & $3$ & $(1)$ & $2$ & 54.347 & IV & $2$ & $(2)$ & $8$ & 58.397 & III & $8$ & $(2, 2)$ & $2$ \\ 
46.248 & IV & $2$ & $(1)$ & $4$ & 51.298 & IV & $7$ & $(2, 2)$ & $2$ & 54.348 & IV & $3$ & $(2)$ & $4$ & 58.398 & III & $5$ & $(2)$ & $2$ \\ 
47.249 & I & $9$ & $(2, 2, 2, 4)$ & $2$ & 51.299 & IV & $3$ & $(2)$ & $4$ & 54.349 & IV & $2$ & $(2)$ & $8$ & 58.399 & III & $1$ & $(1)$ & $4$ \\ 
47.250 & II & $9$ & $(2, 2, 2, 4)$ & $2$ & 51.300 & IV & $4$ & $(2, 2)$ & $4$ & 54.350 & IV & $4$ & $(2)$ & $4$ & 58.400 & IV & $4$ & $(2)$ & $4$ \\ 
\hline \hline 
\end{tabular}

%% file: spinful_OrthoIII.tex
\begin{tabular}{cc|ccc||cc|ccc||cc|ccc||cc|ccc} 
\hline \hline 
58.401 & IV & $4$ & $(2)$ & $4$ & 62.442 & II & $3$ & $(4)$ & $8$ & 65.483 & III & $1$ & $(1)$ & $2$ & 69.524 & III & $9$ & $(2, 2)$ & $1$ \\ 
58.402 & IV & $5$ & $(2)$ & $4$ & 62.443 & III & $1$ & $(1)$ & $4$ & 65.484 & III & $1$ & $(1)$ & $2$ & 69.525 & III & $1$ & $(1)$ & $2$ \\ 
58.403 & IV & $5$ & $(2)$ & $4$ & 62.444 & III & $1$ & $(1)$ & $4$ & 65.485 & III & $12$ & $(2, 2, 2)$ & $1$ & 69.526 & IV & $5$ & $(4)$ & $2$ \\ 
58.404 & IV & $8$ & $(2)$ & $2$ & 62.445 & III & $2$ & $(1)$ & $4$ & 65.486 & III & $10$ & $(2, 2)$ & $1$ & 70.527 & I & $3$ & $(2)$ & $4$ \\ 
59.405 & I & $3$ & $(2)$ & $4$ & 62.446 & III & $3$ & $(2)$ & $4$ & 65.487 & III & $1$ & $(1)$ & $2$ & 70.528 & II & $3$ & $(4)$ & $4$ \\ 
59.406 & II & $3$ & $(2, 4)$ & $4$ & 62.447 & III & $3$ & $(2)$ & $4$ & 65.488 & IV & $4$ & $(4)$ & $4$ & 70.529 & III & $2$ & $(1)$ & $2$ \\ 
59.407 & III & $2$ & $(1)$ & $2$ & 62.448 & III & $4$ & $(2)$ & $4$ & 65.489 & IV & $6$ & $(2, 4)$ & $2$ & 70.530 & III & $4$ & $(2)$ & $2$ \\ 
59.408 & III & $1$ & $(1)$ & $4$ & 62.449 & III & $1$ & $(1)$ & $4$ & 65.490 & IV & $5$ & $(4)$ & $4$ & 70.531 & III & $1$ & $(1)$ & $4$ \\ 
59.409 & III & $5$ & $(2, 2)$ & $2$ & 62.450 & IV & $2$ & $(2)$ & $4$ & 66.491 & I & $7$ & $(2)$ & $2$ & 70.532 & IV & $2$ & $(2)$ & $4$ \\ 
59.410 & III & $4$ & $(2)$ & $2$ & 62.451 & IV & $3$ & $(2)$ & $4$ & 66.492 & II & $5$ & $(2, 4)$ & $4$ & 71.533 & I & $6$ & $(4)$ & $2$ \\ 
59.411 & III & $3$ & $(1)$ & $2$ & 62.452 & IV & $2$ & $(2)$ & $8$ & 66.493 & III & $2$ & $(1)$ & $2$ & 71.534 & II & $6$ & $(2, 4)$ & $2$ \\ 
59.412 & IV & $3$ & $(2)$ & $4$ & 62.453 & IV & $2$ & $(2)$ & $4$ & 66.494 & III & $1$ & $(1)$ & $2$ & 71.535 & III & $1$ & $(1)$ & $2$ \\ 
59.413 & IV & $2$ & $(2)$ & $8$ & 62.454 & IV & $3$ & $(2)$ & $4$ & 66.495 & III & $9$ & $(2, 2)$ & $2$ & 71.536 & III & $9$ & $(2, 2)$ & $1$ \\ 
59.414 & IV & $2$ & $(2)$ & $8$ & 62.455 & IV & $2$ & $(2)$ & $8$ & 66.496 & III & $6$ & $(2, 2)$ & $2$ & 71.537 & III & $1$ & $(1)$ & $2$ \\ 
59.415 & IV & $3$ & $(2)$ & $4$ & 62.456 & IV & $3$ & $(2)$ & $4$ & 66.497 & III & $1$ & $(1)$ & $4$ & 71.538 & IV & $4$ & $(4)$ & $4$ \\ 
59.416 & IV & $2$ & $(2)$ & $4$ & 63.457 & I & $5$ & $(2)$ & $2$ & 66.498 & IV & $5$ & $(2)$ & $2$ & 72.539 & I & $4$ & $(2)$ & $2$ \\ 
60.417 & I & $4$ & $(2)$ & $4$ & 63.458 & II & $4$ & $(2, 4)$ & $4$ & 66.499 & IV & $4$ & $(2)$ & $4$ & 72.540 & II & $4$ & $(2, 4)$ & $4$ \\ 
60.418 & II & $3$ & $(4)$ & $8$ & 63.459 & III & $2$ & $(1)$ & $2$ & 66.500 & IV & $6$ & $(2)$ & $2$ & 72.541 & III & $2$ & $(1)$ & $2$ \\ 
60.419 & III & $1$ & $(1)$ & $4$ & 63.460 & III & $1$ & $(1)$ & $4$ & 67.501 & I & $5$ & $(2)$ & $2$ & 72.542 & III & $1$ & $(1)$ & $2$ \\ 
60.420 & III & $2$ & $(1)$ & $4$ & 63.461 & III & $1$ & $(1)$ & $2$ & 67.502 & II & $5$ & $(2, 2, 4)$ & $4$ & 72.543 & III & $7$ & $(2, 2)$ & $2$ \\ 
60.421 & III & $1$ & $(1)$ & $4$ & 63.462 & III & $5$ & $(2)$ & $2$ & 67.503 & III & $1$ & $(1)$ & $2$ & 72.544 & III & $5$ & $(2, 2)$ & $2$ \\ 
60.422 & III & $3$ & $(2)$ & $4$ & 63.463 & III & $6$ & $(2, 2)$ & $2$ & 67.504 & III & $2$ & $(1)$ & $2$ & 72.545 & III & $1$ & $(1)$ & $4$ \\ 
60.423 & III & $3$ & $(2)$ & $4$ & 63.464 & III & $5$ & $(2, 2)$ & $2$ & 67.505 & III & $7$ & $(2, 2)$ & $2$ & 72.546 & IV & $4$ & $(2)$ & $2$ \\ 
60.424 & III & $4$ & $(2, 2)$ & $4$ & 63.465 & III & $2$ & $(1)$ & $2$ & 67.506 & III & $8$ & $(2, 2)$ & $2$ & 72.547 & IV & $3$ & $(2)$ & $4$ \\ 
60.425 & III & $2$ & $(1)$ & $4$ & 63.466 & IV & $5$ & $(2)$ & $2$ & 67.507 & III & $2$ & $(1)$ & $2$ & 73.548 & I & $3$ & $(2)$ & $4$ \\ 
60.426 & IV & $3$ & $(2)$ & $8$ & 63.467 & IV & $3$ & $(2)$ & $4$ & 67.508 & IV & $3$ & $(2)$ & $4$ & 73.549 & II & $3$ & $(2, 4)$ & $8$ \\ 
60.427 & IV & $2$ & $(2)$ & $8$ & 63.468 & IV & $4$ & $(2)$ & $4$ & 67.509 & IV & $5$ & $(2)$ & $2$ & 73.550 & III & $1$ & $(1)$ & $4$ \\ 
60.428 & IV & $3$ & $(2)$ & $4$ & 64.469 & I & $5$ & $(2)$ & $2$ & 67.510 & IV & $3$ & $(2)$ & $4$ & 73.551 & III & $4$ & $(2, 2)$ & $4$ \\ 
60.429 & IV & $2$ & $(2)$ & $8$ & 64.470 & II & $4$ & $(2, 4)$ & $4$ & 68.511 & I & $3$ & $(2)$ & $4$ & 73.552 & III & $4$ & $(1)$ & $4$ \\ 
60.430 & IV & $3$ & $(2)$ & $4$ & 64.471 & III & $1$ & $(1)$ & $4$ & 68.512 & II & $3$ & $(2, 4)$ & $4$ & 73.553 & IV & $3$ & $(2)$ & $4$ \\ 
60.431 & IV & $4$ & $(2)$ & $4$ & 64.472 & III & $1$ & $(1)$ & $4$ & 68.513 & III & $2$ & $(1)$ & $2$ & 74.554 & I & $7$ & $(2)$ & $2$ \\ 
60.432 & IV & $4$ & $(2)$ & $4$ & 64.473 & III & $1$ & $(1)$ & $4$ & 68.514 & III & $2$ & $(1)$ & $2$ & 74.555 & II & $5$ & $(2, 4)$ & $4$ \\ 
61.433 & I & $3$ & $(2)$ & $4$ & 64.474 & III & $4$ & $(2)$ & $2$ & 68.515 & III & $5$ & $(2, 2)$ & $2$ & 74.556 & III & $1$ & $(1)$ & $2$ \\ 
61.434 & II & $3$ & $(4)$ & $8$ & 64.475 & III & $6$ & $(2, 2)$ & $2$ & 68.516 & III & $4$ & $(2, 2)$ & $2$ & 74.557 & III & $1$ & $(1)$ & $4$ \\ 
61.435 & III & $1$ & $(1)$ & $8$ & 64.476 & III & $5$ & $(2, 2)$ & $2$ & 68.517 & III & $2$ & $(1)$ & $4$ & 74.558 & III & $6$ & $(2, 2)$ & $2$ \\ 
61.436 & III & $3$ & $(2)$ & $4$ & 64.477 & III & $2$ & $(1)$ & $4$ & 68.518 & IV & $3$ & $(2)$ & $4$ & 74.559 & III & $7$ & $(2, 2)$ & $2$ \\ 
61.437 & III & $1$ & $(1)$ & $8$ & 64.478 & IV & $4$ & $(2)$ & $4$ & 68.519 & IV & $3$ & $(2)$ & $4$ & 74.560 & III & $2$ & $(1)$ & $2$ \\ 
61.438 & IV & $2$ & $(2)$ & $8$ & 64.479 & IV & $3$ & $(2)$ & $4$ & 68.520 & IV & $3$ & $(2)$ & $4$ & 74.561 & IV & $5$ & $(2)$ & $4$ \\ 
61.439 & IV & $3$ & $(2)$ & $4$ & 64.480 & IV & $5$ & $(2)$ & $2$ & 69.521 & I & $6$ & $(4)$ & $2$ & 74.562 & IV & $5$ & $(2)$ & $2$ \\ 
61.440 & IV & $2$ & $(2)$ & $8$ & 65.481 & I & $8$ & $(2, 4)$ & $2$ & 69.522 & II & $6$ & $(2, 2, 4)$ & $2$ & ~ & ~ & ~ & ~ \\ 
62.441 & I & $3$ & $(2)$ & $4$ & 65.482 & II & $7$ & $(2, 2, 4)$ & $2$ & 69.523 & III & $1$ & $(1)$ & $2$ & ~ & ~ & ~ & ~ \\ 
\hline \hline 
\end{tabular}

%% file: spinful_OrthoIII_foot.tex
\newlength{\tabLspinfulOrthoIII} 
\settowidth{\tabLspinfulOrthoIII}{\input{spinful_OrthoIII}} 
\begin{minipage}{\tabLspinfulOrthoIII} 
\begin{flushleft} 
{\footnotesize $d$: Rank of the band structure group $\{{\rm BS}\}$\\ 
$X_{\rm BS}$: Symmetry-based indicators of band topology\\ 
$\nu_{\rm BS}$: Set of $\nu$ bands are symmetry-forbidden from being isolated by band gaps if $\nu \not \in \nu_{\rm BS}\, \mathbb Z$ }
\end{flushleft}\end{minipage}

%% file: spinful_Tetra.tex
\begin{tabular}{cc|ccc||cc|ccc||cc|ccc||cc|ccc} 
\hline \hline 
\multicolumn{2}{c|}{MSG} & $d$ & $X_{\rm BS}$ & $\nu_{\rm BS}$ & \multicolumn{2}{c|}{MSG} & $d$ & $X_{\rm BS}$ & $\nu_{\rm BS}$ & \multicolumn{2}{c|}{MSG} & $d$ & $X_{\rm BS}$ & $\nu_{\rm BS}$ & \multicolumn{2}{c|}{MSG} & $d$ & $X_{\rm BS}$ & $\nu_{\rm BS}$\\ 
\hline 
75.1 & I & $8$ & $(4)$ & $1$ & 84.51 & I & $13$ & $(2, 4)$ & $2$ & 90.101 & IV & $4$ & $(1)$ & $2$ & 97.151 & I & $2$ & $(1)$ & $2$ \\ 
75.2 & II & $3$ & $(1)$ & $2$ & 84.52 & II & $7$ & $(2, 4)$ & $4$ & 90.102 & IV & $4$ & $(1)$ & $2$ & 97.152 & II & $2$ & $(1)$ & $2$ \\ 
75.3 & III & $2$ & $(1)$ & $2$ & 84.53 & III & $7$ & $(2)$ & $2$ & 91.103 & I & $4$ & $(1)$ & $4$ & 97.153 & III & $1$ & $(1)$ & $2$ \\ 
75.4 & IV & $3$ & $(1)$ & $2$ & 84.54 & III & $1$ & $(1)$ & $4$ & 91.104 & II & $1$ & $(1)$ & $8$ & 97.154 & III & $5$ & $(2)$ & $1$ \\ 
75.5 & IV & $4$ & $(2)$ & $2$ & 84.55 & III & $5$ & $(2)$ & $2$ & 91.105 & III & $3$ & $(1)$ & $4$ & 97.155 & III & $1$ & $(1)$ & $2$ \\ 
75.6 & IV & $4$ & $(1)$ & $2$ & 84.56 & IV & $6$ & $(4)$ & $4$ & 91.106 & III & $1$ & $(1)$ & $4$ & 97.156 & IV & $2$ & $(1)$ & $2$ \\ 
76.7 & I & $1$ & $(1)$ & $4$ & 84.57 & IV & $6$ & $(4)$ & $4$ & 91.107 & III & $2$ & $(1)$ & $4$ & 98.157 & I & $2$ & $(1)$ & $4$ \\ 
76.8 & II & $1$ & $(1)$ & $8$ & 84.58 & IV & $7$ & $(4)$ & $4$ & 91.108 & IV & $4$ & $(1)$ & $4$ & 98.158 & II & $1$ & $(1)$ & $4$ \\ 
76.9 & III & $1$ & $(1)$ & $4$ & 85.59 & I & $11$ & $(2, 4)$ & $2$ & 91.109 & IV & $2$ & $(1)$ & $8$ & 98.159 & III & $3$ & $(1)$ & $2$ \\ 
76.10 & IV & $1$ & $(1)$ & $4$ & 85.60 & II & $6$ & $(2, 4)$ & $4$ & 91.110 & IV & $3$ & $(1)$ & $4$ & 98.160 & III & $2$ & $(1)$ & $2$ \\ 
76.11 & IV & $1$ & $(1)$ & $8$ & 85.61 & III & $3$ & $(2)$ & $4$ & 92.111 & I & $2$ & $(1)$ & $4$ & 98.161 & III & $1$ & $(1)$ & $4$ \\ 
76.12 & IV & $1$ & $(1)$ & $4$ & 85.62 & III & $4$ & $(1)$ & $2$ & 92.112 & II & $1$ & $(1)$ & $8$ & 98.162 & IV & $2$ & $(1)$ & $4$ \\ 
77.13 & I & $4$ & $(2)$ & $2$ & 85.63 & III & $6$ & $(2)$ & $2$ & 92.113 & III & $1$ & $(1)$ & $4$ & 99.163 & I & $3$ & $(1)$ & $2$ \\ 
77.14 & II & $1$ & $(1)$ & $4$ & 85.64 & IV & $5$ & $(2)$ & $4$ & 92.114 & III & $1$ & $(1)$ & $4$ & 99.164 & II & $3$ & $(1)$ & $2$ \\ 
77.15 & III & $2$ & $(1)$ & $2$ & 85.65 & IV & $6$ & $(2)$ & $4$ & 92.115 & III & $2$ & $(1)$ & $4$ & 99.165 & III & $2$ & $(1)$ & $2$ \\ 
77.16 & IV & $1$ & $(1)$ & $4$ & 85.66 & IV & $5$ & $(2)$ & $4$ & 92.116 & IV & $2$ & $(1)$ & $4$ & 99.166 & III & $1$ & $(1)$ & $2$ \\ 
77.17 & IV & $2$ & $(2)$ & $4$ & 86.67 & I & $9$ & $(2, 2)$ & $2$ & 92.117 & IV & $1$ & $(1)$ & $8$ & 99.167 & III & $8$ & $(4)$ & $1$ \\ 
77.18 & IV & $2$ & $(1)$ & $4$ & 86.68 & II & $5$ & $(2, 4)$ & $4$ & 92.118 & IV & $2$ & $(1)$ & $4$ & 99.168 & IV & $3$ & $(1)$ & $4$ \\ 
78.19 & I & $1$ & $(1)$ & $4$ & 86.69 & III & $3$ & $(2)$ & $4$ & 93.119 & I & $1$ & $(1)$ & $4$ & 99.169 & IV & $2$ & $(1)$ & $4$ \\ 
78.20 & II & $1$ & $(1)$ & $8$ & 86.70 & III & $2$ & $(1)$ & $4$ & 93.120 & II & $1$ & $(1)$ & $4$ & 99.170 & IV & $2$ & $(1)$ & $4$ \\ 
78.21 & III & $1$ & $(1)$ & $4$ & 86.71 & III & $6$ & $(2)$ & $2$ & 93.121 & III & $1$ & $(1)$ & $2$ & 100.171 & I & $4$ & $(1)$ & $2$ \\ 
78.22 & IV & $1$ & $(1)$ & $4$ & 86.72 & IV & $4$ & $(2)$ & $4$ & 93.122 & III & $4$ & $(2)$ & $2$ & 100.172 & II & $2$ & $(1)$ & $4$ \\ 
78.23 & IV & $1$ & $(1)$ & $8$ & 86.73 & IV & $5$ & $(2)$ & $4$ & 93.123 & III & $2$ & $(1)$ & $2$ & 100.173 & III & $1$ & $(1)$ & $4$ \\ 
78.24 & IV & $1$ & $(1)$ & $4$ & 86.74 & IV & $4$ & $(2)$ & $4$ & 93.124 & IV & $1$ & $(1)$ & $4$ & 100.174 & III & $2$ & $(1)$ & $2$ \\ 
79.25 & I & $5$ & $(2)$ & $1$ & 87.75 & I & $16$ & $(4, 4)$ & $1$ & 93.125 & IV & $1$ & $(1)$ & $4$ & 100.175 & III & $5$ & $(4)$ & $2$ \\ 
79.26 & II & $2$ & $(1)$ & $2$ & 87.76 & II & $9$ & $(2, 8)$ & $2$ & 93.126 & IV & $1$ & $(1)$ & $4$ & 100.176 & IV & $2$ & $(1)$ & $4$ \\ 
79.27 & III & $1$ & $(1)$ & $2$ & 87.77 & III & $5$ & $(4)$ & $2$ & 94.127 & I & $2$ & $(1)$ & $4$ & 100.177 & IV & $4$ & $(1)$ & $2$ \\ 
79.28 & IV & $2$ & $(1)$ & $2$ & 87.78 & III & $2$ & $(1)$ & $2$ & 94.128 & II & $1$ & $(1)$ & $4$ & 100.178 & IV & $4$ & $(1)$ & $4$ \\ 
80.29 & I & $2$ & $(1)$ & $2$ & 87.79 & III & $5$ & $(2)$ & $2$ & 94.129 & III & $2$ & $(1)$ & $2$ & 101.179 & I & $1$ & $(1)$ & $4$ \\ 
80.30 & II & $1$ & $(1)$ & $4$ & 87.80 & IV & $8$ & $(4)$ & $2$ & 94.130 & III & $3$ & $(2)$ & $2$ & 101.180 & II & $1$ & $(1)$ & $4$ \\ 
80.31 & III & $2$ & $(1)$ & $2$ & 88.81 & I & $8$ & $(2, 2)$ & $2$ & 94.131 & III & $1$ & $(1)$ & $4$ & 101.181 & III & $2$ & $(1)$ & $4$ \\ 
80.32 & IV & $1$ & $(1)$ & $4$ & 88.82 & II & $5$ & $(4)$ & $4$ & 94.132 & IV & $1$ & $(1)$ & $4$ & 101.182 & III & $1$ & $(1)$ & $2$ \\ 
81.33 & I & $12$ & $(2, 2, 4)$ & $1$ & 88.83 & III & $3$ & $(2)$ & $4$ & 94.133 & IV & $2$ & $(1)$ & $4$ & 101.183 & III & $4$ & $(2)$ & $2$ \\ 
81.34 & II & $5$ & $(2)$ & $2$ & 88.84 & III & $1$ & $(1)$ & $4$ & 94.134 & IV & $2$ & $(1)$ & $4$ & 101.184 & IV & $1$ & $(1)$ & $4$ \\ 
81.35 & III & $2$ & $(1)$ & $2$ & 88.85 & III & $6$ & $(2)$ & $2$ & 95.135 & I & $4$ & $(1)$ & $4$ & 101.185 & IV & $1$ & $(1)$ & $4$ \\ 
81.36 & IV & $5$ & $(2)$ & $2$ & 88.86 & IV & $4$ & $(2)$ & $4$ & 95.136 & II & $1$ & $(1)$ & $8$ & 101.186 & IV & $1$ & $(1)$ & $8$ \\ 
81.37 & IV & $6$ & $(2, 2)$ & $2$ & 89.87 & I & $3$ & $(1)$ & $2$ & 95.137 & III & $3$ & $(1)$ & $4$ & 102.187 & I & $2$ & $(1)$ & $4$ \\ 
81.38 & IV & $6$ & $(2)$ & $2$ & 89.88 & II & $3$ & $(1)$ & $2$ & 95.138 & III & $1$ & $(1)$ & $4$ & 102.188 & II & $1$ & $(1)$ & $4$ \\ 
82.39 & I & $11$ & $(2, 2, 2)$ & $1$ & 89.89 & III & $1$ & $(1)$ & $2$ & 95.139 & III & $2$ & $(1)$ & $4$ & 102.189 & III & $1$ & $(1)$ & $4$ \\ 
82.40 & II & $5$ & $(2)$ & $2$ & 89.90 & III & $8$ & $(4)$ & $1$ & 95.140 & IV & $4$ & $(1)$ & $4$ & 102.190 & III & $2$ & $(1)$ & $2$ \\ 
82.41 & III & $1$ & $(1)$ & $2$ & 89.91 & III & $2$ & $(1)$ & $2$ & 95.141 & IV & $2$ & $(1)$ & $8$ & 102.191 & III & $3$ & $(2)$ & $2$ \\ 
82.42 & IV & $5$ & $(2)$ & $2$ & 89.92 & IV & $3$ & $(1)$ & $2$ & 95.142 & IV & $3$ & $(1)$ & $4$ & 102.192 & IV & $1$ & $(1)$ & $8$ \\ 
83.43 & I & $24$ & $(4, 4, 4)$ & $1$ & 89.93 & IV & $2$ & $(1)$ & $4$ & 96.143 & I & $2$ & $(1)$ & $4$ & 102.193 & IV & $2$ & $(1)$ & $4$ \\ 
83.44 & II & $13$ & $(2, 4, 8)$ & $2$ & 89.94 & IV & $2$ & $(1)$ & $4$ & 96.144 & II & $1$ & $(1)$ & $8$ & 102.194 & IV & $2$ & $(1)$ & $4$ \\ 
83.45 & III & $8$ & $(2, 4)$ & $2$ & 90.95 & I & $4$ & $(1)$ & $2$ & 96.145 & III & $1$ & $(1)$ & $4$ & 103.195 & I & $3$ & $(1)$ & $2$ \\ 
83.46 & III & $3$ & $(1)$ & $2$ & 90.96 & II & $2$ & $(1)$ & $4$ & 96.146 & III & $1$ & $(1)$ & $4$ & 103.196 & II & $3$ & $(1)$ & $4$ \\ 
83.47 & III & $5$ & $(2)$ & $2$ & 90.97 & III & $2$ & $(1)$ & $2$ & 96.147 & III & $2$ & $(1)$ & $4$ & 103.197 & III & $2$ & $(1)$ & $4$ \\ 
83.48 & IV & $11$ & $(4)$ & $2$ & 90.98 & III & $5$ & $(4)$ & $2$ & 96.148 & IV & $2$ & $(1)$ & $4$ & 103.198 & III & $1$ & $(1)$ & $4$ \\ 
83.49 & IV & $12$ & $(4, 4)$ & $2$ & 90.99 & III & $1$ & $(1)$ & $4$ & 96.149 & IV & $1$ & $(1)$ & $8$ & 103.199 & III & $8$ & $(4)$ & $2$ \\ 
83.50 & IV & $12$ & $(4)$ & $2$ & 90.100 & IV & $2$ & $(1)$ & $4$ & 96.150 & IV & $2$ & $(1)$ & $4$ & 103.200 & IV & $3$ & $(1)$ & $2$ \\ 
\hline \hline 
\end{tabular}

%% file: spinful_TetraII.tex
\begin{tabular}{cc|ccc||cc|ccc||cc|ccc||cc|ccc} 
\hline \hline 
103.201 & IV & $2$ & $(1)$ & $4$ & 111.251 & I & $5$ & $(2)$ & $2$ & 117.301 & III & $1$ & $(1)$ & $4$ & 124.351 & I & $11$ & $(4)$ & $2$ \\ 
103.202 & IV & $2$ & $(1)$ & $4$ & 111.252 & II & $5$ & $(2)$ & $2$ & 117.302 & III & $2$ & $(1)$ & $2$ & 124.352 & II & $8$ & $(2, 8)$ & $4$ \\ 
104.203 & I & $4$ & $(1)$ & $2$ & 111.253 & III & $2$ & $(1)$ & $2$ & 117.303 & III & $7$ & $(2, 2, 4)$ & $2$ & 124.353 & III & $3$ & $(1)$ & $2$ \\ 
104.204 & II & $2$ & $(1)$ & $4$ & 111.254 & III & $1$ & $(1)$ & $2$ & 117.304 & IV & $3$ & $(2)$ & $4$ & 124.354 & III & $5$ & $(4)$ & $4$ \\ 
104.205 & III & $1$ & $(1)$ & $4$ & 111.255 & III & $12$ & $(2, 2, 4)$ & $1$ & 117.305 & IV & $6$ & $(2)$ & $2$ & 124.355 & III & $4$ & $(4)$ & $4$ \\ 
104.206 & III & $2$ & $(1)$ & $4$ & 111.256 & IV & $3$ & $(2)$ & $4$ & 117.306 & IV & $4$ & $(2)$ & $4$ & 124.356 & III & $3$ & $(2)$ & $4$ \\ 
104.207 & III & $5$ & $(2)$ & $2$ & 111.257 & IV & $3$ & $(2)$ & $4$ & 118.307 & I & $6$ & $(2)$ & $2$ & 124.357 & III & $16$ & $(4, 4)$ & $2$ \\ 
104.208 & IV & $2$ & $(1)$ & $4$ & 111.258 & IV & $3$ & $(2)$ & $4$ & 118.308 & II & $3$ & $(2)$ & $4$ & 124.358 & III & $3$ & $(2)$ & $4$ \\ 
104.209 & IV & $4$ & $(2)$ & $4$ & 112.259 & I & $5$ & $(2)$ & $2$ & 118.309 & III & $1$ & $(1)$ & $4$ & 124.359 & III & $3$ & $(1)$ & $4$ \\ 
104.210 & IV & $4$ & $(1)$ & $2$ & 112.260 & II & $3$ & $(2)$ & $4$ & 118.310 & III & $2$ & $(1)$ & $2$ & 124.360 & IV & $11$ & $(4)$ & $2$ \\ 
105.211 & I & $1$ & $(1)$ & $4$ & 112.261 & III & $2$ & $(1)$ & $2$ & 118.311 & III & $7$ & $(2, 2, 2)$ & $2$ & 124.361 & IV & $6$ & $(4)$ & $4$ \\ 
105.212 & II & $1$ & $(1)$ & $4$ & 112.262 & III & $1$ & $(1)$ & $4$ & 118.312 & IV & $3$ & $(2)$ & $4$ & 124.362 & IV & $7$ & $(4)$ & $4$ \\ 
105.213 & III & $2$ & $(1)$ & $2$ & 112.263 & III & $8$ & $(2, 2, 2)$ & $2$ & 118.313 & IV & $4$ & $(2)$ & $4$ & 125.363 & I & $6$ & $(2)$ & $4$ \\ 
105.214 & III & $1$ & $(1)$ & $4$ & 112.264 & IV & $5$ & $(2)$ & $2$ & 118.314 & IV & $6$ & $(2)$ & $2$ & 125.364 & II & $6$ & $(2, 4)$ & $4$ \\ 
105.215 & III & $4$ & $(2)$ & $2$ & 112.265 & IV & $3$ & $(2)$ & $4$ & 119.315 & I & $5$ & $(2)$ & $2$ & 125.365 & III & $4$ & $(1)$ & $2$ \\ 
105.216 & IV & $1$ & $(1)$ & $4$ & 112.266 & IV & $3$ & $(2)$ & $4$ & 119.316 & II & $5$ & $(2)$ & $2$ & 125.366 & III & $3$ & $(2)$ & $4$ \\ 
105.217 & IV & $1$ & $(1)$ & $8$ & 113.267 & I & $6$ & $(2)$ & $2$ & 119.317 & III & $1$ & $(1)$ & $2$ & 125.367 & III & $3$ & $(2)$ & $4$ \\ 
105.218 & IV & $1$ & $(1)$ & $4$ & 113.268 & II & $3$ & $(2)$ & $4$ & 119.318 & III & $1$ & $(1)$ & $2$ & 125.368 & III & $3$ & $(2)$ & $4$ \\ 
106.219 & I & $2$ & $(1)$ & $4$ & 113.269 & III & $1$ & $(1)$ & $4$ & 119.319 & III & $11$ & $(2, 2, 2)$ & $1$ & 125.369 & III & $11$ & $(2, 4)$ & $2$ \\ 
106.220 & II & $1$ & $(1)$ & $8$ & 113.270 & III & $2$ & $(1)$ & $2$ & 119.320 & IV & $3$ & $(2)$ & $4$ & 125.370 & III & $6$ & $(2)$ & $2$ \\ 
106.221 & III & $1$ & $(1)$ & $4$ & 113.271 & III & $7$ & $(2, 2, 4)$ & $2$ & 120.321 & I & $5$ & $(2)$ & $2$ & 125.371 & III & $2$ & $(1)$ & $4$ \\ 
106.222 & III & $2$ & $(1)$ & $4$ & 113.272 & IV & $3$ & $(2)$ & $4$ & 120.322 & II & $3$ & $(2)$ & $4$ & 125.372 & IV & $4$ & $(2)$ & $4$ \\ 
106.223 & III & $3$ & $(2)$ & $4$ & 113.273 & IV & $6$ & $(2)$ & $2$ & 120.323 & III & $1$ & $(1)$ & $4$ & 125.373 & IV & $6$ & $(2)$ & $4$ \\ 
106.224 & IV & $1$ & $(1)$ & $4$ & 113.274 & IV & $4$ & $(2)$ & $4$ & 120.324 & III & $1$ & $(1)$ & $2$ & 125.374 & IV & $4$ & $(2)$ & $8$ \\ 
106.225 & IV & $2$ & $(1)$ & $4$ & 114.275 & I & $6$ & $(2)$ & $2$ & 120.325 & III & $7$ & $(2, 2, 2)$ & $2$ & 126.375 & I & $5$ & $(2)$ & $4$ \\ 
106.226 & IV & $2$ & $(1)$ & $4$ & 114.276 & II & $3$ & $(2)$ & $4$ & 120.326 & IV & $5$ & $(2)$ & $2$ & 126.376 & II & $4$ & $(4)$ & $4$ \\ 
107.227 & I & $2$ & $(1)$ & $2$ & 114.277 & III & $1$ & $(1)$ & $4$ & 121.327 & I & $5$ & $(2)$ & $2$ & 126.377 & III & $4$ & $(1)$ & $2$ \\ 
107.228 & II & $2$ & $(1)$ & $2$ & 114.278 & III & $2$ & $(1)$ & $4$ & 121.328 & II & $4$ & $(2)$ & $2$ & 126.378 & III & $2$ & $(2)$ & $4$ \\ 
107.229 & III & $1$ & $(1)$ & $2$ & 114.279 & III & $7$ & $(2, 2, 2)$ & $2$ & 121.329 & III & $1$ & $(1)$ & $2$ & 126.379 & III & $2$ & $(2)$ & $4$ \\ 
107.230 & III & $1$ & $(1)$ & $2$ & 114.280 & IV & $3$ & $(2)$ & $4$ & 121.330 & III & $1$ & $(1)$ & $2$ & 126.380 & III & $3$ & $(2)$ & $4$ \\ 
107.231 & III & $5$ & $(2)$ & $1$ & 114.281 & IV & $4$ & $(2)$ & $4$ & 121.331 & III & $9$ & $(2, 2, 2)$ & $1$ & 126.381 & III & $8$ & $(2, 2)$ & $2$ \\ 
107.232 & IV & $2$ & $(1)$ & $4$ & 114.282 & IV & $6$ & $(2)$ & $2$ & 121.332 & IV & $4$ & $(2)$ & $2$ & 126.382 & III & $4$ & $(2)$ & $4$ \\ 
108.233 & I & $2$ & $(1)$ & $2$ & 115.283 & I & $5$ & $(2)$ & $2$ & 122.333 & I & $7$ & $(2)$ & $2$ & 126.383 & III & $2$ & $(1)$ & $4$ \\ 
108.234 & II & $2$ & $(1)$ & $4$ & 115.284 & II & $5$ & $(2)$ & $2$ & 122.334 & II & $3$ & $(2)$ & $4$ & 126.384 & IV & $5$ & $(2)$ & $4$ \\ 
108.235 & III & $1$ & $(1)$ & $4$ & 115.285 & III & $2$ & $(1)$ & $2$ & 122.335 & III & $1$ & $(1)$ & $4$ & 126.385 & IV & $4$ & $(2)$ & $8$ \\ 
108.236 & III & $1$ & $(1)$ & $2$ & 115.286 & III & $1$ & $(1)$ & $2$ & 122.336 & III & $2$ & $(1)$ & $4$ & 126.386 & IV & $5$ & $(2)$ & $4$ \\ 
108.237 & III & $5$ & $(4)$ & $2$ & 115.287 & III & $12$ & $(2, 2, 4)$ & $1$ & 122.337 & III & $6$ & $(2, 2)$ & $2$ & 127.387 & I & $12$ & $(4, 4)$ & $2$ \\ 
108.238 & IV & $2$ & $(1)$ & $2$ & 115.288 & IV & $3$ & $(2)$ & $4$ & 122.338 & IV & $4$ & $(2)$ & $4$ & 127.388 & II & $8$ & $(4, 8)$ & $4$ \\ 
109.239 & I & $1$ & $(1)$ & $4$ & 115.289 & IV & $3$ & $(2)$ & $4$ & 123.339 & I & $13$ & $(2, 4, 8)$ & $2$ & 127.389 & III & $2$ & $(1)$ & $4$ \\ 
109.240 & II & $1$ & $(1)$ & $4$ & 115.290 & IV & $3$ & $(2)$ & $4$ & 123.340 & II & $13$ & $(2, 4, 8)$ & $2$ & 127.390 & III & $5$ & $(2, 4)$ & $4$ \\ 
109.241 & III & $2$ & $(1)$ & $2$ & 116.291 & I & $5$ & $(2)$ & $2$ & 123.341 & III & $3$ & $(1)$ & $2$ & 127.391 & III & $6$ & $(2, 4)$ & $2$ \\ 
109.242 & III & $1$ & $(1)$ & $4$ & 116.292 & II & $3$ & $(2)$ & $4$ & 123.342 & III & $8$ & $(2, 4)$ & $2$ & 127.392 & III & $3$ & $(2)$ & $4$ \\ 
109.243 & III & $2$ & $(1)$ & $2$ & 116.293 & III & $2$ & $(1)$ & $4$ & 123.343 & III & $7$ & $(2, 2, 4)$ & $2$ & 127.393 & III & $15$ & $(4, 4, 4)$ & $2$ \\ 
109.244 & IV & $1$ & $(1)$ & $8$ & 116.294 & III & $1$ & $(1)$ & $2$ & 123.344 & III & $5$ & $(2)$ & $2$ & 127.394 & III & $3$ & $(2)$ & $4$ \\ 
110.245 & I & $1$ & $(1)$ & $4$ & 116.295 & III & $8$ & $(2, 2, 2)$ & $2$ & 123.345 & III & $24$ & $(4, 4, 4)$ & $1$ & 127.395 & III & $2$ & $(1)$ & $4$ \\ 
110.246 & II & $1$ & $(1)$ & $8$ & 116.296 & IV & $5$ & $(2)$ & $2$ & 123.346 & III & $5$ & $(2)$ & $2$ & 127.396 & IV & $6$ & $(4)$ & $4$ \\ 
110.247 & III & $2$ & $(1)$ & $4$ & 116.297 & IV & $3$ & $(2)$ & $4$ & 123.347 & III & $3$ & $(1)$ & $2$ & 127.397 & IV & $12$ & $(4, 4)$ & $2$ \\ 
110.248 & III & $1$ & $(1)$ & $4$ & 116.298 & IV & $3$ & $(2)$ & $4$ & 123.348 & IV & $8$ & $(2, 8)$ & $4$ & 127.398 & IV & $8$ & $(4)$ & $4$ \\ 
110.249 & III & $2$ & $(2)$ & $4$ & 117.299 & I & $6$ & $(2)$ & $2$ & 123.349 & IV & $8$ & $(4, 8)$ & $4$ & 128.399 & I & $12$ & $(4)$ & $2$ \\ 
110.250 & IV & $1$ & $(1)$ & $4$ & 117.300 & II & $3$ & $(2)$ & $4$ & 123.350 & IV & $7$ & $(8)$ & $4$ & 128.400 & II & $7$ & $(8)$ & $4$ \\ 
\hline \hline 
\end{tabular}

%% file: spinful_TetraIII.tex
\begin{tabular}{cc|ccc||cc|ccc||cc|ccc||cc|ccc} 
\hline \hline 
128.401 & III & $2$ & $(1)$ & $4$ & 131.444 & IV & $6$ & $(2, 4)$ & $4$ & 135.487 & III & $4$ & $(4)$ & $4$ & 138.530 & IV & $3$ & $(2)$ & $8$ \\ 
128.402 & III & $4$ & $(4)$ & $4$ & 131.445 & IV & $4$ & $(4)$ & $8$ & 135.488 & III & $3$ & $(2)$ & $4$ & 139.531 & I & $9$ & $(8)$ & $2$ \\ 
128.403 & III & $5$ & $(4)$ & $4$ & 131.446 & IV & $5$ & $(4)$ & $4$ & 135.489 & III & $8$ & $(2, 4)$ & $4$ & 139.532 & II & $9$ & $(2, 8)$ & $2$ \\ 
128.404 & III & $3$ & $(2)$ & $4$ & 132.447 & I & $6$ & $(4)$ & $4$ & 135.490 & III & $3$ & $(2)$ & $4$ & 139.533 & III & $2$ & $(1)$ & $2$ \\ 
128.405 & III & $13$ & $(4, 4)$ & $2$ & 132.448 & II & $6$ & $(2, 4)$ & $4$ & 135.491 & III & $1$ & $(1)$ & $8$ & 139.534 & III & $5$ & $(4)$ & $2$ \\ 
128.406 & III & $3$ & $(2)$ & $4$ & 132.449 & III & $1$ & $(1)$ & $4$ & 135.492 & IV & $4$ & $(4)$ & $4$ & 139.535 & III & $5$ & $(4)$ & $2$ \\ 
128.407 & III & $2$ & $(1)$ & $4$ & 132.450 & III & $5$ & $(4)$ & $4$ & 135.493 & IV & $6$ & $(4)$ & $4$ & 139.536 & III & $5$ & $(2)$ & $2$ \\ 
128.408 & IV & $7$ & $(4)$ & $4$ & 132.451 & III & $5$ & $(2)$ & $2$ & 135.494 & IV & $6$ & $(4)$ & $4$ & 139.537 & III & $16$ & $(4, 4)$ & $1$ \\ 
128.409 & IV & $8$ & $(4)$ & $4$ & 132.452 & III & $3$ & $(2)$ & $4$ & 136.495 & I & $7$ & $(4)$ & $4$ & 139.538 & III & $4$ & $(2)$ & $2$ \\ 
128.410 & IV & $12$ & $(4)$ & $2$ & 132.453 & III & $12$ & $(2, 4)$ & $2$ & 136.496 & II & $5$ & $(4)$ & $4$ & 139.539 & III & $2$ & $(1)$ & $2$ \\ 
129.411 & I & $6$ & $(2)$ & $4$ & 132.454 & III & $5$ & $(2)$ & $2$ & 136.497 & III & $1$ & $(1)$ & $4$ & 139.540 & IV & $7$ & $(8)$ & $4$ \\ 
129.412 & II & $6$ & $(2, 4)$ & $4$ & 132.455 & III & $1$ & $(1)$ & $4$ & 136.498 & III & $4$ & $(4)$ & $4$ & 140.541 & I & $8$ & $(4)$ & $2$ \\ 
129.413 & III & $2$ & $(1)$ & $4$ & 132.456 & IV & $6$ & $(4)$ & $4$ & 136.499 & III & $6$ & $(2)$ & $2$ & 140.542 & II & $7$ & $(2, 8)$ & $4$ \\ 
129.414 & III & $3$ & $(2)$ & $4$ & 132.457 & IV & $4$ & $(4)$ & $4$ & 136.500 & III & $3$ & $(2)$ & $4$ & 140.543 & III & $2$ & $(1)$ & $2$ \\ 
129.415 & III & $3$ & $(2)$ & $4$ & 132.458 & IV & $4$ & $(4)$ & $8$ & 136.501 & III & $9$ & $(2, 4)$ & $2$ & 140.544 & III & $4$ & $(4)$ & $4$ \\ 
129.416 & III & $6$ & $(2)$ & $2$ & 133.459 & I & $4$ & $(2)$ & $4$ & 136.502 & III & $3$ & $(2)$ & $4$ & 140.545 & III & $4$ & $(4)$ & $2$ \\ 
129.417 & III & $11$ & $(2, 4)$ & $2$ & 133.460 & II & $3$ & $(4)$ & $8$ & 136.503 & III & $1$ & $(1)$ & $4$ & 140.546 & III & $3$ & $(2)$ & $4$ \\ 
129.418 & III & $3$ & $(2)$ & $4$ & 133.461 & III & $2$ & $(1)$ & $4$ & 136.504 & IV & $4$ & $(4)$ & $8$ & 140.547 & III & $13$ & $(4, 4)$ & $2$ \\ 
129.419 & III & $4$ & $(1)$ & $2$ & 133.462 & III & $2$ & $(2)$ & $4$ & 136.505 & IV & $6$ & $(4)$ & $4$ & 140.548 & III & $4$ & $(2)$ & $2$ \\ 
129.420 & IV & $4$ & $(2)$ & $8$ & 133.463 & III & $2$ & $(2)$ & $4$ & 136.506 & IV & $7$ & $(4)$ & $4$ & 140.549 & III & $2$ & $(1)$ & $4$ \\ 
129.421 & IV & $6$ & $(2)$ & $4$ & 133.464 & III & $3$ & $(2)$ & $4$ & 137.507 & I & $4$ & $(2)$ & $4$ & 140.550 & IV & $8$ & $(4)$ & $2$ \\ 
129.422 & IV & $4$ & $(2)$ & $4$ & 133.465 & III & $6$ & $(2, 2)$ & $4$ & 137.508 & II & $4$ & $(4)$ & $4$ & 141.551 & I & $6$ & $(2)$ & $4$ \\ 
130.423 & I & $5$ & $(2)$ & $4$ & 133.466 & III & $4$ & $(2)$ & $4$ & 137.509 & III & $1$ & $(1)$ & $4$ & 141.552 & II & $5$ & $(4)$ & $4$ \\ 
130.424 & II & $4$ & $(4)$ & $8$ & 133.467 & III & $1$ & $(1)$ & $8$ & 137.510 & III & $2$ & $(2)$ & $4$ & 141.553 & III & $1$ & $(1)$ & $4$ \\ 
130.425 & III & $2$ & $(1)$ & $4$ & 133.468 & IV & $4$ & $(2)$ & $4$ & 137.511 & III & $2$ & $(2)$ & $4$ & 141.554 & III & $3$ & $(2)$ & $4$ \\ 
130.426 & III & $2$ & $(2)$ & $8$ & 133.469 & IV & $4$ & $(2)$ & $4$ & 137.512 & III & $6$ & $(2)$ & $2$ & 141.555 & III & $4$ & $(2)$ & $4$ \\ 
130.427 & III & $2$ & $(2)$ & $4$ & 133.470 & IV & $3$ & $(2)$ & $8$ & 137.513 & III & $8$ & $(2, 2)$ & $2$ & 141.556 & III & $6$ & $(2)$ & $2$ \\ 
130.428 & III & $4$ & $(2)$ & $4$ & 134.471 & I & $5$ & $(2)$ & $4$ & 137.514 & III & $3$ & $(2)$ & $4$ & 141.557 & III & $8$ & $(2, 2)$ & $2$ \\ 
130.429 & III & $8$ & $(2, 4)$ & $4$ & 134.472 & II & $5$ & $(2, 4)$ & $4$ & 137.515 & III & $2$ & $(1)$ & $4$ & 141.558 & III & $3$ & $(2)$ & $4$ \\ 
130.430 & III & $3$ & $(2)$ & $4$ & 134.473 & III & $2$ & $(1)$ & $4$ & 137.516 & IV & $3$ & $(2)$ & $8$ & 141.559 & III & $1$ & $(1)$ & $4$ \\ 
130.431 & III & $4$ & $(1)$ & $4$ & 134.474 & III & $3$ & $(2)$ & $4$ & 137.517 & IV & $3$ & $(2)$ & $8$ & 141.560 & IV & $4$ & $(2)$ & $8$ \\ 
130.432 & IV & $5$ & $(2)$ & $4$ & 134.475 & III & $3$ & $(2)$ & $4$ & 137.518 & IV & $4$ & $(2)$ & $4$ & 142.561 & I & $4$ & $(2)$ & $4$ \\ 
130.433 & IV & $4$ & $(2)$ & $4$ & 134.476 & III & $3$ & $(2)$ & $4$ & 138.519 & I & $5$ & $(2)$ & $4$ & 142.562 & II & $3$ & $(4)$ & $8$ \\ 
130.434 & IV & $5$ & $(2)$ & $4$ & 134.477 & III & $9$ & $(2, 2)$ & $2$ & 138.520 & II & $4$ & $(4)$ & $8$ & 142.563 & III & $1$ & $(1)$ & $4$ \\ 
131.435 & I & $7$ & $(2, 4)$ & $4$ & 134.478 & III & $6$ & $(2)$ & $2$ & 138.521 & III & $1$ & $(1)$ & $4$ & 142.564 & III & $2$ & $(2)$ & $8$ \\ 
131.436 & II & $7$ & $(2, 4)$ & $4$ & 134.479 & III & $1$ & $(1)$ & $4$ & 138.522 & III & $3$ & $(2)$ & $4$ & 142.565 & III & $2$ & $(2)$ & $4$ \\ 
131.437 & III & $1$ & $(1)$ & $4$ & 134.480 & IV & $3$ & $(2)$ & $8$ & 138.523 & III & $3$ & $(2)$ & $4$ & 142.566 & III & $5$ & $(2)$ & $4$ \\ 
131.438 & III & $7$ & $(2)$ & $2$ & 134.481 & IV & $4$ & $(2)$ & $4$ & 138.524 & III & $4$ & $(2)$ & $4$ & 142.567 & III & $5$ & $(2, 2)$ & $4$ \\ 
131.439 & III & $5$ & $(2, 4)$ & $4$ & 134.482 & IV & $4$ & $(2)$ & $4$ & 138.525 & III & $7$ & $(2, 2)$ & $4$ & 142.568 & III & $3$ & $(2)$ & $4$ \\ 
131.440 & III & $5$ & $(2)$ & $2$ & 135.483 & I & $6$ & $(4)$ & $4$ & 138.526 & III & $3$ & $(2)$ & $4$ & 142.569 & III & $2$ & $(1)$ & $8$ \\ 
131.441 & III & $13$ & $(2, 4)$ & $2$ & 135.484 & II & $4$ & $(4)$ & $8$ & 138.527 & III & $2$ & $(1)$ & $4$ & 142.570 & IV & $4$ & $(2)$ & $4$ \\ 
131.442 & III & $3$ & $(2)$ & $4$ & 135.485 & III & $1$ & $(1)$ & $4$ & 138.528 & IV & $4$ & $(2)$ & $4$ & ~ & ~ & ~ & ~ \\ 
131.443 & III & $1$ & $(1)$ & $4$ & 135.486 & III & $4$ & $(2)$ & $4$ & 138.529 & IV & $5$ & $(2)$ & $4$ & ~ & ~ & ~ & ~ \\ 
\hline \hline 
\end{tabular}

%% file: spinful_TetraIII_foot.tex
\newlength{\tabLspinfulTetraIII} 
\settowidth{\tabLspinfulTetraIII}{\input{spinful_TetraIII}} 
\begin{minipage}{\tabLspinfulTetraIII} 
\begin{flushleft} 
{\footnotesize $d$: Rank of the band structure group $\{{\rm BS}\}$\\ 
$X_{\rm BS}$: Symmetry-based indicators of band topology\\ 
$\nu_{\rm BS}$: Set of $\nu$ bands are symmetry-forbidden from being isolated by band gaps if $\nu \not \in \nu_{\rm BS}\, \mathbb Z$ }
\end{flushleft}\end{minipage}

%% file: spinful_Hexa.tex
\begin{tabular}{cc|ccc||cc|ccc||cc|ccc||cc|ccc} 
\hline \hline 
\multicolumn{2}{c|}{MSG} & $d$ & $X_{\rm BS}$ & $\nu_{\rm BS}$ & \multicolumn{2}{c|}{MSG} & $d$ & $X_{\rm BS}$ & $\nu_{\rm BS}$ & \multicolumn{2}{c|}{MSG} & $d$ & $X_{\rm BS}$ & $\nu_{\rm BS}$ & \multicolumn{2}{c|}{MSG} & $d$ & $X_{\rm BS}$ & $\nu_{\rm BS}$\\ 
\hline 
143.1 & I & $7$ & $(3)$ & $1$ & 156.51 & III & $7$ & $(3)$ & $1$ & 166.101 & III & $11$ & $(2, 4)$ & $1$ & 177.151 & III & $5$ & $(1)$ & $1$ \\ 
143.2 & II & $4$ & $(1)$ & $2$ & 156.52 & IV & $4$ & $(1)$ & $2$ & 166.102 & IV & $6$ & $(2)$ & $2$ & 177.152 & III & $6$ & $(1)$ & $1$ \\ 
143.3 & IV & $4$ & $(1)$ & $2$ & 157.53 & I & $5$ & $(1)$ & $1$ & 167.103 & I & $7$ & $(2)$ & $2$ & 177.153 & III & $9$ & $(6)$ & $1$ \\ 
144.4 & I & $1$ & $(1)$ & $3$ & 157.54 & II & $3$ & $(1)$ & $2$ & 167.104 & II & $5$ & $(4)$ & $4$ & 177.154 & IV & $4$ & $(1)$ & $2$ \\ 
144.5 & II & $1$ & $(1)$ & $6$ & 157.55 & III & $5$ & $(3)$ & $1$ & 167.105 & III & $2$ & $(1)$ & $2$ & 178.155 & I & $3$ & $(1)$ & $6$ \\ 
144.6 & IV & $1$ & $(1)$ & $6$ & 157.56 & IV & $3$ & $(1)$ & $2$ & 167.106 & III & $3$ & $(1)$ & $2$ & 178.156 & II & $1$ & $(1)$ & $12$ \\ 
145.7 & I & $1$ & $(1)$ & $3$ & 158.57 & I & $4$ & $(1)$ & $2$ & 167.107 & III & $7$ & $(4)$ & $2$ & 178.157 & III & $2$ & $(1)$ & $6$ \\ 
145.8 & II & $1$ & $(1)$ & $6$ & 158.58 & II & $4$ & $(1)$ & $4$ & 167.108 & IV & $6$ & $(2)$ & $2$ & 178.158 & III & $2$ & $(1)$ & $6$ \\ 
145.9 & IV & $1$ & $(1)$ & $6$ & 158.59 & III & $7$ & $(3)$ & $2$ & 168.109 & I & $9$ & $(6)$ & $1$ & 178.159 & III & $1$ & $(1)$ & $6$ \\ 
146.10 & I & $3$ & $(1)$ & $1$ & 158.60 & IV & $4$ & $(1)$ & $2$ & 168.110 & II & $4$ & $(1)$ & $2$ & 178.160 & IV & $3$ & $(1)$ & $6$ \\ 
146.11 & II & $2$ & $(1)$ & $2$ & 159.61 & I & $4$ & $(1)$ & $2$ & 168.111 & III & $4$ & $(1)$ & $1$ & 179.161 & I & $3$ & $(1)$ & $6$ \\ 
146.12 & IV & $2$ & $(1)$ & $2$ & 159.62 & II & $3$ & $(1)$ & $4$ & 168.112 & IV & $4$ & $(1)$ & $2$ & 179.162 & II & $1$ & $(1)$ & $12$ \\ 
147.13 & I & $13$ & $(2, 12)$ & $1$ & 159.63 & III & $5$ & $(3)$ & $2$ & 169.113 & I & $1$ & $(1)$ & $6$ & 179.163 & III & $2$ & $(1)$ & $6$ \\ 
147.14 & II & $9$ & $(2, 4)$ & $2$ & 159.64 & IV & $3$ & $(1)$ & $2$ & 169.114 & II & $1$ & $(1)$ & $12$ & 179.164 & III & $2$ & $(1)$ & $6$ \\ 
147.15 & III & $4$ & $(1)$ & $2$ & 160.65 & I & $3$ & $(1)$ & $1$ & 169.115 & III & $1$ & $(1)$ & $6$ & 179.165 & III & $1$ & $(1)$ & $6$ \\ 
147.16 & IV & $7$ & $(2)$ & $2$ & 160.66 & II & $2$ & $(1)$ & $2$ & 169.116 & IV & $1$ & $(1)$ & $6$ & 179.166 & IV & $3$ & $(1)$ & $6$ \\ 
148.17 & I & $11$ & $(2, 4)$ & $1$ & 160.67 & III & $3$ & $(1)$ & $1$ & 170.117 & I & $1$ & $(1)$ & $6$ & 180.167 & I & $1$ & $(1)$ & $6$ \\ 
148.18 & II & $8$ & $(2, 4)$ & $2$ & 160.68 & IV & $2$ & $(1)$ & $2$ & 170.118 & II & $1$ & $(1)$ & $12$ & 180.168 & II & $1$ & $(1)$ & $6$ \\ 
148.19 & III & $2$ & $(1)$ & $2$ & 161.69 & I & $2$ & $(1)$ & $2$ & 170.119 & III & $1$ & $(1)$ & $6$ & 180.169 & III & $3$ & $(1)$ & $3$ \\ 
148.20 & IV & $6$ & $(2)$ & $2$ & 161.70 & II & $2$ & $(1)$ & $4$ & 170.120 & IV & $1$ & $(1)$ & $6$ & 180.170 & III & $3$ & $(1)$ & $3$ \\ 
149.21 & I & $6$ & $(1)$ & $1$ & 161.71 & III & $3$ & $(1)$ & $2$ & 171.121 & I & $3$ & $(2)$ & $3$ & 180.171 & III & $3$ & $(2)$ & $3$ \\ 
149.22 & II & $4$ & $(1)$ & $2$ & 161.72 & IV & $2$ & $(1)$ & $2$ & 171.122 & II & $1$ & $(1)$ & $6$ & 180.172 & IV & $1$ & $(1)$ & $6$ \\ 
149.23 & III & $7$ & $(3)$ & $1$ & 162.73 & I & $12$ & $(2)$ & $1$ & 171.123 & III & $1$ & $(1)$ & $3$ & 181.173 & I & $1$ & $(1)$ & $6$ \\ 
149.24 & IV & $5$ & $(1)$ & $2$ & 162.74 & II & $9$ & $(2, 4)$ & $2$ & 171.124 & IV & $1$ & $(1)$ & $6$ & 181.174 & II & $1$ & $(1)$ & $6$ \\ 
150.25 & I & $6$ & $(1)$ & $1$ & 162.75 & III & $4$ & $(1)$ & $2$ & 172.125 & I & $3$ & $(2)$ & $3$ & 181.175 & III & $3$ & $(1)$ & $3$ \\ 
150.26 & II & $3$ & $(1)$ & $2$ & 162.76 & III & $3$ & $(1)$ & $2$ & 172.126 & II & $1$ & $(1)$ & $6$ & 181.176 & III & $3$ & $(1)$ & $3$ \\ 
150.27 & III & $5$ & $(3)$ & $1$ & 162.77 & III & $13$ & $(2, 12)$ & $1$ & 172.127 & III & $1$ & $(1)$ & $3$ & 181.177 & III & $3$ & $(2)$ & $3$ \\ 
150.28 & IV & $4$ & $(1)$ & $2$ & 162.78 & IV & $7$ & $(2)$ & $2$ & 172.128 & IV & $1$ & $(1)$ & $6$ & 181.178 & IV & $1$ & $(1)$ & $6$ \\ 
151.29 & I & $3$ & $(1)$ & $3$ & 163.79 & I & $8$ & $(2)$ & $2$ & 173.129 & I & $5$ & $(3)$ & $2$ & 182.179 & I & $5$ & $(1)$ & $2$ \\ 
151.30 & II & $1$ & $(1)$ & $6$ & 163.80 & II & $6$ & $(4)$ & $4$ & 173.130 & II & $3$ & $(1)$ & $4$ & 182.180 & II & $3$ & $(1)$ & $4$ \\ 
151.31 & III & $1$ & $(1)$ & $3$ & 163.81 & III & $4$ & $(1)$ & $2$ & 173.131 & III & $4$ & $(1)$ & $2$ & 182.181 & III & $4$ & $(1)$ & $2$ \\ 
151.32 & IV & $2$ & $(1)$ & $6$ & 163.82 & III & $4$ & $(1)$ & $2$ & 173.132 & IV & $3$ & $(1)$ & $2$ & 182.182 & III & $5$ & $(1)$ & $2$ \\ 
152.33 & I & $3$ & $(1)$ & $3$ & 163.83 & III & $9$ & $(12)$ & $2$ & 174.133 & I & $21$ & $(3, 3, 3)$ & $1$ & 182.183 & III & $5$ & $(3)$ & $2$ \\ 
152.34 & II & $1$ & $(1)$ & $6$ & 163.84 & IV & $7$ & $(2)$ & $2$ & 174.134 & II & $10$ & $(3, 3)$ & $2$ & 182.184 & IV & $5$ & $(1)$ & $2$ \\ 
152.35 & III & $1$ & $(1)$ & $3$ & 164.85 & I & $12$ & $(2)$ & $1$ & 174.135 & III & $4$ & $(1)$ & $1$ & 183.185 & I & $4$ & $(1)$ & $2$ \\ 
152.36 & IV & $2$ & $(1)$ & $6$ & 164.86 & II & $9$ & $(2, 4)$ & $2$ & 174.136 & IV & $11$ & $(3)$ & $2$ & 183.186 & II & $4$ & $(1)$ & $2$ \\ 
153.37 & I & $3$ & $(1)$ & $3$ & 164.87 & III & $3$ & $(1)$ & $2$ & 175.137 & I & $27$ & $(6, 6, 6)$ & $1$ & 183.187 & III & $5$ & $(1)$ & $1$ \\ 
153.38 & II & $1$ & $(1)$ & $6$ & 164.88 & III & $4$ & $(1)$ & $2$ & 175.138 & II & $14$ & $(6, 12)$ & $2$ & 183.188 & III & $4$ & $(1)$ & $1$ \\ 
153.39 & III & $1$ & $(1)$ & $3$ & 164.89 & III & $13$ & $(2, 12)$ & $1$ & 175.139 & III & $10$ & $(3, 3)$ & $2$ & 183.189 & III & $9$ & $(6)$ & $1$ \\ 
153.40 & IV & $2$ & $(1)$ & $6$ & 164.90 & IV & $7$ & $(2)$ & $2$ & 175.140 & III & $4$ & $(1)$ & $2$ & 183.190 & IV & $4$ & $(1)$ & $4$ \\ 
154.41 & I & $3$ & $(1)$ & $3$ & 165.91 & I & $8$ & $(2)$ & $2$ & 175.141 & III & $9$ & $(2, 4)$ & $1$ & 184.191 & I & $4$ & $(1)$ & $2$ \\ 
154.42 & II & $1$ & $(1)$ & $6$ & 165.92 & II & $6$ & $(4)$ & $4$ & 175.142 & IV & $13$ & $(6)$ & $2$ & 184.192 & II & $4$ & $(1)$ & $4$ \\ 
154.43 & III & $1$ & $(1)$ & $3$ & 165.93 & III & $3$ & $(1)$ & $2$ & 176.143 & I & $16$ & $(3, 6)$ & $2$ & 184.193 & III & $4$ & $(1)$ & $2$ \\ 
154.44 & IV & $2$ & $(1)$ & $6$ & 165.94 & III & $5$ & $(1)$ & $2$ & 176.144 & II & $9$ & $(12)$ & $4$ & 184.194 & III & $3$ & $(1)$ & $2$ \\ 
155.45 & I & $4$ & $(1)$ & $1$ & 165.95 & III & $9$ & $(12)$ & $2$ & 176.145 & III & $11$ & $(3)$ & $2$ & 184.195 & III & $9$ & $(6)$ & $2$ \\ 
155.46 & II & $2$ & $(1)$ & $2$ & 165.96 & IV & $7$ & $(2)$ & $2$ & 176.146 & III & $3$ & $(1)$ & $2$ & 184.196 & IV & $4$ & $(1)$ & $2$ \\ 
155.47 & III & $3$ & $(1)$ & $1$ & 166.97 & I & $11$ & $(2)$ & $1$ & 176.147 & III & $7$ & $(4)$ & $2$ & 185.197 & I & $3$ & $(1)$ & $2$ \\ 
155.48 & IV & $3$ & $(1)$ & $2$ & 166.98 & II & $8$ & $(2, 4)$ & $2$ & 176.148 & IV & $9$ & $(6)$ & $2$ & 185.198 & II & $3$ & $(1)$ & $4$ \\ 
156.49 & I & $5$ & $(1)$ & $1$ & 166.99 & III & $2$ & $(1)$ & $2$ & 177.149 & I & $4$ & $(1)$ & $2$ & 185.199 & III & $5$ & $(1)$ & $2$ \\ 
156.50 & II & $4$ & $(1)$ & $2$ & 166.100 & III & $2$ & $(1)$ & $2$ & 177.150 & II & $4$ & $(1)$ & $2$ & 185.200 & III & $3$ & $(1)$ & $2$ \\ 
\hline \hline 
\end{tabular}

%% file: spinful_HexaII.tex
\begin{tabular}{cc|ccc||cc|ccc||cc|ccc||cc|ccc} 
\hline \hline 
185.201 & III & $5$ & $(3)$ & $2$ & 188.219 & III & $14$ & $(3, 3)$ & $2$ & 191.237 & III & $7$ & $(3, 3)$ & $2$ & 193.255 & III & $3$ & $(1)$ & $2$ \\ 
185.202 & IV & $3$ & $(1)$ & $2$ & 188.220 & IV & $12$ & $(3)$ & $2$ & 191.238 & III & $12$ & $(2)$ & $1$ & 193.256 & III & $7$ & $(3)$ & $4$ \\ 
186.203 & I & $3$ & $(1)$ & $2$ & 189.221 & I & $10$ & $(3, 3)$ & $2$ & 191.239 & III & $12$ & $(2)$ & $1$ & 193.257 & III & $8$ & $(3)$ & $2$ \\ 
186.204 & II & $3$ & $(1)$ & $4$ & 189.222 & II & $7$ & $(3, 3)$ & $2$ & 191.240 & III & $27$ & $(6, 6, 6)$ & $1$ & 193.258 & III & $8$ & $(2)$ & $2$ \\ 
186.205 & III & $4$ & $(1)$ & $2$ & 189.223 & III & $4$ & $(1)$ & $1$ & 191.241 & III & $4$ & $(1)$ & $2$ & 193.259 & III & $8$ & $(2)$ & $2$ \\ 
186.206 & III & $4$ & $(1)$ & $2$ & 189.224 & III & $5$ & $(1)$ & $1$ & 191.242 & IV & $9$ & $(12)$ & $4$ & 193.260 & III & $14$ & $(3, 6)$ & $2$ \\ 
186.207 & III & $5$ & $(3)$ & $2$ & 189.225 & III & $15$ & $(3, 3, 3)$ & $1$ & 192.243 & I & $13$ & $(6)$ & $2$ & 193.261 & III & $4$ & $(1)$ & $2$ \\ 
186.208 & IV & $3$ & $(1)$ & $2$ & 189.226 & IV & $6$ & $(3)$ & $2$ & 192.244 & II & $9$ & $(12)$ & $4$ & 193.262 & IV & $10$ & $(6)$ & $2$ \\ 
187.209 & I & $10$ & $(3, 3)$ & $2$ & 190.227 & I & $12$ & $(3)$ & $2$ & 192.245 & III & $4$ & $(1)$ & $2$ & 194.263 & I & $10$ & $(6)$ & $2$ \\ 
187.210 & II & $10$ & $(3, 3)$ & $2$ & 190.228 & II & $6$ & $(3)$ & $4$ & 192.246 & III & $8$ & $(3)$ & $2$ & 194.264 & II & $9$ & $(12)$ & $4$ \\ 
187.211 & III & $6$ & $(1)$ & $1$ & 190.229 & III & $3$ & $(1)$ & $2$ & 192.247 & III & $7$ & $(3)$ & $2$ & 194.265 & III & $3$ & $(1)$ & $2$ \\ 
187.212 & III & $5$ & $(1)$ & $1$ & 190.230 & III & $4$ & $(1)$ & $2$ & 192.248 & III & $7$ & $(2)$ & $2$ & 194.266 & III & $11$ & $(3)$ & $2$ \\ 
187.213 & III & $21$ & $(3, 3, 3)$ & $1$ & 190.231 & III & $12$ & $(3, 3)$ & $2$ & 192.249 & III & $7$ & $(2)$ & $2$ & 194.267 & III & $6$ & $(3)$ & $4$ \\ 
187.214 & IV & $7$ & $(3)$ & $2$ & 190.232 & IV & $9$ & $(3)$ & $2$ & 192.250 & III & $18$ & $(6, 6)$ & $2$ & 194.268 & III & $8$ & $(2)$ & $2$ \\ 
188.215 & I & $12$ & $(3)$ & $2$ & 191.233 & I & $14$ & $(6, 12)$ & $2$ & 192.251 & III & $4$ & $(1)$ & $4$ & 194.269 & III & $8$ & $(2)$ & $2$ \\ 
188.216 & II & $7$ & $(3)$ & $4$ & 191.234 & II & $14$ & $(6, 12)$ & $2$ & 192.252 & IV & $13$ & $(6)$ & $2$ & 194.270 & III & $16$ & $(3, 6)$ & $2$ \\ 
188.217 & III & $5$ & $(1)$ & $2$ & 191.235 & III & $4$ & $(1)$ & $2$ & 193.253 & I & $10$ & $(6)$ & $2$ & 194.271 & III & $4$ & $(1)$ & $2$ \\ 
188.218 & III & $4$ & $(1)$ & $2$ & 191.236 & III & $10$ & $(3, 3)$ & $2$ & 193.254 & II & $8$ & $(12)$ & $4$ & 194.272 & IV & $9$ & $(6)$ & $2$ \\ 
\hline \hline 
\end{tabular}

%% file: spinful_HexaII_foot.tex
\newlength{\tabLspinfulHexaII} 
\settowidth{\tabLspinfulHexaII}{\input{spinful_HexaII}} 
\begin{minipage}{\tabLspinfulHexaII} 
\begin{flushleft} 
{\footnotesize $d$: Rank of the band structure group $\{{\rm BS}\}$\\ 
$X_{\rm BS}$: Symmetry-based indicators of band topology\\ 
$\nu_{\rm BS}$: Set of $\nu$ bands are symmetry-forbidden from being isolated by band gaps if $\nu \not \in \nu_{\rm BS}\, \mathbb Z$ }
\end{flushleft}\end{minipage}

%% file: spinful_Cubic.tex
\begin{tabular}{cc|ccc||cc|ccc||cc|ccc||cc|ccc} 
\hline \hline 
\multicolumn{2}{c|}{MSG} & $d$ & $X_{\rm BS}$ & $\nu_{\rm BS}$ & \multicolumn{2}{c|}{MSG} & $d$ & $X_{\rm BS}$ & $\nu_{\rm BS}$ & \multicolumn{2}{c|}{MSG} & $d$ & $X_{\rm BS}$ & $\nu_{\rm BS}$ & \multicolumn{2}{c|}{MSG} & $d$ & $X_{\rm BS}$ & $\nu_{\rm BS}$\\ 
\hline 
195.1 & I & $3$ & $(1)$ & $2$ & 206.39 & III & $3$ & $(1)$ & $4^* $ & 216.77 & IV & $4$ & $(2)$ & $4$ & 224.115 & IV & $6$ & $(2)$ & $4$ \\ 
195.2 & II & $2$ & $(1)$ & $2$ & 207.40 & I & $4$ & $(1)$ & $2$ & 217.78 & I & $6$ & $(2)$ & $2$ & 225.116 & I & $11$ & $(8)$ & $2$ \\ 
195.3 & IV & $2$ & $(1)$ & $4$ & 207.41 & II & $4$ & $(1)$ & $2$ & 217.79 & II & $5$ & $(2)$ & $2$ & 225.117 & II & $11$ & $(8)$ & $2$ \\ 
196.4 & I & $3$ & $(1)$ & $2$ & 207.42 & III & $3$ & $(1)$ & $2$ & 217.80 & III & $3$ & $(1)$ & $2$ & 225.118 & III & $5$ & $(2)$ & $2$ \\ 
196.5 & II & $2$ & $(1)$ & $2$ & 207.43 & IV & $3$ & $(1)$ & $4$ & 218.81 & I & $6$ & $(2)$ & $2^* $ & 225.119 & III & $10$ & $(4)$ & $2$ \\ 
196.6 & IV & $2$ & $(1)$ & $2^* $ & 208.44 & I & $2$ & $(1)$ & $4$ & 218.82 & II & $4$ & $(2)$ & $4$ & 225.120 & III & $3$ & $(1)$ & $2$ \\ 
197.7 & I & $3$ & $(1)$ & $2$ & 208.45 & II & $2$ & $(1)$ & $4$ & 218.83 & III & $3$ & $(1)$ & $4$ & 225.121 & IV & $8$ & $(8)$ & $4$ \\ 
197.8 & II & $2$ & $(1)$ & $2$ & 208.46 & III & $3$ & $(1)$ & $2^* $ & 218.84 & IV & $4$ & $(2)$ & $4$ & 226.122 & I & $10$ & $(4)$ & $2^* $ \\ 
198.9 & I & $3$ & $(1)$ & $4$ & 208.47 & IV & $2$ & $(1)$ & $4$ & 219.85 & I & $6$ & $(2)$ & $2^* $ & 226.123 & II & $8$ & $(8)$ & $4$ \\ 
198.10 & II & $2$ & $(1)$ & $8$ & 209.48 & I & $3$ & $(1)$ & $2$ & 219.86 & II & $4$ & $(2)$ & $4$ & 226.124 & III & $5$ & $(2)$ & $2^* $ \\ 
198.11 & IV & $2$ & $(1)$ & $4^* $ & 209.49 & II & $3$ & $(1)$ & $2$ & 219.87 & III & $3$ & $(1)$ & $4$ & 226.125 & III & $7$ & $(4)$ & $4$ \\ 
199.12 & I & $4$ & $(1)$ & $2^* $ & 209.50 & III & $3$ & $(1)$ & $2$ & 219.88 & IV & $6$ & $(2)$ & $2^* $ & 226.126 & III & $3$ & $(1)$ & $4$ \\ 
199.13 & II & $2$ & $(1)$ & $4^* $ & 209.51 & IV & $3$ & $(1)$ & $2^* $ & 220.89 & I & $7$ & $(2)$ & $2^* $ & 226.127 & IV & $10$ & $(4)$ & $2^* $ \\ 
200.14 & I & $11$ & $(2, 4)$ & $2$ & 210.52 & I & $3$ & $(1)$ & $4$ & 220.90 & II & $4$ & $(2)$ & $4^* $ & 227.128 & I & $9$ & $(2)$ & $4$ \\ 
200.15 & II & $8$ & $(2, 4)$ & $2$ & 210.53 & II & $2$ & $(1)$ & $4$ & 220.91 & III & $3$ & $(1)$ & $4^* $ & 227.129 & II & $8$ & $(4)$ & $4$ \\ 
200.16 & III & $2$ & $(1)$ & $2$ & 210.54 & III & $3$ & $(1)$ & $4$ & 221.92 & I & $14$ & $(4, 8)$ & $2$ & 227.130 & III & $4$ & $(2)$ & $4$ \\ 
200.17 & IV & $6$ & $(4)$ & $4$ & 210.55 & IV & $3$ & $(1)$ & $4^* $ & 221.93 & II & $14$ & $(4, 8)$ & $2$ & 227.131 & III & $9$ & $(2)$ & $4$ \\ 
201.18 & I & $9$ & $(2)$ & $4$ & 211.56 & I & $3$ & $(1)$ & $2$ & 221.94 & III & $6$ & $(2)$ & $2$ & 227.132 & III & $2$ & $(1)$ & $4$ \\ 
201.19 & II & $6$ & $(2, 4)$ & $4$ & 211.57 & II & $3$ & $(1)$ & $2$ & 221.95 & III & $11$ & $(2, 4)$ & $2$ & 227.133 & IV & $6$ & $(2)$ & $8$ \\ 
201.20 & III & $2$ & $(1)$ & $4$ & 211.58 & III & $3$ & $(1)$ & $2$ & 221.96 & III & $4$ & $(1)$ & $2$ & 228.134 & I & $7$ & $(2)$ & $4^* $ \\ 
201.21 & IV & $5$ & $(2)$ & $4$ & 212.59 & I & $3$ & $(1)$ & $4$ & 221.97 & IV & $8$ & $(8)$ & $4$ & 228.135 & II & $5$ & $(4)$ & $8$ \\ 
202.22 & I & $10$ & $(4)$ & $2$ & 212.60 & II & $2$ & $(1)$ & $8$ & 222.98 & I & $8$ & $(2)$ & $4$ & 228.136 & III & $4$ & $(2)$ & $4^* $ \\ 
202.23 & II & $7$ & $(4)$ & $2$ & 212.61 & III & $3$ & $(1)$ & $4$ & 222.99 & II & $6$ & $(4)$ & $4$ & 228.137 & III & $6$ & $(2)$ & $8$ \\ 
202.24 & III & $2$ & $(1)$ & $2$ & 212.62 & IV & $3$ & $(1)$ & $4^* $ & 222.100 & III & $4$ & $(2)$ & $4$ & 228.138 & III & $3$ & $(1)$ & $8$ \\ 
202.25 & IV & $6$ & $(4)$ & $2^* $ & 213.63 & I & $3$ & $(1)$ & $4$ & 222.101 & III & $6$ & $(2)$ & $4$ & 228.139 & IV & $7$ & $(2)$ & $4^* $ \\ 
203.26 & I & $9$ & $(2)$ & $4$ & 213.64 & II & $2$ & $(1)$ & $8$ & 222.102 & III & $3$ & $(1)$ & $4$ & 229.140 & I & $11$ & $(8)$ & $2$ \\ 
203.27 & II & $6$ & $(4)$ & $4$ & 213.65 & III & $3$ & $(1)$ & $4$ & 222.103 & IV & $8$ & $(2)$ & $4$ & 229.141 & II & $11$ & $(2, 8)$ & $2$ \\ 
203.28 & III & $2$ & $(1)$ & $4$ & 213.66 & IV & $3$ & $(1)$ & $4^* $ & 223.104 & I & $8$ & $(4)$ & $4$ & 229.142 & III & $6$ & $(2)$ & $2$ \\ 
203.29 & IV & $5$ & $(2)$ & $4^* $ & 214.67 & I & $2$ & $(1)$ & $4$ & 223.105 & II & $7$ & $(4)$ & $4$ & 229.143 & III & $10$ & $(4)$ & $2$ \\ 
204.30 & I & $10$ & $(4)$ & $2$ & 214.68 & II & $2$ & $(1)$ & $4^* $ & 223.106 & III & $6$ & $(2)$ & $2^* $ & 229.144 & III & $3$ & $(1)$ & $2$ \\ 
204.31 & II & $7$ & $(2, 4)$ & $2$ & 214.69 & III & $4$ & $(1)$ & $2^* $ & 223.107 & III & $7$ & $(4)$ & $4$ & 230.145 & I & $7$ & $(2)$ & $4^* $ \\ 
204.32 & III & $2$ & $(1)$ & $2$ & 215.70 & I & $6$ & $(2)$ & $2$ & 223.108 & III & $2$ & $(1)$ & $4$ & 230.146 & II & $5$ & $(4)$ & $8^* $ \\ 
205.33 & I & $9$ & $(2)$ & $4$ & 215.71 & II & $6$ & $(2)$ & $2$ & 223.109 & IV & $7$ & $(4)$ & $4$ & 230.147 & III & $5$ & $(2)$ & $4^* $ \\ 
205.34 & II & $6$ & $(4)$ & $8$ & 215.72 & III & $3$ & $(1)$ & $2$ & 224.110 & I & $8$ & $(2)$ & $4$ & 230.148 & III & $6$ & $(2)$ & $4^* $ \\ 
205.35 & III & $2$ & $(1)$ & $8$ & 215.73 & IV & $4$ & $(2)$ & $4$ & 224.111 & II & $8$ & $(2, 4)$ & $4$ & 230.149 & III & $2$ & $(1)$ & $8$ \\ 
205.36 & IV & $5$ & $(2)$ & $8$ & 216.74 & I & $6$ & $(2)$ & $2$ & 224.112 & III & $4$ & $(2)$ & $4$ & ~ & ~ & ~ & ~ \\ 
206.37 & I & $9$ & $(2)$ & $4$ & 216.75 & II & $6$ & $(2)$ & $2$ & 224.113 & III & $9$ & $(2)$ & $4$ & ~ & ~ & ~ & ~ \\ 
206.38 & II & $6$ & $(2, 4)$ & $8$ & 216.76 & III & $3$ & $(1)$ & $2$ & 224.114 & III & $2$ & $(1)$ & $4$ & ~ & ~ & ~ & ~ \\ 
\hline \hline 
\end{tabular}

%% file: spinful_Cubic_foot.tex
\newlength{\tabLspinfulCubic} 
\settowidth{\tabLspinfulCubic}{\input{spinful_Cubic}} 
\begin{minipage}{\tabLspinfulCubic} 
\begin{flushleft} 
{\footnotesize $d$: Rank of the band structure group $\{{\rm BS}\}$\\ 
$X_{\rm BS}$: Symmetry-based indicators of band topology\\ 
$\nu_{\rm BS}$: Set of $\nu$ bands are symmetry-forbidden from being isolated by band gaps if $\nu \not \in \nu_{\rm BS}\, \mathbb Z$ }\\ 
$*$: Exhibiting exceptional filling pattern; see Table \ref{tab:spinfulnuEx} 
\end{flushleft}\end{minipage}

%% file: spinful_MLG.tex
\begin{tabular}{cc|ccc||cc|ccc||cc|ccc||cc|ccc} 
\hline \hline 
\multicolumn{2}{c|}{MSG} & $d$ & $X_{\rm BS}$ & $\nu_{\rm BS}$ & \multicolumn{2}{c|}{MSG} & $d$ & $X_{\rm BS}$ & $\nu_{\rm BS}$ & \multicolumn{2}{c|}{MSG} & $d$ & $X_{\rm BS}$ & $\nu_{\rm BS}$ & \multicolumn{2}{c|}{MSG} & $d$ & $X_{\rm BS}$ & $\nu_{\rm BS}$\\ 
\hline 
1.1 (1) & I & 1 & (1) & 1 & 6.19 (1) & II & 1 & (1) & 2 & 11.52 (1) & III & 2 & (1) & 2 & 16.5 & IV & 1 & (1) & 4 \\ 
1.1 (2) & I & 1 & (1) & 1 & 6.19 (2) & II & 1 & (1) & 2 & 11.52 (2) & III & 2 & (1) & 2 & 17.7 (1) & I & 4 & (1) & 2 \\ 
1.1 (3) & I & 1 & (1) & 1 & 6.19 (3) & II & 1 & (1) & 2 & 11.53 (1) & III & 1 & (1) & 2 & 17.7 (2) & I & 4 & (1) & 2 \\ 
1.2 (1) & II & 1 & (1) & 2 & 6.20 (1) & III & 1 & (1) & 1 & 11.53 (2) & III & 1 & (1) & 2 & 17.8 (1) & II & 1 & (1) & 4 \\ 
1.2 (2) & II & 1 & (1) & 2 & 6.20 (2) & III & 1 & (1) & 1 & 11.54 (1) & III & 3 & (2) & 2 & 17.8 (2) & II & 1 & (1) & 4 \\ 
1.2 (3) & II & 1 & (1) & 2 & 6.20 (3) & III & 1 & (1) & 1 & 11.54 (2) & III & 3 & (2) & 2 & 17.9 (1) & III & 1 & (1) & 2 \\ 
1.3 (1) & IV & 1 & (1) & 2 & 6.21 (1) & IV & 1 & (1) & 2 & 11.55 & IV & 2 & (1) & 4 & 17.9 (2) & III & 1 & (1) & 2 \\ 
1.3 (2) & IV & 1 & (1) & 2 & 6.21 (2) & IV & 1 & (1) & 2 & 11.56 (1) & IV & 3 & (1) & 2 & 17.10 (1) & III & 3 & (2) & 2 \\ 
2.4 (1) & I & 5 & (2) & 1 & 6.22 (1) & IV & 2 & (1) & 2 & 11.56 (2) & IV & 3 & (1) & 2 & 17.10 (2) & III & 2 & (1) & 2 \\ 
2.4 (2) & I & 5 & (2) & 1 & 6.22 (2) & IV & 2 & (1) & 2 & 11.57 & IV & 2 & (1) & 4 & 17.11 & IV & 2 & (1) & 4 \\ 
2.4 (3) & I & 5 & (2) & 1 & 6.23 & IV & 2 & (1) & 2 & 12.58 & I & 6 & (1) & 1 & 17.12 (1) & IV & 4 & (1) & 2 \\ 
2.5 (1) & II & 5 & (2) & 2 & 7.24 (1) & I & 1 & (1) & 2 & 12.59 & II & 4 & (2) & 2 & 17.12 (2) & IV & 4 & (1) & 2 \\ 
2.5 (2) & II & 5 & (2) & 2 & 7.24 (2) & I & 1 & (1) & 2 & 12.60 & III & 1 & (1) & 2 & 17.13 & IV & 3 & (1) & 2 \\ 
2.5 (3) & II & 5 & (2) & 2 & 7.25 (1) & II & 1 & (1) & 4 & 12.61 & III & 1 & (1) & 2 & 18.16 & I & 3 & (1) & 2 \\ 
2.6 (1) & III & 1 & (1) & 2 & 7.25 (2) & II & 1 & (1) & 4 & 12.62 & III & 4 & (2) & 1 & 18.17 & II & 1 & (1) & 4 \\ 
2.6 (2) & III & 1 & (1) & 2 & 7.26 (1) & III & 1 & (1) & 2 & 12.64 & IV & 3 & (1) & 2 & 18.18 & III & 3 & (2) & 2 \\ 
2.6 (3) & III & 1 & (1) & 2 & 7.26 (2) & III & 1 & (1) & 2 & 13.65 (1) & I & 4 & (1) & 2 & 18.19 & III & 1 & (1) & 2 \\ 
2.7 (1) & IV & 3 & (1) & 2 & 7.27 & IV & 1 & (1) & 4 & 13.65 (2) & I & 5 & (2) & 2 & 18.20 & IV & 2 & (1) & 4 \\ 
2.7 (2) & IV & 3 & (1) & 2 & 7.28 (1) & IV & 1 & (1) & 2 & 13.66 (1) & II & 3 & (2) & 4 & 18.23 & IV & 3 & (1) & 2 \\ 
3.1 (1) & I & 3 & (1) & 1 & 7.28 (2) & IV & 1 & (1) & 2 & 13.66 (2) & II & 3 & (2) & 4 & 21.38 & I & 2 & (1) & 2 \\ 
3.1 (2) & I & 5 & (2) & 1 & 7.29 & IV & 1 & (1) & 4 & 13.67 (1) & III & 1 & (1) & 2 & 21.39 & II & 1 & (1) & 2 \\ 
3.1 (3) & I & 3 & (1) & 1 & 7.31 & IV & 1 & (1) & 2 & 13.67 (2) & III & 1 & (1) & 2 & 21.40 & III & 4 & (2) & 1 \\ 
3.2 (1) & II & 1 & (1) & 2 & 8.32 & I & 2 & (1) & 1 & 13.68 (1) & III & 2 & (1) & 2 & 21.41 & III & 2 & (1) & 1 \\ 
3.2 (2) & II & 1 & (1) & 2 & 8.33 & II & 1 & (1) & 2 & 13.68 (2) & III & 3 & (1) & 2 & 21.43 & IV & 2 & (1) & 2 \\ 
3.2 (3) & II & 1 & (1) & 2 & 8.34 & III & 1 & (1) & 1 & 13.69 (1) & III & 3 & (2) & 2 & 25.57 (1) & I & 1 & (1) & 2 \\ 
3.3 (1) & III & 1 & (1) & 1 & 8.36 & IV & 1 & (1) & 2 & 13.69 (2) & III & 3 & (2) & 2 & 25.57 (2) & I & 1 & (1) & 2 \\ 
3.3 (2) & III & 1 & (1) & 1 & 10.42 (1) & I & 9 & (1) & 1 & 13.70 & IV & 3 & (1) & 4 & 25.57 (3) & I & 1 & (1) & 2 \\ 
3.3 (3) & III & 1 & (1) & 1 & 10.42 (2) & I & 10 & (2, 2) & 1 & 13.71 & IV & 2 & (1) & 4 & 25.58 (1) & II & 1 & (1) & 2 \\ 
3.4 (1) & IV & 3 & (2) & 2 & 10.42 (3) & I & 9 & (1) & 1 & 13.72 (1) & IV & 3 & (1) & 2 & 25.58 (2) & II & 1 & (1) & 2 \\ 
3.4 (2) & IV & 2 & (1) & 2 & 10.43 (1) & II & 5 & (2) & 2 & 13.72 (2) & IV & 3 & (1) & 2 & 25.58 (3) & II & 1 & (1) & 2 \\ 
3.5 (1) & IV & 1 & (1) & 2 & 10.43 (2) & II & 5 & (2) & 2 & 13.73 & IV & 2 & (1) & 4 & 25.59 (1) & III & 3 & (1) & 1 \\ 
3.5 (2) & IV & 1 & (1) & 2 & 10.43 (3) & II & 5 & (2) & 2 & 14.75 & I & 3 & (1) & 2 & 25.59 (2) & III & 2 & (1) & 1 \\ 
3.6 & IV & 2 & (1) & 2 & 10.44 (1) & III & 1 & (1) & 2 & 14.76 & II & 3 & (2) & 4 & 25.59 (3) & III & 3 & (1) & 1 \\ 
4.7 (1) & I & 1 & (1) & 2 & 10.44 (2) & III & 1 & (1) & 2 & 14.77 & III & 1 & (1) & 4 & 25.60 (1) & III & 3 & (1) & 1 \\ 
4.7 (2) & I & 1 & (1) & 2 & 10.44 (3) & III & 1 & (1) & 2 & 14.78 & III & 1 & (1) & 4 & 25.60 (2) & III & 3 & (1) & 1 \\ 
4.8 (1) & II & 1 & (1) & 4 & 10.45 (1) & III & 1 & (1) & 2 & 14.79 & III & 3 & (2) & 2 & 25.60 (3) & III & 5 & (2) & 1 \\ 
4.8 (2) & II & 1 & (1) & 4 & 10.45 (2) & III & 1 & (1) & 2 & 14.81 & IV & 2 & (1) & 4 & 25.61 (1) & IV & 1 & (1) & 4 \\ 
4.9 (1) & III & 1 & (1) & 2 & 10.45 (3) & III & 1 & (1) & 2 & 14.82 & IV & 2 & (1) & 4 & 25.61 (2) & IV & 1 & (1) & 4 \\ 
4.9 (2) & III & 1 & (1) & 2 & 10.46 (1) & III & 5 & (2) & 1 & 14.83 & IV & 3 & (1) & 2 & 25.62 (1) & IV & 1 & (1) & 2 \\ 
4.10 & IV & 1 & (1) & 4 & 10.46 (2) & III & 5 & (2) & 1 & 16.1 (1) & I & 1 & (1) & 2 & 25.62 (2) & IV & 1 & (1) & 2 \\ 
4.11 (1) & IV & 1 & (1) & 2 & 10.46 (3) & III & 5 & (2) & 1 & 16.1 (2) & I & 1 & (1) & 2 & 25.63 & IV & 1 & (1) & 4 \\ 
4.11 (2) & IV & 1 & (1) & 2 & 10.47 (1) & IV & 5 & (2) & 2 & 16.1 (3) & I & 1 & (1) & 2 & 25.64 & IV & 1 & (1) & 4 \\ 
4.12 & IV & 1 & (1) & 2 & 10.47 (2) & IV & 4 & (1) & 2 & 16.2 (1) & II & 1 & (1) & 2 & 26.66 (1) & I & 1 & (1) & 2 \\ 
5.13 & I & 2 & (1) & 1 & 10.48 (1) & IV & 4 & (1) & 2 & 16.2 (2) & II & 1 & (1) & 2 & 26.66 (2) & I & 1 & (1) & 2 \\ 
5.14 & II & 1 & (1) & 2 & 10.48 (2) & IV & 4 & (1) & 2 & 16.2 (3) & II & 1 & (1) & 2 & 26.67 (1) & II & 1 & (1) & 4 \\ 
5.15 & III & 1 & (1) & 1 & 10.49 & IV & 5 & (1) & 2 & 16.3 (1) & III & 3 & (1) & 1 & 26.67 (2) & II & 1 & (1) & 4 \\ 
5.17 & IV & 1 & (1) & 2 & 11.50 (1) & I & 4 & (1) & 2 & 16.3 (2) & III & 3 & (1) & 1 & 26.68 (1) & III & 1 & (1) & 2 \\ 
6.18 (1) & I & 3 & (1) & 1 & 11.50 (2) & I & 4 & (1) & 2 & 16.3 (3) & III & 5 & (2) & 1 & 26.68 (2) & III & 1 & (1) & 2 \\ 
6.18 (2) & I & 2 & (1) & 1 & 11.51 (1) & II & 3 & (2) & 4 & 16.4 (1) & IV & 1 & (1) & 2 & 26.69 (1) & III & 2 & (1) & 2 \\ 
6.18 (3) & I & 3 & (1) & 1 & 11.51 (2) & II & 3 & (2) & 4 & 16.4 (2) & IV & 1 & (1) & 2 & 26.69 (2) & III & 3 & (1) & 2 \\ 
\hline \hline 
\end{tabular}

%% file: spinful_MLGII.tex
\begin{tabular}{cc|ccc||cc|ccc||cc|ccc||cc|ccc} 
\hline \hline 
26.70 (1) & III & 1 & (1) & 2 & 30.118 & IV & 1 & (1) & 4 & 47.255 & IV & 3 & (2) & 4 & 51.302 & IV & 3 & (2) & 4 \\ 
26.70 (2) & III & 1 & (1) & 2 & 30.119 & IV & 2 & (1) & 2 & 49.265 (1) & I & 4 & (1) & 2 & 51.303 & IV & 3 & (1) & 4 \\ 
26.71 & IV & 1 & (1) & 4 & 31.123 & I & 2 & (1) & 2 & 49.265 (2) & I & 4 & (1) & 2 & 53.321 & I & 6 & (1) & 2 \\ 
26.72 & IV & 1 & (1) & 4 & 31.124 & II & 1 & (1) & 4 & 49.266 (1) & II & 3 & (2) & 4 & 53.322 & II & 3 & (2) & 4 \\ 
26.73 (1) & IV & 1 & (1) & 2 & 31.125 & III & 1 & (1) & 2 & 49.266 (2) & II & 3 & (2) & 4 & 53.323 & III & 1 & (1) & 4 \\ 
26.73 (2) & IV & 1 & (1) & 2 & 31.126 & III & 2 & (1) & 2 & 49.267 (1) & III & 3 & (1) & 2 & 53.324 & III & 2 & (1) & 4 \\ 
26.74 & IV & 1 & (1) & 2 & 31.127 & III & 1 & (1) & 2 & 49.267 (2) & III & 2 & (1) & 2 & 53.325 & III & 1 & (1) & 4 \\ 
26.75 & IV & 1 & (1) & 4 & 31.128 & IV & 2 & (1) & 4 & 49.268 (1) & III & 1 & (1) & 2 & 53.326 & III & 3 & (1) & 2 \\ 
27.78 (1) & I & 1 & (1) & 2 & 31.130 & IV & 1 & (1) & 4 & 49.268 (2) & III & 1 & (1) & 2 & 53.327 & III & 5 & (1) & 2 \\ 
27.78 (2) & I & 1 & (1) & 2 & 31.132 & IV & 2 & (1) & 2 & 49.269 (1) & III & 6 & (1) & 2 & 53.328 & III & 4 & (2) & 2 \\ 
27.79 (1) & II & 1 & (1) & 4 & 32.135 & I & 3 & (1) & 2 & 49.269 (2) & III & 6 & (1) & 2 & 53.329 & III & 2 & (1) & 4 \\ 
27.79 (2) & II & 1 & (1) & 4 & 32.136 & II & 1 & (1) & 4 & 49.270 (1) & III & 4 & (1) & 2 & 53.330 & IV & 4 & (1) & 4 \\ 
27.80 (1) & III & 1 & (1) & 2 & 32.137 & III & 1 & (1) & 2 & 49.270 (2) & III & 5 & (2) & 2 & 53.332 & IV & 4 & (1) & 4 \\ 
27.80 (2) & III & 1 & (1) & 2 & 32.138 & III & 3 & (2) & 2 & 49.271 (1) & III & 1 & (1) & 4 & 53.334 & IV & 5 & (1) & 2 \\ 
27.81 (1) & III & 3 & (1) & 2 & 32.140 & IV & 2 & (1) & 4 & 49.271 (2) & III & 1 & (1) & 4 & 54.337 & I & 3 & (1) & 4 \\ 
27.81 (2) & III & 3 & (1) & 2 & 32.141 & IV & 3 & (1) & 2 & 49.272 & IV & 2 & (1) & 4 & 54.338 & II & 2 & (2) & 8 \\ 
27.82 (1) & IV & 1 & (1) & 2 & 35.165 & I & 2 & (1) & 2 & 49.273 (1) & IV & 4 & (1) & 2 & 54.339 & III & 1 & (1) & 4 \\ 
27.82 (2) & IV & 1 & (1) & 2 & 35.166 & II & 1 & (1) & 2 & 49.273 (2) & IV & 4 & (1) & 2 & 54.340 & III & 2 & (1) & 4 \\ 
27.83 & IV & 1 & (1) & 4 & 35.167 & III & 2 & (1) & 1 & 49.274 & IV & 3 & (1) & 4 & 54.341 & III & 1 & (1) & 4 \\ 
27.85 & IV & 1 & (1) & 4 & 35.168 & III & 4 & (2) & 1 & 50.277 & I & 2 & (1) & 4 & 54.342 & III & 3 & (1) & 4 \\ 
28.87 (1) & I & 3 & (1) & 2 & 35.170 & IV & 2 & (1) & 2 & 50.278 & II & 2 & (2) & 4 & 54.343 & III & 2 & (1) & 4 \\ 
28.87 (2) & I & 4 & (1) & 2 & 38.187 & I & 1 & (1) & 2 & 50.279 & III & 2 & (1) & 2 & 54.344 & III & 3 & (2) & 4 \\ 
28.88 (1) & II & 1 & (1) & 4 & 38.188 & II & 1 & (1) & 2 & 50.280 & III & 3 & (1) & 2 & 54.345 & III & 3 & (1) & 4 \\ 
28.88 (2) & II & 1 & (1) & 4 & 38.189 & III & 2 & (1) & 1 & 50.281 & III & 4 & (2) & 2 & 54.346 & IV & 3 & (1) & 4 \\ 
28.89 (1) & III & 1 & (1) & 2 & 38.190 & III & 2 & (1) & 1 & 50.282 & III & 3 & (1) & 2 & 54.348 & IV & 2 & (1) & 4 \\ 
28.89 (2) & III & 1 & (1) & 2 & 38.191 & III & 2 & (1) & 1 & 50.283 & III & 1 & (1) & 4 & 54.350 & IV & 3 & (1) & 4 \\ 
28.90 (1) & III & 2 & (1) & 2 & 38.193 & IV & 1 & (1) & 2 & 50.284 & IV & 2 & (1) & 4 & 55.353 & I & 5 & (2) & 2 \\ 
28.90 (2) & III & 2 & (1) & 2 & 39.195 & I & 1 & (1) & 2 & 50.287 & IV & 2 & (1) & 4 & 55.354 & II & 3 & (2) & 4 \\ 
28.91 (1) & III & 2 & (1) & 2 & 39.196 & II & 1 & (1) & 4 & 51.289 (1) & I & 5 & (2) & 2 & 55.355 & III & 1 & (1) & 4 \\ 
28.91 (2) & III & 3 & (2) & 2 & 39.197 & III & 2 & (1) & 2 & 51.289 (2) & I & 4 & (1) & 2 & 55.356 & III & 1 & (1) & 4 \\ 
28.92 (1) & IV & 3 & (1) & 2 & 39.198 & III & 1 & (1) & 2 & 51.290 (1) & II & 3 & (2) & 4 & 55.357 & III & 6 & (2, 2) & 2 \\ 
28.92 (2) & IV & 4 & (1) & 2 & 39.199 & III & 2 & (1) & 2 & 51.290 (2) & II & 3 & (2) & 4 & 55.358 & III & 3 & (1) & 2 \\ 
28.93 & IV & 2 & (1) & 4 & 39.201 & IV & 1 & (1) & 2 & 51.291 (1) & III & 1 & (1) & 2 & 55.359 & III & 1 & (1) & 4 \\ 
28.94 & IV & 1 & (1) & 4 & 47.249 (1) & I & 5 & (2) & 2 & 51.291 (2) & III & 1 & (1) & 2 & 55.360 & IV & 3 & (2) & 4 \\ 
28.96 & IV & 3 & (1) & 4 & 47.249 (2) & I & 5 & (2) & 2 & 51.292 (1) & III & 2 & (1) & 2 & 55.363 & IV & 5 & (2) & 2 \\ 
28.97 & IV & 3 & (1) & 2 & 47.249 (3) & I & 5 & (2) & 2 & 51.292 (2) & III & 2 & (1) & 2 & 57.377 & I & 3 & (1) & 4 \\ 
29.99 & I & 1 & (1) & 4 & 47.250 (1) & II & 5 & (2) & 2 & 51.293 (1) & III & 1 & (1) & 4 & 57.378 & II & 2 & (2) & 8 \\ 
29.100 & II & 1 & (1) & 8 & 47.250 (2) & II & 5 & (2) & 2 & 51.293 (2) & III & 1 & (1) & 4 & 57.379 & III & 3 & (1) & 4 \\ 
29.101 & III & 1 & (1) & 4 & 47.250 (3) & II & 5 & (2) & 2 & 51.294 (1) & III & 4 & (1) & 2 & 57.380 & III & 1 & (1) & 4 \\ 
29.102 & III & 1 & (1) & 4 & 47.251 (1) & III & 1 & (1) & 2 & 51.294 (2) & III & 5 & (2) & 2 & 57.381 & III & 1 & (1) & 4 \\ 
29.103 & III & 1 & (1) & 4 & 47.251 (2) & III & 1 & (1) & 2 & 51.295 (1) & III & 4 & (1) & 2 & 57.382 & III & 3 & (1) & 4 \\ 
29.104 & IV & 1 & (1) & 4 & 47.251 (3) & III & 1 & (1) & 2 & 51.295 (2) & III & 4 & (1) & 2 & 57.383 & III & 3 & (2) & 4 \\ 
29.106 & IV & 1 & (1) & 4 & 47.252 (1) & III & 9 & (1) & 1 & 51.296 (1) & III & 6 & (2, 2) & 2 & 57.384 & III & 2 & (1) & 4 \\ 
29.108 & IV & 1 & (1) & 4 & 47.252 (2) & III & 9 & (1) & 1 & 51.296 (2) & III & 6 & (1) & 2 & 57.385 & III & 2 & (1) & 4 \\ 
30.111 & I & 2 & (1) & 2 & 47.252 (3) & III & 10 & (2, 2) & 1 & 51.297 (1) & III & 2 & (1) & 2 & 57.387 & IV & 2 & (1) & 4 \\ 
30.112 & II & 1 & (1) & 4 & 47.253 (1) & III & 1 & (1) & 2 & 51.297 (2) & III & 3 & (1) & 2 & 57.388 & IV & 3 & (1) & 4 \\ 
30.113 & III & 1 & (1) & 2 & 47.253 (2) & III & 1 & (1) & 2 & 51.298 (1) & IV & 5 & (2) & 2 & 57.389 & IV & 3 & (1) & 4 \\ 
30.114 & III & 1 & (1) & 2 & 47.253 (3) & III & 1 & (1) & 2 & 51.298 (2) & IV & 4 & (1) & 2 & 59.405 & I & 2 & (1) & 4 \\ 
30.115 & III & 2 & (1) & 2 & 47.254 (1) & IV & 3 & (2) & 4 & 51.299 & IV & 2 & (1) & 4 & 59.406 & II & 2 & (2) & 4 \\ 
30.117 & IV & 2 & (1) & 4 & 47.254 (2) & IV & 3 & (2) & 4 & 51.300 & IV & 3 & (2) & 4 & 59.407 & III & 2 & (1) & 2 \\ 
\hline \hline 
\end{tabular}

%% file: spinful_MLGIII.tex
\begin{tabular}{cc|ccc||cc|ccc||cc|ccc||cc|ccc} 
\hline \hline 
59.408 & III & 1 & (1) & 4 & 90.97 & III & 2 & (1) & 2 & 125.363 & I & 4 & (1) & 4 & 162.76 & III & 3 & (1) & 2 \\ 
59.409 & III & 4 & (2) & 2 & 90.98 & III & 5 & (4) & 2 & 125.364 & II & 4 & (2) & 4 & 162.77 & III & 9 & (6) & 1 \\ 
59.410 & III & 3 & (1) & 2 & 90.99 & III & 1 & (1) & 4 & 125.365 & III & 4 & (1) & 2 & 164.85 & I & 8 & (1) & 1 \\ 
59.411 & III & 3 & (1) & 2 & 90.101 & IV & 4 & (1) & 2 & 125.366 & III & 2 & (1) & 4 & 164.86 & II & 6 & (2) & 2 \\ 
59.412 & IV & 2 & (1) & 4 & 99.163 & I & 3 & (1) & 2 & 125.367 & III & 2 & (1) & 4 & 164.87 & III & 3 & (1) & 2 \\ 
59.415 & IV & 2 & (1) & 4 & 99.164 & II & 3 & (1) & 2 & 125.368 & III & 2 & (1) & 4 & 164.88 & III & 4 & (1) & 2 \\ 
65.481 & I & 5 & (2) & 2 & 99.165 & III & 2 & (1) & 2 & 125.369 & III & 8 & (4) & 2 & 164.89 & III & 9 & (6) & 1 \\ 
65.482 & II & 4 & (2) & 2 & 99.166 & III & 1 & (1) & 2 & 125.370 & III & 4 & (1) & 2 & 168.109 & I & 9 & (6) & 1 \\ 
65.483 & III & 1 & (1) & 2 & 99.167 & III & 8 & (4) & 1 & 125.371 & III & 2 & (1) & 4 & 168.110 & II & 4 & (1) & 2 \\ 
65.484 & III & 1 & (1) & 2 & 99.169 & IV & 2 & (1) & 4 & 125.373 & IV & 4 & (1) & 4 & 168.111 & III & 4 & (1) & 1 \\ 
65.485 & III & 8 & (2, 2) & 1 & 100.171 & I & 4 & (1) & 2 & 127.387 & I & 8 & (4) & 2 & 174.133 & I & 14 & (3, 3) & 1 \\ 
65.486 & III & 6 & (1) & 1 & 100.172 & II & 2 & (1) & 4 & 127.388 & II & 5 & (4) & 4 & 174.134 & II & 7 & (3) & 2 \\ 
65.487 & III & 1 & (1) & 2 & 100.173 & III & 1 & (1) & 4 & 127.389 & III & 2 & (1) & 4 & 174.135 & III & 4 & (1) & 1 \\ 
65.489 & IV & 4 & (2) & 2 & 100.174 & III & 2 & (1) & 2 & 127.390 & III & 3 & (2) & 4 & 175.137 & I & 18 & (6, 6) & 1 \\ 
67.501 & I & 3 & (1) & 2 & 100.175 & III & 5 & (4) & 2 & 127.391 & III & 4 & (2) & 2 & 175.138 & II & 9 & (6) & 2 \\ 
67.502 & II & 3 & (2) & 4 & 100.177 & IV & 4 & (1) & 2 & 127.392 & III & 2 & (1) & 4 & 175.139 & III & 7 & (3) & 2 \\ 
67.503 & III & 1 & (1) & 2 & 111.251 & I & 3 & (1) & 2 & 127.393 & III & 10 & (4, 4) & 2 & 175.140 & III & 4 & (1) & 2 \\ 
67.504 & III & 2 & (1) & 2 & 111.252 & II & 3 & (1) & 2 & 127.394 & III & 2 & (1) & 4 & 175.141 & III & 6 & (2) & 1 \\ 
67.505 & III & 5 & (2) & 2 & 111.253 & III & 2 & (1) & 2 & 127.395 & III & 2 & (1) & 4 & 177.149 & I & 4 & (1) & 2 \\ 
67.506 & III & 5 & (1) & 2 & 111.254 & III & 1 & (1) & 2 & 127.397 & IV & 8 & (4) & 2 & 177.150 & II & 4 & (1) & 2 \\ 
67.507 & III & 2 & (1) & 2 & 111.255 & III & 8 & (4) & 1 & 129.411 & I & 4 & (1) & 4 & 177.151 & III & 4 & (1) & 1 \\ 
67.509 & IV & 3 & (1) & 2 & 111.257 & IV & 2 & (1) & 4 & 129.412 & II & 4 & (2) & 4 & 177.152 & III & 5 & (1) & 1 \\ 
75.1 & I & 8 & (4) & 1 & 113.267 & I & 4 & (1) & 2 & 129.413 & III & 2 & (1) & 4 & 177.153 & III & 9 & (6) & 1 \\ 
75.2 & II & 3 & (1) & 2 & 113.268 & II & 2 & (1) & 4 & 129.414 & III & 2 & (1) & 4 & 183.185 & I & 4 & (1) & 2 \\ 
75.3 & III & 2 & (1) & 2 & 113.269 & III & 1 & (1) & 4 & 129.415 & III & 2 & (1) & 4 & 183.186 & II & 4 & (1) & 2 \\ 
75.5 & IV & 4 & (2) & 2 & 113.270 & III & 2 & (1) & 2 & 129.416 & III & 4 & (1) & 2 & 183.187 & III & 5 & (1) & 1 \\ 
81.33 & I & 8 & (4) & 1 & 113.271 & III & 5 & (4) & 2 & 129.417 & III & 8 & (4) & 2 & 183.188 & III & 4 & (1) & 1 \\ 
81.34 & II & 3 & (1) & 2 & 113.273 & IV & 4 & (1) & 2 & 129.418 & III & 2 & (1) & 4 & 183.189 & III & 9 & (6) & 1 \\ 
81.35 & III & 2 & (1) & 2 & 115.283 & I & 3 & (1) & 2 & 129.419 & III & 4 & (1) & 2 & 187.209 & I & 7 & (3) & 2 \\ 
81.37 & IV & 4 & (2) & 2 & 115.284 & II & 3 & (1) & 2 & 129.421 & IV & 4 & (1) & 4 & 187.210 & II & 7 & (3) & 2 \\ 
83.43 & I & 16 & (4, 4) & 1 & 115.285 & III & 2 & (1) & 2 & 143.1 & I & 7 & (3) & 1 & 187.211 & III & 5 & (1) & 1 \\ 
83.44 & II & 8 & (4) & 2 & 115.286 & III & 1 & (1) & 2 & 143.2 & II & 4 & (1) & 2 & 187.212 & III & 5 & (1) & 1 \\ 
83.45 & III & 5 & (2) & 2 & 115.287 & III & 8 & (4) & 1 & 147.13 & I & 9 & (6) & 1 & 187.213 & III & 14 & (3, 3) & 1 \\ 
83.46 & III & 3 & (1) & 2 & 115.289 & IV & 2 & (1) & 4 & 147.14 & II & 6 & (2) & 2 & 189.221 & I & 7 & (3) & 2 \\ 
83.47 & III & 3 & (1) & 2 & 117.299 & I & 4 & (1) & 2 & 147.15 & III & 4 & (1) & 2 & 189.222 & II & 5 & (3) & 2 \\ 
83.49 & IV & 8 & (4) & 2 & 117.300 & II & 2 & (1) & 4 & 149.21 & I & 5 & (1) & 1 & 189.223 & III & 4 & (1) & 1 \\ 
85.59 & I & 8 & (4) & 2 & 117.301 & III & 1 & (1) & 4 & 149.22 & II & 4 & (1) & 2 & 189.224 & III & 4 & (1) & 1 \\ 
85.60 & II & 4 & (2) & 4 & 117.302 & III & 2 & (1) & 2 & 149.23 & III & 7 & (3) & 1 & 189.225 & III & 10 & (3, 3) & 1 \\ 
85.61 & III & 2 & (1) & 4 & 117.303 & III & 5 & (4) & 2 & 150.25 & I & 5 & (1) & 1 & 191.233 & I & 9 & (6) & 2 \\ 
85.62 & III & 4 & (1) & 2 & 117.305 & IV & 4 & (1) & 2 & 150.26 & II & 3 & (1) & 2 & 191.234 & II & 9 & (6) & 2 \\ 
85.63 & III & 4 & (1) & 2 & 123.339 & I & 8 & (4) & 2 & 150.27 & III & 5 & (3) & 1 & 191.235 & III & 4 & (1) & 2 \\ 
85.65 & IV & 4 & (1) & 4 & 123.340 & II & 8 & (4) & 2 & 156.49 & I & 5 & (1) & 1 & 191.236 & III & 7 & (3) & 2 \\ 
89.87 & I & 3 & (1) & 2 & 123.341 & III & 3 & (1) & 2 & 156.50 & II & 4 & (1) & 2 & 191.237 & III & 5 & (3) & 2 \\ 
89.88 & II & 3 & (1) & 2 & 123.342 & III & 5 & (2) & 2 & 156.51 & III & 7 & (3) & 1 & 191.238 & III & 8 & (1) & 1 \\ 
89.89 & III & 1 & (1) & 2 & 123.343 & III & 4 & (2) & 2 & 157.53 & I & 5 & (1) & 1 & 191.239 & III & 8 & (1) & 1 \\ 
89.90 & III & 8 & (4) & 1 & 123.344 & III & 3 & (1) & 2 & 157.54 & II & 3 & (1) & 2 & 191.240 & III & 18 & (6, 6) & 1 \\ 
89.91 & III & 2 & (1) & 2 & 123.345 & III & 16 & (4, 4) & 1 & 157.55 & III & 5 & (3) & 1 & 191.241 & III & 4 & (1) & 2 \\ 
89.93 & IV & 2 & (1) & 4 & 123.346 & III & 3 & (1) & 2 & 162.73 & I & 8 & (1) & 1 & ~ & ~ & ~ & ~ \\ 
90.95 & I & 4 & (1) & 2 & 123.347 & III & 3 & (1) & 2 & 162.74 & II & 6 & (2) & 2 & ~ & ~ & ~ & ~ \\ 
90.96 & II & 2 & (1) & 4 & 123.349 & IV & 5 & (4) & 4 & 162.75 & III & 4 & (1) & 2 & ~ & ~ & ~ & ~ \\ 
\hline \hline 
\end{tabular}

%% file: spinful_MLGIII_foot.tex
\newlength{\tabLspinfulMLGIII} 
\settowidth{\tabLspinfulMLGIII}{\input{spinful_MLGIII}} 
\begin{minipage}{\tabLspinfulMLGIII} 
\begin{flushleft} 
{\footnotesize $d$: Rank of the band structure group $\{{\rm BS}\}$\\ 
$X_{\rm BS}$: Symmetry-based indicators of band topology\\ 
$\nu_{\rm BS}$: Set of $\nu$ bands can only be isolated by band gaps if $\nu \in \nu_{\rm BS}\, \mathbb Z$ }
\end{flushleft}
\end{minipage}

%% file: spinful_nuSkip.tex
\begin{tabular}{cc|cc||cc|cc||cc|cc||cc|cc} 
\hline \hline 
\multicolumn{2}{c|}{MSG} & $\{ \nu \}_{\rm AI}$ & $\{ \nu \}_{\rm BS}$ & \multicolumn{2}{c|}{MSG} & $\{ \nu \}_{\rm AI}$ & $\{ \nu \}_{\rm BS}$ & \multicolumn{2}{c|}{MSG} & $\{ \nu \}_{\rm AI}$ & $\{ \nu \}_{\rm BS}$ & \multicolumn{2}{c|}{MSG} & $\{ \nu \}_{\rm AI}$ & $\{ \nu \}_{\rm BS}$\\ 
\hline 
196.6 & IV & $2\mathbb N \setminus \{2\}$ & $2\mathbb N $ & 209.51 & IV & $2\mathbb N \setminus \{2\}$ & $2\mathbb N $ & 219.88 & IV & $2\mathbb N \setminus \{2\}$ & $2\mathbb N $ & 228.134 & I & $4\mathbb N \setminus \{4\}$ & $4\mathbb N $ \\ 
198.11 & IV & $4\mathbb N \setminus \{4\}$ & $4\mathbb N $ & 210.55 & IV & $4\mathbb N \setminus \{4\}$ & $4\mathbb N $ & 220.89 & I & $2\mathbb N \setminus \{2, 4, 10\}$ & $2\mathbb N \setminus \{2\}$ & 228.136 & III & $4\mathbb N \setminus \{4\}$ & $4\mathbb N $ \\ 
199.12 & I & $2\mathbb N \setminus \{2\}$ & -- & 212.62 & IV & $4\mathbb N \setminus \{4\}$ & $4\mathbb N $ & 220.90 & II & $4\mathbb N \setminus \{4, 8, 20\}$ & $4\mathbb N \setminus \{4\}$ & 228.139 & IV & $4\mathbb N \setminus \{4\}$ & $4\mathbb N $ \\ 
199.13 & II & $4\mathbb N \setminus \{4\}$ & $4\mathbb N $ & 213.66 & IV & $4\mathbb N \setminus \{4\}$ & $4\mathbb N $ & 220.91 & III & $4\mathbb N \setminus \{4\}$ & -- & 230.145 & I & $4\mathbb N \setminus \{4\}$ & $4\mathbb N $ \\ 
202.25 & IV & $2\mathbb N \setminus \{2\}$ & $2\mathbb N $ & 214.68 & II & $4\mathbb N \setminus \{4\}$ & $4\mathbb N $ & 223.106 & III & $2\mathbb N \setminus \{2\}$ & $2\mathbb N $ & 230.146 & II & $8\mathbb N \setminus \{8\}$ & $8\mathbb N $ \\ 
203.29 & IV & $4\mathbb N \setminus \{4\}$ & $4\mathbb N $ & 214.69 & III & $2\mathbb N \setminus \{2\}$ & -- & 226.122 & I & $2\mathbb N \setminus \{2\}$ & $2\mathbb N $ & 230.147 & III & $4\mathbb N \setminus \{4\}$ & -- \\ 
206.39 & III & $4\mathbb N \setminus \{4\}$ & -- & 218.81 & I & $2\mathbb N \setminus \{2\}$ & $2\mathbb N $ & 226.124 & III & $2\mathbb N \setminus \{2\}$ & $2\mathbb N $ & 230.148 & III & $4\mathbb N \setminus \{4\}$ & -- \\ 
208.46 & III & $2\mathbb N \setminus \{2\}$ & $2\mathbb N $ & 219.85 & I & $2\mathbb N \setminus \{2\}$ & $2\mathbb N $ & 226.127 & IV & $2\mathbb N \setminus \{2\}$ & $2\mathbb N $ & ~ & ~ & ~ \\ 
\hline \hline 
\end{tabular}

%% file: spinful_nuSkip_foot.tex
\newlength{\tabLspinfulnuSkip} 
\settowidth{\tabLspinfulnuSkip}{\input{spinful_nuSkip}} 
\begin{minipage}{\tabLspinfulnuSkip} 
\begin{flushleft} 
{\footnotesize 
$\mathbb N$: The set of natural numbers\\ 
$\{ \nu \}_{\rm AI}$: Set of fillings for which physical atomic insulators are possible\\ 
$\{ \nu \}_{\rm BS}$: Set of fillings for which physical band structure are possible; a dash (--) indicates $\{ \nu \}_{\rm BS} = \{ \nu \}_{\rm AI}$ for that MSG}
\end{flushleft} 
\end{minipage}

%% file: spinful_pfeSM.tex
\begin{tabular}{cc|cc||cc|cc||cc|cc||cc|cc||cc|cc||cc|cc} 
\hline \hline 
\multicolumn{2}{c|}{MSG} & $\nu_{\rm P}$ & FS & \multicolumn{2}{c|}{MSG} & $\nu_{\rm P}$ & FS & \multicolumn{2}{c|}{MSG} & $\nu_{\rm P}$ & FS & \multicolumn{2}{c|}{MSG} & $\nu_{\rm P}$ & FS & \multicolumn{2}{c|}{MSG} & $\nu_{\rm P}$ & FS & \multicolumn{2}{c|}{MSG} & $\nu_{\rm P}$ & FS\\ 
\hline 
1.2 & II & $1$ & $2$ & 59.415 & IV & $2$ & $4$ & 100.173 & III & $2$ & $4$ & 125.366 & III & $2$ & $4$ & 139.540 & IV & $2$ & $4$ & 195.3 & IV & $2$ & $4$ \\ 
3.2 & II & $1$ & $2$ & 63.468 & IV & $2$ & $4$ & 101.180 & II & $2$ & $4$ & 125.368 & III & $2$ & $4$ & 140.542 & II & $2$ & $4$ & 196.4 & I & $1$ & $2$ \\ 
5.14 & II & $1$ & $2$ & 66.492 & II & $2$ & $4$ & 101.181 & III & $2$ & $4$ & 125.371 & III & $2$ & $4$ & 140.544 & III & $2$ & $4$ & 196.5 & II & $1$ & $2$ \\ 
6.19 & II & $1$ & $2$ & 66.497 & III & $2$ & $4$ & 101.184 & IV & $2$ & $4$ & 125.374 & IV & $4$ & $8$ & 140.546 & III & $2$ & $4$ & 197.7 & I & $1$ & $2$ \\ 
8.33 & II & $1$ & $2$ & 67.502 & II & $2$ & $4$ & 102.188 & II & $2$ & $4$ & 126.375 & I & $2$ & $4$ & 140.549 & III & $2$ & $4$ & 197.8 & II & $1$ & $2$ \\ 
11.51 & II & $2$ & $4$ & 68.512 & II & $2$ & $4$ & 102.189 & III & $2$ & $4$ & 126.376 & II & $2$ & $4$ & 141.551 & I & $2$ & $4$ & 201.19 & II & $2$ & $4$ \\ 
11.57 & IV & $2$ & $4$ & 68.517 & III & $2$ & $4$ & 102.194 & IV & $2$ & $4$ & 126.378 & III & $2$ & $4$ & 141.552 & II & $2$ & $4$ & 201.20 & III & $2$ & $4$ \\ 
13.66 & II & $2$ & $4$ & 68.520 & IV & $2$ & $4$ & 103.196 & II & $2$ & $4$ & 126.379 & III & $2$ & $4$ & 141.555 & III & $2$ & $4$ & 203.27 & II & $2$ & $4$ \\ 
13.73 & IV & $2$ & $4$ & 70.528 & II & $2$ & $4$ & 103.197 & III & $2$ & $4$ & 126.380 & III & $2$ & $4$ & 141.558 & III & $2$ & $4$ & 203.28 & III & $2$ & $4$ \\ 
14.76 & II & $2$ & $4$ & 70.531 & III & $2$ & $4$ & 103.198 & III & $2$ & $4$ & 126.382 & III & $2$ & $4$ & 141.559 & III & $2$ & $4$ & 207.40 & I & $1$ & $2$ \\ 
15.86 & II & $2$ & $4$ & 71.538 & IV & $2$ & $4$ & 104.204 & II & $2$ & $4$ & 126.383 & III & $2$ & $4$ & 143.2 & II & $1$ & $2$ & 207.41 & II & $1$ & $2$ \\ 
16.1 & I & $1$ & $2$ & 72.540 & II & $2$ & $4$ & 104.205 & III & $2$ & $4$ & 126.385 & IV & $4$ & $8$ & 144.5 & II & $3$ & $2$ & 207.42 & III & $1$ & $2$ \\ 
16.2 & II & $1$ & $2$ & 72.545 & III & $2$ & $4$ & 104.206 & III & $2$ & $4$ & 127.392 & III & $2$ & $4$ & 145.8 & II & $3$ & $2$ & 207.43 & IV & $2$ & $4$ \\ 
16.5 & IV & $2$ & $4$ & 74.555 & II & $2$ & $4$ & 105.212 & II & $2$ & $4$ & 127.395 & III & $2$ & $4$ & 146.11 & II & $1$ & $2$ & 208.47 & IV & $2$ & $4$ \\ 
16.6 & IV & $2$ & $4$ & 75.2 & II & $1$ & $2$ & 105.214 & III & $2$ & $4$ & 128.403 & III & $2$ & $4$ & 149.22 & II & $1$ & $2$ & 209.48 & I & $1$ & $2$ \\ 
18.17 & II & $2$ & $4$ & 75.3 & III & $1$ & $2$ & 105.216 & IV & $2$ & $4$ & 128.404 & III & $2$ & $4$ & 150.26 & II & $1$ & $2$ & 209.49 & II & $1$ & $2$ \\ 
21.38 & I & $1$ & $2$ & 79.26 & II & $1$ & $2$ & 105.218 & IV & $2$ & $4$ & 128.407 & III & $2$ & $4$ & 151.30 & II & $3$ & $2$ & 209.50 & III & $1$ & $2$ \\ 
21.39 & II & $1$ & $2$ & 79.27 & III & $1$ & $2$ & 107.232 & IV & $2$ & $4$ & 129.411 & I & $2$ & $4$ & 152.34 & II & $3$ & $2$ & 211.56 & I & $1$ & $2$ \\ 
22.45 & I & $1$ & $2$ & 84.52 & II & $2$ & $4$ & 108.234 & II & $2$ & $4$ & 129.420 & IV & $4$ & $8$ & 153.38 & II & $3$ & $2$ & 211.57 & II & $1$ & $2$ \\ 
22.46 & II & $1$ & $2$ & 84.56 & IV & $2$ & $4$ & 108.235 & III & $2$ & $4$ & 129.421 & IV & $2$ & $4$ & 154.42 & II & $3$ & $2$ & 211.58 & III & $1$ & $2$ \\ 
23.49 & I & $1$ & $2$ & 84.58 & IV & $2$ & $4$ & 109.240 & II & $2$ & $4$ & 130.424 & II & $4$ & $8$ & 155.46 & II & $1$ & $2$ & 218.82 & II & $2$ & $4$ \\ 
23.50 & II & $1$ & $2$ & 85.60 & II & $2$ & $4$ & 109.242 & III & $2$ & $4$ & 131.435 & I & $2$ & $4$ & 156.50 & II & $1$ & $2$ & 218.83 & III & $2$ & $4$ \\ 
25.61 & IV & $2$ & $4$ & 85.65 & IV & $2$ & $4$ & 111.256 & IV & $2$ & $4$ & 131.436 & II & $2$ & $4$ & 157.54 & II & $1$ & $2$ & 220.90 & II & $2$ & $4$ \\ 
25.63 & IV & $2$ & $4$ & 85.66 & IV & $2$ & $4$ & 112.260 & II & $2$ & $4$ & 131.439 & III & $2$ & $4$ & 160.66 & II & $1$ & $2$ & 220.91 & III & $2$ & $4$ \\ 
25.64 & IV & $2$ & $4$ & 86.68 & II & $2$ & $4$ & 112.262 & III & $2$ & $4$ & 131.442 & III & $2$ & $4$ & 163.80 & II & $2$ & $4$ & 222.98 & I & $2$ & $4$ \\ 
25.65 & IV & $2$ & $4$ & 86.74 & IV & $2$ & $4$ & 113.274 & IV & $2$ & $4$ & 131.443 & III & $2$ & $4$ & 165.92 & II & $2$ & $4$ & 222.99 & II & $2$ & $4$ \\ 
26.67 & II & $2$ & $4$ & 88.82 & II & $2$ & $4$ & 114.278 & III & $2$ & $4$ & 131.445 & IV & $4$ & $8$ & 167.104 & II & $2$ & $4$ & 222.100 & III & $2$ & $4$ \\ 
27.79 & II & $2$ & $4$ & 89.87 & I & $1$ & $2$ & 115.288 & IV & $2$ & $4$ & 132.447 & I & $2$ & $4$ & 168.110 & II & $1$ & $2$ & 222.101 & III & $2$ & $4$ \\ 
30.112 & II & $2$ & $4$ & 89.88 & II & $1$ & $2$ & 115.289 & IV & $2$ & $4$ & 132.448 & II & $2$ & $4$ & 171.122 & II & $3$ & $2$ & 222.102 & III & $2$ & $4$ \\ 
31.124 & II & $2$ & $4$ & 89.89 & III & $1$ & $2$ & 115.290 & IV & $2$ & $4$ & 132.450 & III & $2$ & $4$ & 172.126 & II & $3$ & $2$ & 223.104 & I & $2$ & $4$ \\ 
32.136 & II & $2$ & $4$ & 89.91 & III & $1$ & $2$ & 116.292 & II & $2$ & $4$ & 132.452 & III & $2$ & $4$ & 174.134 & II & $1$ & $2$ & 223.105 & II & $2$ & $4$ \\ 
34.157 & II & $2$ & $4$ & 89.93 & IV & $2$ & $4$ & 116.293 & III & $2$ & $4$ & 132.455 & III & $2$ & $4$ & 176.144 & II & $2$ & $4$ & 223.107 & III & $2$ & $4$ \\ 
35.169 & IV & $2$ & $4$ & 89.94 & IV & $2$ & $4$ & 117.300 & II & $2$ & $4$ & 132.458 & IV & $4$ & $8$ & 177.149 & I & $1$ & $2$ & 223.108 & III & $2$ & $4$ \\ 
35.171 & IV & $2$ & $4$ & 90.96 & II & $2$ & $4$ & 117.301 & III & $2$ & $4$ & 134.471 & I & $2$ & $4$ & 177.150 & II & $1$ & $2$ & 224.110 & I & $2$ & $4$ \\ 
36.173 & II & $2$ & $4$ & 90.99 & III & $2$ & $4$ & 118.308 & II & $2$ & $4$ & 134.472 & II & $2$ & $4$ & 180.167 & I & $3$ & $2$ & 224.111 & II & $2$ & $4$ \\ 
37.181 & II & $2$ & $4$ & 93.126 & IV & $2$ & $4$ & 118.309 & III & $2$ & $4$ & 134.474 & III & $2$ & $4$ & 180.168 & II & $3$ & $2$ & 224.112 & III & $2$ & $4$ \\ 
43.225 & II & $2$ & $4$ & 94.128 & II & $2$ & $4$ & 122.334 & II & $2$ & $4$ & 134.476 & III & $2$ & $4$ & 181.173 & I & $3$ & $2$ & 224.114 & III & $2$ & $4$ \\ 
48.258 & II & $2$ & $4$ & 94.131 & III & $2$ & $4$ & 122.336 & III & $2$ & $4$ & 134.479 & III & $2$ & $4$ & 181.174 & II & $3$ & $2$ & 225.121 & IV & $2$ & $4$ \\ 
48.261 & III & $2$ & $4$ & 97.151 & I & $1$ & $2$ & 124.352 & II & $2$ & $4$ & 135.484 & II & $4$ & $8$ & 183.190 & IV & $2$ & $4$ & 226.123 & II & $2$ & $4$ \\ 
49.266 & II & $2$ & $4$ & 97.152 & II & $1$ & $2$ & 124.354 & III & $2$ & $4$ & 136.495 & I & $2$ & $4$ & 184.192 & II & $2$ & $4$ & 226.125 & III & $2$ & $4$ \\ 
49.271 & III & $2$ & $4$ & 97.153 & III & $1$ & $2$ & 124.355 & III & $2$ & $4$ & 136.500 & III & $2$ & $4$ & 185.198 & II & $2$ & $4$ & 226.126 & III & $2$ & $4$ \\ 
50.278 & II & $2$ & $4$ & 97.155 & III & $1$ & $2$ & 124.356 & III & $2$ & $4$ & 136.503 & III & $2$ & $4$ & 186.204 & II & $2$ & $4$ & 227.128 & I & $2$ & $4$ \\ 
50.283 & III & $2$ & $4$ & 99.168 & IV & $2$ & $4$ & 124.358 & III & $2$ & $4$ & 136.504 & IV & $4$ & $8$ & 192.244 & II & $2$ & $4$ & 227.129 & II & $2$ & $4$ \\ 
53.329 & III & $2$ & $4$ & 99.169 & IV & $2$ & $4$ & 124.359 & III & $2$ & $4$ & 137.507 & I & $2$ & $4$ & 192.251 & III & $2$ & $4$ & 227.130 & III & $2$ & $4$ \\ 
55.359 & III & $2$ & $4$ & 99.170 & IV & $2$ & $4$ & 125.363 & I & $2$ & $4$ & 137.511 & III & $2$ & $4$ & 195.1 & I & $1$ & $2$ & 227.132 & III & $2$ & $4$ \\ 
58.399 & III & $2$ & $4$ & 100.172 & II & $2$ & $4$ & 125.364 & II & $2$ & $4$ & 137.515 & III & $2$ & $4$ & 195.2 & II & $1$ & $2$ & ~ & ~ & ~ & ~ \\ 
\hline \hline 
\end{tabular}

%% file: spinful_pfeSM_foot.tex
\newlength{\tabLspinfulpfeSM} 
\settowidth{\tabLspinfulpfeSM}{\input{spinful_pfeSM}} 
\begin{minipage}{\tabLspinfulpfeSM} 
\begin{flushleft} 
{\footnotesize 
$\nu_{\rm P}$: The physically achievable fillings are quantized in units of $\nu_{\rm P}$\\
FS: Dimension of the irrep(s) dissected by the Fermi energy when the physical filling is an odd-integer-multiple of $\nu_{\rm P}$
}
\end{flushleft} 
\end{minipage}

%% file: spinful_mfeSM.tex
\begin{tabular}{cc|c||cc|c||cc|c||cc|c||cc|c||cc|c} 
\hline \hline 
\multicolumn{2}{c|}{MSG} & $\nu_{\rm P}$ & \multicolumn{2}{c|}{MSG} & $\nu_{\rm P}$ & \multicolumn{2}{c|}{MSG} & $\nu_{\rm P}$ & \multicolumn{2}{c|}{MSG} & $\nu_{\rm P}$ & \multicolumn{2}{c|}{MSG} & $\nu_{\rm P}$ & \multicolumn{2}{c|}{MSG} & $\nu_{\rm P}$\\ 
\hline 
4.8 & II & $2$ & 53.325 & III & $2$ & 77.16 & IV & $2$ & 102.192 & IV & $4$ & 138.520 & II & $4$ & 206.38 & II & $4$ \\ 
7.25 & II & $2$ & 54.338 & II & $4$ & 77.18 & IV & $2$ & 105.211 & I & $2$ & 138.530 & IV & $4$ & 208.44 & I & $2$ \\ 
9.38 & II & $2$ & 54.349 & IV & $4$ & 78.20 & II & $4$ & 105.217 & IV & $4$ & 141.553 & III & $2$ & 208.45 & II & $2$ \\ 
14.77 & III & $2$ & 56.366 & II & $4$ & 80.30 & II & $2$ & 106.220 & II & $4$ & 141.554 & III & $2$ & 210.52 & I & $2$ \\ 
14.78 & III & $2$ & 57.378 & II & $4$ & 84.54 & III & $2$ & 109.239 & I & $2$ & 141.560 & IV & $4$ & 210.53 & II & $2$ \\ 
17.8 & II & $2$ & 57.390 & IV & $4$ & 85.61 & III & $2$ & 109.244 & IV & $4$ & 142.562 & II & $4$ & 210.54 & III & $2$ \\ 
19.26 & II & $4$ & 59.405 & I & $2$ & 86.69 & III & $2$ & 110.246 & II & $4$ & 142.564 & III & $4$ & 212.60 & II & $4$ \\ 
20.32 & II & $2$ & 59.413 & IV & $4$ & 86.70 & III & $2$ & 119.320 & IV & $2$ & 142.569 & III & $4$ & 213.64 & II & $4$ \\ 
24.54 & II & $2$ & 59.414 & IV & $4$ & 88.83 & III & $2$ & 120.322 & II & $2$ & 158.58 & II & $2$ & 214.67 & I & $2$ \\ 
28.88 & II & $2$ & 60.418 & II & $4$ & 88.84 & III & $2$ & 120.323 & III & $2$ & 159.62 & II & $2$ & 214.68 & II & $2$ \\ 
29.100 & II & $4$ & 60.429 & IV & $4$ & 91.104 & II & $4$ & 122.335 & III & $2$ & 161.70 & II & $2$ & 216.77 & IV & $2$ \\ 
33.145 & II & $4$ & 61.434 & II & $4$ & 92.112 & II & $4$ & 125.367 & III & $2$ & 169.114 & II & $6$ & 219.86 & II & $2$ \\ 
38.194 & IV & $2$ & 61.435 & III & $4$ & 93.119 & I & $2$ & 129.415 & III & $2$ & 170.118 & II & $6$ & 219.87 & III & $2$ \\ 
39.196 & II & $2$ & 61.437 & III & $4$ & 93.120 & II & $2$ & 130.426 & III & $4$ & 173.130 & II & $2$ & 224.113 & III & $2$ \\ 
40.204 & II & $2$ & 62.442 & II & $4$ & 93.124 & IV & $2$ & 131.437 & III & $2$ & 178.156 & II & $6$ & 227.131 & III & $2$ \\ 
41.212 & II & $2$ & 62.455 & IV & $4$ & 94.127 & I & $2$ & 132.449 & III & $2$ & 179.162 & II & $6$ & 227.133 & IV & $4$ \\ 
44.233 & IV & $2$ & 64.471 & III & $2$ & 94.134 & IV & $2$ & 133.460 & II & $4$ & 182.180 & II & $2$ & 228.135 & II & $4$ \\ 
45.236 & II & $2$ & 64.473 & III & $2$ & 95.136 & II & $4$ & 133.467 & III & $4$ & 188.216 & II & $2$ & 228.137 & III & $4$ \\ 
46.242 & II & $2$ & 64.477 & III & $2$ & 96.144 & II & $4$ & 133.470 & IV & $4$ & 190.228 & II & $2$ & 228.138 & III & $4$ \\ 
48.257 & I & $2$ & 68.511 & I & $2$ & 98.157 & I & $2$ & 134.473 & III & $2$ & 198.10 & II & $4$ & 230.146 & II & $4$ \\ 
50.277 & I & $2$ & 70.527 & I & $2$ & 98.158 & II & $2$ & 134.475 & III & $2$ & 199.13 & II & $2$ & 230.149 & III & $4$ \\ 
50.286 & IV & $4$ & 73.549 & II & $4$ & 98.161 & III & $2$ & 134.480 & IV & $4$ & 201.18 & I & $2$ & ~ & ~ & ~ \\ 
50.288 & IV & $4$ & 74.557 & III & $2$ & 101.179 & I & $2$ & 135.491 & III & $4$ & 203.26 & I & $2$ & ~ & ~ & ~ \\ 
52.306 & II & $4$ & 76.8 & II & $4$ & 101.186 & IV & $4$ & 137.516 & IV & $4$ & 205.34 & II & $4$ & ~ & ~ & ~ \\ 
53.323 & III & $2$ & 77.14 & II & $2$ & 102.187 & I & $2$ & 137.517 & IV & $4$ & 205.35 & III & $4$ & ~ & ~ & ~ \\ 
\hline \hline 
\end{tabular}

%% file: spinful_mfeSM_foot.tex
\newlength{\tabLspinfulmfeSM} 
\settowidth{\tabLspinfulmfeSM}{\input{spinful_mfeSM}} 
\begin{minipage}{\tabLspinfulmfeSM} 
\begin{flushleft} 
{\footnotesize 
$\nu_{\rm P}$: The physically achievable fillings are quantized in units of $\nu_{\rm P}$
}
\end{flushleft} 
\end{minipage}

%% file: spinful_feM.tex
\begin{tabular}{cc|cc||cc|cc||cc|cc||cc|cc||cc|cc||cc|cc} 
\hline \hline 
\multicolumn{2}{c|}{MSG} & $\nu_{\rm P}$ & Deg & \multicolumn{2}{c|}{MSG} & $\nu_{\rm P}$ & Deg & \multicolumn{2}{c|}{MSG} & $\nu_{\rm P}$ & Deg & \multicolumn{2}{c|}{MSG} & $\nu_{\rm P}$ & Deg & \multicolumn{2}{c|}{MSG} & $\nu_{\rm P}$ & Deg & \multicolumn{2}{c|}{MSG} & $\nu_{\rm P}$ & Deg\\ 
\hline 
2.5 & II & $1$ & $3$ & 59.406 & II & $2$ & $1$ & 99.166 & III & $1$ & $1$ & 123.350 & IV & $2$ & $1$ & 147.14 & II & $1$ & $3$ & 202.22 & I & $1$ & $1$ \\ 
2.6 & III & $1$ & $3$ & 59.408 & III & $2$ & $1$ & 107.227 & I & $1$ & $1$ & 125.373 & IV & $2$ & $1$ & 147.15 & III & $1$ & $3$ & 202.23 & II & $1$ & $3$ \\ 
10.43 & II & $1$ & $3$ & 59.416 & IV & $2$ & $1$ & 107.228 & II & $1$ & $1$ & 126.386 & IV & $2$ & $1$ & 148.18 & II & $1$ & $3$ & 202.24 & III & $1$ & $3$ \\ 
10.44 & III & $1$ & $3$ & 63.458 & II & $2$ & $1$ & 107.229 & III & $1$ & $1$ & 127.388 & II & $2$ & $1$ & 148.19 & III & $1$ & $3$ & 204.30 & I & $1$ & $1$ \\ 
10.45 & III & $1$ & $3$ & 63.460 & III & $2$ & $1$ & 107.230 & III & $1$ & $1$ & 127.389 & III & $2$ & $1$ & 162.74 & II & $1$ & $3$ & 204.31 & II & $1$ & $3$ \\ 
12.59 & II & $1$ & $3$ & 64.470 & II & $2$ & $1$ & 111.251 & I & $1$ & $1$ & 127.390 & III & $2$ & $1$ & 162.75 & III & $1$ & $3$ & 204.32 & III & $1$ & $3$ \\ 
12.60 & III & $1$ & $3$ & 64.472 & III & $2$ & $1$ & 111.252 & II & $1$ & $1$ & 127.394 & III & $2$ & $1$ & 162.76 & III & $1$ & $3$ & 215.70 & I & $1$ & $1$ \\ 
12.61 & III & $1$ & $3$ & 65.481 & I & $1$ & $1$ & 111.253 & III & $1$ & $1$ & 128.400 & II & $2$ & $1$ & 164.86 & II & $1$ & $3$ & 215.71 & II & $1$ & $1$ \\ 
25.57 & I & $1$ & $1$ & 65.482 & II & $1$ & $3$ & 111.254 & III & $1$ & $1$ & 128.401 & III & $2$ & $1$ & 164.87 & III & $1$ & $3$ & 215.72 & III & $1$ & $1$ \\ 
25.58 & II & $1$ & $1$ & 65.483 & III & $1$ & $3$ & 111.257 & IV & $2$ & $1$ & 128.402 & III & $2$ & $1$ & 164.88 & III & $1$ & $3$ & 215.73 & IV & $2$ & $1$ \\ 
35.165 & I & $1$ & $1$ & 65.484 & III & $1$ & $3$ & 111.258 & IV & $2$ & $1$ & 128.406 & III & $2$ & $1$ & 166.98 & II & $1$ & $3$ & 216.74 & I & $1$ & $1$ \\ 
35.166 & II & $1$ & $1$ & 65.487 & III & $1$ & $3$ & 112.266 & IV & $2$ & $1$ & 129.412 & II & $2$ & $1$ & 166.99 & III & $1$ & $3$ & 216.75 & II & $1$ & $1$ \\ 
38.187 & I & $1$ & $1$ & 65.488 & IV & $2$ & $1$ & 113.268 & II & $2$ & $1$ & 129.413 & III & $2$ & $1$ & 166.100 & III & $1$ & $3$ & 216.76 & III & $1$ & $1$ \\ 
38.188 & II & $1$ & $1$ & 65.490 & IV & $2$ & $1$ & 113.269 & III & $2$ & $1$ & 129.414 & III & $2$ & $1$ & 175.138 & II & $1$ & $3$ & 217.78 & I & $1$ & $1$ \\ 
42.219 & I & $1$ & $1$ & 67.510 & IV & $2$ & $1$ & 114.276 & II & $2$ & $1$ & 129.418 & III & $2$ & $1$ & 175.139 & III & $1$ & $3$ & 217.79 & II & $1$ & $1$ \\ 
42.220 & II & $1$ & $1$ & 69.521 & I & $1$ & $1$ & 114.277 & III & $2$ & $1$ & 129.422 & IV & $2$ & $1$ & 175.140 & III & $1$ & $3$ & 217.80 & III & $1$ & $1$ \\ 
44.229 & I & $1$ & $1$ & 69.522 & II & $1$ & $3$ & 115.283 & I & $1$ & $1$ & 131.444 & IV & $2$ & $1$ & 183.185 & I & $1$ & $1$ & 218.84 & IV & $2$ & $1$ \\ 
44.230 & II & $1$ & $1$ & 69.523 & III & $1$ & $3$ & 115.284 & II & $1$ & $1$ & 131.446 & IV & $2$ & $1$ & 183.186 & II & $1$ & $1$ & 221.92 & I & $1$ & $1$ \\ 
47.249 & I & $1$ & $1$ & 69.525 & III & $1$ & $3$ & 115.285 & III & $1$ & $1$ & 132.456 & IV & $2$ & $1$ & 187.209 & I & $1$ & $1$ & 221.93 & II & $1$ & $3$ \\ 
47.250 & II & $1$ & $3$ & 71.533 & I & $1$ & $1$ & 115.286 & III & $1$ & $1$ & 134.482 & IV & $2$ & $1$ & 187.210 & II & $1$ & $1$ & 221.94 & III & $1$ & $3$ \\ 
47.251 & III & $1$ & $3$ & 71.534 & II & $1$ & $3$ & 119.315 & I & $1$ & $1$ & 136.496 & II & $2$ & $1$ & 189.221 & I & $1$ & $1$ & 221.95 & III & $1$ & $1$ \\ 
47.253 & III & $1$ & $3$ & 71.535 & III & $1$ & $3$ & 119.316 & II & $1$ & $1$ & 136.497 & III & $2$ & $1$ & 189.222 & II & $1$ & $1$ & 221.96 & III & $1$ & $3$ \\ 
47.254 & IV & $2$ & $1$ & 71.537 & III & $1$ & $3$ & 119.317 & III & $1$ & $1$ & 136.498 & III & $2$ & $1$ & 191.233 & I & $1$ & $1$ & 221.97 & IV & $2$ & $1$ \\ 
47.255 & IV & $2$ & $1$ & 81.34 & II & $1$ & $1$ & 119.318 & III & $1$ & $1$ & 136.502 & III & $2$ & $1$ & 191.234 & II & $1$ & $3$ & 222.103 & IV & $2$ & $1$ \\ 
47.256 & IV & $2$ & $1$ & 81.35 & III & $1$ & $1$ & 121.327 & I & $1$ & $1$ & 136.506 & IV & $2$ & $1$ & 191.235 & III & $1$ & $3$ & 223.109 & IV & $2$ & $1$ \\ 
48.264 & IV & $2$ & $1$ & 82.40 & II & $1$ & $1$ & 121.328 & II & $1$ & $1$ & 137.508 & II & $2$ & $1$ & 191.236 & III & $1$ & $3$ & 224.115 & IV & $2$ & $1$ \\ 
50.287 & IV & $2$ & $1$ & 82.41 & III & $1$ & $1$ & 121.329 & III & $1$ & $1$ & 137.509 & III & $2$ & $1$ & 191.237 & III & $1$ & $3$ & 225.116 & I & $1$ & $1$ \\ 
51.290 & II & $2$ & $1$ & 83.44 & II & $1$ & $3$ & 121.330 & III & $1$ & $1$ & 137.510 & III & $2$ & $1$ & 191.241 & III & $1$ & $3$ & 225.117 & II & $1$ & $3$ \\ 
51.293 & III & $2$ & $1$ & 83.45 & III & $1$ & $1$ & 123.339 & I & $1$ & $1$ & 137.514 & III & $2$ & $1$ & 191.242 & IV & $2$ & $1$ & 225.118 & III & $1$ & $3$ \\ 
51.302 & IV & $2$ & $1$ & 83.46 & III & $1$ & $3$ & 123.340 & II & $1$ & $3$ & 137.518 & IV & $2$ & $1$ & 193.254 & II & $2$ & $1$ & 225.119 & III & $1$ & $1$ \\ 
53.322 & II & $2$ & $1$ & 83.47 & III & $1$ & $3$ & 123.341 & III & $1$ & $3$ & 139.531 & I & $1$ & $1$ & 193.256 & III & $2$ & $1$ & 225.120 & III & $1$ & $3$ \\ 
53.324 & III & $2$ & $1$ & 87.76 & II & $1$ & $3$ & 123.342 & III & $1$ & $1$ & 139.532 & II & $1$ & $3$ & 194.264 & II & $2$ & $1$ & 229.140 & I & $1$ & $1$ \\ 
55.354 & II & $2$ & $1$ & 87.77 & III & $1$ & $1$ & 123.343 & III & $1$ & $1$ & 139.533 & III & $1$ & $3$ & 194.267 & III & $2$ & $1$ & 229.141 & II & $1$ & $3$ \\ 
55.355 & III & $2$ & $1$ & 87.78 & III & $1$ & $3$ & 123.344 & III & $1$ & $3$ & 139.534 & III & $1$ & $1$ & 200.14 & I & $1$ & $1$ & 229.142 & III & $1$ & $3$ \\ 
55.356 & III & $2$ & $1$ & 87.79 & III & $1$ & $3$ & 123.346 & III & $1$ & $3$ & 139.535 & III & $1$ & $1$ & 200.15 & II & $1$ & $3$ & 229.143 & III & $1$ & $1$ \\ 
58.394 & II & $2$ & $1$ & 99.163 & I & $1$ & $1$ & 123.347 & III & $1$ & $3$ & 139.536 & III & $1$ & $3$ & 200.16 & III & $1$ & $3$ & 229.144 & III & $1$ & $3$ \\ 
58.395 & III & $2$ & $1$ & 99.164 & II & $1$ & $1$ & 123.348 & IV & $2$ & $1$ & 139.538 & III & $1$ & $3$ & 200.17 & IV & $2$ & $1$ & ~ & ~ & ~ & ~ \\ 
58.396 & III & $2$ & $1$ & 99.165 & III & $1$ & $1$ & 123.349 & IV & $2$ & $1$ & 139.539 & III & $1$ & $3$ & 201.21 & IV & $2$ & $1$ & ~ & ~ & ~ & ~ \\ 
\hline \hline 
\end{tabular}

%% file: spinful_feM_foot.tex
\newlength{\tabLspinfulfeM} 
\settowidth{\tabLspinfulfeM}{\input{spinful_feM}} 
\begin{minipage}{\tabLspinfulfeM} 
\begin{flushleft} 
{\footnotesize 
$\nu_{\rm P}$: The physically achievable fillings are quantized in units of $\nu_{\rm P}$\\
Deg: Dimension of the manifold on which band sticking occurs, which enforces metallic behavior at fillings which are odd-integer multiples of $\nu_{\rm P}$
}
\end{flushleft} 
\end{minipage}

%% file: SOC_feSM.tex
\begin{tabular}{c|cc||c|cc||c|cc||c|cc||c|cc||c|cc} 
\hline \hline 
SG & $\nu_{\rm P}$ & FS & SG & $\nu_{\rm P}$ & FS & SG & $\nu_{\rm P}$ & FS & SG & $\nu_{\rm P}$ & FS & SG & $\nu_{\rm P}$ & FS & SG & $\nu_{\rm P}$ & FS\\ 
\hline 
1 & $1$ & M$_3$ $\rightarrow$ p & 40 & $2$ & M$_1$ $\rightarrow$ m & 79 & $1$ & M$_3$ $\rightarrow$ p & 118 & $2$ & M$_1$ $\rightarrow$ p & 157 & $1$ & M$_3$ $\rightarrow$ p & 196 & $1$ & M$_3$ $\rightarrow$ p \\ 
2 & $1$ & M$_3$ $\rightarrow$ M$_3$ & 41 & $2$ & M$_1$ $\rightarrow$ m & 80 & $2$ & p $\rightarrow$ m & 119 & $1$ & M$_3$ $\rightarrow$ M$_1$ & 158 & $2$ & M$_1$ $\rightarrow$ m & 197 & $1$ & M$_3$ $\rightarrow$ p \\ 
3 & $1$ & M$_3$ $\rightarrow$ p & 42 & $1$ & M$_3$ $\rightarrow$ M$_1$ & 81 & $1$ & M$_3$ $\rightarrow$ M$_1$ & 120 & $2$ & M$_1$ $\rightarrow$ m & 159 & $2$ & M$_1$ $\rightarrow$ m & 198 & $4$ & p $\rightarrow$ m \\ 
4 & $2$ & M$_2$ $\rightarrow$ m & 43 & $2$ & M$_1$ $\rightarrow$ p & 82 & $1$ & M$_3$ $\rightarrow$ M$_1$ & 121 & $1$ & M$_3$ $\rightarrow$ M$_1$ & 160 & $1$ & M$_3$ $\rightarrow$ p & 199 & $2$ & p $\rightarrow$ m \\ 
5 & $1$ & M$_3$ $\rightarrow$ p & 44 & $1$ & M$_3$ $\rightarrow$ M$_1$ & 83 & $1$ & M$_3$ $\rightarrow$ M$_3$ & 122 & $2$ & M$_1$ $\rightarrow$ p & 161 & $2$ & M$_1$ $\rightarrow$ m & 200 & $1$ & M$_3$ $\rightarrow$ M$_3$ \\ 
6 & $1$ & M$_3$ $\rightarrow$ p & 45 & $2$ & M$_1$ $\rightarrow$ m & 84 & $2$ & p $\rightarrow$ p & 123 & $1$ & M$_3$ $\rightarrow$ M$_3$ & 162 & $1$ & M$_3$ $\rightarrow$ M$_3$ & 201 & $2$ & M$_1$ $\rightarrow$ p \\ 
7 & $2$ & M$_1$ $\rightarrow$ m & 46 & $2$ & M$_1$ $\rightarrow$ m & 85 & $2$ & M$_1$ $\rightarrow$ p & 124 & $2$ & M$_1$ $\rightarrow$ p & 163 & $2$ & M$_1$ $\rightarrow$ p & 202 & $1$ & M$_3$ $\rightarrow$ M$_3$ \\ 
8 & $1$ & M$_3$ $\rightarrow$ p & 47 & $1$ & M$_3$ $\rightarrow$ M$_3$ & 86 & $2$ & M$_1$ $\rightarrow$ p & 125 & $2$ & M$_1$ $\rightarrow$ p & 164 & $1$ & M$_3$ $\rightarrow$ M$_3$ & 203 & $2$ & M$_1$ $\rightarrow$ p \\ 
9 & $2$ & M$_1$ $\rightarrow$ m & 48 & $2$ & M$_1$ $\rightarrow$ p & 87 & $1$ & M$_3$ $\rightarrow$ M$_3$ & 126 & $2$ & M$_1$ $\rightarrow$ p & 165 & $2$ & M$_1$ $\rightarrow$ p & 204 & $1$ & M$_3$ $\rightarrow$ M$_3$ \\ 
10 & $1$ & M$_3$ $\rightarrow$ M$_3$ & 49 & $2$ & M$_1$ $\rightarrow$ p & 88 & $2$ & M$_1$ $\rightarrow$ p & 127 & $2$ & M$_2$ $\rightarrow$ M$_1$ & 166 & $1$ & M$_3$ $\rightarrow$ M$_3$ & 205 & $4$ & M$_1$ $\rightarrow$ m \\ 
11 & $2$ & M$_2$ $\rightarrow$ p & 50 & $2$ & M$_1$ $\rightarrow$ p & 89 & $1$ & M$_3$ $\rightarrow$ p & 128 & $2$ & M$_2$ $\rightarrow$ M$_1$ & 167 & $2$ & M$_1$ $\rightarrow$ p & 206 & $4$ & p $\rightarrow$ m \\ 
12 & $1$ & M$_3$ $\rightarrow$ M$_3$ & 51 & $2$ & M$_2$ $\rightarrow$ M$_1$ & 90 & $2$ & M$_2$ $\rightarrow$ p & 129 & $2$ & M$_2$ $\rightarrow$ M$_1$ & 168 & $1$ & M$_3$ $\rightarrow$ p & 207 & $1$ & M$_3$ $\rightarrow$ p \\ 
13 & $2$ & M$_1$ $\rightarrow$ p & 52 & $4$ & p $\rightarrow$ m & 91 & $4$ & m $\rightarrow$ m & 130 & $4$ & p $\rightarrow$ p & 169 & $6$ & M$_2$ $\rightarrow$ m & 208 & $2$ & m $\rightarrow$ m \\ 
14 & $2$ & M$_2$ $\rightarrow$ p & 53 & $2$ & M$_2$ $\rightarrow$ M$_1$ & 92 & $4$ & p $\rightarrow$ m & 131 & $2$ & M$_1$ $\rightarrow$ p & 170 & $6$ & M$_2$ $\rightarrow$ m & 209 & $1$ & M$_3$ $\rightarrow$ p \\ 
15 & $2$ & M$_1$ $\rightarrow$ p & 54 & $4$ & p $\rightarrow$ m & 93 & $2$ & m $\rightarrow$ m & 132 & $2$ & M$_1$ $\rightarrow$ p & 171 & $3$ & M$_3$ $\rightarrow$ p & 210 & $2$ & p $\rightarrow$ m \\ 
16 & $1$ & M$_3$ $\rightarrow$ p & 55 & $2$ & M$_2$ $\rightarrow$ M$_1$ & 94 & $2$ & M$_2$ $\rightarrow$ p & 133 & $4$ & m $\rightarrow$ m & 172 & $3$ & M$_3$ $\rightarrow$ p & 211 & $1$ & M$_3$ $\rightarrow$ p \\ 
17 & $2$ & M$_2$ $\rightarrow$ m & 56 & $4$ & p $\rightarrow$ m & 95 & $4$ & m $\rightarrow$ m & 134 & $2$ & M$_1$ $\rightarrow$ p & 173 & $2$ & M$_2$ $\rightarrow$ m & 212 & $4$ & p $\rightarrow$ m \\ 
18 & $2$ & M$_2$ $\rightarrow$ p & 57 & $4$ & M$_1$ $\rightarrow$ m & 96 & $4$ & p $\rightarrow$ m & 135 & $4$ & p $\rightarrow$ p & 174 & $1$ & M$_3$ $\rightarrow$ p & 213 & $4$ & p $\rightarrow$ m \\ 
19 & $4$ & p $\rightarrow$ m & 58 & $2$ & M$_2$ $\rightarrow$ M$_1$ & 97 & $1$ & M$_3$ $\rightarrow$ p & 136 & $2$ & M$_2$ $\rightarrow$ M$_1$ & 175 & $1$ & M$_3$ $\rightarrow$ M$_3$ & 214 & $2$ & p $\rightarrow$ m \\ 
20 & $2$ & M$_2$ $\rightarrow$ m & 59 & $2$ & M$_2$ $\rightarrow$ M$_1$ & 98 & $2$ & p $\rightarrow$ m & 137 & $2$ & M$_2$ $\rightarrow$ M$_1$ & 176 & $2$ & M$_2$ $\rightarrow$ p & 215 & $1$ & M$_3$ $\rightarrow$ M$_1$ \\ 
21 & $1$ & M$_3$ $\rightarrow$ p & 60 & $4$ & M$_1$ $\rightarrow$ m & 99 & $1$ & M$_3$ $\rightarrow$ M$_1$ & 138 & $4$ & p $\rightarrow$ m & 177 & $1$ & M$_3$ $\rightarrow$ p & 216 & $1$ & M$_3$ $\rightarrow$ M$_1$ \\ 
22 & $1$ & M$_3$ $\rightarrow$ p & 61 & $4$ & M$_1$ $\rightarrow$ m & 100 & $2$ & M$_1$ $\rightarrow$ p & 139 & $1$ & M$_3$ $\rightarrow$ M$_3$ & 178 & $6$ & M$_2$ $\rightarrow$ m & 217 & $1$ & M$_3$ $\rightarrow$ M$_1$ \\ 
23 & $1$ & M$_3$ $\rightarrow$ p & 62 & $4$ & M$_1$ $\rightarrow$ m & 101 & $2$ & M$_1$ $\rightarrow$ p & 140 & $2$ & M$_1$ $\rightarrow$ p & 179 & $6$ & M$_2$ $\rightarrow$ m & 218 & $2$ & M$_1$ $\rightarrow$ p \\ 
24 & $2$ & p $\rightarrow$ m & 63 & $2$ & M$_2$ $\rightarrow$ M$_1$ & 102 & $2$ & M$_1$ $\rightarrow$ p & 141 & $2$ & M$_1$ $\rightarrow$ p & 180 & $3$ & M$_3$ $\rightarrow$ p & 219 & $2$ & M$_1$ $\rightarrow$ m \\ 
25 & $1$ & M$_3$ $\rightarrow$ M$_1$ & 64 & $2$ & M$_2$ $\rightarrow$ M$_1$ & 103 & $2$ & M$_1$ $\rightarrow$ p & 142 & $4$ & p $\rightarrow$ m & 181 & $3$ & M$_3$ $\rightarrow$ p & 220 & $2$ & M$_1$ $\rightarrow$ p \\ 
26 & $2$ & M$_2$ $\rightarrow$ p & 65 & $1$ & M$_3$ $\rightarrow$ M$_3$ & 104 & $2$ & M$_1$ $\rightarrow$ p & 143 & $1$ & M$_3$ $\rightarrow$ p & 182 & $2$ & M$_2$ $\rightarrow$ m & 221 & $1$ & M$_3$ $\rightarrow$ M$_3$ \\ 
27 & $2$ & M$_1$ $\rightarrow$ p & 66 & $2$ & M$_1$ $\rightarrow$ p & 105 & $2$ & M$_1$ $\rightarrow$ p & 144 & $3$ & M$_3$ $\rightarrow$ p & 183 & $1$ & M$_3$ $\rightarrow$ M$_1$ & 222 & $2$ & M$_1$ $\rightarrow$ p \\ 
28 & $2$ & M$_1$ $\rightarrow$ m & 67 & $2$ & M$_1$ $\rightarrow$ p & 106 & $4$ & m $\rightarrow$ m & 145 & $3$ & M$_3$ $\rightarrow$ p & 184 & $2$ & M$_1$ $\rightarrow$ p & 223 & $2$ & M$_1$ $\rightarrow$ p \\ 
29 & $4$ & p $\rightarrow$ m & 68 & $2$ & M$_1$ $\rightarrow$ p & 107 & $1$ & M$_3$ $\rightarrow$ M$_1$ & 146 & $1$ & M$_3$ $\rightarrow$ p & 185 & $2$ & M$_2$ $\rightarrow$ p & 224 & $2$ & M$_1$ $\rightarrow$ p \\ 
30 & $2$ & M$_1$ $\rightarrow$ p & 69 & $1$ & M$_3$ $\rightarrow$ M$_3$ & 108 & $2$ & M$_1$ $\rightarrow$ p & 147 & $1$ & M$_3$ $\rightarrow$ M$_3$ & 186 & $2$ & M$_2$ $\rightarrow$ p & 225 & $1$ & M$_3$ $\rightarrow$ M$_3$ \\ 
31 & $2$ & M$_2$ $\rightarrow$ p & 70 & $2$ & M$_1$ $\rightarrow$ p & 109 & $2$ & M$_1$ $\rightarrow$ p & 148 & $1$ & M$_3$ $\rightarrow$ M$_3$ & 187 & $1$ & M$_3$ $\rightarrow$ M$_1$ & 226 & $2$ & M$_1$ $\rightarrow$ p \\ 
32 & $2$ & M$_1$ $\rightarrow$ p & 71 & $1$ & M$_3$ $\rightarrow$ M$_3$ & 110 & $4$ & p $\rightarrow$ m & 149 & $1$ & M$_3$ $\rightarrow$ p & 188 & $2$ & M$_1$ $\rightarrow$ m & 227 & $2$ & M$_1$ $\rightarrow$ p \\ 
33 & $4$ & p $\rightarrow$ m & 72 & $2$ & M$_1$ $\rightarrow$ p & 111 & $1$ & M$_3$ $\rightarrow$ M$_1$ & 150 & $1$ & M$_3$ $\rightarrow$ p & 189 & $1$ & M$_3$ $\rightarrow$ M$_1$ & 228 & $4$ & p $\rightarrow$ m \\ 
34 & $2$ & M$_1$ $\rightarrow$ p & 73 & $4$ & p $\rightarrow$ m & 112 & $2$ & M$_1$ $\rightarrow$ p & 151 & $3$ & M$_3$ $\rightarrow$ p & 190 & $2$ & M$_1$ $\rightarrow$ m & 229 & $1$ & M$_3$ $\rightarrow$ M$_3$ \\ 
35 & $1$ & M$_3$ $\rightarrow$ M$_1$ & 74 & $2$ & M$_1$ $\rightarrow$ p & 113 & $2$ & M$_2$ $\rightarrow$ M$_1$ & 152 & $3$ & M$_3$ $\rightarrow$ p & 191 & $1$ & M$_3$ $\rightarrow$ M$_3$ & 230 & $4$ & p $\rightarrow$ m \\ 
36 & $2$ & M$_2$ $\rightarrow$ p & 75 & $1$ & M$_3$ $\rightarrow$ p & 114 & $2$ & M$_2$ $\rightarrow$ M$_1$ & 153 & $3$ & M$_3$ $\rightarrow$ p & 192 & $2$ & M$_1$ $\rightarrow$ p & ~ & ~ & ~ \\ 
37 & $2$ & M$_1$ $\rightarrow$ p & 76 & $4$ & m $\rightarrow$ m & 115 & $1$ & M$_3$ $\rightarrow$ M$_1$ & 154 & $3$ & M$_3$ $\rightarrow$ p & 193 & $2$ & M$_2$ $\rightarrow$ M$_1$ & ~ & ~ & ~ \\ 
38 & $1$ & M$_3$ $\rightarrow$ M$_1$ & 77 & $2$ & m $\rightarrow$ m & 116 & $2$ & M$_1$ $\rightarrow$ p & 155 & $1$ & M$_3$ $\rightarrow$ p & 194 & $2$ & M$_2$ $\rightarrow$ M$_1$ & ~ & ~ & ~ \\ 
39 & $2$ & M$_1$ $\rightarrow$ m & 78 & $4$ & m $\rightarrow$ m & 117 & $2$ & M$_1$ $\rightarrow$ p & 156 & $1$ & M$_3$ $\rightarrow$ p & 195 & $1$ & M$_3$ $\rightarrow$ p & ~ & ~ & ~ \\ 
\hline \hline 
\end{tabular}

%% file: SOC_feSM_foot.tex
\newlength{\tabLSOCfeSM} 
\settowidth{\tabLSOCfeSM}{\input{SOC_feSM}} 
\begin{minipage}{\tabLSOCfeSM} 
\begin{flushleft} 
{\footnotesize 
SG: Space group describing the spatial symmetry of a material with time-reversal symmetry\\
$\nu_{\rm P}$: The physically achievable fillings are quantized in units of $\nu_{\rm P}$\\
FS: Fermiology for a material with a filling being an odd-integer multiple of $\nu_{\rm P}$, dubbed a filling-enforced (semi-)metal\\
$M_d$: Filling-enforced metals due to band sticking on $d$-dimensional manifolds\\
$p$: Filling-enforced semimetals with pinned, nodal-point Fermi surfaces\\
$m$: Filling-enforced semimetals with movable nodal Fermi surfaces
}
\end{flushleft} 
\end{minipage}

%% file: spinless_Tri.tex
\begin{tabular}{cc|ccc||cc|ccc||cc|ccc||cc|ccc} 
\hline \hline 
\multicolumn{2}{c|}{MSG} & $d$ & $X_{\rm BS}$ & $\nu_{\rm BS}$ & \multicolumn{2}{c|}{MSG} & $d$ & $X_{\rm BS}$ & $\nu_{\rm BS}$ & \multicolumn{2}{c|}{MSG} & $d$ & $X_{\rm BS}$ & $\nu_{\rm BS}$ & \multicolumn{2}{c|}{MSG} & $d$ & $X_{\rm BS}$ & $\nu_{\rm BS}$\\ 
\hline 
1.1 & I & $1$ & $(1)$ & $1$ & 1.3 & IV & $1$ & $(1)$ & $2$ & 2.5 & II & $9$ & $(2, 2, 2, 4)$ & $1$ & 2.7 & IV & $5$ & $(2, 2, 4)$ & $2$ \\ 
1.2 & II & $1$ & $(1)$ & $1$ & 2.4 & I & $9$ & $(2, 2, 2, 4)$ & $1$ & 2.6 & III & $1$ & $(1)$ & $1$ & ~ & ~ & ~ & ~ \\ 
\hline \hline 
\end{tabular}

%% file: spinless_Tri_foot.tex
\newlength{\tabLspinlessTri} 
\settowidth{\tabLspinlessTri}{\input{spinless_Tri}} 
\begin{minipage}{\tabLspinlessTri} 
\begin{flushleft} 
{\footnotesize $d$: Rank of the band structure group $\{{\rm BS}\}$\\ 
$X_{\rm BS}$: Symmetry-based indicators of band topology\\ 
$\nu_{\rm BS}$: Set of $\nu$ bands are symmetry-forbidden from being isolated by band gaps if $\nu \not \in \nu_{\rm BS}\, \mathbb Z$ }
\end{flushleft}\end{minipage}

%% file: spinless_Mono.tex
\begin{tabular}{cc|ccc||cc|ccc||cc|ccc||cc|ccc} 
\hline \hline 
\multicolumn{2}{c|}{MSG} & $d$ & $X_{\rm BS}$ & $\nu_{\rm BS}$ & \multicolumn{2}{c|}{MSG} & $d$ & $X_{\rm BS}$ & $\nu_{\rm BS}$ & \multicolumn{2}{c|}{MSG} & $d$ & $X_{\rm BS}$ & $\nu_{\rm BS}$ & \multicolumn{2}{c|}{MSG} & $d$ & $X_{\rm BS}$ & $\nu_{\rm BS}$\\ 
\hline 
3.1 & I & $5$ & $(2)$ & $1$ & 7.24 & I & $1$ & $(1)$ & $2$ & 10.47 & IV & $9$ & $(2, 2, 2)$ & $2$ & 13.70 & IV & $4$ & $(2, 2)$ & $4$ \\ 
3.2 & II & $5$ & $(2)$ & $1$ & 7.25 & II & $1$ & $(1)$ & $2$ & 10.48 & IV & $10$ & $(2, 2)$ & $2$ & 13.71 & IV & $5$ & $(2, 2)$ & $4$ \\ 
3.3 & III & $1$ & $(1)$ & $1$ & 7.26 & III & $1$ & $(1)$ & $2$ & 10.49 & IV & $8$ & $(2, 2)$ & $2$ & 13.72 & IV & $7$ & $(2, 2)$ & $2$ \\ 
3.4 & IV & $3$ & $(2)$ & $2$ & 7.27 & IV & $1$ & $(1)$ & $4$ & 11.50 & I & $6$ & $(2)$ & $2$ & 13.73 & IV & $5$ & $(2)$ & $2$ \\ 
3.5 & IV & $5$ & $(2)$ & $2$ & 7.28 & IV & $1$ & $(1)$ & $2$ & 11.51 & II & $6$ & $(2)$ & $2$ & 13.74 & IV & $4$ & $(2)$ & $4$ \\ 
3.6 & IV & $3$ & $(1)$ & $2$ & 7.29 & IV & $1$ & $(1)$ & $4$ & 11.52 & III & $2$ & $(1)$ & $2$ & 14.75 & I & $5$ & $(2)$ & $2$ \\ 
4.7 & I & $1$ & $(1)$ & $2$ & 7.30 & IV & $1$ & $(1)$ & $4$ & 11.53 & III & $1$ & $(1)$ & $2$ & 14.76 & II & $5$ & $(2)$ & $2$ \\ 
4.8 & II & $1$ & $(1)$ & $2$ & 7.31 & IV & $1$ & $(1)$ & $2$ & 11.54 & III & $5$ & $(2, 2, 4)$ & $2$ & 14.77 & III & $1$ & $(1)$ & $2$ \\ 
4.9 & III & $1$ & $(1)$ & $2$ & 8.32 & I & $2$ & $(1)$ & $1$ & 11.55 & IV & $4$ & $(2)$ & $4$ & 14.78 & III & $1$ & $(1)$ & $2$ \\ 
4.10 & IV & $1$ & $(1)$ & $4$ & 8.33 & II & $2$ & $(1)$ & $1$ & 11.56 & IV & $6$ & $(2)$ & $2$ & 14.79 & III & $5$ & $(2, 4)$ & $2$ \\ 
4.11 & IV & $1$ & $(1)$ & $2$ & 8.34 & III & $1$ & $(1)$ & $1$ & 11.57 & IV & $4$ & $(2)$ & $2$ & 14.80 & IV & $3$ & $(2)$ & $4$ \\ 
4.12 & IV & $1$ & $(1)$ & $2$ & 8.35 & IV & $2$ & $(1)$ & $2$ & 12.58 & I & $10$ & $(2, 2)$ & $1$ & 14.81 & IV & $3$ & $(2)$ & $4$ \\ 
5.13 & I & $3$ & $(1)$ & $1$ & 8.36 & IV & $2$ & $(1)$ & $2$ & 12.59 & II & $10$ & $(2, 2)$ & $1$ & 14.82 & IV & $3$ & $(2)$ & $4$ \\ 
5.14 & II & $3$ & $(1)$ & $1$ & 9.37 & I & $1$ & $(1)$ & $2$ & 12.60 & III & $2$ & $(1)$ & $1$ & 14.83 & IV & $5$ & $(2)$ & $2$ \\ 
5.15 & III & $1$ & $(1)$ & $1$ & 9.38 & II & $1$ & $(1)$ & $2$ & 12.61 & III & $3$ & $(1)$ & $1$ & 14.84 & IV & $3$ & $(2)$ & $4$ \\ 
5.16 & IV & $2$ & $(1)$ & $2$ & 9.39 & III & $1$ & $(1)$ & $2$ & 12.62 & III & $7$ & $(2, 2, 4)$ & $1$ & 15.85 & I & $6$ & $(2, 2)$ & $2$ \\ 
5.17 & IV & $3$ & $(2)$ & $2$ & 9.40 & IV & $1$ & $(1)$ & $2$ & 12.63 & IV & $6$ & $(2, 2)$ & $2$ & 15.86 & II & $6$ & $(2, 2)$ & $2$ \\ 
6.18 & I & $3$ & $(1)$ & $1$ & 9.41 & IV & $1$ & $(2)$ & $2^* $ & 12.64 & IV & $8$ & $(2, 2)$ & $2$ & 15.87 & III & $1$ & $(1)$ & $2$ \\ 
6.19 & II & $3$ & $(1)$ & $1$ & 10.42 & I & $15$ & $(2, 2, 2)$ & $1$ & 13.65 & I & $7$ & $(2, 2)$ & $2$ & 15.88 & III & $2$ & $(1)$ & $2$ \\ 
6.20 & III & $1$ & $(1)$ & $1$ & 10.43 & II & $15$ & $(2, 2, 2)$ & $1$ & 13.66 & II & $7$ & $(2, 2)$ & $2$ & 15.89 & III & $5$ & $(2, 4)$ & $2$ \\ 
6.21 & IV & $3$ & $(1)$ & $2$ & 10.44 & III & $3$ & $(1)$ & $1$ & 13.67 & III & $1$ & $(1)$ & $2$ & 15.90 & IV & $5$ & $(2, 2)$ & $2$ \\ 
6.22 & IV & $2$ & $(1)$ & $2$ & 10.45 & III & $5$ & $(2)$ & $1$ & 13.68 & III & $3$ & $(2)$ & $2$ & 15.91 & IV & $4$ & $(2, 2, 2)$ & $2^* $ \\ 
6.23 & IV & $2$ & $(1)$ & $2$ & 10.46 & III & $9$ & $(2, 2, 2, 4)$ & $1$ & 13.69 & III & $5$ & $(2, 2, 4)$ & $2$ & ~ & ~ & ~ & ~ \\ 
\hline \hline 
\end{tabular}

%% file: spinless_Mono_foot.tex
\newlength{\tabLspinlessMono} 
\settowidth{\tabLspinlessMono}{\input{spinless_Mono}} 
\begin{minipage}{\tabLspinlessMono} 
\begin{flushleft} 
{\footnotesize $d$: Rank of the band structure group $\{{\rm BS}\}$\\ 
$X_{\rm BS}$: Symmetry-based indicators of band topology\\ 
$\nu_{\rm BS}$: Set of $\nu$ bands are symmetry-forbidden from being isolated by band gaps if $\nu \not \in \nu_{\rm BS}\, \mathbb Z$ }\\ 
$*$: Exhibiting exceptional filling pattern; see Table \ref{tab:spinlessnuEx} 
\end{flushleft}\end{minipage}

%% file: spinless_Ortho.tex
\begin{tabular}{cc|ccc||cc|ccc||cc|ccc||cc|ccc} 
\hline \hline 
\multicolumn{2}{c|}{MSG} & $d$ & $X_{\rm BS}$ & $\nu_{\rm BS}$ & \multicolumn{2}{c|}{MSG} & $d$ & $X_{\rm BS}$ & $\nu_{\rm BS}$ & \multicolumn{2}{c|}{MSG} & $d$ & $X_{\rm BS}$ & $\nu_{\rm BS}$ & \multicolumn{2}{c|}{MSG} & $d$ & $X_{\rm BS}$ & $\nu_{\rm BS}$\\ 
\hline 
16.1 & I & $13$ & $(1)$ & $1$ & 23.51 & III & $3$ & $(1)$ & $1$ & 29.101 & III & $1$ & $(1)$ & $4$ & 33.151 & IV & $1$ & $(1)$ & $4$ \\ 
16.2 & II & $13$ & $(1)$ & $1$ & 23.52 & IV & $5$ & $(2)$ & $2$ & 29.102 & III & $1$ & $(1)$ & $4$ & 33.152 & IV & $1$ & $(1)$ & $4$ \\ 
16.3 & III & $5$ & $(2)$ & $1$ & 24.53 & I & $4$ & $(1)$ & $2$ & 29.103 & III & $1$ & $(1)$ & $4$ & 33.153 & IV & $1$ & $(1)$ & $4$ \\ 
16.4 & IV & $9$ & $(2)$ & $2$ & 24.54 & II & $4$ & $(1)$ & $2$ & 29.104 & IV & $1$ & $(1)$ & $4$ & 33.154 & IV & $1$ & $(1)$ & $4$ \\ 
16.5 & IV & $7$ & $(2)$ & $2$ & 24.55 & III & $2$ & $(1)$ & $2$ & 29.105 & IV & $1$ & $(2)$ & $4^* $ & 33.155 & IV & $1$ & $(1)$ & $4$ \\ 
16.6 & IV & $7$ & $(2)$ & $2$ & 24.56 & IV & $4$ & $(2)$ & $2$ & 29.106 & IV & $1$ & $(1)$ & $4$ & 34.156 & I & $3$ & $(1)$ & $2$ \\ 
17.7 & I & $5$ & $(1)$ & $2$ & 25.57 & I & $9$ & $(1)$ & $1$ & 29.107 & IV & $1$ & $(1)$ & $4$ & 34.157 & II & $3$ & $(1)$ & $2$ \\ 
17.8 & II & $5$ & $(1)$ & $2$ & 25.58 & II & $9$ & $(1)$ & $1$ & 29.108 & IV & $1$ & $(1)$ & $4$ & 34.158 & III & $1$ & $(1)$ & $2$ \\ 
17.9 & III & $1$ & $(1)$ & $2$ & 25.59 & III & $3$ & $(1)$ & $1$ & 29.109 & IV & $1$ & $(1)$ & $4$ & 34.159 & III & $3$ & $(1)$ & $2$ \\ 
17.10 & III & $3$ & $(2)$ & $2$ & 25.60 & III & $5$ & $(2)$ & $1$ & 29.110 & IV & $1$ & $(1)$ & $4$ & 34.160 & IV & $2$ & $(1)$ & $4$ \\ 
17.11 & IV & $4$ & $(2)$ & $4$ & 25.61 & IV & $9$ & $(1)$ & $2$ & 30.111 & I & $3$ & $(1)$ & $2$ & 34.161 & IV & $3$ & $(2)$ & $4$ \\ 
17.12 & IV & $5$ & $(1)$ & $2$ & 25.62 & IV & $6$ & $(1)$ & $2$ & 30.112 & II & $3$ & $(1)$ & $2$ & 34.162 & IV & $2$ & $(1)$ & $4$ \\ 
17.13 & IV & $4$ & $(1)$ & $2$ & 25.63 & IV & $5$ & $(1)$ & $2$ & 30.113 & III & $1$ & $(1)$ & $2$ & 34.163 & IV & $3$ & $(2)$ & $4$ \\ 
17.14 & IV & $3$ & $(2)$ & $4$ & 25.64 & IV & $6$ & $(1)$ & $2$ & 30.114 & III & $1$ & $(1)$ & $2$ & 34.164 & IV & $3$ & $(1)$ & $2$ \\ 
17.15 & IV & $3$ & $(1)$ & $4$ & 25.65 & IV & $5$ & $(1)$ & $2$ & 30.115 & III & $3$ & $(1)$ & $2$ & 35.165 & I & $6$ & $(1)$ & $1$ \\ 
18.16 & I & $3$ & $(1)$ & $2$ & 26.66 & I & $3$ & $(1)$ & $2$ & 30.116 & IV & $2$ & $(1)$ & $4$ & 35.166 & II & $6$ & $(1)$ & $1$ \\ 
18.17 & II & $3$ & $(1)$ & $2$ & 26.67 & II & $3$ & $(1)$ & $2$ & 30.117 & IV & $3$ & $(2)$ & $4$ & 35.167 & III & $2$ & $(1)$ & $1$ \\ 
18.18 & III & $3$ & $(2)$ & $2$ & 26.68 & III & $1$ & $(1)$ & $2$ & 30.118 & IV & $3$ & $(2)$ & $4$ & 35.168 & III & $4$ & $(2)$ & $1$ \\ 
18.19 & III & $1$ & $(1)$ & $2$ & 26.69 & III & $3$ & $(1)$ & $2$ & 30.119 & IV & $3$ & $(1)$ & $2$ & 35.169 & IV & $6$ & $(1)$ & $2$ \\ 
18.20 & IV & $2$ & $(1)$ & $4$ & 26.70 & III & $1$ & $(1)$ & $2$ & 30.120 & IV & $2$ & $(2)$ & $4$ & 35.170 & IV & $5$ & $(1)$ & $2$ \\ 
18.21 & IV & $3$ & $(2)$ & $4$ & 26.71 & IV & $2$ & $(1)$ & $4$ & 30.121 & IV & $2$ & $(1)$ & $4$ & 35.171 & IV & $5$ & $(1)$ & $2$ \\ 
18.22 & IV & $2$ & $(1)$ & $4$ & 26.72 & IV & $3$ & $(1)$ & $4$ & 30.122 & IV & $2$ & $(1)$ & $4$ & 36.172 & I & $2$ & $(1)$ & $2$ \\ 
18.23 & IV & $3$ & $(1)$ & $2$ & 26.73 & IV & $3$ & $(1)$ & $2$ & 31.123 & I & $2$ & $(1)$ & $2$ & 36.173 & II & $2$ & $(1)$ & $2$ \\ 
18.24 & IV & $3$ & $(1)$ & $2$ & 26.74 & IV & $3$ & $(1)$ & $2$ & 31.124 & II & $2$ & $(1)$ & $2$ & 36.174 & III & $1$ & $(1)$ & $2$ \\ 
19.25 & I & $1$ & $(1)$ & $4$ & 26.75 & IV & $2$ & $(1)$ & $4$ & 31.125 & III & $1$ & $(1)$ & $2$ & 36.175 & III & $2$ & $(1)$ & $2$ \\ 
19.26 & II & $1$ & $(1)$ & $4$ & 26.76 & IV & $2$ & $(1)$ & $4$ & 31.126 & III & $2$ & $(1)$ & $2$ & 36.176 & III & $1$ & $(1)$ & $2$ \\ 
19.27 & III & $1$ & $(1)$ & $4$ & 26.77 & IV & $2$ & $(1)$ & $4$ & 31.127 & III & $1$ & $(1)$ & $2$ & 36.177 & IV & $2$ & $(1)$ & $2$ \\ 
19.28 & IV & $1$ & $(1)$ & $4$ & 27.78 & I & $5$ & $(2)$ & $2$ & 31.128 & IV & $2$ & $(1)$ & $4$ & 36.178 & IV & $2$ & $(1)$ & $4$ \\ 
19.29 & IV & $1$ & $(1)$ & $4$ & 27.79 & II & $5$ & $(2)$ & $2$ & 31.129 & IV & $2$ & $(1)$ & $4$ & 36.179 & IV & $2$ & $(1)$ & $2$ \\ 
19.30 & IV & $1$ & $(1)$ & $4$ & 27.80 & III & $1$ & $(1)$ & $2$ & 31.130 & IV & $2$ & $(1)$ & $4$ & 37.180 & I & $4$ & $(2)$ & $2$ \\ 
20.31 & I & $3$ & $(1)$ & $2$ & 27.81 & III & $5$ & $(2)$ & $2$ & 31.131 & IV & $2$ & $(1)$ & $4$ & 37.181 & II & $4$ & $(2)$ & $2$ \\ 
20.32 & II & $3$ & $(1)$ & $2$ & 27.82 & IV & $5$ & $(2)$ & $2$ & 31.132 & IV & $2$ & $(1)$ & $2$ & 37.182 & III & $1$ & $(1)$ & $2$ \\ 
20.33 & III & $1$ & $(1)$ & $2$ & 27.83 & IV & $3$ & $(2)$ & $4$ & 31.133 & IV & $2$ & $(1)$ & $4$ & 37.183 & III & $4$ & $(2)$ & $2$ \\ 
20.34 & III & $2$ & $(1)$ & $2$ & 27.84 & IV & $3$ & $(2)$ & $4$ & 31.134 & IV & $2$ & $(1)$ & $2$ & 37.184 & IV & $4$ & $(2)$ & $2$ \\ 
20.35 & IV & $3$ & $(1)$ & $2$ & 27.85 & IV & $3$ & $(2)$ & $4$ & 32.135 & I & $3$ & $(1)$ & $2$ & 37.185 & IV & $3$ & $(2, 2)$ & $2^* $ \\ 
20.36 & IV & $3$ & $(2)$ & $4$ & 27.86 & IV & $3$ & $(2)$ & $4$ & 32.136 & II & $3$ & $(1)$ & $2$ & 37.186 & IV & $3$ & $(2)$ & $2$ \\ 
20.37 & IV & $3$ & $(1)$ & $2$ & 28.87 & I & $4$ & $(1)$ & $2$ & 32.137 & III & $1$ & $(1)$ & $2$ & 38.187 & I & $6$ & $(1)$ & $1$ \\ 
21.38 & I & $8$ & $(1)$ & $1$ & 28.88 & II & $4$ & $(1)$ & $2$ & 32.138 & III & $3$ & $(2)$ & $2$ & 38.188 & II & $6$ & $(1)$ & $1$ \\ 
21.39 & II & $8$ & $(1)$ & $1$ & 28.89 & III & $1$ & $(1)$ & $2$ & 32.139 & IV & $3$ & $(2)$ & $4$ & 38.189 & III & $2$ & $(1)$ & $1$ \\ 
21.40 & III & $4$ & $(2)$ & $1$ & 28.90 & III & $2$ & $(1)$ & $2$ & 32.140 & IV & $2$ & $(1)$ & $4$ & 38.190 & III & $3$ & $(1)$ & $1$ \\ 
21.41 & III & $3$ & $(1)$ & $1$ & 28.91 & III & $3$ & $(2)$ & $2$ & 32.141 & IV & $3$ & $(1)$ & $2$ & 38.191 & III & $3$ & $(1)$ & $1$ \\ 
21.42 & IV & $6$ & $(2)$ & $2$ & 28.92 & IV & $4$ & $(1)$ & $2$ & 32.142 & IV & $2$ & $(1)$ & $4$ & 38.192 & IV & $4$ & $(1)$ & $2$ \\ 
21.43 & IV & $7$ & $(2)$ & $2$ & 28.93 & IV & $3$ & $(1)$ & $4$ & 32.143 & IV & $3$ & $(1)$ & $4$ & 38.193 & IV & $6$ & $(1)$ & $2$ \\ 
21.44 & IV & $5$ & $(2)$ & $2$ & 28.94 & IV & $4$ & $(1)$ & $4$ & 33.144 & I & $1$ & $(1)$ & $4$ & 38.194 & IV & $4$ & $(1)$ & $2$ \\ 
22.45 & I & $7$ & $(1)$ & $1$ & 28.95 & IV & $3$ & $(1)$ & $4$ & 33.145 & II & $1$ & $(1)$ & $4$ & 39.195 & I & $4$ & $(1)$ & $2$ \\ 
22.46 & II & $7$ & $(1)$ & $1$ & 28.96 & IV & $4$ & $(1)$ & $4$ & 33.146 & III & $1$ & $(1)$ & $4$ & 39.196 & II & $4$ & $(1)$ & $2$ \\ 
22.47 & III & $3$ & $(1)$ & $1$ & 28.97 & IV & $3$ & $(1)$ & $2$ & 33.147 & III & $1$ & $(1)$ & $4$ & 39.197 & III & $2$ & $(1)$ & $2$ \\ 
22.48 & IV & $7$ & $(2)$ & $2$ & 28.98 & IV & $3$ & $(1)$ & $4$ & 33.148 & III & $1$ & $(1)$ & $4$ & 39.198 & III & $1$ & $(1)$ & $2$ \\ 
23.49 & I & $7$ & $(1)$ & $1$ & 29.99 & I & $1$ & $(1)$ & $4$ & 33.149 & IV & $1$ & $(1)$ & $4$ & 39.199 & III & $3$ & $(2)$ & $2$ \\ 
23.50 & II & $7$ & $(1)$ & $1$ & 29.100 & II & $1$ & $(1)$ & $4$ & 33.150 & IV & $1$ & $(1)$ & $8$ & 39.200 & IV & $3$ & $(1)$ & $4$ \\ 
\hline \hline 
\end{tabular}

%% file: spinless_OrthoII.tex
\begin{tabular}{cc|ccc||cc|ccc||cc|ccc||cc|ccc} 
\hline \hline 
39.201 & IV & $4$ & $(1)$ & $2$ & 47.251 & III & $9$ & $(1)$ & $1$ & 51.301 & IV & $7$ & $(1)$ & $4$ & 54.351 & IV & $4$ & $(2)$ & $4$ \\ 
39.202 & IV & $3$ & $(1)$ & $2$ & 47.252 & III & $15$ & $(2, 2, 2)$ & $1$ & 51.302 & IV & $9$ & $(1)$ & $2$ & 54.352 & IV & $4$ & $(2)$ & $8$ \\ 
40.203 & I & $3$ & $(1)$ & $2$ & 47.253 & III & $13$ & $(1)$ & $1$ & 51.303 & IV & $8$ & $(1)$ & $4$ & 55.353 & I & $9$ & $(1)$ & $2$ \\ 
40.204 & II & $3$ & $(1)$ & $2$ & 47.254 & IV & $18$ & $(1)$ & $2$ & 51.304 & IV & $7$ & $(1)$ & $4$ & 55.354 & II & $9$ & $(1)$ & $2$ \\ 
40.205 & III & $1$ & $(1)$ & $2$ & 47.255 & IV & $15$ & $(1)$ & $2$ & 52.305 & I & $5$ & $(2)$ & $4$ & 55.355 & III & $3$ & $(1)$ & $2$ \\ 
40.206 & III & $2$ & $(1)$ & $2$ & 47.256 & IV & $14$ & $(1)$ & $2$ & 52.306 & II & $5$ & $(2)$ & $4$ & 55.356 & III & $3$ & $(1)$ & $2$ \\ 
40.207 & III & $2$ & $(1)$ & $2$ & 48.257 & I & $9$ & $(2)$ & $2$ & 52.307 & III & $2$ & $(1)$ & $4$ & 55.357 & III & $9$ & $(2, 2, 2)$ & $2$ \\ 
40.208 & IV & $3$ & $(1)$ & $2$ & 48.258 & II & $9$ & $(2)$ & $2$ & 52.308 & III & $1$ & $(1)$ & $4$ & 55.358 & III & $5$ & $(2)$ & $2$ \\ 
40.209 & IV & $3$ & $(1)$ & $4$ & 48.259 & III & $3$ & $(1)$ & $2$ & 52.309 & III & $2$ & $(1)$ & $4$ & 55.359 & III & $3$ & $(1)$ & $2$ \\ 
40.210 & IV & $3$ & $(1)$ & $2$ & 48.260 & III & $5$ & $(2)$ & $2$ & 52.310 & III & $4$ & $(2)$ & $4$ & 55.360 & IV & $6$ & $(1)$ & $4$ \\ 
41.211 & I & $2$ & $(1)$ & $2$ & 48.261 & III & $7$ & $(2)$ & $2$ & 52.311 & III & $4$ & $(2)$ & $4$ & 55.361 & IV & $6$ & $(1)$ & $4$ \\ 
41.212 & II & $2$ & $(1)$ & $2$ & 48.262 & IV & $6$ & $(2, 2)$ & $4$ & 52.312 & III & $3$ & $(2)$ & $4$ & 55.362 & IV & $5$ & $(1)$ & $4$ \\ 
41.213 & III & $1$ & $(1)$ & $2$ & 48.263 & IV & $7$ & $(2, 2)$ & $4$ & 52.313 & III & $3$ & $(1)$ & $4$ & 55.363 & IV & $9$ & $(1)$ & $2$ \\ 
41.214 & III & $1$ & $(1)$ & $2$ & 48.264 & IV & $8$ & $(2)$ & $2$ & 52.314 & IV & $5$ & $(2, 2)$ & $4$ & 55.364 & IV & $6$ & $(1)$ & $4$ \\ 
41.215 & III & $2$ & $(2)$ & $2$ & 49.265 & I & $14$ & $(2)$ & $2$ & 52.315 & IV & $4$ & $(2)$ & $4$ & 56.365 & I & $5$ & $(2)$ & $4$ \\ 
41.216 & IV & $2$ & $(1)$ & $4$ & 49.266 & II & $14$ & $(2)$ & $2$ & 52.316 & IV & $4$ & $(2, 2)$ & $8$ & 56.366 & II & $5$ & $(2)$ & $4$ \\ 
41.217 & IV & $2$ & $(2)$ & $2^* $ & 49.267 & III & $4$ & $(1)$ & $2$ & 52.317 & IV & $5$ & $(2)$ & $4$ & 56.367 & III & $1$ & $(1)$ & $4$ \\ 
41.218 & IV & $2$ & $(1)$ & $2$ & 49.268 & III & $5$ & $(2)$ & $2$ & 52.318 & IV & $4$ & $(2)$ & $4$ & 56.368 & III & $3$ & $(2)$ & $4$ \\ 
42.219 & I & $5$ & $(1)$ & $1$ & 49.269 & III & $10$ & $(2, 2)$ & $2$ & 52.319 & IV & $4$ & $(2, 2)$ & $4$ & 56.369 & III & $5$ & $(2, 2)$ & $4$ \\ 
42.220 & II & $5$ & $(1)$ & $1$ & 49.270 & III & $7$ & $(2, 2)$ & $2$ & 52.320 & IV & $5$ & $(2)$ & $4$ & 56.370 & III & $3$ & $(2)$ & $4$ \\ 
42.221 & III & $2$ & $(1)$ & $1$ & 49.271 & III & $9$ & $(2)$ & $2$ & 53.321 & I & $9$ & $(2)$ & $2$ & 56.371 & III & $3$ & $(2)$ & $4$ \\ 
42.222 & III & $3$ & $(2)$ & $1$ & 49.272 & IV & $9$ & $(2)$ & $4$ & 53.322 & II & $9$ & $(2)$ & $2$ & 56.372 & IV & $3$ & $(2)$ & $8$ \\ 
42.223 & IV & $5$ & $(1)$ & $2$ & 49.273 & IV & $14$ & $(2)$ & $2$ & 53.323 & III & $3$ & $(1)$ & $2$ & 56.373 & IV & $4$ & $(2)$ & $4$ \\ 
43.224 & I & $2$ & $(1)$ & $2$ & 49.274 & IV & $9$ & $(2)$ & $4$ & 53.324 & III & $3$ & $(1)$ & $2$ & 56.374 & IV & $4$ & $(2)$ & $4$ \\ 
43.225 & II & $2$ & $(1)$ & $2$ & 49.275 & IV & $8$ & $(2)$ & $4$ & 53.325 & III & $2$ & $(1)$ & $2$ & 56.375 & IV & $5$ & $(2)$ & $4$ \\ 
43.226 & III & $1$ & $(1)$ & $2$ & 49.276 & IV & $8$ & $(2)$ & $4$ & 53.326 & III & $5$ & $(2)$ & $2$ & 56.376 & IV & $4$ & $(2)$ & $4$ \\ 
43.227 & III & $2$ & $(1)$ & $2$ & 50.277 & I & $9$ & $(2)$ & $2$ & 53.327 & III & $8$ & $(2, 2)$ & $2$ & 57.377 & I & $5$ & $(1)$ & $4$ \\ 
43.228 & IV & $2$ & $(2)$ & $2^* $ & 50.278 & II & $9$ & $(2)$ & $2$ & 53.328 & III & $6$ & $(2, 2)$ & $2$ & 57.378 & II & $5$ & $(1)$ & $4$ \\ 
44.229 & I & $5$ & $(1)$ & $1$ & 50.279 & III & $3$ & $(1)$ & $2$ & 53.329 & III & $4$ & $(1)$ & $2$ & 57.379 & III & $3$ & $(1)$ & $4$ \\ 
44.230 & II & $5$ & $(1)$ & $1$ & 50.280 & III & $3$ & $(1)$ & $2$ & 53.330 & IV & $7$ & $(2)$ & $4$ & 57.380 & III & $2$ & $(1)$ & $4$ \\ 
44.231 & III & $2$ & $(1)$ & $1$ & 50.281 & III & $5$ & $(2, 2)$ & $2$ & 53.331 & IV & $6$ & $(2)$ & $4$ & 57.381 & III & $1$ & $(1)$ & $4$ \\ 
44.232 & III & $3$ & $(1)$ & $1$ & 50.282 & III & $5$ & $(2)$ & $2$ & 53.332 & IV & $7$ & $(2)$ & $4$ & 57.382 & III & $4$ & $(2)$ & $4$ \\ 
44.233 & IV & $5$ & $(1)$ & $2$ & 50.283 & III & $7$ & $(2)$ & $2$ & 53.333 & IV & $6$ & $(2)$ & $4$ & 57.383 & III & $4$ & $(2, 2)$ & $4$ \\ 
44.234 & IV & $4$ & $(1)$ & $2$ & 50.284 & IV & $8$ & $(2, 2)$ & $4$ & 53.334 & IV & $9$ & $(2)$ & $2$ & 57.384 & III & $3$ & $(2)$ & $4$ \\ 
45.235 & I & $3$ & $(2)$ & $2$ & 50.285 & IV & $6$ & $(2, 2)$ & $4$ & 53.335 & IV & $6$ & $(2)$ & $4$ & 57.385 & III & $2$ & $(1)$ & $4$ \\ 
45.236 & II & $3$ & $(1)$ & $2$ & 50.286 & IV & $5$ & $(2, 2)$ & $4$ & 53.336 & IV & $6$ & $(2)$ & $4$ & 57.386 & IV & $4$ & $(1)$ & $8$ \\ 
45.237 & III & $1$ & $(1)$ & $2$ & 50.287 & IV & $9$ & $(2)$ & $2$ & 54.337 & I & $6$ & $(2)$ & $4$ & 57.387 & IV & $5$ & $(1)$ & $4$ \\ 
45.238 & III & $3$ & $(2)$ & $2$ & 50.288 & IV & $6$ & $(2)$ & $4$ & 54.338 & II & $6$ & $(2)$ & $4$ & 57.388 & IV & $5$ & $(1)$ & $4$ \\ 
45.239 & IV & $3$ & $(2)$ & $2$ & 51.289 & I & $12$ & $(1)$ & $2$ & 54.339 & III & $1$ & $(1)$ & $4$ & 57.389 & IV & $5$ & $(1)$ & $4$ \\ 
45.240 & IV & $2$ & $(2)$ & $2^* $ & 51.290 & II & $12$ & $(1)$ & $2$ & 54.340 & III & $2$ & $(1)$ & $4$ & 57.390 & IV & $4$ & $(1)$ & $4$ \\ 
46.241 & I & $3$ & $(1)$ & $2$ & 51.291 & III & $3$ & $(1)$ & $2$ & 54.341 & III & $3$ & $(2)$ & $4$ & 57.391 & IV & $4$ & $(1)$ & $4$ \\ 
46.242 & II & $3$ & $(1)$ & $2$ & 51.292 & III & $4$ & $(1)$ & $2$ & 54.342 & III & $5$ & $(2, 2)$ & $4$ & 57.392 & IV & $4$ & $(1)$ & $4$ \\ 
46.243 & III & $1$ & $(1)$ & $2$ & 51.293 & III & $6$ & $(1)$ & $2$ & 54.343 & III & $3$ & $(2)$ & $4$ & 58.393 & I & $8$ & $(2)$ & $2$ \\ 
46.244 & III & $2$ & $(1)$ & $2$ & 51.294 & III & $7$ & $(2, 2)$ & $2$ & 54.344 & III & $4$ & $(2, 2)$ & $4$ & 58.394 & II & $8$ & $(2)$ & $2$ \\ 
46.245 & III & $2$ & $(1)$ & $2$ & 51.295 & III & $6$ & $(2)$ & $2$ & 54.345 & III & $4$ & $(2)$ & $4$ & 58.395 & III & $2$ & $(1)$ & $2$ \\ 
46.246 & IV & $3$ & $(1)$ & $2$ & 51.296 & III & $9$ & $(2, 2, 2)$ & $2$ & 54.346 & IV & $6$ & $(2)$ & $4$ & 58.396 & III & $3$ & $(1)$ & $2$ \\ 
46.247 & IV & $3$ & $(1)$ & $2$ & 51.297 & III & $5$ & $(1)$ & $2$ & 54.347 & IV & $4$ & $(2, 2)$ & $4^* $ & 58.397 & III & $8$ & $(2, 2)$ & $2$ \\ 
46.248 & IV & $3$ & $(1)$ & $4$ & 51.298 & IV & $12$ & $(1)$ & $2$ & 54.348 & IV & $6$ & $(2)$ & $4$ & 58.398 & III & $5$ & $(2)$ & $2$ \\ 
47.249 & I & $27$ & $(1)$ & $1$ & 51.299 & IV & $8$ & $(1)$ & $4$ & 54.349 & IV & $4$ & $(2)$ & $4$ & 58.399 & III & $3$ & $(1)$ & $2$ \\ 
47.250 & II & $27$ & $(1)$ & $1$ & 51.300 & IV & $9$ & $(1)$ & $4$ & 54.350 & IV & $6$ & $(2)$ & $4$ & 58.400 & IV & $5$ & $(2)$ & $4$ \\ 
\hline \hline 
\end{tabular}

%% file: spinless_OrthoIII.tex
\begin{tabular}{cc|ccc||cc|ccc||cc|ccc||cc|ccc} 
\hline \hline
58.401 & IV & $6$ & $(2)$ & $4$ & 62.442 & II & $4$ & $(1)$ & $4$ & 65.483 & III & $6$ & $(1)$ & $1$ & 69.524 & III & $9$ & $(2, 2)$ & $1$ \\ 
58.402 & IV & $5$ & $(2)$ & $4$ & 62.443 & III & $2$ & $(1)$ & $4$ & 65.484 & III & $6$ & $(1)$ & $1$ & 69.525 & III & $7$ & $(2)$ & $1$ \\ 
58.403 & IV & $6$ & $(2)$ & $4$ & 62.444 & III & $1$ & $(1)$ & $4$ & 65.485 & III & $12$ & $(2, 2, 2)$ & $1$ & 69.526 & IV & $14$ & $(1)$ & $2$ \\ 
58.404 & IV & $8$ & $(2)$ & $2$ & 62.445 & III & $2$ & $(1)$ & $4$ & 65.486 & III & $10$ & $(2, 2)$ & $1$ & 70.527 & I & $6$ & $(2)$ & $2$ \\ 
59.405 & I & $7$ & $(1)$ & $2$ & 62.446 & III & $3$ & $(2)$ & $4$ & 65.487 & III & $8$ & $(1)$ & $1$ & 70.528 & II & $6$ & $(2)$ & $2$ \\ 
59.406 & II & $7$ & $(1)$ & $2$ & 62.447 & III & $3$ & $(2)$ & $4$ & 65.488 & IV & $12$ & $(1)$ & $2$ & 70.529 & III & $2$ & $(1)$ & $2$ \\ 
59.407 & III & $2$ & $(1)$ & $2$ & 62.448 & III & $4$ & $(2)$ & $4$ & 65.489 & IV & $15$ & $(1)$ & $2$ & 70.530 & III & $4$ & $(2)$ & $2$ \\ 
59.408 & III & $5$ & $(1)$ & $2$ & 62.449 & III & $1$ & $(1)$ & $4$ & 65.490 & IV & $11$ & $(1)$ & $2$ & 70.531 & III & $4$ & $(1)$ & $2$ \\ 
59.409 & III & $5$ & $(2, 2)$ & $2$ & 62.450 & IV & $3$ & $(1)$ & $4$ & 66.491 & I & $11$ & $(2)$ & $2$ & 70.532 & IV & $5$ & $(2, 2)$ & $2^* $ \\ 
59.410 & III & $4$ & $(2)$ & $2$ & 62.451 & IV & $4$ & $(1)$ & $4$ & 66.492 & II & $11$ & $(2)$ & $2$ & 71.533 & I & $15$ & $(1)$ & $1$ \\ 
59.411 & III & $3$ & $(1)$ & $2$ & 62.452 & IV & $3$ & $(1)$ & $8$ & 66.493 & III & $3$ & $(1)$ & $2$ & 71.534 & II & $15$ & $(1)$ & $1$ \\ 
59.412 & IV & $6$ & $(1)$ & $4$ & 62.453 & IV & $3$ & $(1)$ & $4$ & 66.494 & III & $4$ & $(2)$ & $2$ & 71.535 & III & $5$ & $(1)$ & $1$ \\ 
59.413 & IV & $6$ & $(1)$ & $4$ & 62.454 & IV & $4$ & $(1)$ & $4$ & 66.495 & III & $9$ & $(2, 2)$ & $2$ & 71.536 & III & $9$ & $(2, 2)$ & $1$ \\ 
59.414 & IV & $5$ & $(1)$ & $4$ & 62.455 & IV & $3$ & $(1)$ & $4$ & 66.496 & III & $6$ & $(2, 2)$ & $2$ & 71.537 & III & $7$ & $(1)$ & $1$ \\ 
59.415 & IV & $7$ & $(1)$ & $2$ & 62.456 & IV & $4$ & $(1)$ & $4$ & 66.497 & III & $6$ & $(2)$ & $2$ & 71.538 & IV & $11$ & $(1)$ & $2$ \\ 
59.416 & IV & $6$ & $(1)$ & $2$ & 63.457 & I & $8$ & $(1)$ & $2$ & 66.498 & IV & $10$ & $(2)$ & $2$ & 72.539 & I & $9$ & $(1)$ & $2$ \\ 
60.417 & I & $4$ & $(2)$ & $4$ & 63.458 & II & $8$ & $(1)$ & $2$ & 66.499 & IV & $8$ & $(2)$ & $4$ & 72.540 & II & $9$ & $(1)$ & $2$ \\ 
60.418 & II & $4$ & $(2)$ & $4$ & 63.459 & III & $3$ & $(1)$ & $2$ & 66.500 & IV & $9$ & $(2)$ & $2$ & 72.541 & III & $3$ & $(1)$ & $2$ \\ 
60.419 & III & $1$ & $(1)$ & $4$ & 63.460 & III & $4$ & $(1)$ & $2$ & 67.501 & I & $13$ & $(1)$ & $2$ & 72.542 & III & $3$ & $(2)$ & $2$ \\ 
60.420 & III & $2$ & $(1)$ & $4$ & 63.461 & III & $2$ & $(1)$ & $2$ & 67.502 & II & $13$ & $(1)$ & $2$ & 72.543 & III & $7$ & $(2, 2)$ & $2$ \\ 
60.421 & III & $1$ & $(1)$ & $4$ & 63.462 & III & $5$ & $(2)$ & $2$ & 67.503 & III & $4$ & $(1)$ & $2$ & 72.544 & III & $5$ & $(2, 2)$ & $2$ \\ 
60.422 & III & $3$ & $(2)$ & $4$ & 63.463 & III & $6$ & $(2, 2)$ & $2$ & 67.504 & III & $5$ & $(1)$ & $2$ & 72.545 & III & $5$ & $(2)$ & $2$ \\ 
60.423 & III & $3$ & $(2)$ & $4$ & 63.464 & III & $5$ & $(2, 2)$ & $2$ & 67.505 & III & $7$ & $(2, 2)$ & $2$ & 72.546 & IV & $9$ & $(1)$ & $2$ \\ 
60.424 & III & $4$ & $(2, 2)$ & $4$ & 63.465 & III & $3$ & $(1)$ & $2$ & 67.506 & III & $8$ & $(2, 2)$ & $2$ & 72.547 & IV & $7$ & $(1)$ & $4$ \\ 
60.425 & III & $2$ & $(1)$ & $4$ & 63.466 & IV & $8$ & $(1)$ & $2$ & 67.507 & III & $7$ & $(2)$ & $2$ & 73.548 & I & $6$ & $(2)$ & $2^* $ \\ 
60.426 & IV & $3$ & $(2, 2)$ & $4^* $ & 63.467 & IV & $7$ & $(1)$ & $4$ & 67.508 & IV & $9$ & $(1)$ & $4$ & 73.549 & II & $6$ & $(1)$ & $4$ \\ 
60.427 & IV & $3$ & $(2)$ & $8$ & 63.468 & IV & $7$ & $(1)$ & $2$ & 67.509 & IV & $13$ & $(1)$ & $2$ & 73.550 & III & $2$ & $(2)$ & $2^* $ \\ 
60.428 & IV & $4$ & $(2)$ & $4$ & 64.469 & I & $7$ & $(1)$ & $2$ & 67.510 & IV & $9$ & $(1)$ & $2$ & 73.551 & III & $4$ & $(2, 2)$ & $4$ \\ 
60.429 & IV & $3$ & $(2)$ & $4$ & 64.470 & II & $7$ & $(1)$ & $2$ & 68.511 & I & $7$ & $(2)$ & $2$ & 73.552 & III & $4$ & $(2, 2)$ & $4$ \\ 
60.430 & IV & $3$ & $(2)$ & $4$ & 64.471 & III & $2$ & $(1)$ & $2$ & 68.512 & II & $7$ & $(2)$ & $2$ & 73.553 & IV & $6$ & $(2)$ & $2^* $ \\ 
60.431 & IV & $4$ & $(2)$ & $4$ & 64.472 & III & $3$ & $(1)$ & $2$ & 68.513 & III & $2$ & $(1)$ & $2$ & 74.554 & I & $10$ & $(1)$ & $2$ \\ 
60.432 & IV & $4$ & $(2)$ & $4$ & 64.473 & III & $2$ & $(1)$ & $2$ & 68.514 & III & $3$ & $(2)$ & $2$ & 74.555 & II & $10$ & $(1)$ & $2$ \\ 
61.433 & I & $3$ & $(2)$ & $4$ & 64.474 & III & $4$ & $(2)$ & $2$ & 68.515 & III & $5$ & $(2, 2)$ & $2$ & 74.556 & III & $3$ & $(1)$ & $2$ \\ 
61.434 & II & $3$ & $(1)$ & $4$ & 64.475 & III & $6$ & $(2, 2)$ & $2$ & 68.516 & III & $4$ & $(2, 2)$ & $2$ & 74.557 & III & $4$ & $(1)$ & $2$ \\ 
61.435 & III & $1$ & $(1)$ & $4$ & 64.476 & III & $5$ & $(2, 2)$ & $2$ & 68.517 & III & $5$ & $(2, 2)$ & $2$ & 74.558 & III & $6$ & $(2, 2)$ & $2$ \\ 
61.436 & III & $3$ & $(2)$ & $4$ & 64.477 & III & $3$ & $(2)$ & $2$ & 68.518 & IV & $7$ & $(2)$ & $4$ & 74.559 & III & $7$ & $(2, 2)$ & $2$ \\ 
61.437 & III & $1$ & $(1)$ & $4$ & 64.478 & IV & $6$ & $(1)$ & $4$ & 68.519 & IV & $7$ & $(2, 2)$ & $2^* $ & 74.560 & III & $4$ & $(1)$ & $2$ \\ 
61.438 & IV & $2$ & $(2)$ & $4^* $ & 64.479 & IV & $6$ & $(1)$ & $4$ & 68.520 & IV & $7$ & $(2)$ & $2$ & 74.561 & IV & $8$ & $(1)$ & $4$ \\ 
61.439 & IV & $3$ & $(2)$ & $4$ & 64.480 & IV & $7$ & $(1)$ & $2$ & 69.521 & I & $15$ & $(1)$ & $1$ & 74.562 & IV & $9$ & $(1)$ & $2$ \\ 
61.440 & IV & $2$ & $(2)$ & $4^* $ & 65.481 & I & $18$ & $(1)$ & $1$ & 69.522 & II & $15$ & $(1)$ & $1$ & ~ & ~ & ~ & ~ \\ 
62.441 & I & $4$ & $(1)$ & $4$ & 65.482 & II & $18$ & $(1)$ & $1$ & 69.523 & III & $5$ & $(1)$ & $1$ & ~ & ~ & ~ & ~ \\ 
\hline \hline 
\end{tabular}

%% file: spinless_OrthoIII_foot.tex
\newlength{\tabLspinlessOrthoIII} 
\settowidth{\tabLspinlessOrthoIII}{\input{spinless_OrthoIII}} 
\begin{minipage}{\tabLspinlessOrthoIII} 
\begin{flushleft} 
{\footnotesize $d$: Rank of the band structure group $\{{\rm BS}\}$\\ 
$X_{\rm BS}$: Symmetry-based indicators of band topology\\ 
$\nu_{\rm BS}$: Set of $\nu$ bands are symmetry-forbidden from being isolated by band gaps if $\nu \not \in \nu_{\rm BS}\, \mathbb Z$ \\
$*$: Exhibiting exceptional filling pattern; see Table \ref{tab:spinlessnuEx} 
}
\end{flushleft}\end{minipage}

%% file: spinless_Tetra.tex
\begin{tabular}{cc|ccc||cc|ccc||cc|ccc||cc|ccc} 
\hline \hline 
\multicolumn{2}{c|}{MSG} & $d$ & $X_{\rm BS}$ & $\nu_{\rm BS}$ & \multicolumn{2}{c|}{MSG} & $d$ & $X_{\rm BS}$ & $\nu_{\rm BS}$ & \multicolumn{2}{c|}{MSG} & $d$ & $X_{\rm BS}$ & $\nu_{\rm BS}$ & \multicolumn{2}{c|}{MSG} & $d$ & $X_{\rm BS}$ & $\nu_{\rm BS}$\\ 
\hline 
75.1 & I & $8$ & $(4)$ & $1$ & 84.51 & I & $13$ & $(2, 4)$ & $2$ & 90.101 & IV & $7$ & $(1)$ & $2$ & 97.151 & I & $8$ & $(1)$ & $1$ \\ 
75.2 & II & $6$ & $(2)$ & $1$ & 84.52 & II & $11$ & $(2, 2)$ & $2$ & 90.102 & IV & $6$ & $(2)$ & $2$ & 97.152 & II & $8$ & $(1)$ & $1$ \\ 
75.3 & III & $4$ & $(1)$ & $1$ & 84.53 & III & $9$ & $(2)$ & $2$ & 91.103 & I & $4$ & $(1)$ & $4$ & 97.153 & III & $5$ & $(1)$ & $1$ \\ 
75.4 & IV & $6$ & $(2)$ & $2$ & 84.54 & III & $4$ & $(2)$ & $2$ & 91.104 & II & $4$ & $(1)$ & $4$ & 97.154 & III & $5$ & $(2)$ & $1$ \\ 
75.5 & IV & $5$ & $(2)$ & $2$ & 84.55 & III & $8$ & $(2, 2, 2)$ & $2$ & 91.105 & III & $3$ & $(1)$ & $4$ & 97.155 & III & $5$ & $(1)$ & $1$ \\ 
75.6 & IV & $5$ & $(2)$ & $2$ & 84.56 & IV & $10$ & $(2)$ & $2$ & 91.106 & III & $1$ & $(1)$ & $4$ & 97.156 & IV & $7$ & $(1)$ & $2$ \\ 
76.7 & I & $1$ & $(1)$ & $4$ & 84.57 & IV & $8$ & $(2, 2)$ & $4$ & 91.107 & III & $2$ & $(1)$ & $4$ & 98.157 & I & $5$ & $(1)$ & $2$ \\ 
76.8 & II & $1$ & $(1)$ & $4$ & 84.58 & IV & $9$ & $(2)$ & $2$ & 91.108 & IV & $4$ & $(1)$ & $4$ & 98.158 & II & $5$ & $(1)$ & $2$ \\ 
76.9 & III & $1$ & $(1)$ & $4$ & 85.59 & I & $11$ & $(2, 4)$ & $2$ & 91.109 & IV & $3$ & $(2)$ & $4^* $ & 98.159 & III & $3$ & $(1)$ & $2$ \\ 
76.10 & IV & $1$ & $(1)$ & $4$ & 85.60 & II & $8$ & $(2)$ & $2$ & 91.110 & IV & $3$ & $(1)$ & $4$ & 98.160 & III & $2$ & $(1)$ & $2$ \\ 
76.11 & IV & $1$ & $(2)$ & $4^* $ & 85.61 & III & $5$ & $(1)$ & $2$ & 92.111 & I & $2$ & $(1)$ & $4$ & 98.161 & III & $4$ & $(1)$ & $2$ \\ 
76.12 & IV & $1$ & $(1)$ & $4$ & 85.62 & III & $5$ & $(2)$ & $2$ & 92.112 & II & $2$ & $(1)$ & $4$ & 98.162 & IV & $5$ & $(2)$ & $2^* $ \\ 
77.13 & I & $4$ & $(2)$ & $2$ & 85.63 & III & $7$ & $(2, 2, 2)$ & $2$ & 92.113 & III & $1$ & $(1)$ & $4$ & 99.163 & I & $9$ & $(1)$ & $1$ \\ 
77.14 & II & $4$ & $(2)$ & $2$ & 85.64 & IV & $7$ & $(2)$ & $4$ & 92.114 & III & $1$ & $(1)$ & $4$ & 99.164 & II & $9$ & $(1)$ & $1$ \\ 
77.15 & III & $4$ & $(2)$ & $2$ & 85.65 & IV & $8$ & $(2)$ & $2$ & 92.115 & III & $2$ & $(1)$ & $4$ & 99.165 & III & $5$ & $(1)$ & $1$ \\ 
77.16 & IV & $4$ & $(1)$ & $2$ & 85.66 & IV & $7$ & $(2)$ & $2$ & 92.116 & IV & $2$ & $(1)$ & $4$ & 99.166 & III & $6$ & $(1)$ & $1$ \\ 
77.17 & IV & $3$ & $(2, 2)$ & $2^* $ & 86.67 & I & $9$ & $(2, 2)$ & $2$ & 92.117 & IV & $2$ & $(2)$ & $4^* $ & 99.167 & III & $8$ & $(4)$ & $1$ \\ 
77.18 & IV & $3$ & $(1)$ & $2$ & 86.68 & II & $7$ & $(2)$ & $2$ & 92.118 & IV & $2$ & $(1)$ & $4$ & 99.168 & IV & $9$ & $(1)$ & $2$ \\ 
78.19 & I & $1$ & $(1)$ & $4$ & 86.69 & III & $5$ & $(2)$ & $2$ & 93.119 & I & $10$ & $(1)$ & $2$ & 99.169 & IV & $6$ & $(1)$ & $2$ \\ 
78.20 & II & $1$ & $(1)$ & $4$ & 86.70 & III & $3$ & $(2)$ & $2$ & 93.120 & II & $10$ & $(1)$ & $2$ & 99.170 & IV & $6$ & $(1)$ & $2$ \\ 
78.21 & III & $1$ & $(1)$ & $4$ & 86.71 & III & $7$ & $(2, 2, 2, 2)$ & $2$ & 93.121 & III & $8$ & $(2)$ & $2$ & 100.171 & I & $6$ & $(1)$ & $2$ \\ 
78.22 & IV & $1$ & $(1)$ & $4$ & 86.72 & IV & $6$ & $(1)$ & $4$ & 93.122 & III & $4$ & $(2)$ & $2$ & 100.172 & II & $5$ & $(1)$ & $2$ \\ 
78.23 & IV & $1$ & $(2)$ & $4^* $ & 86.73 & IV & $7$ & $(2, 2)$ & $2^* $ & 93.123 & III & $6$ & $(2)$ & $2$ & 100.173 & III & $4$ & $(1)$ & $2$ \\ 
78.24 & IV & $1$ & $(1)$ & $4$ & 86.74 & IV & $6$ & $(1)$ & $2$ & 93.124 & IV & $7$ & $(1)$ & $2$ & 100.174 & III & $3$ & $(1)$ & $2$ \\ 
79.25 & I & $5$ & $(2)$ & $1$ & 87.75 & I & $16$ & $(4, 4)$ & $1$ & 93.125 & IV & $7$ & $(2)$ & $2^* $ & 100.175 & III & $5$ & $(4)$ & $2$ \\ 
79.26 & II & $4$ & $(1)$ & $1$ & 87.76 & II & $12$ & $(2, 2)$ & $1$ & 93.126 & IV & $6$ & $(1)$ & $2$ & 100.176 & IV & $5$ & $(1)$ & $4$ \\ 
79.27 & III & $3$ & $(1)$ & $1$ & 87.77 & III & $8$ & $(1)$ & $1$ & 94.127 & I & $5$ & $(1)$ & $2$ & 100.177 & IV & $6$ & $(1)$ & $2$ \\ 
79.28 & IV & $4$ & $(1)$ & $2$ & 87.78 & III & $4$ & $(2)$ & $1$ & 94.128 & II & $5$ & $(1)$ & $2$ & 100.178 & IV & $6$ & $(1)$ & $4$ \\ 
80.29 & I & $2$ & $(1)$ & $2$ & 87.79 & III & $7$ & $(2, 2)$ & $1$ & 94.129 & III & $3$ & $(2)$ & $2$ & 101.179 & I & $5$ & $(1)$ & $2$ \\ 
80.30 & II & $2$ & $(1)$ & $2$ & 87.80 & IV & $11$ & $(2, 2)$ & $2$ & 94.130 & III & $3$ & $(2)$ & $2$ & 101.180 & II & $5$ & $(1)$ & $2$ \\ 
80.31 & III & $2$ & $(1)$ & $2$ & 88.81 & I & $8$ & $(2, 2)$ & $2$ & 94.131 & III & $5$ & $(2)$ & $2$ & 101.181 & III & $5$ & $(1)$ & $2$ \\ 
80.32 & IV & $2$ & $(2)$ & $2^* $ & 88.82 & II & $6$ & $(2)$ & $2$ & 94.132 & IV & $4$ & $(1)$ & $4$ & 101.182 & III & $4$ & $(2)$ & $2$ \\ 
81.33 & I & $12$ & $(2, 2, 4)$ & $1$ & 88.83 & III & $4$ & $(2)$ & $2$ & 94.133 & IV & $5$ & $(2)$ & $2^* $ & 101.183 & III & $4$ & $(2)$ & $2$ \\ 
81.34 & II & $8$ & $(2, 2)$ & $1$ & 88.84 & III & $2$ & $(1)$ & $2$ & 94.134 & IV & $4$ & $(1)$ & $2$ & 101.184 & IV & $5$ & $(1)$ & $2$ \\ 
81.35 & III & $4$ & $(1)$ & $1$ & 88.85 & III & $6$ & $(2, 2)$ & $2$ & 95.135 & I & $4$ & $(1)$ & $4$ & 101.185 & IV & $4$ & $(1)$ & $4$ \\ 
81.36 & IV & $8$ & $(2, 2)$ & $2$ & 88.86 & IV & $5$ & $(2, 2)$ & $2^* $ & 95.136 & II & $4$ & $(1)$ & $4$ & 101.186 & IV & $4$ & $(1)$ & $4$ \\ 
81.37 & IV & $7$ & $(2, 2, 2)$ & $2$ & 89.87 & I & $12$ & $(1)$ & $1$ & 95.137 & III & $3$ & $(1)$ & $4$ & 102.187 & I & $4$ & $(1)$ & $2$ \\ 
81.38 & IV & $7$ & $(2, 2)$ & $2$ & 89.88 & II & $12$ & $(1)$ & $1$ & 95.138 & III & $1$ & $(1)$ & $4$ & 102.188 & II & $4$ & $(1)$ & $2$ \\ 
82.39 & I & $11$ & $(2, 2, 2)$ & $1$ & 89.89 & III & $8$ & $(1)$ & $1$ & 95.139 & III & $2$ & $(1)$ & $4$ & 102.189 & III & $4$ & $(1)$ & $2$ \\ 
82.40 & II & $7$ & $(2)$ & $1$ & 89.90 & III & $8$ & $(4)$ & $1$ & 95.140 & IV & $4$ & $(1)$ & $4$ & 102.190 & III & $3$ & $(2)$ & $2$ \\ 
82.41 & III & $3$ & $(1)$ & $1$ & 89.91 & III & $6$ & $(1)$ & $1$ & 95.141 & IV & $3$ & $(2)$ & $4^* $ & 102.191 & III & $3$ & $(2)$ & $2$ \\ 
82.42 & IV & $7$ & $(2, 2)$ & $2$ & 89.92 & IV & $9$ & $(2)$ & $2$ & 95.142 & IV & $3$ & $(1)$ & $4$ & 102.192 & IV & $4$ & $(1)$ & $4$ \\ 
83.43 & I & $24$ & $(4, 4, 4)$ & $1$ & 89.93 & IV & $8$ & $(1)$ & $2$ & 96.143 & I & $2$ & $(1)$ & $4$ & 102.193 & IV & $4$ & $(1)$ & $4$ \\ 
83.44 & II & $18$ & $(2, 2, 2)$ & $1$ & 89.94 & IV & $7$ & $(2)$ & $2$ & 96.144 & II & $2$ & $(1)$ & $4$ & 102.194 & IV & $4$ & $(1)$ & $2$ \\ 
83.45 & III & $12$ & $(1)$ & $1$ & 90.95 & I & $7$ & $(1)$ & $2$ & 96.145 & III & $1$ & $(1)$ & $4$ & 103.195 & I & $6$ & $(2)$ & $2$ \\ 
83.46 & III & $6$ & $(2)$ & $1$ & 90.96 & II & $6$ & $(1)$ & $2$ & 96.146 & III & $1$ & $(1)$ & $4$ & 103.196 & II & $6$ & $(2)$ & $2$ \\ 
83.47 & III & $8$ & $(2, 2)$ & $1$ & 90.97 & III & $3$ & $(1)$ & $2$ & 96.147 & III & $2$ & $(1)$ & $4$ & 103.197 & III & $4$ & $(1)$ & $2$ \\ 
83.48 & IV & $14$ & $(2, 4)$ & $2$ & 90.98 & III & $5$ & $(4)$ & $2$ & 96.148 & IV & $2$ & $(1)$ & $4$ & 103.198 & III & $4$ & $(1)$ & $2$ \\ 
83.49 & IV & $15$ & $(2, 2, 2)$ & $2$ & 90.99 & III & $5$ & $(1)$ & $2$ & 96.149 & IV & $2$ & $(2)$ & $4^* $ & 103.199 & III & $8$ & $(4)$ & $2$ \\ 
83.50 & IV & $13$ & $(2, 4)$ & $2$ & 90.100 & IV & $5$ & $(2)$ & $4$ & 96.150 & IV & $2$ & $(1)$ & $4$ & 103.200 & IV & $6$ & $(2)$ & $2$ \\ 
\hline \hline 
\end{tabular}

%% file: spinless_TetraII.tex
\begin{tabular}{cc|ccc||cc|ccc||cc|ccc||cc|ccc} 
\hline \hline 
103.201 & IV & $4$ & $(1)$ & $4$ & 111.251 & I & $13$ & $(1)$ & $1$ & 117.301 & III & $5$ & $(1)$ & $2$ & 124.351 & I & $17$ & $(4)$ & $2$ \\ 
103.202 & IV & $4$ & $(2)$ & $4$ & 111.252 & II & $13$ & $(1)$ & $1$ & 117.302 & III & $3$ & $(1)$ & $2$ & 124.352 & II & $15$ & $(2)$ & $2$ \\ 
104.203 & I & $5$ & $(2)$ & $2$ & 111.253 & III & $5$ & $(1)$ & $1$ & 117.303 & III & $7$ & $(2, 2, 4)$ & $2$ & 124.353 & III & $6$ & $(2)$ & $2$ \\ 
104.204 & II & $4$ & $(1)$ & $2$ & 111.254 & III & $8$ & $(1)$ & $1$ & 117.304 & IV & $6$ & $(2)$ & $4$ & 124.354 & III & $9$ & $(1)$ & $2$ \\ 
104.205 & III & $3$ & $(1)$ & $2$ & 111.255 & III & $12$ & $(2, 2, 4)$ & $1$ & 117.305 & IV & $9$ & $(2)$ & $2$ & 124.355 & III & $10$ & $(1)$ & $2$ \\ 
104.206 & III & $3$ & $(1)$ & $2$ & 111.256 & IV & $9$ & $(1)$ & $2$ & 117.306 & IV & $6$ & $(2)$ & $4$ & 124.356 & III & $8$ & $(1)$ & $2$ \\ 
104.207 & III & $5$ & $(2)$ & $2$ & 111.257 & IV & $8$ & $(1)$ & $2$ & 118.307 & I & $9$ & $(2)$ & $2$ & 124.357 & III & $16$ & $(4, 4)$ & $2$ \\ 
104.208 & IV & $4$ & $(2)$ & $4$ & 111.258 & IV & $8$ & $(1)$ & $2$ & 118.308 & II & $7$ & $(1)$ & $2$ & 124.358 & III & $7$ & $(1)$ & $2$ \\ 
104.209 & IV & $5$ & $(2)$ & $4$ & 112.259 & I & $12$ & $(2)$ & $2$ & 118.309 & III & $5$ & $(1)$ & $2$ & 124.359 & III & $9$ & $(2)$ & $2$ \\ 
104.210 & IV & $5$ & $(2)$ & $2$ & 112.260 & II & $10$ & $(1)$ & $2$ & 118.310 & III & $3$ & $(1)$ & $2$ & 124.360 & IV & $17$ & $(4)$ & $2$ \\ 
105.211 & I & $6$ & $(1)$ & $2$ & 112.261 & III & $4$ & $(2)$ & $2$ & 118.311 & III & $7$ & $(2, 2, 2)$ & $2$ & 124.361 & IV & $11$ & $(2)$ & $4$ \\ 
105.212 & II & $6$ & $(1)$ & $2$ & 112.262 & III & $8$ & $(1)$ & $2$ & 118.312 & IV & $6$ & $(2)$ & $4$ & 124.362 & IV & $11$ & $(4)$ & $4$ \\ 
105.213 & III & $4$ & $(2)$ & $2$ & 112.263 & III & $8$ & $(2, 2, 2)$ & $2$ & 118.313 & IV & $7$ & $(2)$ & $4$ & 125.363 & I & $13$ & $(1)$ & $2$ \\ 
105.214 & III & $6$ & $(1)$ & $2$ & 112.264 & IV & $10$ & $(2)$ & $2$ & 118.314 & IV & $8$ & $(2)$ & $2$ & 125.364 & II & $13$ & $(1)$ & $2$ \\ 
105.215 & III & $4$ & $(2)$ & $2$ & 112.265 & IV & $7$ & $(2)$ & $2^* $ & 119.315 & I & $10$ & $(1)$ & $1$ & 125.365 & III & $6$ & $(1)$ & $2$ \\ 
105.216 & IV & $6$ & $(1)$ & $2$ & 112.266 & IV & $7$ & $(2)$ & $2$ & 119.316 & II & $10$ & $(1)$ & $1$ & 125.366 & III & $8$ & $(1)$ & $2$ \\ 
105.217 & IV & $4$ & $(1)$ & $4$ & 113.267 & I & $8$ & $(1)$ & $2$ & 119.317 & III & $5$ & $(1)$ & $1$ & 125.367 & III & $7$ & $(1)$ & $2$ \\ 
105.218 & IV & $4$ & $(1)$ & $2$ & 113.268 & II & $6$ & $(1)$ & $2$ & 119.318 & III & $4$ & $(1)$ & $1$ & 125.368 & III & $8$ & $(1)$ & $2$ \\ 
106.219 & I & $3$ & $(2)$ & $2^* $ & 113.269 & III & $4$ & $(1)$ & $2$ & 119.319 & III & $11$ & $(2, 2, 2)$ & $1$ & 125.369 & III & $11$ & $(2, 4)$ & $2$ \\ 
106.220 & II & $3$ & $(1)$ & $4$ & 113.270 & III & $3$ & $(1)$ & $2$ & 119.320 & IV & $8$ & $(1)$ & $2$ & 125.370 & III & $9$ & $(2)$ & $2$ \\ 
106.221 & III & $3$ & $(2)$ & $4$ & 113.271 & III & $7$ & $(2, 2, 4)$ & $2$ & 120.321 & I & $9$ & $(2)$ & $2$ & 125.371 & III & $8$ & $(1)$ & $2$ \\ 
106.222 & III & $3$ & $(2)$ & $2^* $ & 113.272 & IV & $6$ & $(1)$ & $4$ & 120.322 & II & $7$ & $(1)$ & $2$ & 125.372 & IV & $9$ & $(1)$ & $4$ \\ 
106.223 & III & $3$ & $(2)$ & $4$ & 113.273 & IV & $8$ & $(1)$ & $2$ & 120.323 & III & $5$ & $(1)$ & $2$ & 125.373 & IV & $13$ & $(1)$ & $2$ \\ 
106.224 & IV & $3$ & $(1)$ & $4$ & 113.274 & IV & $6$ & $(1)$ & $2$ & 120.324 & III & $3$ & $(2)$ & $2$ & 125.374 & IV & $9$ & $(1)$ & $4$ \\ 
106.225 & IV & $3$ & $(2)$ & $2^* $ & 114.275 & I & $7$ & $(2)$ & $2$ & 120.325 & III & $7$ & $(2, 2, 2)$ & $2$ & 126.375 & I & $10$ & $(2)$ & $2$ \\ 
106.226 & IV & $3$ & $(1)$ & $4$ & 114.276 & II & $5$ & $(1)$ & $2$ & 120.326 & IV & $9$ & $(2)$ & $2$ & 126.376 & II & $9$ & $(1)$ & $2$ \\ 
107.227 & I & $6$ & $(1)$ & $1$ & 114.277 & III & $3$ & $(2)$ & $2$ & 121.327 & I & $10$ & $(1)$ & $1$ & 126.377 & III & $5$ & $(2)$ & $2$ \\ 
107.228 & II & $6$ & $(1)$ & $1$ & 114.278 & III & $3$ & $(1)$ & $2$ & 121.328 & II & $9$ & $(1)$ & $1$ & 126.378 & III & $5$ & $(1)$ & $2$ \\ 
107.229 & III & $4$ & $(1)$ & $1$ & 114.279 & III & $7$ & $(2, 2, 2)$ & $2$ & 121.329 & III & $4$ & $(1)$ & $1$ & 126.379 & III & $6$ & $(1)$ & $2$ \\ 
107.230 & III & $4$ & $(1)$ & $1$ & 114.280 & IV & $5$ & $(2)$ & $4$ & 121.330 & III & $5$ & $(1)$ & $1$ & 126.380 & III & $7$ & $(1)$ & $2$ \\ 
107.231 & III & $5$ & $(2)$ & $1$ & 114.281 & IV & $5$ & $(2)$ & $2^* $ & 121.331 & III & $9$ & $(2, 2, 2)$ & $1$ & 126.381 & III & $8$ & $(2, 2)$ & $2$ \\ 
107.232 & IV & $6$ & $(1)$ & $2$ & 114.282 & IV & $7$ & $(2)$ & $2$ & 121.332 & IV & $8$ & $(1)$ & $2$ & 126.382 & III & $6$ & $(2)$ & $2$ \\ 
108.233 & I & $5$ & $(1)$ & $2$ & 115.283 & I & $12$ & $(1)$ & $1$ & 122.333 & I & $7$ & $(2)$ & $2$ & 126.383 & III & $7$ & $(1)$ & $2$ \\ 
108.234 & II & $5$ & $(1)$ & $2$ & 115.284 & II & $12$ & $(1)$ & $1$ & 122.334 & II & $5$ & $(1)$ & $2$ & 126.384 & IV & $9$ & $(2)$ & $4$ \\ 
108.235 & III & $4$ & $(1)$ & $2$ & 115.285 & III & $6$ & $(1)$ & $1$ & 122.335 & III & $2$ & $(1)$ & $2$ & 126.385 & IV & $8$ & $(2)$ & $4$ \\ 
108.236 & III & $3$ & $(1)$ & $2$ & 115.286 & III & $6$ & $(1)$ & $1$ & 122.336 & III & $3$ & $(1)$ & $2$ & 126.386 & IV & $10$ & $(2)$ & $2$ \\ 
108.237 & III & $5$ & $(4)$ & $2$ & 115.287 & III & $12$ & $(2, 2, 4)$ & $1$ & 122.337 & III & $6$ & $(2, 2)$ & $2$ & 127.387 & I & $18$ & $(1)$ & $2$ \\ 
108.238 & IV & $5$ & $(1)$ & $2$ & 115.288 & IV & $9$ & $(1)$ & $2$ & 122.338 & IV & $5$ & $(2)$ & $2^* $ & 127.388 & II & $15$ & $(1)$ & $2$ \\ 
109.239 & I & $3$ & $(1)$ & $2$ & 115.289 & IV & $8$ & $(1)$ & $2$ & 123.339 & I & $27$ & $(1)$ & $1$ & 127.389 & III & $5$ & $(1)$ & $2$ \\ 
109.240 & II & $3$ & $(1)$ & $2$ & 115.290 & IV & $7$ & $(1)$ & $2$ & 123.340 & II & $27$ & $(1)$ & $1$ & 127.390 & III & $12$ & $(1)$ & $2$ \\ 
109.241 & III & $2$ & $(1)$ & $2$ & 116.291 & I & $10$ & $(2)$ & $2$ & 123.341 & III & $9$ & $(1)$ & $1$ & 127.391 & III & $9$ & $(1)$ & $2$ \\ 
109.242 & III & $3$ & $(1)$ & $2$ & 116.292 & II & $8$ & $(1)$ & $2$ & 123.342 & III & $15$ & $(1)$ & $1$ & 127.392 & III & $6$ & $(1)$ & $2$ \\ 
109.243 & III & $2$ & $(1)$ & $2$ & 116.293 & III & $6$ & $(1)$ & $2$ & 123.343 & III & $18$ & $(1)$ & $1$ & 127.393 & III & $15$ & $(4, 4, 4)$ & $2$ \\ 
109.244 & IV & $3$ & $(1)$ & $4$ & 116.294 & III & $4$ & $(2)$ & $2$ & 123.344 & III & $13$ & $(1)$ & $1$ & 127.394 & III & $7$ & $(1)$ & $2$ \\ 
110.245 & I & $2$ & $(2)$ & $2^* $ & 116.295 & III & $8$ & $(2, 2, 2)$ & $2$ & 123.345 & III & $24$ & $(4, 4, 4)$ & $1$ & 127.395 & III & $6$ & $(1)$ & $2$ \\ 
110.246 & II & $2$ & $(1)$ & $4$ & 116.296 & IV & $9$ & $(2)$ & $2$ & 123.346 & III & $12$ & $(1)$ & $1$ & 127.396 & IV & $11$ & $(1)$ & $4$ \\ 
110.247 & III & $2$ & $(2)$ & $4$ & 116.297 & IV & $7$ & $(2)$ & $4$ & 123.347 & III & $12$ & $(1)$ & $1$ & 127.397 & IV & $18$ & $(1)$ & $2$ \\ 
110.248 & III & $2$ & $(2)$ & $2^* $ & 116.298 & IV & $6$ & $(2)$ & $4$ & 123.348 & IV & $18$ & $(1)$ & $2$ & 127.398 & IV & $12$ & $(1)$ & $4$ \\ 
110.249 & III & $2$ & $(2)$ & $4$ & 117.299 & I & $9$ & $(2)$ & $2$ & 123.349 & IV & $18$ & $(1)$ & $2$ & 128.399 & I & $14$ & $(4)$ & $2$ \\ 
110.250 & IV & $2$ & $(2)$ & $2^* $ & 117.300 & II & $7$ & $(1)$ & $2$ & 123.350 & IV & $15$ & $(1)$ & $2$ & 128.400 & II & $11$ & $(2)$ & $2$ \\ 
\hline \hline 
\end{tabular}

%% file: spinless_TetraIII.tex
\begin{tabular}{cc|ccc||cc|ccc||cc|ccc||cc|ccc} 
\hline \hline 
128.401 & III & $4$ & $(2)$ & $2$ & 131.444 & IV & $15$ & $(1)$ & $2$ & 135.487 & III & $6$ & $(1)$ & $4$ & 138.530 & IV & $7$ & $(1)$ & $4$ \\ 
128.402 & III & $8$ & $(1)$ & $2$ & 131.445 & IV & $11$ & $(1)$ & $4$ & 135.488 & III & $5$ & $(2)$ & $4$ & 139.531 & I & $18$ & $(1)$ & $1$ \\ 
128.403 & III & $7$ & $(1)$ & $2$ & 131.446 & IV & $12$ & $(1)$ & $2$ & 135.489 & III & $8$ & $(2, 4)$ & $4$ & 139.532 & II & $18$ & $(1)$ & $1$ \\ 
128.404 & III & $5$ & $(1)$ & $2$ & 132.447 & I & $15$ & $(1)$ & $2$ & 135.490 & III & $6$ & $(2)$ & $2^* $ & 139.533 & III & $6$ & $(1)$ & $1$ \\ 
128.405 & III & $13$ & $(4, 4)$ & $2$ & 132.448 & II & $15$ & $(1)$ & $2$ & 135.491 & III & $4$ & $(1)$ & $4$ & 139.534 & III & $11$ & $(1)$ & $1$ \\ 
128.406 & III & $6$ & $(1)$ & $2$ & 132.449 & III & $5$ & $(1)$ & $2$ & 135.492 & IV & $8$ & $(1)$ & $4$ & 139.535 & III & $11$ & $(1)$ & $1$ \\ 
128.407 & III & $5$ & $(1)$ & $2$ & 132.450 & III & $11$ & $(1)$ & $2$ & 135.493 & IV & $9$ & $(1)$ & $4$ & 139.536 & III & $10$ & $(1)$ & $1$ \\ 
128.408 & IV & $10$ & $(4)$ & $4$ & 132.451 & III & $10$ & $(2)$ & $2$ & 135.494 & IV & $9$ & $(1)$ & $4$ & 139.537 & III & $16$ & $(4, 4)$ & $1$ \\ 
128.409 & IV & $11$ & $(2)$ & $4$ & 132.452 & III & $9$ & $(1)$ & $2$ & 136.495 & I & $12$ & $(1)$ & $2$ & 139.538 & III & $9$ & $(1)$ & $1$ \\ 
128.410 & IV & $14$ & $(4)$ & $2$ & 132.453 & III & $12$ & $(2, 4)$ & $2$ & 136.496 & II & $11$ & $(1)$ & $2$ & 139.539 & III & $8$ & $(1)$ & $1$ \\ 
129.411 & I & $12$ & $(1)$ & $2$ & 132.454 & III & $10$ & $(2)$ & $2$ & 136.497 & III & $4$ & $(1)$ & $2$ & 139.540 & IV & $15$ & $(1)$ & $2$ \\ 
129.412 & II & $12$ & $(1)$ & $2$ & 132.455 & III & $8$ & $(1)$ & $2$ & 136.498 & III & $10$ & $(1)$ & $2$ & 140.541 & I & $15$ & $(1)$ & $2$ \\ 
129.413 & III & $6$ & $(1)$ & $2$ & 132.456 & IV & $14$ & $(1)$ & $2$ & 136.499 & III & $7$ & $(2)$ & $2$ & 140.542 & II & $14$ & $(1)$ & $2$ \\ 
129.414 & III & $8$ & $(1)$ & $2$ & 132.457 & IV & $11$ & $(1)$ & $4$ & 136.500 & III & $6$ & $(1)$ & $2$ & 140.543 & III & $5$ & $(1)$ & $2$ \\ 
129.415 & III & $6$ & $(1)$ & $2$ & 132.458 & IV & $10$ & $(1)$ & $4$ & 136.501 & III & $9$ & $(2, 4)$ & $2$ & 140.544 & III & $10$ & $(1)$ & $2$ \\ 
129.416 & III & $8$ & $(1)$ & $2$ & 133.459 & I & $9$ & $(2)$ & $2^* $ & 136.502 & III & $7$ & $(2)$ & $2$ & 140.545 & III & $8$ & $(1)$ & $2$ \\ 
129.417 & III & $11$ & $(2, 4)$ & $2$ & 133.460 & II & $8$ & $(1)$ & $4$ & 136.503 & III & $5$ & $(1)$ & $2$ & 140.546 & III & $7$ & $(1)$ & $2$ \\ 
129.418 & III & $8$ & $(1)$ & $2$ & 133.461 & III & $3$ & $(2)$ & $2^* $ & 136.504 & IV & $9$ & $(1)$ & $4$ & 140.547 & III & $13$ & $(4, 4)$ & $2$ \\ 
129.419 & III & $7$ & $(1)$ & $2$ & 133.462 & III & $5$ & $(2)$ & $4$ & 136.505 & IV & $11$ & $(1)$ & $4$ & 140.548 & III & $8$ & $(1)$ & $2$ \\ 
129.420 & IV & $9$ & $(1)$ & $4$ & 133.463 & III & $6$ & $(2)$ & $2^* $ & 136.506 & IV & $11$ & $(1)$ & $2$ & 140.549 & III & $7$ & $(2)$ & $2$ \\ 
129.421 & IV & $12$ & $(1)$ & $2$ & 133.464 & III & $7$ & $(2)$ & $4$ & 137.507 & I & $8$ & $(1)$ & $2$ & 140.550 & IV & $15$ & $(1)$ & $2$ \\ 
129.422 & IV & $9$ & $(1)$ & $2$ & 133.465 & III & $6$ & $(2, 2)$ & $4$ & 137.508 & II & $8$ & $(1)$ & $2$ & 141.551 & I & $9$ & $(1)$ & $2$ \\ 
130.423 & I & $8$ & $(2)$ & $4$ & 133.466 & III & $6$ & $(2, 2)$ & $2^* $ & 137.509 & III & $4$ & $(1)$ & $2$ & 141.552 & II & $9$ & $(1)$ & $2$ \\ 
130.424 & II & $7$ & $(2)$ & $4$ & 133.467 & III & $6$ & $(1)$ & $4$ & 137.510 & III & $5$ & $(2)$ & $2$ & 141.553 & III & $3$ & $(1)$ & $2$ \\ 
130.425 & III & $4$ & $(2)$ & $4$ & 133.468 & IV & $8$ & $(1)$ & $4$ & 137.511 & III & $5$ & $(1)$ & $2$ & 141.554 & III & $5$ & $(2)$ & $2$ \\ 
130.426 & III & $5$ & $(1)$ & $4$ & 133.469 & IV & $9$ & $(2)$ & $2^* $ & 137.512 & III & $7$ & $(2)$ & $2$ & 141.555 & III & $6$ & $(1)$ & $2$ \\ 
130.427 & III & $4$ & $(1)$ & $4$ & 133.470 & IV & $7$ & $(1)$ & $4$ & 137.513 & III & $8$ & $(2, 2)$ & $2$ & 141.556 & III & $7$ & $(2)$ & $2$ \\ 
130.428 & III & $5$ & $(1)$ & $4$ & 134.471 & I & $12$ & $(1)$ & $2$ & 137.514 & III & $7$ & $(1)$ & $2$ & 141.557 & III & $8$ & $(2, 2)$ & $2$ \\ 
130.429 & III & $8$ & $(2, 4)$ & $4$ & 134.472 & II & $12$ & $(1)$ & $2$ & 137.515 & III & $4$ & $(1)$ & $2$ & 141.558 & III & $6$ & $(1)$ & $2$ \\ 
130.430 & III & $6$ & $(1)$ & $4$ & 134.473 & III & $4$ & $(1)$ & $2$ & 137.516 & IV & $7$ & $(1)$ & $4$ & 141.559 & III & $4$ & $(1)$ & $2$ \\ 
130.431 & III & $6$ & $(2)$ & $4$ & 134.474 & III & $8$ & $(1)$ & $2$ & 137.517 & IV & $7$ & $(1)$ & $4$ & 141.560 & IV & $7$ & $(1)$ & $4$ \\ 
130.432 & IV & $8$ & $(2)$ & $4$ & 134.475 & III & $7$ & $(2)$ & $2$ & 137.518 & IV & $8$ & $(1)$ & $2$ & 142.561 & I & $7$ & $(2)$ & $2^* $ \\ 
130.433 & IV & $7$ & $(1)$ & $4$ & 134.476 & III & $8$ & $(1)$ & $2$ & 138.519 & I & $10$ & $(1)$ & $4$ & 142.562 & II & $6$ & $(1)$ & $4$ \\ 
130.434 & IV & $8$ & $(2)$ & $4$ & 134.477 & III & $9$ & $(2, 2)$ & $2$ & 138.520 & II & $9$ & $(1)$ & $4$ & 142.563 & III & $2$ & $(2)$ & $2^* $ \\ 
131.435 & I & $18$ & $(1)$ & $2$ & 134.478 & III & $9$ & $(2, 2)$ & $2$ & 138.521 & III & $4$ & $(1)$ & $4$ & 142.564 & III & $4$ & $(2)$ & $4$ \\ 
131.436 & II & $18$ & $(1)$ & $2$ & 134.479 & III & $7$ & $(1)$ & $2$ & 138.522 & III & $8$ & $(1)$ & $4$ & 142.565 & III & $4$ & $(2)$ & $2^* $ \\ 
131.437 & III & $6$ & $(1)$ & $2$ & 134.480 & IV & $8$ & $(1)$ & $4$ & 138.523 & III & $5$ & $(2)$ & $4$ & 142.566 & III & $5$ & $(2, 2)$ & $4$ \\ 
131.438 & III & $10$ & $(2)$ & $2$ & 134.481 & IV & $10$ & $(1)$ & $4$ & 138.524 & III & $6$ & $(1)$ & $4$ & 142.567 & III & $5$ & $(2, 2)$ & $4$ \\ 
131.439 & III & $15$ & $(1)$ & $2$ & 134.482 & IV & $10$ & $(1)$ & $2$ & 138.525 & III & $7$ & $(2, 2)$ & $4$ & 142.568 & III & $5$ & $(2)$ & $2^* $ \\ 
131.440 & III & $12$ & $(2)$ & $2$ & 135.483 & I & $9$ & $(1)$ & $4$ & 138.526 & III & $7$ & $(2)$ & $4$ & 142.569 & III & $4$ & $(2)$ & $4$ \\ 
131.441 & III & $13$ & $(2, 4)$ & $2$ & 135.484 & II & $8$ & $(1)$ & $4$ & 138.527 & III & $5$ & $(1)$ & $4$ & 142.570 & IV & $7$ & $(2)$ & $2^* $ \\ 
131.442 & III & $9$ & $(1)$ & $2$ & 135.485 & III & $3$ & $(2)$ & $2^* $ & 138.528 & IV & $8$ & $(1)$ & $4$ & ~ & ~ & ~ & ~ \\ 
131.443 & III & $9$ & $(1)$ & $2$ & 135.486 & III & $7$ & $(2)$ & $4$ & 138.529 & IV & $10$ & $(1)$ & $4$ & ~ & ~ & ~ & ~ \\ 
\hline \hline 
\end{tabular}

%% file: spinless_TetraIII_foot.tex
\newlength{\tabLspinlessTetraIII} 
\settowidth{\tabLspinlessTetraIII}{\input{spinless_TetraIII}} 
\begin{minipage}{\tabLspinlessTetraIII} 
\begin{flushleft} 
{\footnotesize $d$: Rank of the band structure group $\{{\rm BS}\}$\\ 
$X_{\rm BS}$: Symmetry-based indicators of band topology\\ 
$\nu_{\rm BS}$: Set of $\nu$ bands are symmetry-forbidden from being isolated by band gaps if $\nu \not \in \nu_{\rm BS}\, \mathbb Z$ }\\ 
$*$: Exhibiting exceptional filling pattern; see Table \ref{tab:spinlessnuEx} 
\end{flushleft}\end{minipage}

%% file: spinless_Hexa.tex
\begin{tabular}{cc|ccc||cc|ccc||cc|ccc||cc|ccc} 
\hline \hline 
\multicolumn{2}{c|}{MSG} & $d$ & $X_{\rm BS}$ & $\nu_{\rm BS}$ & \multicolumn{2}{c|}{MSG} & $d$ & $X_{\rm BS}$ & $\nu_{\rm BS}$ & \multicolumn{2}{c|}{MSG} & $d$ & $X_{\rm BS}$ & $\nu_{\rm BS}$ & \multicolumn{2}{c|}{MSG} & $d$ & $X_{\rm BS}$ & $\nu_{\rm BS}$\\ 
\hline 
143.1 & I & $7$ & $(3)$ & $1$ & 156.51 & III & $7$ & $(3)$ & $1$ & 166.101 & III & $11$ & $(2, 4)$ & $1$ & 177.151 & III & $5$ & $(1)$ & $1$ \\ 
143.2 & II & $4$ & $(1)$ & $1$ & 156.52 & IV & $5$ & $(1)$ & $2$ & 166.102 & IV & $7$ & $(2)$ & $2$ & 177.152 & III & $6$ & $(1)$ & $1$ \\ 
143.3 & IV & $4$ & $(1)$ & $2$ & 157.53 & I & $5$ & $(1)$ & $1$ & 167.103 & I & $7$ & $(2)$ & $2$ & 177.153 & III & $9$ & $(6)$ & $1$ \\ 
144.4 & I & $1$ & $(1)$ & $3$ & 157.54 & II & $4$ & $(1)$ & $1$ & 167.104 & II & $6$ & $(2)$ & $2$ & 177.154 & IV & $8$ & $(2)$ & $2$ \\ 
144.5 & II & $1$ & $(1)$ & $3$ & 157.55 & III & $5$ & $(3)$ & $1$ & 167.105 & III & $2$ & $(1)$ & $2$ & 178.155 & I & $3$ & $(1)$ & $6$ \\ 
144.6 & IV & $1$ & $(1)$ & $6$ & 157.56 & IV & $4$ & $(1)$ & $2$ & 167.106 & III & $3$ & $(1)$ & $2$ & 178.156 & II & $3$ & $(1)$ & $6$ \\ 
145.7 & I & $1$ & $(1)$ & $3$ & 158.57 & I & $4$ & $(1)$ & $2$ & 167.107 & III & $7$ & $(4)$ & $2$ & 178.157 & III & $2$ & $(1)$ & $6$ \\ 
145.8 & II & $1$ & $(1)$ & $3$ & 158.58 & II & $4$ & $(1)$ & $2$ & 167.108 & IV & $7$ & $(2)$ & $2$ & 178.158 & III & $2$ & $(1)$ & $6$ \\ 
145.9 & IV & $1$ & $(1)$ & $6$ & 158.59 & III & $7$ & $(3)$ & $2$ & 168.109 & I & $9$ & $(6)$ & $1$ & 178.159 & III & $1$ & $(1)$ & $6$ \\ 
146.10 & I & $3$ & $(1)$ & $1$ & 158.60 & IV & $4$ & $(1)$ & $2$ & 168.110 & II & $6$ & $(2)$ & $1$ & 178.160 & IV & $3$ & $(1)$ & $6$ \\ 
146.11 & II & $2$ & $(1)$ & $1$ & 159.61 & I & $4$ & $(1)$ & $2$ & 168.111 & III & $4$ & $(1)$ & $1$ & 179.161 & I & $3$ & $(1)$ & $6$ \\ 
146.12 & IV & $2$ & $(1)$ & $2$ & 159.62 & II & $3$ & $(1)$ & $2$ & 168.112 & IV & $6$ & $(2)$ & $2$ & 179.162 & II & $3$ & $(1)$ & $6$ \\ 
147.13 & I & $13$ & $(2, 12)$ & $1$ & 159.63 & III & $5$ & $(3)$ & $2$ & 169.113 & I & $1$ & $(1)$ & $6$ & 179.163 & III & $2$ & $(1)$ & $6$ \\ 
147.14 & II & $9$ & $(2, 4)$ & $1$ & 159.64 & IV & $3$ & $(1)$ & $2$ & 169.114 & II & $1$ & $(1)$ & $6$ & 179.164 & III & $2$ & $(1)$ & $6$ \\ 
147.15 & III & $4$ & $(1)$ & $1$ & 160.65 & I & $3$ & $(1)$ & $1$ & 169.115 & III & $1$ & $(1)$ & $6$ & 179.165 & III & $1$ & $(1)$ & $6$ \\ 
147.16 & IV & $7$ & $(4)$ & $2$ & 160.66 & II & $3$ & $(1)$ & $1$ & 169.116 & IV & $1$ & $(1)$ & $6$ & 179.166 & IV & $3$ & $(1)$ & $6$ \\ 
148.17 & I & $11$ & $(2, 4)$ & $1$ & 160.67 & III & $3$ & $(1)$ & $1$ & 170.117 & I & $1$ & $(1)$ & $6$ & 180.167 & I & $7$ & $(1)$ & $3$ \\ 
148.18 & II & $8$ & $(2, 4)$ & $1$ & 160.68 & IV & $3$ & $(1)$ & $2$ & 170.118 & II & $1$ & $(1)$ & $6$ & 180.168 & II & $7$ & $(1)$ & $3$ \\ 
148.19 & III & $2$ & $(1)$ & $1$ & 161.69 & I & $2$ & $(1)$ & $2$ & 170.119 & III & $1$ & $(1)$ & $6$ & 180.169 & III & $3$ & $(1)$ & $3$ \\ 
148.20 & IV & $6$ & $(4)$ & $2$ & 161.70 & II & $2$ & $(1)$ & $2$ & 170.120 & IV & $1$ & $(1)$ & $6$ & 180.170 & III & $3$ & $(1)$ & $3$ \\ 
149.21 & I & $6$ & $(1)$ & $1$ & 161.71 & III & $3$ & $(1)$ & $2$ & 171.121 & I & $3$ & $(2)$ & $3$ & 180.171 & III & $3$ & $(2)$ & $3$ \\ 
149.22 & II & $6$ & $(1)$ & $1$ & 161.72 & IV & $2$ & $(1)$ & $2$ & 171.122 & II & $3$ & $(2)$ & $3$ & 180.172 & IV & $5$ & $(2)$ & $6$ \\ 
149.23 & III & $7$ & $(3)$ & $1$ & 162.73 & I & $12$ & $(2)$ & $1$ & 171.123 & III & $1$ & $(1)$ & $3$ & 181.173 & I & $7$ & $(1)$ & $3$ \\ 
149.24 & IV & $5$ & $(1)$ & $2$ & 162.74 & II & $12$ & $(2)$ & $1$ & 171.124 & IV & $3$ & $(2)$ & $6$ & 181.174 & II & $7$ & $(1)$ & $3$ \\ 
150.25 & I & $6$ & $(1)$ & $1$ & 162.75 & III & $5$ & $(1)$ & $1$ & 172.125 & I & $3$ & $(2)$ & $3$ & 181.175 & III & $3$ & $(1)$ & $3$ \\ 
150.26 & II & $5$ & $(1)$ & $1$ & 162.76 & III & $5$ & $(1)$ & $1$ & 172.126 & II & $3$ & $(2)$ & $3$ & 181.176 & III & $3$ & $(1)$ & $3$ \\ 
150.27 & III & $5$ & $(3)$ & $1$ & 162.77 & III & $13$ & $(2, 12)$ & $1$ & 172.127 & III & $1$ & $(1)$ & $3$ & 181.177 & III & $3$ & $(2)$ & $3$ \\ 
150.28 & IV & $4$ & $(1)$ & $2$ & 162.78 & IV & $8$ & $(2)$ & $2$ & 172.128 & IV & $3$ & $(2)$ & $6$ & 181.178 & IV & $5$ & $(2)$ & $6$ \\ 
151.29 & I & $3$ & $(1)$ & $3$ & 163.79 & I & $8$ & $(2)$ & $2$ & 173.129 & I & $5$ & $(3)$ & $2$ & 182.179 & I & $5$ & $(1)$ & $2$ \\ 
151.30 & II & $3$ & $(1)$ & $3$ & 163.80 & II & $7$ & $(2)$ & $2$ & 173.130 & II & $3$ & $(1)$ & $2$ & 182.180 & II & $5$ & $(1)$ & $2$ \\ 
151.31 & III & $1$ & $(1)$ & $3$ & 163.81 & III & $4$ & $(1)$ & $2$ & 173.131 & III & $4$ & $(1)$ & $2$ & 182.181 & III & $4$ & $(1)$ & $2$ \\ 
151.32 & IV & $2$ & $(1)$ & $6$ & 163.82 & III & $4$ & $(1)$ & $2$ & 173.132 & IV & $3$ & $(1)$ & $2$ & 182.182 & III & $5$ & $(1)$ & $2$ \\ 
152.33 & I & $3$ & $(1)$ & $3$ & 163.83 & III & $9$ & $(12)$ & $2$ & 174.133 & I & $21$ & $(3, 3, 3)$ & $1$ & 182.183 & III & $5$ & $(3)$ & $2$ \\ 
152.34 & II & $3$ & $(1)$ & $3$ & 163.84 & IV & $8$ & $(2)$ & $2$ & 174.134 & II & $12$ & $(1)$ & $1$ & 182.184 & IV & $5$ & $(1)$ & $2$ \\ 
152.35 & III & $1$ & $(1)$ & $3$ & 164.85 & I & $12$ & $(2)$ & $1$ & 174.135 & III & $4$ & $(1)$ & $1$ & 183.185 & I & $8$ & $(1)$ & $1$ \\ 
152.36 & IV & $2$ & $(1)$ & $6$ & 164.86 & II & $12$ & $(2)$ & $1$ & 174.136 & IV & $11$ & $(3)$ & $2$ & 183.186 & II & $8$ & $(1)$ & $1$ \\ 
153.37 & I & $3$ & $(1)$ & $3$ & 164.87 & III & $4$ & $(1)$ & $1$ & 175.137 & I & $27$ & $(6, 6, 6)$ & $1$ & 183.187 & III & $5$ & $(1)$ & $1$ \\ 
153.38 & II & $3$ & $(1)$ & $3$ & 164.88 & III & $6$ & $(1)$ & $1$ & 175.138 & II & $18$ & $(2, 2, 2)$ & $1$ & 183.188 & III & $4$ & $(1)$ & $1$ \\ 
153.39 & III & $1$ & $(1)$ & $3$ & 164.89 & III & $13$ & $(2, 12)$ & $1$ & 175.139 & III & $12$ & $(1)$ & $1$ & 183.189 & III & $9$ & $(6)$ & $1$ \\ 
153.40 & IV & $2$ & $(1)$ & $6$ & 164.90 & IV & $8$ & $(2)$ & $2$ & 175.140 & III & $6$ & $(2)$ & $1$ & 183.190 & IV & $8$ & $(1)$ & $2$ \\ 
154.41 & I & $3$ & $(1)$ & $3$ & 165.91 & I & $8$ & $(2)$ & $2$ & 175.141 & III & $9$ & $(2, 4)$ & $1$ & 184.191 & I & $6$ & $(2)$ & $2$ \\ 
154.42 & II & $3$ & $(1)$ & $3$ & 165.92 & II & $7$ & $(2)$ & $2$ & 175.142 & IV & $15$ & $(2, 6)$ & $2$ & 184.192 & II & $6$ & $(2)$ & $2$ \\ 
154.43 & III & $1$ & $(1)$ & $3$ & 165.93 & III & $3$ & $(1)$ & $2$ & 176.143 & I & $16$ & $(3, 6)$ & $2$ & 184.193 & III & $4$ & $(1)$ & $2$ \\ 
154.44 & IV & $2$ & $(1)$ & $6$ & 165.94 & III & $5$ & $(1)$ & $2$ & 176.144 & II & $10$ & $(2)$ & $2$ & 184.194 & III & $3$ & $(1)$ & $2$ \\ 
155.45 & I & $4$ & $(1)$ & $1$ & 165.95 & III & $9$ & $(12)$ & $2$ & 176.145 & III & $11$ & $(3)$ & $2$ & 184.195 & III & $9$ & $(6)$ & $2$ \\ 
155.46 & II & $4$ & $(1)$ & $1$ & 165.96 & IV & $8$ & $(2)$ & $2$ & 176.146 & III & $3$ & $(1)$ & $2$ & 184.196 & IV & $6$ & $(2)$ & $2$ \\ 
155.47 & III & $3$ & $(1)$ & $1$ & 166.97 & I & $11$ & $(2)$ & $1$ & 176.147 & III & $7$ & $(4)$ & $2$ & 185.197 & I & $4$ & $(1)$ & $2$ \\ 
155.48 & IV & $3$ & $(1)$ & $2$ & 166.98 & II & $11$ & $(2)$ & $1$ & 176.148 & IV & $10$ & $(2)$ & $2$ & 185.198 & II & $4$ & $(1)$ & $2$ \\ 
156.49 & I & $5$ & $(1)$ & $1$ & 166.99 & III & $3$ & $(1)$ & $1$ & 177.149 & I & $10$ & $(1)$ & $1$ & 185.199 & III & $5$ & $(1)$ & $2$ \\ 
156.50 & II & $5$ & $(1)$ & $1$ & 166.100 & III & $4$ & $(1)$ & $1$ & 177.150 & II & $10$ & $(1)$ & $1$ & 185.200 & III & $3$ & $(1)$ & $2$ \\ 
\hline \hline 
\end{tabular}

%% file: spinless_HexaII.tex
\begin{tabular}{cc|ccc||cc|ccc||cc|ccc||cc|ccc} 
\hline \hline 
185.201 & III & $5$ & $(3)$ & $2$ & 188.219 & III & $14$ & $(3, 3)$ & $2$ & 191.237 & III & $12$ & $(1)$ & $1$ & 193.255 & III & $4$ & $(1)$ & $2$ \\ 
185.202 & IV & $4$ & $(1)$ & $2$ & 188.220 & IV & $12$ & $(3)$ & $2$ & 191.238 & III & $12$ & $(2)$ & $1$ & 193.256 & III & $10$ & $(1)$ & $2$ \\ 
186.203 & I & $4$ & $(1)$ & $2$ & 189.221 & I & $15$ & $(1)$ & $1$ & 191.239 & III & $12$ & $(2)$ & $1$ & 193.257 & III & $9$ & $(3)$ & $2$ \\ 
186.204 & II & $4$ & $(1)$ & $2$ & 189.222 & II & $12$ & $(1)$ & $1$ & 191.240 & III & $27$ & $(6, 6, 6)$ & $1$ & 193.258 & III & $8$ & $(2)$ & $2$ \\ 
186.205 & III & $4$ & $(1)$ & $2$ & 189.223 & III & $4$ & $(1)$ & $1$ & 191.241 & III & $10$ & $(1)$ & $1$ & 193.259 & III & $8$ & $(2)$ & $2$ \\ 
186.206 & III & $4$ & $(1)$ & $2$ & 189.224 & III & $5$ & $(1)$ & $1$ & 191.242 & IV & $16$ & $(1)$ & $2$ & 193.260 & III & $14$ & $(3, 6)$ & $2$ \\ 
186.207 & III & $5$ & $(3)$ & $2$ & 189.225 & III & $15$ & $(3, 3, 3)$ & $1$ & 192.243 & I & $17$ & $(6)$ & $2$ & 193.261 & III & $5$ & $(1)$ & $2$ \\ 
186.208 & IV & $4$ & $(1)$ & $2$ & 189.226 & IV & $9$ & $(1)$ & $2$ & 192.244 & II & $14$ & $(2)$ & $2$ & 193.262 & IV & $13$ & $(1)$ & $2$ \\ 
187.209 & I & $15$ & $(1)$ & $1$ & 190.227 & I & $12$ & $(3)$ & $2$ & 192.245 & III & $6$ & $(2)$ & $2$ & 194.263 & I & $13$ & $(1)$ & $2$ \\ 
187.210 & II & $15$ & $(1)$ & $1$ & 190.228 & II & $8$ & $(1)$ & $2$ & 192.246 & III & $9$ & $(1)$ & $2$ & 194.264 & II & $13$ & $(1)$ & $2$ \\ 
187.211 & III & $6$ & $(1)$ & $1$ & 190.229 & III & $3$ & $(1)$ & $2$ & 192.247 & III & $8$ & $(1)$ & $2$ & 194.265 & III & $4$ & $(1)$ & $2$ \\ 
187.212 & III & $5$ & $(1)$ & $1$ & 190.230 & III & $4$ & $(1)$ & $2$ & 192.248 & III & $7$ & $(2)$ & $2$ & 194.266 & III & $12$ & $(3)$ & $2$ \\ 
187.213 & III & $21$ & $(3, 3, 3)$ & $1$ & 190.231 & III & $12$ & $(3, 3)$ & $2$ & 192.249 & III & $7$ & $(2)$ & $2$ & 194.267 & III & $9$ & $(1)$ & $2$ \\ 
187.214 & IV & $10$ & $(1)$ & $2$ & 190.232 & IV & $9$ & $(3)$ & $2$ & 192.250 & III & $18$ & $(6, 6)$ & $2$ & 194.268 & III & $8$ & $(2)$ & $2$ \\ 
188.215 & I & $12$ & $(3)$ & $2$ & 191.233 & I & $24$ & $(1)$ & $1$ & 192.251 & III & $8$ & $(2)$ & $2$ & 194.269 & III & $8$ & $(2)$ & $2$ \\ 
188.216 & II & $9$ & $(1)$ & $2$ & 191.234 & II & $24$ & $(1)$ & $1$ & 192.252 & IV & $17$ & $(6)$ & $2$ & 194.270 & III & $16$ & $(3, 6)$ & $2$ \\ 
188.217 & III & $5$ & $(1)$ & $2$ & 191.235 & III & $8$ & $(1)$ & $1$ & 193.253 & I & $13$ & $(1)$ & $2$ & 194.271 & III & $5$ & $(1)$ & $2$ \\ 
188.218 & III & $4$ & $(1)$ & $2$ & 191.236 & III & $15$ & $(1)$ & $1$ & 193.254 & II & $12$ & $(1)$ & $2$ & 194.272 & IV & $12$ & $(1)$ & $2$ \\ 
\hline \hline 
\end{tabular}

%% file: spinless_HexaII_foot.tex
\newlength{\tabLspinlessHexaII} 
\settowidth{\tabLspinlessHexaII}{\input{spinless_HexaII}} 
\begin{minipage}{\tabLspinlessHexaII} 
\begin{flushleft} 
{\footnotesize $d$: Rank of the band structure group $\{{\rm BS}\}$\\ 
$X_{\rm BS}$: Symmetry-based indicators of band topology\\ 
$\nu_{\rm BS}$: Set of $\nu$ bands are symmetry-forbidden from being isolated by band gaps if $\nu \not \in \nu_{\rm BS}\, \mathbb Z$ }
\end{flushleft}\end{minipage}

%% file: spinless_Cubic.tex
\begin{tabular}{cc|ccc||cc|ccc||cc|ccc||cc|ccc} 
\hline \hline 
\multicolumn{2}{c|}{MSG} & $d$ & $X_{\rm BS}$ & $\nu_{\rm BS}$ & \multicolumn{2}{c|}{MSG} & $d$ & $X_{\rm BS}$ & $\nu_{\rm BS}$ & \multicolumn{2}{c|}{MSG} & $d$ & $X_{\rm BS}$ & $\nu_{\rm BS}$ & \multicolumn{2}{c|}{MSG} & $d$ & $X_{\rm BS}$ & $\nu_{\rm BS}$\\ 
\hline 
195.1 & I & $7$ & $(1)$ & $1$ & 206.39 & III & $3$ & $(1)$ & $4$ & 216.77 & IV & $7$ & $(1)$ & $2$ & 224.115 & IV & $10$ & $(1)$ & $2$ \\ 
195.2 & II & $6$ & $(1)$ & $1$ & 207.40 & I & $9$ & $(1)$ & $1$ & 217.78 & I & $9$ & $(1)$ & $1$ & 225.116 & I & $17$ & $(1)$ & $1$ \\ 
195.3 & IV & $4$ & $(2)$ & $2$ & 207.41 & II & $9$ & $(1)$ & $1$ & 217.79 & II & $8$ & $(1)$ & $1$ & 225.117 & II & $17$ & $(1)$ & $1$ \\ 
196.4 & I & $5$ & $(1)$ & $1$ & 207.42 & III & $6$ & $(1)$ & $1$ & 217.80 & III & $5$ & $(1)$ & $1$ & 225.118 & III & $8$ & $(1)$ & $1$ \\ 
196.5 & II & $4$ & $(1)$ & $1$ & 207.43 & IV & $6$ & $(2)$ & $2$ & 218.81 & I & $9$ & $(2)$ & $2$ & 225.119 & III & $13$ & $(1)$ & $1$ \\ 
196.6 & IV & $4$ & $(2)$ & $2$ & 208.44 & I & $7$ & $(1)$ & $2$ & 218.82 & II & $7$ & $(1)$ & $2$ & 225.120 & III & $7$ & $(1)$ & $1$ \\ 
197.7 & I & $5$ & $(1)$ & $1$ & 208.45 & II & $7$ & $(1)$ & $2$ & 218.83 & III & $6$ & $(1)$ & $2$ & 225.121 & IV & $13$ & $(1)$ & $2$ \\ 
197.8 & II & $4$ & $(1)$ & $1$ & 208.46 & III & $6$ & $(2)$ & $2$ & 218.84 & IV & $6$ & $(2)$ & $2$ & 226.122 & I & $14$ & $(1)$ & $2$ \\ 
198.9 & I & $3$ & $(1)$ & $4$ & 208.47 & IV & $5$ & $(1)$ & $2$ & 219.85 & I & $8$ & $(2)$ & $2$ & 226.123 & II & $12$ & $(1)$ & $2$ \\ 
198.10 & II & $2$ & $(1)$ & $4$ & 209.48 & I & $7$ & $(1)$ & $1$ & 219.86 & II & $6$ & $(1)$ & $2$ & 226.124 & III & $7$ & $(1)$ & $2$ \\ 
198.11 & IV & $2$ & $(1)$ & $4^* $ & 209.49 & II & $7$ & $(1)$ & $1$ & 219.87 & III & $5$ & $(1)$ & $2$ & 226.125 & III & $10$ & $(1)$ & $2$ \\ 
199.12 & I & $4$ & $(1)$ & $2^* $ & 209.50 & III & $5$ & $(1)$ & $1$ & 219.88 & IV & $8$ & $(2)$ & $2$ & 226.126 & III & $6$ & $(2)$ & $2$ \\ 
199.13 & II & $3$ & $(1)$ & $2^* $ & 209.51 & IV & $6$ & $(1)$ & $2$ & 220.89 & I & $7$ & $(2)$ & $2^* $ & 226.127 & IV & $14$ & $(1)$ & $2$ \\ 
200.14 & I & $17$ & $(1)$ & $1$ & 210.52 & I & $4$ & $(1)$ & $2$ & 220.90 & II & $5$ & $(1)$ & $2^* $ & 227.128 & I & $11$ & $(1)$ & $2$ \\ 
200.15 & II & $14$ & $(1)$ & $1$ & 210.53 & II & $4$ & $(1)$ & $2$ & 220.91 & III & $4$ & $(1)$ & $2^* $ & 227.129 & II & $11$ & $(1)$ & $2$ \\ 
200.16 & III & $6$ & $(1)$ & $1$ & 210.54 & III & $4$ & $(1)$ & $2$ & 221.92 & I & $22$ & $(1)$ & $1$ & 227.130 & III & $6$ & $(1)$ & $2$ \\ 
200.17 & IV & $9$ & $(1)$ & $2$ & 210.55 & IV & $4$ & $(2)$ & $2^* $ & 221.93 & II & $22$ & $(1)$ & $1$ & 227.131 & III & $10$ & $(2)$ & $2$ \\ 
201.18 & I & $11$ & $(2)$ & $2$ & 211.56 & I & $7$ & $(1)$ & $1$ & 221.94 & III & $10$ & $(1)$ & $1$ & 227.132 & III & $4$ & $(1)$ & $2$ \\ 
201.19 & II & $8$ & $(2)$ & $2$ & 211.57 & II & $7$ & $(1)$ & $1$ & 221.95 & III & $16$ & $(1)$ & $1$ & 227.133 & IV & $8$ & $(1)$ & $4$ \\ 
201.20 & III & $4$ & $(2)$ & $2$ & 211.58 & III & $5$ & $(1)$ & $1$ & 221.96 & III & $9$ & $(1)$ & $1$ & 228.134 & I & $9$ & $(2)$ & $2^* $ \\ 
201.21 & IV & $7$ & $(2)$ & $2$ & 212.59 & I & $3$ & $(1)$ & $4$ & 221.97 & IV & $13$ & $(1)$ & $2$ & 228.135 & II & $7$ & $(1)$ & $4$ \\ 
202.22 & I & $13$ & $(1)$ & $1$ & 212.60 & II & $3$ & $(1)$ & $4$ & 222.98 & I & $11$ & $(2)$ & $2$ & 228.136 & III & $5$ & $(2)$ & $2^* $ \\ 
202.23 & II & $10$ & $(1)$ & $1$ & 212.61 & III & $3$ & $(1)$ & $4$ & 222.99 & II & $9$ & $(1)$ & $2$ & 228.137 & III & $7$ & $(2)$ & $4$ \\ 
202.24 & III & $4$ & $(2)$ & $1$ & 212.62 & IV & $3$ & $(1)$ & $4^* $ & 222.100 & III & $6$ & $(1)$ & $2$ & 228.138 & III & $4$ & $(2)$ & $4$ \\ 
202.25 & IV & $9$ & $(1)$ & $2$ & 213.63 & I & $3$ & $(1)$ & $4$ & 222.101 & III & $8$ & $(1)$ & $2$ & 228.139 & IV & $9$ & $(2)$ & $2^* $ \\ 
203.26 & I & $10$ & $(2)$ & $2$ & 213.64 & II & $3$ & $(1)$ & $4$ & 222.102 & III & $6$ & $(1)$ & $2$ & 229.140 & I & $17$ & $(1)$ & $1$ \\ 
203.27 & II & $7$ & $(2)$ & $2$ & 213.65 & III & $3$ & $(1)$ & $4$ & 222.103 & IV & $11$ & $(2)$ & $2$ & 229.141 & II & $17$ & $(1)$ & $1$ \\ 
203.28 & III & $3$ & $(1)$ & $2$ & 213.66 & IV & $3$ & $(1)$ & $4^* $ & 223.104 & I & $13$ & $(1)$ & $2$ & 229.142 & III & $9$ & $(1)$ & $1$ \\ 
203.29 & IV & $6$ & $(2, 2)$ & $2^* $ & 214.67 & I & $5$ & $(1)$ & $2^* $ & 223.105 & II & $12$ & $(1)$ & $2$ & 229.143 & III & $13$ & $(1)$ & $1$ \\ 
204.30 & I & $13$ & $(1)$ & $1$ & 214.68 & II & $5$ & $(1)$ & $2^* $ & 223.106 & III & $9$ & $(2)$ & $2$ & 229.144 & III & $7$ & $(1)$ & $1$ \\ 
204.31 & II & $10$ & $(1)$ & $1$ & 214.69 & III & $4$ & $(1)$ & $2^* $ & 223.107 & III & $11$ & $(1)$ & $2$ & 230.145 & I & $9$ & $(2)$ & $2^* $ \\ 
204.32 & III & $4$ & $(1)$ & $1$ & 215.70 & I & $10$ & $(1)$ & $1$ & 223.108 & III & $6$ & $(1)$ & $2$ & 230.146 & II & $7$ & $(1)$ & $4^* $ \\ 
205.33 & I & $9$ & $(2)$ & $4$ & 215.71 & II & $10$ & $(1)$ & $1$ & 223.109 & IV & $11$ & $(1)$ & $2$ & 230.147 & III & $5$ & $(2)$ & $4^* $ \\ 
205.34 & II & $6$ & $(1)$ & $4$ & 215.72 & III & $6$ & $(1)$ & $1$ & 224.110 & I & $13$ & $(1)$ & $2$ & 230.148 & III & $7$ & $(2)$ & $2^* $ \\ 
205.35 & III & $2$ & $(1)$ & $4$ & 215.73 & IV & $7$ & $(1)$ & $2$ & 224.111 & II & $13$ & $(1)$ & $2$ & 230.149 & III & $4$ & $(1)$ & $4^* $ \\ 
205.36 & IV & $5$ & $(2)$ & $4^* $ & 216.74 & I & $9$ & $(1)$ & $1$ & 224.112 & III & $7$ & $(1)$ & $2$ & ~ & ~ & ~ & ~ \\ 
206.37 & I & $10$ & $(2)$ & $2^* $ & 216.75 & II & $9$ & $(1)$ & $1$ & 224.113 & III & $11$ & $(2)$ & $2$ & ~ & ~ & ~ & ~ \\ 
206.38 & II & $7$ & $(1)$ & $4$ & 216.76 & III & $5$ & $(1)$ & $1$ & 224.114 & III & $6$ & $(1)$ & $2$ & ~ & ~ & ~ & ~ \\ 
\hline \hline 
\end{tabular}

%% file: spinless_Cubic_foot.tex
\newlength{\tabLspinlessCubic} 
\settowidth{\tabLspinlessCubic}{\input{spinless_Cubic}} 
\begin{minipage}{\tabLspinlessCubic} 
\begin{flushleft} 
{\footnotesize $d$: Rank of the band structure group $\{{\rm BS}\}$\\ 
$X_{\rm BS}$: Symmetry-based indicators of band topology\\ 
$\nu_{\rm BS}$: Set of $\nu$ bands are symmetry-forbidden from being isolated by band gaps if $\nu \not \in \nu_{\rm BS}\, \mathbb Z$ }\\ 
$*$: Exhibiting exceptional filling pattern; see Table \ref{tab:spinlessnuEx} 
\end{flushleft}\end{minipage}

%% file: spinless_nuSkip.tex
\begin{tabular}{cc|cc||cc|cc||cc|cc||cc|cc} 
\hline \hline 
\multicolumn{2}{c|}{MSG} & $\{ \nu \}_{\rm AI}$ & $\{ \nu \}_{\rm BS}$ & \multicolumn{2}{c|}{MSG} & $\{ \nu \}_{\rm AI}$ & $\{ \nu \}_{\rm BS}$ & \multicolumn{2}{c|}{MSG} & $\{ \nu \}_{\rm AI}$ & $\{ \nu \}_{\rm BS}$ & \multicolumn{2}{c|}{MSG} & $\{ \nu \}_{\rm AI}$ & $\{ \nu \}_{\rm BS}$\\ 
\hline 
9.41 & IV & $4\mathbb N $ & $2\mathbb N $ & 80.32 & IV & $4\mathbb N $ & $2\mathbb N $ & 133.459 & I & $4\mathbb N $ & $2\mathbb N \setminus \{2\}$ & 212.62 & IV & $4\mathbb N \setminus \{4\}$ & $4\mathbb N $ \\ 
15.91 & IV & $4\mathbb N $ & $2\mathbb N $ & 86.73 & IV & $4\mathbb N $ & $2\mathbb N $ & 133.461 & III & $4\mathbb N $ & $2\mathbb N $ & 213.66 & IV & $4\mathbb N \setminus \{4\}$ & $4\mathbb N $ \\ 
29.105 & IV & $8\mathbb N $ & $4\mathbb N $ & 88.86 & IV & $4\mathbb N $ & $2\mathbb N $ & 133.463 & III & $4\mathbb N $ & $2\mathbb N \setminus \{2\}$ & 214.67 & I & $2\mathbb N \setminus \{2\}$ & -- \\ 
37.185 & IV & $4\mathbb N $ & $2\mathbb N $ & 91.109 & IV & $8\mathbb N $ & $4\mathbb N $ & 133.466 & III & $4\mathbb N $ & $2\mathbb N $ & 214.68 & II & $2\mathbb N \setminus \{2\}$ & -- \\ 
41.217 & IV & $4\mathbb N $ & $2\mathbb N $ & 92.117 & IV & $8\mathbb N $ & $4\mathbb N $ & 133.469 & IV & $4\mathbb N $ & $2\mathbb N \setminus \{2\}$ & 214.69 & III & $2\mathbb N \setminus \{2\}$ & -- \\ 
43.228 & IV & $4\mathbb N $ & $2\mathbb N $ & 93.125 & IV & $4\mathbb N $ & $2\mathbb N \setminus \{2\}$ & 135.485 & III & $4\mathbb N $ & $2\mathbb N $ & 220.89 & I & $2\mathbb N \setminus \{2, 4, 10\}$ & $2\mathbb N \setminus \{2\}$ \\ 
45.240 & IV & $4\mathbb N $ & $2\mathbb N $ & 94.133 & IV & $4\mathbb N $ & $2\mathbb N $ & 135.490 & III & $4\mathbb N $ & $2\mathbb N \setminus \{2\}$ & 220.90 & II & $2\mathbb N \setminus \{2, 4, 10\}$ & -- \\ 
54.347 & IV & $8\mathbb N $ & $4\mathbb N $ & 95.141 & IV & $8\mathbb N $ & $4\mathbb N $ & 142.561 & I & $4\mathbb N $ & $2\mathbb N \setminus \{2\}$ & 220.91 & III & $2\mathbb N \setminus \{2, 4, 10\}$ & -- \\ 
60.426 & IV & $8\mathbb N $ & $4\mathbb N $ & 96.149 & IV & $8\mathbb N $ & $4\mathbb N $ & 142.563 & III & $4\mathbb N $ & $2\mathbb N $ & 228.134 & I & $4\mathbb N $ & $2\mathbb N \setminus \{2\}$ \\ 
61.438 & IV & $8\mathbb N $ & $4\mathbb N $ & 98.162 & IV & $4\mathbb N $ & $2\mathbb N $ & 142.565 & III & $4\mathbb N $ & $2\mathbb N \setminus \{2\}$ & 228.136 & III & $4\mathbb N $ & $2\mathbb N \setminus \{2\}$ \\ 
61.440 & IV & $8\mathbb N $ & $4\mathbb N $ & 106.219 & I & $4\mathbb N $ & $2\mathbb N $ & 142.568 & III & $4\mathbb N $ & $2\mathbb N \setminus \{2\}$ & 228.139 & IV & $4\mathbb N $ & $2\mathbb N \setminus \{2\}$ \\ 
68.519 & IV & $4\mathbb N $ & $2\mathbb N $ & 106.222 & III & $4\mathbb N $ & $2\mathbb N $ & 142.570 & IV & $4\mathbb N $ & $2\mathbb N \setminus \{2\}$ & 230.145 & I & $4\mathbb N \setminus \{4\}$ & $2\mathbb N \setminus \{2, 4\}$ \\ 
70.532 & IV & $4\mathbb N $ & $2\mathbb N $ & 106.225 & IV & $4\mathbb N $ & $2\mathbb N $ & 198.11 & IV & $4\mathbb N \setminus \{4\}$ & $4\mathbb N $ & 230.146 & II & $4\mathbb N \setminus \{4\}$ & -- \\ 
73.548 & I & $4\mathbb N $ & $2\mathbb N \setminus \{2\}$ & 110.245 & I & $4\mathbb N $ & $2\mathbb N $ & 199.12 & I & $2\mathbb N \setminus \{2\}$ & -- & 230.147 & III & $4\mathbb N \setminus \{4\}$ & $4\mathbb N $ \\ 
73.550 & III & $4\mathbb N $ & $2\mathbb N $ & 110.248 & III & $4\mathbb N $ & $2\mathbb N $ & 199.13 & II & $2\mathbb N \setminus \{2\}$ & -- & 230.148 & III & $4\mathbb N \setminus \{4\}$ & $2\mathbb N \setminus \{2, 4, 10\}$ \\ 
73.553 & IV & $4\mathbb N $ & $2\mathbb N \setminus \{2\}$ & 110.250 & IV & $4\mathbb N $ & $2\mathbb N $ & 203.29 & IV & $4\mathbb N $ & $2\mathbb N $ & 230.149 & III & $4\mathbb N \setminus \{4\}$ & -- \\ 
76.11 & IV & $8\mathbb N $ & $4\mathbb N $ & 112.265 & IV & $4\mathbb N $ & $2\mathbb N \setminus \{2\}$ & 205.36 & IV & $8\mathbb N $ & $4\mathbb N $ & ~ & ~ & ~ \\ 
77.17 & IV & $4\mathbb N $ & $2\mathbb N $ & 114.281 & IV & $4\mathbb N $ & $2\mathbb N $ & 206.37 & I & $4\mathbb N $ & $2\mathbb N \setminus \{2\}$ & ~ & ~ & ~ \\ 
78.23 & IV & $8\mathbb N $ & $4\mathbb N $ & 122.338 & IV & $4\mathbb N $ & $2\mathbb N $ & 210.55 & IV & $4\mathbb N $ & $2\mathbb N $ & ~ & ~ & ~ \\ 
\hline \hline 
\end{tabular}

%% file: spinless_nuSkip_foot.tex
\newlength{\tabLspinlessnuSkip} 
\settowidth{\tabLspinlessnuSkip}{\input{spinless_nuSkip}} 
\begin{minipage}{\tabLspinlessnuSkip} 
\begin{flushleft} 
{\footnotesize 
$\mathbb N$: The set of natural numbers\\ 
$\{ \nu \}_{\rm AI}$: Set of fillings for which physical atomic insulators are possible\\ 
$\{ \nu \}_{\rm BS}$: Set of fillings for which physical band structure are possible; a dash (--) indicates $\{ \nu \}_{\rm BS} = \{ \nu \}_{\rm AI}$ for that MSG }
\end{flushleft} 
\end{minipage}